\documentclass[
    12pt, 				% The default document font size, options: 10pt, 11pt, 12pt
    twoside, 			% Two side (alternating margins) for binding by default, uncomment to switch to one side
    openright, 
    french, 
    singlespacing, 		% Single line spacing, alternatives: onehalfspacing or doublespacing or singlespacing (voir \setstretch ci-dessous)
    parskip, 			% Uncomment to add space between paragraphs
]{Configurations/Template_Thesis} % The class file specifying the document structure

\usepackage[utf8x]{inputenc}
\usepackage[T1]{fontenc}
\usepackage[round]{natbib}
\usepackage[french]{babel}
\usepackage{geometry}

\usepackage{calc}
\usepackage{ulem}  % to underline text
\usepackage{pdfpages}  % to integrate pdf pages in the document
\usepackage{hyphenat}  % to specify when hyphenating ('\hyph{}')
\usepackage[autostyle=true]{csquotes}
\usepackage{rotating}

\usepackage{icomma}
\usepackage{amsmath}
\usepackage{amssymb}
\usepackage{stmaryrd}  % for double brackets (\llbracket, \rrbracket...)
\usepackage{upgreek}  % to display greek letters not in italic in math expressions
\usepackage{marvosym}  % to get the euro symbols
\usepackage{tikz}  % To draw arrows on the side of matrices
\newcommand{\tikzmark}[1]{\tikz[overlay, remember picture] \coordinate (#1);}

\usepackage{graphicx}
\usepackage[format=plain,labelfont={bf},labelsep=endash]{caption}
\usepackage{subcaption}
\newcommand{\figwidth}{\textwidth}
\usepackage[pagebackref]{hyperref}

\usepackage{multirow}  % for tables
\renewcommand{\arraystretch}{1.2}

\usepackage[nottoc]{tocbibind}
\usepackage{minitoc}
\newlength{\spaceafterminitoc}
\addto{\captionsfrench}{% Making babel aware of special titles
  
}

\usepackage[acronym,toc,shortcuts]{glossaries}
\makeglossaries
\newacronym{IAU}{IAU}{International Astronomical Union}
\newacronym{FIRST}{FIRST}{Fibered Interferometer foR a Single Telescope}
\newacronym{FIRSTv1}{FIRSTv1}{Fibered Interferometer foR a Single Telescope version~1}
\newacronym{FIRSTv2}{FIRSTv2}{Fibered Interferometer foR a Single Telescope version~2}
\newacronym{LESIA}{LESIA}{Laboratoire d'\'Etudes Spatiales et d'Instrumentation en Astrophysique}
\newacronym{HRA}{HRA}{haut contraste et Haute R\'esolution Angulaire}
\newacronym{SCExAO}{SCExAO}{Subaru Coronagraphic Extreme Adaptive Optics}
\newacronym{CACAO}{CACAO}{Control And Compute for Adaptive Optics}
\newacronym{V2PM}{V2PM}{Visibility To Pixel Matrix}
\newacronym{P2VM}{P2VM}{Pixel To Visibility Matrix}
\newacronym{SVD}{SVD}{Singular Value Decomposition}
\newacronym{CMOS}{CMOS}{Complementary Metal Oxide Semiconductor}
\newacronym{ADU}{ADU}{Analog-Digital Unit}
\newacronym{OPD}{OPD}{Optical Path Difference}
\newacronym{ODL}{ODL}{Optical Delay Line}
\newacronym{MEMS}{MEMS}{Micro-ElectroMechanical Systems}
\newacronym{PSF}{PSF}{Point Spread Function}
\newacronym{UV}{UV}{UltraViolet}
\newacronym{CA}{CA}{Core Accretion}
\newacronym{GI}{GI}{Gravitational Instabilities}
\newacronym{TTS}{TTS}{T Tauri Star}
\newacronym{CTTS}{CTTS}{Classical T Tauri Star}
\newacronym{WTTS}{WTTS}{Weak T Tauri Star}
\newacronym{MUSE}{MUSE}{Multi Unit Spectroscopic Explorer}
\newacronym{VLT}{VLT}{Very Large Telescope}
\newacronym{VLTI}{VLTI}{Very Large Telescope Interferometer}
\newacronym{ELT}{ELT}{Extremely Large Telescope}
\newacronym{NIRC2}{NIRC2}{Near InfraRed Camera 2}
\newacronym{MagAO}{MagAO}{Magellan Adaptive Optics}
\newacronym{MagAO-X}{MagAO-X}{Magellan Adaptive Optics - eXtreme}
\newacronym{LMIRCam}{LMIRCam}{Large binocular telescope Mid-InfraRed Camera}
\newacronym{LBTI}{LBTI}{Large Binocular Telescope Interferometer}
\newacronym{ISIS}{ISIS}{Intermediate-dispersion Spectrograph and Imaging System}
\newacronym{WHT}{WHT}{William Herschel Telescope}
\newacronym{CHARIS}{CHARIS}{Coronagraphic High Angular Resolution Imaging Spectrograph}
\newacronym{SPHERE}{SPHERE}{Spectro-Polarimetic High contrast imager for Exoplanets REsearch}
\newacronym{NACO}{NACO}{Nasmyth Adaptive Optics System - Coude Near Infrared Camera}
\newacronym{GPI}{GPI}{Gemini Planet Imager}
\newacronym{GST}{GST}{Gemini South Telescope}
\newacronym{GAPlanetS}{GAPlanetS}{Giant Accreting Protoplanet Survey}
\newacronym{AMBER}{AMBER}{Astronomical Multi-BEam combineR}
\newacronym{GRAVITY}{GRAVITY}{General Relativity Analysis via VLT InTerferometrY}
\newacronym{SNR}{SNR}{Signal-to-Noise Ratio}
\newacronym{BCD}{BCD}{Beam Commutation Device}
\newacronym{JWST}{JWST}{James Webb Space Telescope}
\newacronym{MIRI}{MIRI}{Mid-Infrared Instrument}
\newacronym{MFD}{MFD}{Mode Field Diameter}
\newacronym{VPH}{VPH}{Volume Phase Holographic}
\newacronym{pyMILK}{pyMILK}{PYthon Multi-purpose Imaging Libraries toolKit}
\newacronym{MILK}{MILK}{Multi-purpose Imaging Libraries toolKit}
\newacronym{SDK}{SDK}{Software Development Kit}
\newacronym{SHM}{SHM}{SHared Memory}
\newacronym{IPAG}{IPAG}{Institut de Plan\'etologie et d'Astrophysique de Grenoble}
\newacronym{RAM}{RAM}{Random Access Memory}
\newacronym{CCD}{CCD}{Charge Coupled Device}
\newacronym{EMCCD}{EMCCD}{Electron Multiplying Charge Coupled Device}
\newacronym{GLINT}{GLINT}{Guided-Light Interferometric Nulling Technology}
\newacronym{AU}{AU}{Astronomical Unit}
\newacronym{UA}{UA}{Unit\'e Astronomique}
\newacronym{TESS}{TESS}{Transiting Exoplanet Survey Satellite}
\newacronym{ADI}{ADI}{Angular Differential Imaging}
\newacronym{SDI}{SDI}{Spectral Differential Imaging}
\newacronym{RDI}{RDI}{Reference Star Differential Imaging}
\newacronym{TMT}{TMT}{Thirty Meter Telescope}
\newacronym{LBTAO}{LBTAO}{Large Binocular Telescope Adaptive Optics}
\newacronym{CHARA}{CHARA}{Center for High Angular Resolution Astronomy}
\newacronym{SAM}{SAM}{Sparse Aperture Masking}
\newacronym{NIRISS}{NIRISS}{Near InfraRed Imager and Slitless Spectrograph}
\newacronym{MICADO}{MICADO}{Multi-ao Imaging CAmera for Deep Observations}
\newacronym{HST}{HST}{Hubble Space Telescope}
\newacronym{NICI}{NICI}{Near-Infrared Coronagraphic Imager}
\newacronym{VAMPIRES}{VAMPIRES}{Visible Aperture Masking Polarimetric Imager for Resolved Exoplanetary Structures}
\newacronym{REACH}{REACH}{Rigorous Exoplanetary Atmosphere Characterization}
\newacronym{MEC}{MEC}{MKID Exoplanet Camera}
\newacronym{PDI}{PDI}{Polarization Differential Imaging}
\newacronym{RHEA}{RHEA}{Replicable High resolution Exoplanet and Asteroseismology}
\newacronym{PIAA}{PIAA}{Phase-Induced Amplitude Apodization}
\newacronym{IR}{IR}{InfraRouge}
\newacronym{WDM}{WDM}{Wavelength Division Multiplexer}
\newacronym{ExAO}{ExAO}{Extreme Adaptive Optics}

\usepackage{titlesec}
\titleformat{\paragraph}{\normalfont\normalsize\bfseries}{\theparagraph}{1em}{}
\titlespacing*{\paragraph}{0pt}{3.25ex plus 1ex minus .2ex}{1.5ex plus .2ex}
\newcommand{\threesubsection}{\paragraph}

\newcommand{\vv}[1]{\overrightarrow{#1}}
\newcommand{\ha}[0]{\text{$\text{H}\upalpha$}}
\newcommand{\Lha}[0]{\text{$\text{L}_{\ha}$}}
\newcommand{\Lacc}[0]{\text{$\text{L}_{\text{acc}}$}}
\newcommand{\Wd}[0]{\text{$\text{W}_{10}$}}
\newcommand{\sk}[0]{\textit{Super K}}
\newcommand{\wiggles}[0]{\textit{wiggles}}
\newcommand{\degree}[0]{^{\circ}}
\newcommand{\um}{$\upmu$m}
\newcommand{\Like}[0]{\mathcal{L}}
\newcommand{\MJ}[0]{\text{$\text{M}_\text{J}$}}
\newcommand{\MS}[0]{\text{$\text{M}_{\odot}$}}

\usepackage{xcolor}

\usepackage{comment} % to comment parts of the text with \begin{comment} \end{comment}

\newenvironment{mytitlepage}
{
}%
{
}

\newcommand{\logoupc}{
	\hspace{-2cm}\raisebox{-2cm}{\includegraphics[width=6cm]{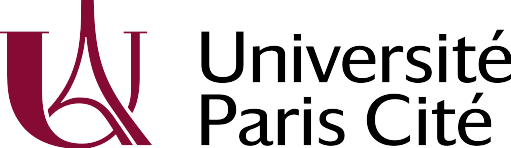}}
}

\newlength{\ongletwidth}
\newlength{\ongletheight}

\newcommand{\ongletfontI}{%
	\fontfamily{pag}\fontseries{b}\fontshape{n}\fontsize{1cm}{1cm}\selectfont\color{white}
}

\newcommand{\ongletfontII}{%
	\fontfamily{pag}\fontseries{b}\fontshape{n}\fontsize{1cm}{1cm}\selectfont\color{white}
} 

\newcommand{\ongletfont}{%	
	\ifnum\pdfstrcmp{\thechapter}{A}=0 % teste si les 2 strings sont indentiques
		\ongletfontII%
	\else\ifnum\pdfstrcmp{\thechapter}{B}=0 % teste si les 2 strings sont indentiques
		\ongletfontII% 
	\else\ifnum\pdfstrcmp{\thechapter}{C}=0 % teste si les 2 strings sont indentiques
		\ongletfontII% 
	\else\ifnum\pdfstrcmp{\thechapter}{D}=0 % teste si les 2 strings sont indentiques
		\ongletfontII% 
	\else\ifnum\pdfstrcmp{\thechapter}{E}=0 % teste si les 2 strings sont indentiques
		\ongletfontII%
	\else\ifnum\pdfstrcmp{\thechapter}{F}=0 % teste si les 2 strings sont indentiques
		\ongletfontII%
	\else\ifnum\pdfstrcmp{\thechapter}{G}=0 % teste si les 2 strings sont indentiques
		\ongletfontII%
	\else\ifnum\pdfstrcmp{\thechapter}{H}=0 % teste si les 2 strings sont indentiques
		\ongletfontII%
	\else\ifnum\pdfstrcmp{\thechapter}{I}=0 % teste si les 2 strings sont indentiques
		\ongletfontII%
	\else\ifnum\pdfstrcmp{\thechapter}{J}=0 % teste si les 2 strings sont indentiques
		\ongletfontII%
	\else\ifnum\pdfstrcmp{\thechapter}{K}=0 % teste si les 2 strings sont indentiques
		\ongletfontII%
	\else\ifnum\pdfstrcmp{\thechapter}{L}=0 % teste si les 2 strings sont indentiques
		\ongletfontII%
	\else\ifnum\pdfstrcmp{\thechapter}{M}=0 % teste si les 2 strings sont indentiques
		\ongletfontII%
	\else\ifnum\pdfstrcmp{\thechapter}{N}=0 % teste si les 2 strings sont indentiques
		\ongletfontII%
	\else
		\ongletfontI%
	\fi\fi\fi\fi\fi\fi\fi\fi\fi\fi\fi\fi\fi\fi
}

\newcommand{\conclusionchapternumber}{6}

\newcommand{\valueforonglet}{%	
	\ifnum\pdfstrcmp{\thechapter}{A}=0 % teste si les 2 strings sont indentiques
		\conclusionchapternumber%
	\else\ifnum\pdfstrcmp{\thechapter}{B}=0 % teste si les 2 strings sont indentiques
		\conclusionchapternumber%
	\else\ifnum\pdfstrcmp{\thechapter}{C}=0 % teste si les 2 strings sont indentiques
		\conclusionchapternumber% 
	\else\ifnum\pdfstrcmp{\thechapter}{D}=0 % teste si les 2 strings sont indentiques
		\conclusionchapternumber% 
	\else\ifnum\pdfstrcmp{\thechapter}{E}=0 % teste si les 2 strings sont indentiques
		\conclusionchapternumber%
	\else\ifnum\pdfstrcmp{\thechapter}{F}=0 % teste si les 2 strings sont indentiques
		\conclusionchapternumber%
	\else\ifnum\pdfstrcmp{\thechapter}{G}=0 % teste si les 2 strings sont indentiques
		\conclusionchapternumber%
	\else\ifnum\pdfstrcmp{\thechapter}{H}=0 % teste si les 2 strings sont indentiques
		\conclusionchapternumber%
	\else\ifnum\pdfstrcmp{\thechapter}{I}=0 % teste si les 2 strings sont indentiques
		\conclusionchapternumber%
	\else\ifnum\pdfstrcmp{\thechapter}{J}=0 % teste si les 2 strings sont indentiques
		\conclusionchapternumber%
	\else\ifnum\pdfstrcmp{\thechapter}{K}=0 % teste si les 2 strings sont indentiques
		\conclusionchapternumber%
	\else\ifnum\pdfstrcmp{\thechapter}{L}=0 % teste si les 2 strings sont indentiques
		\conclusionchapternumber%
	\else\ifnum\pdfstrcmp{\thechapter}{M}=0 % teste si les 2 strings sont indentiques
		\conclusionchapternumber%
	\else\ifnum\pdfstrcmp{\thechapter}{N}=0 % teste si les 2 strings sont indentiques
		\conclusionchapternumber%
	\else
		\thechapter%
	\fi\fi\fi\fi\fi\fi\fi\fi\fi\fi\fi\fi\fi\fi
}

\newcommand{\coulorforonglet}{%	
	\ifnum\pdfstrcmp{\thechapter}{A}=0% teste si les 2 strings sont indentiques
		colorboxchap%
	\else\ifnum\pdfstrcmp{\thechapter}{B}=0% teste si les 2 strings sont indentiques
		colorboxchap% 
	\else\ifnum\pdfstrcmp{\thechapter}{C}=0% teste si les 2 strings sont indentiques
		colorboxchap% 
	\else\ifnum\pdfstrcmp{\thechapter}{D}=0% teste si les 2 strings sont indentiques
		colorboxchap%
	\else\ifnum\pdfstrcmp{\thechapter}{E}=0% teste si les 2 strings sont indentiques
		colorboxchap%
	\else\ifnum\pdfstrcmp{\thechapter}{F}=0% teste si les 2 strings sont indentiques
		colorboxchap%
	\else\ifnum\pdfstrcmp{\thechapter}{G}=0% teste si les 2 strings sont indentiques
		colorboxchap%
	\else\ifnum\pdfstrcmp{\thechapter}{H}=0% teste si les 2 strings sont indentiques
		colorboxchap%
	\else\ifnum\pdfstrcmp{\thechapter}{I}=0% teste si les 2 strings sont indentiques
		colorboxchap%
	\else\ifnum\pdfstrcmp{\thechapter}{J}=0% teste si les 2 strings sont indentiques
		colorboxchap%
	\else\ifnum\pdfstrcmp{\thechapter}{K}=0% teste si les 2 strings sont indentiques
		colorboxchap%
	\else\ifnum\pdfstrcmp{\thechapter}{L}=0% teste si les 2 strings sont indentiques
		colorboxchap%
	\else\ifnum\pdfstrcmp{\thechapter}{M}=0% teste si les 2 strings sont indentiques
		colorboxchap%
	\else\ifnum\pdfstrcmp{\thechapter}{N}=0% teste si les 2 strings sont indentiques
		colorboxchap%
	\else%
		color16_126_179%
	\fi\fi\fi\fi\fi\fi\fi\fi\fi\fi\fi\fi\fi\fi
}

\newcommand{\textforonglet}{%
	\ifnum\pdfstrcmp{\thechapter}{0}=0% teste si les 2 strings sont indentiques
		I% % pour introduction
	\else\ifnum\pdfstrcmp{\thechapter}{6}=0% teste si les 2 strings sont indentiques
		C% pour conclusion (vérifier que c'est le 6e chapitre)
	\else%
		\thechapter%
	\fi\fi
}

\newlength{\ongletvtop}
\newlength{\ongletvpos}

\newlength{\frameboxsize}

\newlength{\espacepaire}

\newcommand{\ongletcote}{%
	\ifnum\pdfstrcmp{\thechapter}{D}=0% teste si les 2 strings sont indentiques
		\setlength{\ongletheight}{1.7cm}
		\setlength{\frameboxsize}{1.5cm}
		\setlength{\espacepaire}{0.1cm}
		\setlength{\ongletvpos}{\ongletvtop-\ongletheight*\real{\valueforonglet}*\real{1.19}}%
	\else\ifnum\pdfstrcmp{\thechapter}{E}=0% teste si les 2 strings sont indentiques
		\setlength{\ongletheight}{1.7cm}
		\setlength{\frameboxsize}{1.5cm}
		\setlength{\espacepaire}{0.1cm}
		\setlength{\ongletvpos}{\ongletvtop-\ongletheight*\real{\valueforonglet}*\real{1.19}}%
	\else%
		\setlength{\ongletheight}{1.5cm}% environ 1.7cm apparent sur PDF A4 (colorbox déborbe framebox)
		\setlength{\frameboxsize}{1.25cm}
		\setlength{\espacepaire}{0.cm}
		\setlength{\ongletvpos}{\ongletvtop-\ongletheight*\real{\valueforonglet}*\real{1.2}}%
	\fi\fi	%
	\ifthenelse{\isodd{\value{page}}}{% IF (Boîte page impaire)
		\makebox[0pt][l]{%
			\hspace{0.42cm}% petit espace étrange
			\hspace{1cm}% margin outer (2.5cm) - 1.5cm = 1cm (bien centré)
			\raisebox{\ongletvpos}[0pt][0pt]{%
				\colorbox{\coulorforonglet}{%
					\parbox[t][\ongletheight][c]{\ongletwidth}{%
						\setlength{\fboxrule}{0pt}% Masquer trait de framebox
						\framebox[\frameboxsize]{%
							\ongletfont\textforonglet%
						}
					}
				}
			}
		}
	}{% ELSE (Boîte page paire)
		\makebox[0pt][r]{%
			\raisebox{\ongletvpos}[0pt][0pt]{%
				\colorbox{\coulorforonglet}{%
					\parbox[t][\ongletheight][c]{\ongletwidth}{%
						\setlength{\fboxrule}{0pt}% Masquer trait de framebox
						\framebox[1.25cm]{%
						}
					}
				}
			}
			\hspace{\espacepaire}% petit espace étrange
			\hspace{1cm}% margin outer (2.5cm) - 1.5cm = 1cm (bien centré)
		}
	}
}

\newcommand{\titreongletfont}{%
	\fontfamily{pag}\fontseries{b}\fontshape{n}\fontsize{0.5cm}{0.6cm}\selectfont\textcolor{\coulorforonglet}%
}

\newlength{\titreongletvpos}

\newcommand{\titreongletcote}{%
	\ifnum\pdfstrcmp{\thechapter}{D}=0% teste si les 2 strings sont indentiques
		\setlength{\titreongletvpos}{\ongletvtop-\ongletheight*\real{\valueforonglet}*\real{1.19}-\ongletheight*\real{1.19}}%
	\else\ifnum\pdfstrcmp{\thechapter}{E}=0% teste si les 2 strings sont indentiques
		\setlength{\titreongletvpos}{\ongletvtop-\ongletheight*\real{\valueforonglet}*\real{1.19}-\ongletheight*\real{1.19}}%
	\else%
		\setlength{\titreongletvpos}{\ongletvtop-\ongletheight*\real{\valueforonglet}*\real{1.2}-\ongletheight*\real{1.2}}%
	\fi\fi%
	\ifthenelse{\isodd{\value{page}}}{% IF (Boîte page impaire)
		\makebox[0pt][l]{%
			\hspace{0cm}% (bien centré sur bord texte avant décallage = 0cm)
			\hspace{1.75cm}% margin outer (2.5cm) - 1.5cm = 1cm + moitié boite (0.75cm)
			\raisebox{\titreongletvpos}[0pt][0pt]{\rotatebox{270}{\titreongletfont\leftmark}}
		}
	}{% ELSE (Boîte page paire)
	}
}

\thesistitle{Caract\'erisation et int\'egration de composants d'optique int\'egr\'ee sur l'interf\'erom\`etre fibr\'e FIRST au t\'elescope Subaru pour l'\'etude des protoplan\`etes en accr\'etion}

\supervisor{Sylvestre Lacour}

\degreen{Docteur}

\author{Kévin Barjot}

\subject{Astronomie et Astrophysique}

\keywordsfr{Haute Résolution Angulaire, haut contraste, interférométrie, accrétion de protoplanètes, photonique}
\keywordsen{High angular resolution, high contrast, interferometry, accreting protoplanets, photonic}

\university{Université Paris Cité}

\school{École doctorale\\Astronomie et Astrophysique d'Ile de France (127)}

\lab{Laboratoire d'\'Etudes Spatiales et d'Instrumentation en Astrophysique}

\addresses{}

\begin{document}

\newcommand*\aap{A\&A}
\let\astap=\aap
\newcommand*\aapr{A\&A~Rev.}
\newcommand*\aaps{A\&AS}
\newcommand*\actaa{Acta Astron.}
\newcommand*\aj{AJ}
\newcommand*\ao{Appl.~Opt.}
\let\applopt\ao
\newcommand*\apj{ApJ}
\newcommand*\apjl{ApJ}
\let\apjlett\apjl
\newcommand*\apjs{ApJS}
\let\apjsupp\apjs
\newcommand*\aplett{Astrophys.~Lett.}
\newcommand*\apspr{Astrophys.~Space~Phys.~Res.}
\newcommand*\apss{Ap\&SS}
\newcommand*\araa{ARA\&A}
\newcommand*\azh{AZh}
\newcommand*\baas{BAAS}
\newcommand*\bac{Bull. astr. Inst. Czechosl.}
\newcommand*\bain{Bull.~Astron.~Inst.~Netherlands}
\newcommand*\caa{Chinese Astron. Astrophys.}
\newcommand*\cjaa{Chinese J. Astron. Astrophys.}
\newcommand*\fcp{Fund.~Cosmic~Phys.}
\newcommand*\gca{Geochim.~Cosmochim.~Acta}
\newcommand*\grl{Geophys.~Res.~Lett.}
\newcommand*\iaucirc{IAU~Circ.}
\newcommand*\icarus{Icarus}
\newcommand*\jcap{J. Cosmology Astropart. Phys.}
\newcommand*\jcp{J.~Chem.~Phys.}
\newcommand*\jgr{J.~Geophys.~Res.}
\newcommand*\jqsrt{J.~Quant.~Spectr.~Rad.~Transf.}
\newcommand*\jrasc{JRASC}
\newcommand*\memras{MmRAS}
\newcommand*\memsai{Mem.~Soc.~Astron.~Italiana}
\newcommand*\mnras{MNRAS}
\newcommand*\na{New A}
\newcommand*\nar{New A Rev.}
\newcommand*\nat{Nature}
\newcommand*\nphysa{Nucl.~Phys.~A}
\newcommand*\pasa{PASA}
\newcommand*\pasj{PASJ}
\newcommand*\pasp{PASP}
\newcommand*\physrep{Phys.~Rep.}
\newcommand*\physscr{Phys.~Scr}
\newcommand*\planss{Planet.~Space~Sci.}
\newcommand*\pra{Phys.~Rev.~A}
\newcommand*\prb{Phys.~Rev.~B}
\newcommand*\prc{Phys.~Rev.~C}
\newcommand*\prd{Phys.~Rev.~D}
\newcommand*\pre{Phys.~Rev.~E}
\newcommand*\prl{Phys.~Rev.~Lett.}
\newcommand*\procspie{Proc.~SPIE}
\newcommand*\qjras{QJRAS}
\newcommand*\rmxaa{Rev. Mexicana Astron. Astrofis.}
\newcommand*\skytel{S\&T}
\newcommand*\solphys{Sol.~Phys.}
\newcommand*\sovast{Soviet~Ast.}
\newcommand*\ssr{Space~Sci.~Rev.}
\newcommand*\zap{ZAp}

\pagenumbering{Alph} % pour avoir bonnes refs pages dans biblio
\KOMAoptions{headsepline=false} % trait sous en-tête
\pagestyle{firstpage}

% Include the title page
%%%%%%%%%%%%%%%%%%%%%%%%%%%%%%%%%%%%%%%%%%%%%%%%%%%%%%%%%%%
% Karine Perraut <karine.perraut@univ-grenoble-alpes.fr>
% Lucas Labadie <labadie@ph1.uni-koeln.de>
% Arthur Vigan <arthur.vigan@lam.fr>
% Anthony Boccaletti <anthony.boccaletti@obspm.fr>
% Daniel Rouan <daniel.rouan@obspm.fr>
%%%%%%%%%%%%%%%%%%%%%%%%%%%%%%%%%%%%%%%%%%%%%%%%%%%%%%%%%%%

\begin{mytitlepage}

	\begin{center}
		
		\vspace*{0.3cm}
		\LARGE\textbf{\univname} \\
		\vspace{0.2cm}
		\Large\textbf{\schoolname} \\
		\vspace{0.3cm}
		\Large\textbf{\labname} \\
		\vspace{1cm}
		\huge{\textbf{\ttitle}} \\
		\vspace{0.8cm}
		\large{Par \authorname} \\
		\vspace{0.6cm}
		\large{Thèse de doctorat en \subjectname} \\
		\vspace{0.6cm}
		\large{Dirigée par Sylvestre Lacour} \\
            \large{Et co-encadrée par Elsa Huby} \\
		
		\normalsize
		\vspace{0.7cm}
		Présentée et soutenue publiquement le 20 avril 2023 \\
		\vspace{0.7cm}
		Devant un jury composé de :
		
		% Taille de l'espace entre les lignes sdu tableau de jury
		\newlength{\espacejury}
		\setlength{\espacejury}{0.1cm}
		
		\small
		
		\noindent\makebox[\textwidth]{%
		\begin{tabular}{llll}
			$M.$ & \textsc{Anthony Boccaletti} &  Directeur de recherche, Observatoire de Paris PSL & Président du jury \\[\espacejury]
			$M.$ & \textsc{Sylvestre Lacour} & Chargé de recherche, CNRS - Observatoire de Paris PSL & Directeur de thèse \\ [\espacejury] 
			$Mme$ & \textsc{Elsa Huby} & Astronome adjointe, Observatoire de Paris PSL & Co-encadrante \\ [\espacejury]
			$M.$ & \textsc{Lucas Labadie} & Professeur, Université de Cologne & Rapporteur \\[\espacejury]
			$Mme$ & \textsc{Karine Perraut} & Astronome, Université Grenoble Alpes  & Rapportrice \\[\espacejury]		
			$M.$ & \textsc{Arthur Vigan}  & Chargé de Recherche, Aix-Marseille Université & Examinateur \\[\espacejury]
			$M.$ & \textsc{Daniel Rouan} & Chercheur émérite, Observatoire de Paris PSL & Invité \\ [\espacejury]
		\end{tabular} }
		
		\vfill
	\end{center}
\end{mytitlepage}
 % ATTENTION : penser à demander une impression sur la tranche

\frontmatter % Use roman page numbering style (i, ii, iii, iv...) for the pre-content pages
\KOMAoptions{headsepline=false} % trait sous en-tête
\pagestyle{frontpage} % Default to the plain heading style until the thesis style is called for the body content

\newpage\thispagestyle{empty}
\null
\newpage\thispagestyle{empty}
\null

\newpage
\thispagestyle{empty}
\chapter*{Remerciements}

\footnotesize Quand on accomplit un projet aussi grand qu'une thèse, l'humilité pousse naturellement aux remerciements.

Je tiens, tout d'abord, à remercier mes merveilleux encadrants de thèse, Elsa Huby et Sylvestre Lacour. Merci à vous de m'avoir fait confiance pour travailer sur le projet FIRST. J'ai eu un grand plaisir à me l'approprier et à travailler dessus avec vous. Merci d'avoir su être présent pendant vos congés parentaux et surtout pendant la pandémie mondiale de la COVID-19. Et un grand merci pour votre relecture consciencieuse de mon manuscrit. J'ai beaucoup appris auprès de vous, j'admire les scientifiques et les humains que vous êtes.

Je remercie chaleureusement tous les membres de mon jury de thèse, Anthony Boccaletti, Karine Perraut, Lucas Labadie, Arthur Vigan, Daniel Rouan qui ont pris le temps de s'intéresser avec sincérité à mon travail de thèse et qui m'ont octroyer le titre de docteur. Je remercie plus particulièrement Karine et Lucas pour votre lecture et vos corrections de mon manuscrit de thèse.

Je remercie grandement toute l'équipe FIRST avec laquelle j'ai beaucoup aimée travailler. Merci Nick Cvetojevic de m'avoir intéressé au sujet de cette thèse en m'accompagnant à son démarrage ainsi que pour toutes tes anecdotes sur l'Australie. Merci Daniel Rouan de m'avoir accordé ton temps pour travailler ensemble. Merci Vincent Lapeyrère pour ton aide au labo. Merci Guillermo Martin pour nos discussions sur la photonique et surtout pour les bons moments de travail qu'on a pus avoir au labo. Merci Manon Lallement pour les bons moments passés ensemble et les bonnes tranches de rigolade. Merci Harry-Dean Kenchington pour ton témoignage sur l'écriture du manuscrit. Un grand merci à Pierre Fédou qui m'a permis de faire la mise à jour de la version 2 vers la version 3 de Python qui, comme tu me l'avais dit, \og c'est comme changer de langage de programmation \fg. et merci pour ta grande bonne humeur. Merci à Franck Marchis et Clément Chalumeau pour leur aide pour faire fonctionner les MEMS IrisAO en Python 3.

Un grand merci à toute l'équipe SCExAO à Hawaii pour votre accueil lors de mes missions ainsi que pour l'environnement de travail passionant que vous avez instauré. Énorme merci à Sébastien Vievard, c'était un réel plaisir de travailler avec toi sur FIRST et merci pour ton encadrement lorsque je suis venu en mission : j'ai réalisé un de mes vieux rêves en venant travailler dans un des plus grands observatoires au monde sur une île incroyable qu'est la Big Island. J'admire l'humain que tu es et le scientifique que tu es en train de devenir. Merci Vincent Déo pour toutes ces discussions en info, pour ton aide sur l'installation de MILK et pyMILK sur l'ordinateur de FIRSTv2 et pour m'avoir emmené voir Mike Love en soirée à Hawaii. Thank you Barnaby Norris for your help on the Andor camera software. Thank you Robin for having me at your house in February 2022, for your great kindness and your hugs.

Je remercie le LESIA au sein duquel j'ai fait ma thèse et qui m'a donné un bon cadre. Merci aux personnels de l'administration qui sont d'une grande aide. Je remercie mon comité de suivi de thèse composé de Carine Briand et de Yann Clénet pour avoir pris le temps de me suivre tout au long de ma thèse. Je remercie aussi Thierry Fouchet et Jérome Rodriguez pour votre investissement et grand soutien que vous vouez aux doctorants. Je remercie tous mes collègues de couloirs qui m'ont offert un excellent environnement professionnel pendant ma thèse et avec qui c'était un plaisir d'aller déjeuner et de prendre des pauses café. Merci aux doctorants et post-doctorants pour les coups à boire. Je remercie plus particulièrement Christian Wilkinson pour la soirée qu'on a faite chez toi et pour ta délicieuse salade. Je remercie chaleuresement Iva Laginja avec qui j'ai des discussions intéressantes et pleines de leçons.

Je voudrais remercier ceux qui ont participé à ce long chemin pour devenir un scientifique : tous mes professeurs aussi loin que je m'en souvienne. Ils m'ont éduqué et appris tellement de choses. Plus particulièrement, mes professeurs de maths et de physique qui m'ont donné ma passion dévorante pour la science et m'ont inculqué la rigueur scientifique.

Je remercie mes tuteurs de stage, Yann Girard, Jaime Dawson, Patrice Martinez et Mathilde Beaulieu avec qui j'ai eu de merveilleux stages. Vous m'avez permis d'exercer la recherche scientifique et m'avez mené à la réussite de ma thèse.

Je remercie chaleureusement l'Uranoscope de l'île de France et tous ses membres, qui sont une deuxième famille pour moi. Vous m'avez tout appris sur la vulgarisation et l'observation astronomique. Grâce à vous j'ai perdu ma timidité en faisant une activité qui faisait sens pour moi. Merci pour tous ces samedis soir au club, pour avoir crû en moi et pour tous les rires.

Je remercie tous mes amis. Vous êtes le Soleil de ma vie. Merci Véro, Élodie, Linda et Chrystelle pour tous ces merveilleux moments à rire et à délirer sous les étoiles. Merci Le Crew pour toutes ces magnifiques soirées. Merci Manou pour ton réconfort et tes discussions. Merci Louis pour ta bonne humeur et les soirées. Merci Riad pour ton amour. Merci Antoine pour tes enseignements et ton bon vivre. Merci Emma pour tous ces merveilleux moments qu'on a passés ensemble, les loongues pauses déj et cafés et pour avoir le même humour que moi. Merci Louis et Tanguy pour tous les délires, les rires et les soirées depuis la fac. Merci Antoine en prépa. Merci Warren à la fac. Merci Gaetan mon frère jumeau. Merci Manon pour m'avoir fait pleurer de rire pendant ma thèse. Merci Yéhudi pour l'amour. Merci Charles de me gâter et de m'avoir emmené à ma première soirée bisounours. Merci à mes meilleures amies de lycée, Victoria, Clémence, Rudia pour notre belle histoire. Merci Glwadys pour les subways. Merci Mathieu pour ta bienveillance et tes lumières. Merci Fériel pour ton écoute pendant nos merveilleuses pauses KFé. Merci Lucas d'être complètement toi-même et de m'avoir tant fait rire. Merci Fabien pour les bons moments passés ensemble surtout pendant les covoiturages pendant le master OSAE. Un grand merci Ophélie pour tout, merci d'être ma soeur et pour tous ces moments délicieux passés avec toi. Merci Valentin mon alter ego depuis le lycée, mon coupain de toujours, je suis tellement heureux qu'on se soit croisé.

Remerciement très spécial à Assia, mon grand amour. Je te remercie d'avoir été à mes côtés tout au long de ma thèse. Sans toi elle aurait été bien plus difficile à traverser. Merci de me faire grandir. Je t'aime de tout mon coeur.

Enfin, je remercie les personnes les plus importantes dans ma vie. Celles qui sont là depuis le début. Je veux remercier ma si belle famille. Merci Maman de nous avoir élevés toute seule avec Nico et de m'avoir fait aller aussi loin dans ma vie. Je t'admire énormément pour ça ! Merci Papa de nous avoir gâtés et de m'avoir inculqué ma grande curiosité et ton humour pourri qui me faire rire chaque jour. Merci à mes beaux-parents José et Valérie qui sont de merveilleuses personnes. Merci à mes petits frères Nicolas et Valentin mon boulou. Merci à mes grands-parents qui agrémentent mon enfance de doux souvenirs. Merci à mes tatas Auréa et Sylvie, merci à mes oncles Mourad et Michel, merci à mes cousins de Paris Émilie, Michel, Mélanie, Elsa et ma merveilleurse Sophie avec qui je passe de bons moments de bonheur qui me font grandir et merci à ma marraine Elisabeth et mon parrain Luis. Merci à mes cousins de Lyon, Élise et mon Vincent adoré. J'ai passé avec vous les meilleurs moments de ma vie et je suis fier de faire partie de notre famille.

% \input{Openings/Dedicace}
% \input{Openings/Quotation}
%----------------------------------------------------------------------------------------
%	FRENCH
%----------------------------------------------------------------------------------------

\newpage
\thispagestyle{empty}
\chapter*{Résumé}

Dans le cadre de ma thèse, j'ai travaillé dans le domaine de l'imagerie haut contraste à haute résolution angulaire (HRA) pour la détection et la caractérisation de compagnon stellaire ou planétaire. Je me suis plus particulièrement intéressé au cas des protoplanètes en cours de formation qui accrètent de la matière, induisant un fort rayonnement à la longueur d'onde de la raie de l'hydrogène \ha. À cette longueur d'onde le contraste est diminué par rapport aux longueurs d'onde InfraRouges (bande K), facilitant sa détection et la mesure de son intensité permettrait d'apporter des contraintes sur ce phénomène à l'oeuvre dans les systèmes planétaires en formation. FIRST (Fibered Interferometer foR a Single Telescope) est un instrument installé sur le banc d'optique adaptative extrême du télescope Subaru (SCExAO) et exploite la technique de masquage et de réarrangement de pupille, qui permet d'atteindre des résolutions angulaires jusqu'à deux fois meilleures que celle du télescope. Pour cela, la pupille d'entrée de l'instrument est sous-divisée en sous-pupilles dont la lumière est injectée dans des fibres optiques monomodes qui appliquent un filtrage spatial du front d'onde supprimant ainsi les aberrations optiques à l'échelle des sous-pupilles. Sur la deuxième version de FIRST (FIRSTv2), la recombinaison de ces faisceaux est effectuée par un composant d'optique intégrée, afin d'augmenter les performances d'imagerie haut contraste. Le développement de la technologie photonique dans le visible est complexe et innovant et l'objectif de ma thèse est d'évaluer ses performances et la faisabilité de son application pour l'imagerie HRA. Cela a nécessité de continuer le développement du banc de test, aussi bien au niveau du montage optique que du logiciel de contrôle, de mettre au point une procédure spécifique d'acquisition des données et de développer un programme de traitement et d'analyse de données.

Une réplique de FIRST a été développée au \ac{LESIA}, afin de pouvoir implémenter, tester et valider les nouveaux développements de composants photoniques avant leur intégration au télescope Subaru. La recombinaison de chaque paire de sous-pupilles est ainsi codée sur une ou deux sorties de la puce photonique (selon la technologie utilisée) qui sont imagées sur la caméra sur quelques pixels. L'échantillonnage correct des franges d'interférences nécessite une modulation temporelle de la différence de marche. Pour cela, nous utilisons un miroir segmenté contrôlable en position, pour changer le déphasage entre les sous-pupilles. Dans un premier temps, j'ai amélioré le logiciel de contrôle de ce banc pour augmenter sa rapidité ainsi que pour permettre l'acquisition de données interférométriques nécessitant le contrôle synchronisé des différents composants pour la modulation des franges (miroir déformable, lignes à retard et caméra).

J'ai ensuite caractérisé les propriétés optiques de deux puces photoniques qui utilisent deux techniques différentes de recombinaison interférométrique de faisceaux, telles que leur transmission, la quantité de fuite du signal entre les différents guides d'onde (appelé \textit{cross-talk}), leur comportement dans les deux polarisations ainsi que le contraste instrumental. Ensuite, j'ai construit un système optique simulant une source binaire sur le banc de test, qui permet d'injecter à la fois une source avec une large bande spectrale pour simuler une étoile et une source avec une bande spectrale étroite pour simuler une exoplanète avec une raie d'émission. Cela a été crucial pour démontrer les capacités de l'instrument à détecter et mesurer le signal du compagnon. Pour mesurer ce dernier, j'ai tiré profit des propriétés de phase différentielle, qui est une grandeur auto-étalonnée. En effet, la mesure de la phase sur une large bande spectrale permet l'étalonnage des mesures de phases par le signal du continuum mettant en évidence un possible signal dans la raie d'intérêt. J'ai ainsi développé un programme de traitement et d'analyse de données permettant d'estimer la phase différentielle ainsi que la visibilité complexe et la clôture de phase et de les ajuster avec un modèle étoile-compagnon. Cela m'a permis de démontrer que FIRSTv2 pouvait détecter un compagnon de type protoplanète à une séparation équivalente à $0.7 \lambda / B$ de la source centrale, avec un contraste atteignant $0.1$.

Une des puces photoniques a été intégrée dans l'instrument FIRST installé sur le banc SCExAO, ce qui a donné lieu à la première lumière de FIRSTv2 le 10 septembre 2021. J'ai ainsi acquis des données sur ciel lors de plusieurs nuits d'observations à distance mais aussi lors d'une mission à Hawaii en février 2022. Cela a été l'occasion pour moi d'intégrer le deuxième composant pour comparaison avec le premier, ainsi que de déployer le logiciel développé au laboratoire. Les résultats du traitement de ces données sur ciel ont permis d'en apprendre plus sur la méthode d'acquisition et de traitement des données interférométriques.

En conclusion, j'ai caractérisé et étudié les performances de la technologie d'optique intégrée dans le cadre de l'imagerie interférométrique à haut contraste et haute résolution dans le visible. J'ai ainsi développé le logiciel de contrôle de FIRSTv2 en laboratoire avant de le déployer sur le banc SCExAO pour sa première lumière. Ensuite, j'ai développé le programme de traitement et d'analyse de données interférométriques pour FIRSTv2, en y incluant un ajustement des observables interférométriques par un modèle de protoplanète présentant une forte raie d'émission dans son spectre. L'objectif est de caractériser les mécanismes d'accrétion de jeunes systèmes exoplanétaires en formation. Enfin, j'ai eu l'occasion de participer à de nombreuses nuits d'observations durant lesquelles j'ai acquis des données sur des cibles simples telles que des binaires d'étoiles à faible contraste compagnon/étoile.

%----------------------------------------------------------------------------------------
%	ENGLISH
%----------------------------------------------------------------------------------------

\newpage
\thispagestyle{empty}
\chapter*{Abstract}

During my thesis, I worked in the context of high contrast and high angular resolution (HRA) imaging for the detection and the characterisation of stellar or planetary companions. I was particularly interested in the case of forming protoplanets which accrete matter, inducing a strong emission at \ha~wavelength. In addition to decreasing the contrast between the star and the protoplanet at this wavelength, facilitating its detection, the measurement of its intensity would constraints this ongoing phenomenon in protoplanetary systems. FIRST (Fibered Interferometer foR a Single Telescope) is an instrument installed on the Subaru coronagraphic extreme adaptive optics (SCExAO) testbed and exploits the pupil masking and remapping technique, which allows to reach up to twice the angular resolution of the telescope. To achieve this, the instrument's entrance pupil is subdivided into sub-pupils whose light is injected into single-mode optical fibres that spatially filter the wavefront, thereby suppressing optical aberrations at the sub-pupil level. In the second version of FIRST (FIRSTv2), the recombination of these beams is performed by an integrated optics component in order to increase the high contrast imaging performance. The development of photonic technology in the visible range is complex and innovative and the objective of my thesis is to evaluate its performance and the feasibility of its application for HRA imaging. In that purpose I conducted the further development of the testbed, both in terms of the optical setup and the control software, developed a specific data acquisition procedure and a data processing and analysis pipeline.

A replica of FIRST has been built at LESIA, in order to implement, test and validate new developments of photonic components before their integration at the Subaru telescope. The recombination of each pair of sub-pupils is encoded on one or two outputs of the photonic chip (depending on the technology used) and imaged on the camera on a few pixels. The correct sampling of the interference fringes requires a temporal modulation of the optical path difference (OPD). For this purpose, we use a segmented mirror controllable in piston and tip/tilt to change the phase shift between the sub-pupils. First, I improved the control software of the testbed to increase its speed and to allow the acquisition of interferometric data requiring the synchronised control of the different components for the modulation of the fringes (deformable mirror, delay lines and camera).

Then I characterised the optical properties of two photonic chips that use two different interferometric beam recombination techniques, such as their transmission, the cross-talk between the waveguides, their behaviour in the two polarisations and the instrumental contrast. I built an optical system simulating a binary source on the testbed, which allows to inject both a source with a wide spectral band to simulate a central star and a source with a narrow spectral band to simulate an exoplanet with an emission line. This was crucial to demonstrate the instrument's ability to detect and measure the companion signal. To measure the latter, I took advantage of the properties of spectral differential phase, which is a self-calibrated quantity. Indeed, the measurement of the phase over a wide spectral band allows the calibration of the phase measurements by the continuum signal highlighting a possible signal in the emission line of interest. Thus, I developed a data processing and analysis program allowing to estimate the differential phase as well as the complex visibility and the closure phase and to fit them with a star-companion model. Therefore I was able to demonstrate that FIRSTv2 could detect a protoplanet-like companion at an equivalent separation of $0.7 \lambda / B$ from the central source, with a contrast of up to $0.1$.

One of the photonic chips was integrated to the FIRST instrument installed on the SCExAO bench, resulting in the first light from FIRSTv2 on 10 September 2021. I acquired on-sky data during several remote observation nights as well as during a mission to Hawaii in February 2022. It was the opportunity for me to integrate the second component for comparison with the first, as well as to deploy the control software developed in the laboratory. The results of the processing of these on-sky data allowed me to learn more about the acquisition and processing methods of interferometric data.

To conclude, I characterised and studied the performance of the integrated optics technology in the context of high contrast and high resolution interferometric imaging in the visible range. I developed the FIRSTv2 control software in the laboratory before deploying it on the SCExAO bench for the first light. Then, I developed the interferometric data processing and analysis program for FIRSTv2, including a fitting of the interferometric observables by a protoplanet model with a strong emission line in its spectrum. The objective is to characterise the accretion mechanisms of young exoplanetary systems in formation. Finally, I have had the opportunity to participate in many observing nights during which I have acquired data on simple targets such as companion-star binaries with low contrast.

%%%%%%%%%%%%%%%%%%%%%%%%%%%%%%%%%%%%%%%%%%%%%%%%%%%%%%%%%%%%%%%%
% Tabel of contents
%%%%%%%%%%%%%%%%%%%%%%%%%%%%%%%%%%%%%%%%%%%%%%%%%%%%%%%%%%%%%%%%

%% Build a per-chapter table of content
\dominitoc
\tableofcontents

%%%%%%%%%%%%%%%%%%%%%%%%%%%%%%%%%%%%%%%%%%%%%%%%%%%%%%%%%%%%%%%%
% Manuscript content
%%%%%%%%%%%%%%%%%%%%%%%%%%%%%%%%%%%%%%%%%%%%%%%%%%%%%%%%%%%%%%%%

\begingroup{
\KOMAoptions{headsepline=true} % trait sous en-tête
% Re-initialize page numbering and set it to arabic numbers
\mainmatter
\pagestyle{thesis} % Return the page headers back to the "thesis" style

%%%%%%%%%%%%%%%%%%%%%%%%%%%%%%%%%%%%%%%%%%%%%%%%%%%%%%%%%%%%%%%%
\chapter{Introduction}
\setcounter{figure}{0}
\setcounter{table}{0}
\setcounter{equation}{0}

\minitoc

\clearpage
Les détections de plusieurs milliers d'exoplanètes sont actuellement confirmées grâce à de multiples méthodes de détection. Les méthodes de détection indirectes sont celles qui ont été les plus fructueuses pour détecter de nouvelles planètes mais, contrairement aux méthodes de détection directes, elle ne permettent pas (ou peu) la caractérisation chimique et astrométrique des exoplanètes. En effet, pour cela il s'agit de concevoir des instruments atteignant des hautes performances en terme de résolution angulaire et de contraste afin de détecter la lumière provenant directement de la planète.

La technique de masquage et de réarrangement de pupille à l'aide de fibres optiques monomodes permet de relever à la fois ces deux défis techniques et de permettre la caractérisation spectroscopique des systèmes observés. L'instrument \ac{FIRST} l'implémentant a montré par le passé ses capacités de haute résolution angulaire pour la caractérisation de binaires d'étoiles \citep{huby2013}. Mon travail de thèse s'inscrit dans la volonté d'améliorer les performances en contraste de cet instrument en intégrant la technologie d'optique intégrée.

Dans cette partie j'introduirai donc le contexte de l'étude des exoplanètes, les solutions techniques existantes et les apports des techniques de masquage et de réarrangement de pupille. Je présenterai le développement de \ac{FIRST} et l'apport de \ac{FIRSTv2}. J'exposerai enfin le contexte de l'objectif scientifique dans lequel ce dernier s'inscrit : l'étude des protoplanètes.

%%%%%%%%%%%%%%%%%%%%%%%%%%%%%%%%
\section{Imagerie haut contraste et haute résolution angulaire pour l'étude des exoplanètes}

%%%%%%%%%%%%%%%%
\subsection{L'étude des exoplanètes}

Depuis la première détection d'une exoplanète, 51 Peg b \citep{mayor1995}, en (l'an de grâce) $1995$, les techniques de détection et de caractérisation de systèmes exoplanétaires n'ont cessé de gagner en sensibilité et de se diversifier. Les détections de plus de $5\,000$ exoplanètes sont confirmées à ce jour et la figure~\ref{fig:ExoplanetDetection} présente une partie de ces détections (celles dont le demi-grand axe a pu être mesuré), pour les quatre techniques de détection par vitesse radiale (cercle orange), par transit (triangle vert), par micro-lentille gravitationnelle (point violet) et par imagerie directe (carré bleu). La masse (en unité de masse de Jupiter) de l'exoplanète est tracée en fonction de son demi-grand axe en \ac{AU} ou \ac{UA}, en français. Les huit planètes du système solaire y sont aussi représentées pour comparaison, par la première lettre de leur nom dans une bulle. Enfin, seul les corps ayant une masse inférieure à $13 \,$\MJ (correspondant à la limite différenciant une planète d'une étoile selon l'\ac{IAU} \citep{lecavelierdesetangs2022}) sont représentés, ce qui induit un plafonnement horizontal des corps détectés en haut du graphique.

\begin{figure}[ht!]
    \centering
    \includegraphics[width=\figwidth]{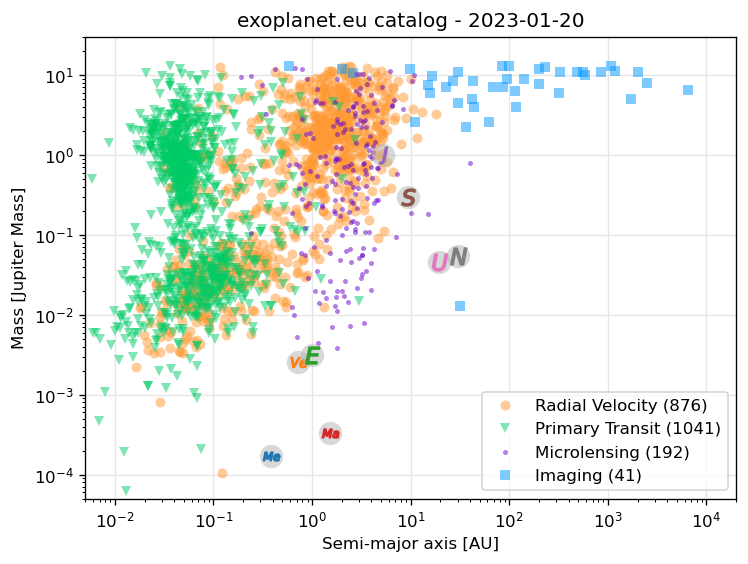}
    \caption[Graphique des exoplanètes détectées à ce jour suivant les quatre techniques de détection principales.]{Graphique de la masse mesurées des exoplanètes détectées à ce jour en fonction de leur demi-grand axe estimé. Les détections sont identifiées par les techniques de détection par vitesse radiale (cercle orange), par transit (triangle vert), par micro-lentille gravitationnelle (point violet) et par imagerie directe (carré bleu). Seules $2\,150$ exoplanètes, dont le demi-grand axe est mesuré, sont affichées. Données tirées de la base de données de \textit{exoplanet.eu} (\url{http://exoplanet.eu/}).}
    \label{fig:ExoplanetDetection}
\end{figure}

On remarque que les deux techniques de transit et de vitesse radiale (méthode qui détecte pour la première fois 51 Peg b) ont permis de détecter la majorité des exoplanètes. Cela s'explique par le fait qu'elles sont les plus simples techniquement à implémenter en comparaison des autres. Les missions spatiales Kepler \citep{borucki2010} et \ac{TESS} \citep{ricker2016} ont ainsi été lancées afin de détecter et étudier de nouvelles exoplanètes par la méthode des transits. Ces deux missions ont permis de détecter plus de $2\,700$ et $300$ (pour $6\,000$ candidats) nouvelles exoplanètes, respectivement.

De plus, on remarque que chaque technique de détection découvre une population d'exoplanètes qui sont homologues en terme de masse et de demi-grand axe, se traduisant sur le graphique par un regroupement des points par technique de détection avec peu de chevauchement. En effet, la méthode des transits mesure la variation d'intensité lumineuse de l'étoile due au passage du compagnon en premier plan, ce qui favorise la détection des systèmes qui ont un petit demi-grand axe car cela augmente le nombre de passages et les chances que le compagnon passe entre l'étoile et la Terre. La méthode des vitesses radiales mesure le déplacement de l'étoile dans la ligne de visée et ce déplacement est d'autant plus élevé que le compagnon est massif ou que le compagnon est plus proche de l'étoile lorsqu'il est moins massif. La méthode d'imagerie consistant en l'observation de la lumière provenant du compagnon, est plus sensible aux systèmes avec une grande séparation et dont l'exoplanète est très massive (voir plus de détails dans la section~\ref{sec:ImagerieDirecte}) : elle réfléchit alors plus de lumière de son étoile ou émet un rayonnement de plus forte intensité lorsque c'est une exoplanète de type Jupiter chaude nouvellement formée. Enfin, la méthode de micro-lentille gravitationnelle qui consiste à détecter l'augmentation du flux lumineux d'une étoile en arrière-plan par l'effet de lentille gravitationnelle du compagnon d'un système exoplanétaire en avant-plan, est sensible pour une séparation idéale avec l'étoile qui ne peut ni être trop petite car on ne pourrait pas discerner l'effet de lentille de l'étoile de celui du compagnon, ni trop grande car les chances qu'à la fois l'étoile et le compagnon passent devant l'étoile d'arrière-plan sont faibles.

Par conséquent, il est nécessaire de garder à l'esprit que l'exploration de l'espace des paramètres des exoplanètes détectées est limitée et biaisée selon la méthode de détection utilisée. Cela motive l'amélioration des technologies existantes et la recherche de nouvelles solutions techniques afin d'accéder à des cas différents d'exoplanètes pour diversifier l'ensemble de nos connaissances sur les systèmes planétaires. L'imagerie directe a encore peu exploré cet espace des paramètres et se révèle très intéressante en ce qu'elle permet l'étude spectroscopique et photométrique des exoplanètes ce qui est un atout majeur dans la recherche de traces de vie extra-terrestre.

%%%%%%%%%%%%%%%%
\subsection{L'imagerie directe des systèmes exoplanétaires}
\label{sec:ImagerieDirecte}

L'imagerie directe d'un système exoplanétaire permet d'une part l'étude du mouvement orbital du compagnon \citep{chauvin2012, wang2018} et d'autre part la mesure spectroscopique de sa lumière. Cette dernière rend possible l'étude de la composition chimique de l'atmosphère de la planète (détection de la présence d'eau ou de molécules organiques comme le méthane ou le monoxyde de carbone), d'inférer la présence de nuages \citep{marley2015} (en ajustant des modèles atmosphériques), la gravité de surface \citep{marley2012} (la présence de nuages dans l'atmosphère et sa chimie sont fortement influencées par la gravité de surface) ou la rotation de l'exoplanète \citep{bryan2020} (à partir de l'élargissement des raies spectrales). Actuellement, seulement une vingtaine d'images directes de systèmes exoplanétaires ont été obtenues (figure~\ref{fig:ExoplanetDetection}) car les performances à atteindre sont ambitieuses. En effet, il s'agit de relever deux défis techniques majeurs : un grand pouvoir de résolution angulaire et un haut contraste.

Pour illustrer ce propos, la figure~\ref{fig:ContrastSeparation} tirée de \cite{mawet2012} présente le contraste de l'intensité lumineuse entre le compagnon et l'étoile centrale (\textit{Planet/Star Contrast}) en fonction de la séparation apparente, pour les planètes du système solaire (dans la partie inférieure du graphique) telles qu'elles seraient observées à une distance de $10 \,$pc (distance typique entre la Terre et les systèmes exoplanétaires) ainsi que pour les exoplanètes imagées du système HR8799 \citep{marois2008} et $\upbeta$ Pic b \citep{lagrange2010} (dans la partie supérieure du graphique). Les domaines de performances de quelques instruments en service et futurs sont représentés par les lignes colorées nommées par leur nom. On remarque que les performances à atteindre pour observer des planètes telles que celles du système solaire sont encore hors d'atteinte de $1-2$ ordres de grandeur par les instruments futurs (en cours de construction) et de $2-4$ ordres de grandeur pour les instruments actuels.

\begin{figure}[ht!]
    \centering
    \includegraphics[width=\figwidth]{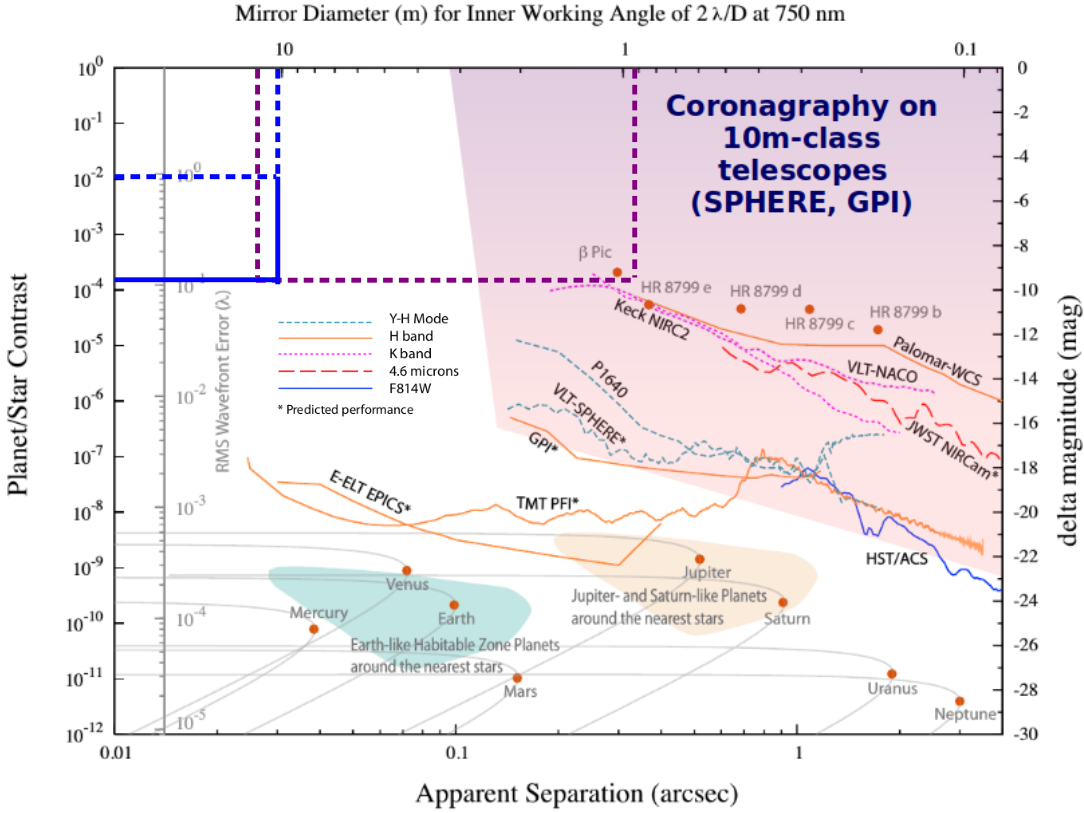}
    \caption[Contraste en fonction de la séparation apparente de certaines exoplanètes imagées et des planètes du système solaire.]{Contraste (\textit{Planet/Star Contrast}) en fonction de la séparation apparente (en arcsec) des exoplanètes du système HR8799 (bcde) et de $\upbeta$ Pic b (dans la partie supérieure) et des planètes du système solaire telles que vues à une distance de $10 \,$pc (dans la partie inférieure). Un deuxième axe des ordonnées, à gauche, indique l'erreur rms (en unité de longueur d'onde) sur le front d'onde à atteindre pour imager un système exoplanétaire au contraste correspondant à la même ordonnée. Un troisième axe des ordonnées, à droite, indique la différence de magnitude entre l'étoile et la planète. Les limites hautes des performances de quelques instruments récents (Keck-NIRC2, VLT-NACO, Palomar-WCS, HST-ACS, GPI, SPHERE, Palomar-P3K-P1640 et JWST-NIRCam) et de futurs instruments (TMT-PFI et E-ELT-EPICS) sont tracées par les lignes colorées (on note que ce graphique, datant de 2012, indique par une étoile certains futurs projets qui sont actuellement en fonctionnement). Les zones verte et bleue indiquent, respectivement, les planètes de type terrestre dans la zone d'habitabilité et les planètes de type Jupiter et Saturne, autour des étoiles les plus proches. La zone de couleur violette dégradée indique le domaine de performances des systèmes coronographiques sur les télescopes de $10 \,$m. Le rectangle en tirets violets indique le domaine de performances de la technique de masquage de pupille sur les télescopes de $10 \,$m. Les rectangles en tirets bleus et en trait continu bleu indiquent les domaines de performance de la technique de masquage de pupille dans le visible des instruments FIRSTv1 et FIRSTv2, respectivement. Adapté de \cite{mawet2012}.}
    \label{fig:ContrastSeparation}
\end{figure}

Le premier défi technique à relever est le haut contraste qui est défini comme le rapport entre l'intensité lumineuse de la planète par celle de l'étoile. Celui-ci atteint $\sim 10^{-3}$ dans le cas des planètes géantes gazeuses qui émettent un rayonnement thermique infrarouge. Ce sont la plupart des systèmes imagés actuellement (ce sont les plus faciles à détecter) mais sont une population restreinte dans la taxonomie des exoplanètes car ce sont des systèmes très jeunes (de l'ordre de la dizaine de millions d'années ou moins). En revanche, le contraste des planètes de type terrestre est de $\sim 10^{-10}$. Ce sont des systèmes plus vieux (de l'ordre de centaines de millions à plusieurs milliards d'années) dans lesquels le compagnon réfléchit la lumière de l'étoile dans le visible et constituent donc un plus grand intérêt dans le cadre de la recherche de traces de vie extra-terrestre. Les systèmes imagés jusque maintenant appartiennent uniquement à la première catégorie.

La technique de la coronographie \citep{lyot1939} a été la plus largement utilisée pour obtenir ces images et son domaine de performance est représenté en dégradé de violet sur la figure~\ref{fig:ContrastSeparation}. Le composant principal de cette technique est un coronographe et il permet de supprimer (par masquage opaque ou par interférence destructive) la lumière incidente de l'étoile centrée sur son axe optique, tout en laissant apparaître la lumière du compagnon décentré par rapport à l'étoile. En revanche, les perturbations atmosphériques qui affectent le front d'onde incident lors d'observations au sol diminuent fortement les performances de cette technique. Sur la même figure, un deuxième axe vertical à gauche indique l'erreur rms maximum du front d'onde afin d'obtenir la performance en contraste correspondante. Il s'agit d'erreurs de l'ordre de $100 \,$nm rms et de $0,01 \,$nm rms dans les gammes de longueur d'onde infrarouge et visible, respectivement. Il est donc nécessaire de corriger les perturbations atmosphériques en amont du coronographe \citep{sivaramakrishnan2001} à l'aide de systèmes d'optique adaptative \citep{rousset1990} et d'optique adaptative extrême, qui permettent d'atteindre aujourd'hui des contrastes en entrée de l'instrument de $10^{-3}$ à $10^{-4,5}$ : e.g. \ac{LBTAO} \citep{esposito2011}, PALM-3000 sur le télescope Hale de l'observatoire Palomar \citep{dekany2013}, \ac{GPI} sur le télescope Gemini-South \citep{macintosh2014}, \ac{SPHERE} \citep{beuzit2019} sur le \ac{VLT}, \ac{SCExAO} \citep{jovanovic2015} sur le télescope Subaru, \ac{MagAO-X} \citep{males2020} sur le télescope Clay de l'observatoire Magellan.

Le deuxième défi est d'atteindre un fort pouvoir de résolution angulaire car il s'agit ici de détecter la lumière d'un point se trouvant à de très petites distances apparentes par rapport à l'étoile : de l'ordre de $10 - 100 \,$mas, ce qui correspond à $\sim 1 \,$ua à une distance de la Terre de $100 - 10 \,$pc. Or la coronographie n'est performante que pour des grandes séparations angulaires, typiquement égales à plus de $4 \uplambda / \text{D}$ de l'étoile centrale (avec $\uplambda$ la longueur d'onde d'observation et D le diamètre du miroir primaire du télescope) car la lumière de l'étoile parvient à fuir aux petites séparations ce qui rend difficile la détection de compagnons trop proches. De nouveaux concepts voient le jour pour améliorer cette performance des coronographes \citep{mawet2012} mais ils parviennent encore difficilement à des séparations de l'ordre de la limite de diffraction $\uplambda / \text{D}$. Le masquage de pupille (voir la section~\ref{sec:PupilMasking}) est une technique permettant d'augmenter le pouvoir de résolution en imagerie jusqu'à $0,5 \uplambda / \text{D}$. Son domaine de performance est tracé sur la figure~\ref{fig:ContrastSeparation} par les rectangles en tirets violets, en tirets bleus et en trait continu bleu, respectivement sur des télescopes de $8 - 10 \,$m dans l'infrarouge, sur \acrshort{FIRSTv1} dans le visible et \ac{FIRSTv2} dans le visible. On note l'intérêt d'observations dans le visible qui permettent de meilleures résolutions angulaires mais aussi les bas contrastes jusqu'alors atteints ($< 10^{-4}$) par cette technique dus à sa nature peu sensible. Comme nous le verrons par la suite, la technique de réarrangement de pupille est une technique permettant d'utiliser un masquage redondant de la pupille afin de contrebalancer ce problème.

Enfin, des programmes de traitement et d'analyse de données élaborés permettent d'augmenter un peu plus les performances en contraste sur les images de $1-2$ ordres de grandeur : e.g. \ac{ADI} \citep{marois2006}, \ac{SDI} \citep{marois2000}, \ac{RDI} \citep{lafreniere2009}. C'est en combinant toutes ces applications instrumentales (optique adaptative, coronographie et traitement de données approfondis) qu'il a été possible d'imager directement quelques systèmes exoplanétaires. Cela a permis d'ouvrir un nouveau domaine de l'étude des exoplanètes qui grandit toujours actuellement et qui a encore beaucoup à faire. Parmi les différentes voies existantes, l'interférométrie est une technique proposant d'atteindre de meilleures performances en résolution angulaire de plusieurs ordres de grandeur et c'est dans ce cadre que s'inscrit mon travail de thèse.

%%%%%%%%%%%%%%%%%%%%%%%%%%%%%%%%
\section{L'instrument FIRST dans le contexte de l'imagerie directe des exoplanètes}

%%%%%%%%%%%%%%%%
\subsection{L'interférométrie}

En reprenant les explications présentées dans la section 1.2.1 de la thèse d'Elsa Huby \citep{huby2013these}, chaque paire de points du front d'onde incident sur un télescope interfère et produit un interférogramme dans le plan focal. Ainsi, la tâche de diffraction dans le plan focal du télescope peut s'interpréter comme la superposition de l'ensemble de ces interférogrammes. Pour un front d'onde provenant d'une source lumineuse astrophysique (à l'infini) non résolue incident sur un télescope avec une pupille d'entrée circulaire, de diamètre D, la tâche de diffraction dans le plan focal est une fonction d'Airy, dont l'expression s'écrit :

\begin{equation}
    \text{I}_{\text{Airy}}(x) = \left( \frac{2 \text{J}_{1} (\uppi \text{D}x / \uplambda \text{f})}{\uppi \text{D}x / \uplambda \text{f}} \right)^2
\end{equation}

\noindent où f est la distance focale du télescope et $\text{J}_{1}$ est la fonction de Bessel de première espèce d'ordre 1.

Le pouvoir de résolution du télescope est défini par la largeur à mi-hauteur de cette tâche de diffraction, qui s'approxime à : $\uplambda / \text{D}$. Cela correspond au plus petit détail discernable par un télescope. De la même manière, la recombinaison interférométrique des deux faisceaux collectés à la sortie de deux télescopes séparés d'une distance B, appelée base, qui observent simultanément une source lumineuse donne un interférogramme équivalent à celui résultant de l'interférence de deux points séparés d'une distance B sur la pupille d'un télescope. Ainsi, la recombinaison de ces deux faisceaux permet l'observation d'une cible astrophysique avec un pouvoir de résolution égal à $\uplambda / \text{B}$ qui peut être augmenté en éloignant les télescopes, évitant ainsi la construction d'un télescope monolithique de diamètre égal à B.

C'est ce principe qui est implémenté actuellement dans les observatoires \ac{CHARA} \citep{tenbrummelaar2005} sur le Mont Wilson, qui peut combiner six télescopes de $1 \,$m de diamètre, avec des bases de $330 \,$m de longueur, dans le visible et l'infrarouge et \ac{VLTI} \citep{haguenauer2012} sur le Cerro Paranal, qui peut combiner quatre télescopes de $\sim 8 \,$m et quatre télescopes de $1,8 \,$m de diamètre, avec des bases de $130 \,$m de longueur, dans l'infrarouge proche et moyen. C'est sur un principe interférométrique similaire que la première image de l'environnement proche du trou noir du centre de la galaxie M87 a été construite \citep{EHTC2019}. Il s'agit dans ce cas-là d'interférométrie hétérodyne, à partir de la combinaison des faisceaux radio de télescopes répartis sur Terre, offrant une base de longueur de son diamètre ($\sim 10^4 \,$km). Mais encore, le futur projet \textit{hypertélescope} \citep{labeyrie2013} a pour ambition de combiner les faisceaux de télescopes espacés jusqu'à $10^5 \,$km dans l'espace pour l'étude des galaxies, l'étude de surfaces stellaires et l'imagerie directe d'exoplanètes.

Ces solutions techniques font partie du domaine de l'interférométrie longue base. La technique que je vais maintenant exposer et qui est implémentée dans le concept de \ac{FIRST} est l'interférométrie à masquage de pupille qui s'utilise sur un unique télescope.

%%%%%%%%%%%%%%%%
\subsection{Les techniques de masquage et réarrangement de pupille}
\label{sec:PupilMasking}

Comme nous l'avons vu dans la partie précédente, la figure mesurée dans le plan focal d'un télescope est la superposition d'une infinité d'interférogrammes. Ainsi, une infinité de paires de points de base égale à $\vv{\text{B}}$ (repéré dans le plan pupille) contribuent à une infinité d'interférogrammes de fréquence spatiale égale à $\vv{\text{B}} / \uplambda$ (avec $\uplambda$ la longueur d'onde d'observation) qui se superposent lorsqu'ils sont imagés. Dans le cas où le front d'onde incident est perturbé par la turbulence atmosphérique, y compris après la correction par un système d'optique adaptative qui laisse des résidus, ces paires de points sont déphasés et les interférogrammes (de même fréquence spatiale) ne se superposent plus. Ces perturbations brouillent les franges ce qui induit une diminution du pouvoir de résolution de l'instrument. 

La technique de masquage de pupille \citep{baldwin1986, haniff1987} propose d'appliquer un masque comportant des trous sur la pupille du télescope afin de sélectionner des paires de sous-pupilles disposées de façon non-redondante. La non-redondance assure une unique contribution de la part des paires de sous-pupilles à chaque interférogramme ce qui empêche le brouillage des franges. Cela a, par exemple, été implémenté sur l'un des télescopes Keck \citep{tuthill2000}. La figure~\ref{fig:KeckPupilMaskingA} présente le masque utilisé, disposant de $21$ trous, superposé au miroir primaire. La figure~\ref{fig:KeckPupilMaskingB} est une image du détecteur obtenue lors d'une observation avec le masque installé. La tâche est la superposition de $21 \times 20 / 2 = 210$ réseaux de franges. Enfin, la figure~\ref{fig:KeckPupilMaskingC} est la transformée de Fourier de l'image précédente, que l'on nomme plan UV des fréquences spatiales. On remarque que ce plan est échantillonné et donne l'information sur $210$ fréquences spatiales indépendantes.

\begin{figure}[ht!]
    \centering
    \begin{subfigure}[t]{0.31\textwidth}
        \centering
        \includegraphics[width=0.9\textwidth]{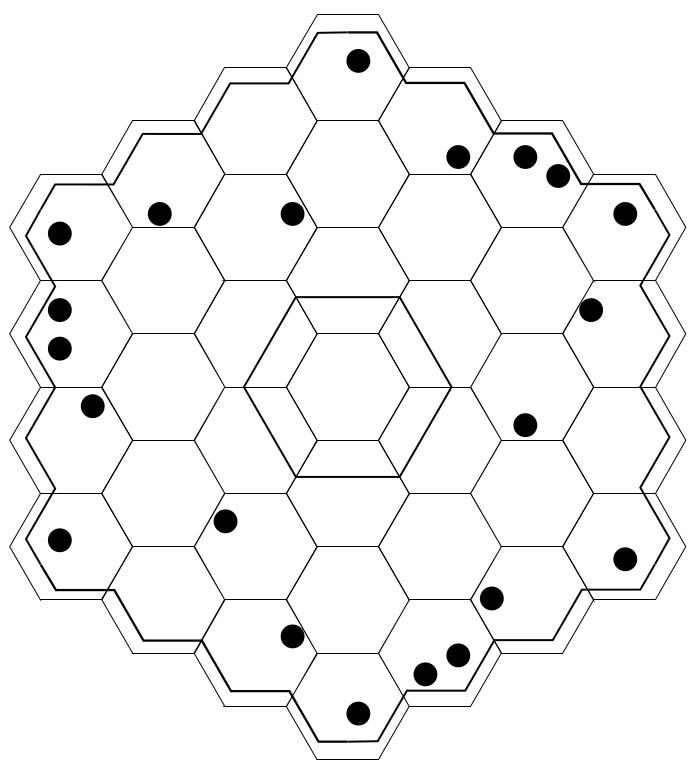}
        \caption{Pupille du télescope avec le masque. Le miroir du télescope est composé d'un pavage de segments hexagonaux et les trous du masque sont représentés par des points noirs.}
        \label{fig:KeckPupilMaskingA}
    \end{subfigure}\hspace{0.01\textwidth}
    \begin{subfigure}[t]{0.31\textwidth}
        \centering
        \includegraphics[width=0.9\textwidth]{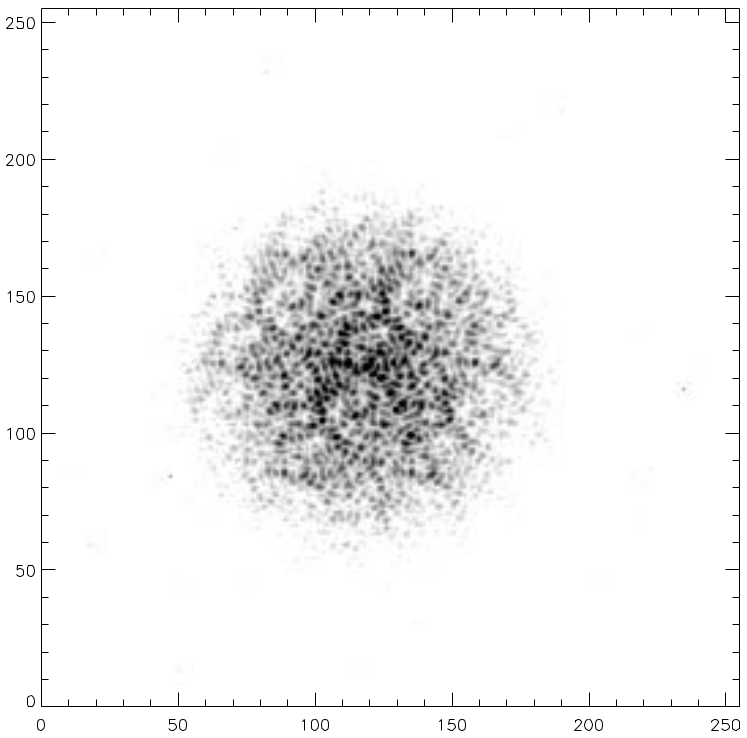}
        \caption{Image des réseaux de franges obtenus sur la caméra.}
        \label{fig:KeckPupilMaskingB}
    \end{subfigure}\hspace{0.01\textwidth}
    \begin{subfigure}[t]{0.31\textwidth}
        \centering
        \includegraphics[width=0.9\textwidth]{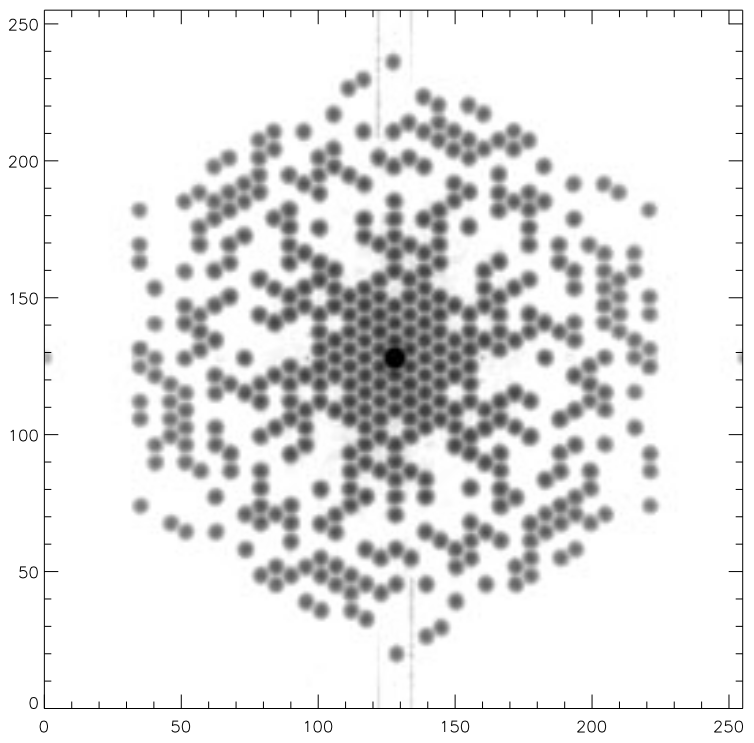}
        \caption{Plan UV obtenu par la transformé de Fourier de l'image de la caméra.}
        \label{fig:KeckPupilMaskingC}
    \end{subfigure}
    \caption[Expérience de masquage de pupille sur l'un des télescope Keck.]{Expérience de masquage de pupille sur l'un des télescope Keck. Figures tirées de \cite{tuthill2000}.}
    \label{fig:KeckPupilMasking}
\end{figure}

Cette technique a été utilisée comme mode d'observation \ac{SAM} sur l'instrument \ac{NACO} \citep{tuthill2010, lacour2011a} du \ac{VLT}, actuellement sur l'instrument \ac{SPHERE} \citep{cheetham2016} du même télescope et sur l'instrument \ac{VAMPIRES} \citep{norris2015} du télescope Subaru, ainsi que sur l'instrument \ac{NIRISS} du télescope spatial \ac{JWST} \citep{sivaramakrishnan2012}. Ce mode est aussi prévu pour l'instrument \ac{MICADO} \citep{lacour2014} du futur télescope \ac{ELT} de $39 \,$m de diamètre.

L'avantage majeur de cette technique est qu'il est possible d'atteindre, après analyse des données, un pouvoir de résolution allant jusqu'à $0,5 \uplambda / \text{D}$ où D étant le diamètre du télescope \citep{lacour2011b}. En revanche son désavantage est sa faible transmission du flux lumineux à cause de l'exploitation partielle de la pupille d'entrée du télescope. Par exemple, le masque de la figure~\ref{fig:KeckPupilMaskingA} a une transmission du flux lumineux de $10\%$.

Afin de contrebalancer ce problème, le réarrangement de pupille est proposé \citep{perrin2006, lacour2007}. Il permet d'exploiter l'entièreté de la pupille du télescope en combinaison avec la technique de masquage de pupille. Pour cela, toute la pupille est divisée en sous-pupilles afin d'être réarrangées de manière non-redondante. Pour ce faire, le flux lumineux des faisceaux de chaque sous-pupille est injecté dans des fibres optiques monomodes. Celles-ci permettent à la fois le réarrangement simple des sous-pupilles et le filtrage du front d'onde de chaque sous-pupille des résidus de phase subsistant après l'optique adaptative. \ac{FIRST} est le premier instrument implémentant cette technique de réarrangement de pupille fibré \citep{kotani2008} pour l'imagerie haut contraste à haute résolution angulaire.

Enfin, l'instrument \ac{GLINT} \citep{martinod2021}, successeur de l'instrument Dragonfly \citep{jovanovic2012}, propose une autre solution technique du réarrangement de pupille, sans fibre optique, en injectant directement les faisceaux dans un composant d'optique intégrée (servant à la recombinaison interférométrique). Les entrées ont alors la même disposition spatiale que celle des sous-pupilles. Les faisceaux sont ensuite guidés dans le composant avant d'être recombinés par paire.

Comme on vient de le voir, ces techniques atteignent un pouvoir de résolution allant jusqu'à $0,5 \uplambda / \text{D}$. Le masquage de pupille sur un télescope unique de $8 \,$m de diamètre permet des mesures sur un compagnon d'étoile avec une séparation de l'ordre de la dizaine de mas. De plus, l'étude faite par \cite{fernandes2019} sur la population des planètes géantes ($0,1 - 20 \text{M}_{\text{J}}$) détectées par la méthode des vitesses radiales et par la mission Kepler, résumée par le graphique de la figure~\ref{fig:Fernandes2019F2}, montre une forte occurrence des exoplanètes avec un demi-grand axe égal à $\sim 2 \,$ua. Or pour des systèmes exoplanétaires à une distance d'environ $100 \,$pc de la Terre (la distance où se trouve le regroupement d'étoiles en formation le plus proche), de telles valeurs de demi-grand axe, représentent une séparation projetée sur le ciel qui vaut $\sim 20 \,$mas.

\begin{figure}[ht!]
    \centering
    \includegraphics[width=\figwidth]{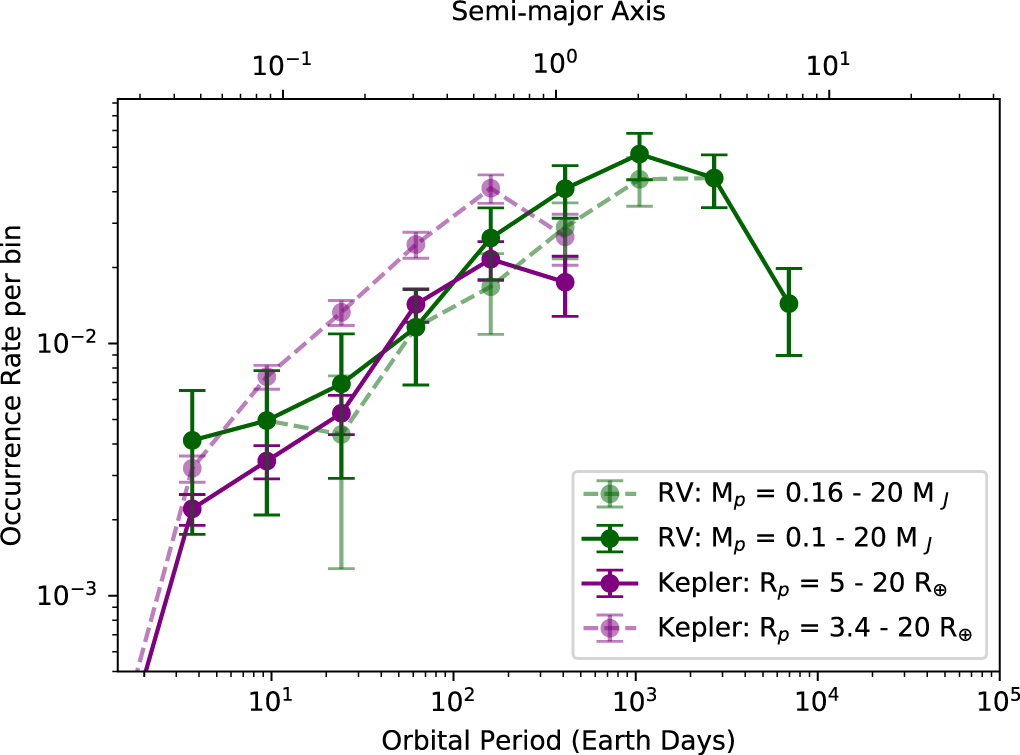}
    \caption[Le nombre d'occurrence des exoplanètes détectées par la méthode des vitesses radiales et par la mission Kepler en fonction de la période orbitale et du demi-grand axe.]{Le nombre d'occurrence des exoplanètes géantes en fonction de la période orbitale (axe du bas) et du demi-grand axe (axe du haut) détectées par la méthode des vitesses radiales (ligne continue verte) et par la mission Kepler (ligne continue violette). Les courbes sont données pour plusieurs intervalles de masse et de rayon (lignes continues ou discontinues). Figure tirée de \cite{fernandes2019}.}
    \label{fig:Fernandes2019F2}
\end{figure}

L'utilisation de cette technique a déjà fait certaines découvertes de compagnon sub-stellaires. Par exemple, le mode \ac{SAM} de l'instrument \ac{NACO} du \ac{VLT} a permis la découverte d'un compagnon autour de l'étoile HD 142527 \citep{biller2012} avec une séparation de $\sim 88 \,$mas, correspondant à un demi-grand axe estimé à $\sim 13 \,$ua. Le compagnon a ensuite été observé et confirmé en \ha~\citep{close2014} et son taux d'accrétion a été estimé. Cette détection ont été effectuée en infrarouge et le développement de la technique de masquage et réarrangement de pupille dans la gamme des longueurs d'onde du visible, sur des télescopes de la classe $10 \,$m de diamètre, permettrait d'augmenter encore les performances en résolution angulaire tout en permettant la caractérisation d'exoplanètes en formation (voir la section~\ref{sec:Protoplanetes}).

%%%%%%%%%%%%%%%%
\subsection{De FIRSTv1 à FIRSTv2}

L'instrument \ac{FIRST} a été développé \citep{kotani2008} pour appliquer le masquage et le réarrangement de pupille fibré combiné avec un spectrographe, dans la gamme de longueurs d'onde du visible, selon le concept développé par Sylvestre Lacour pendant sa thèse \citep{lacour2010} à partir de l'idée proposée par Guy Perrin \citep{perrin2006}. La figure~\ref{fig:FIRSTv1Scheme} présente un schéma de principe de cet instrument. La pupille du télescope est divisée en sous-pupilles représentée par le miroir déformable segmenté (1) dont les segments bleus (au nombre de $9$) sont ceux dont la lumière est injectée dans les fibres optiques monomodes (3) à l'aide d'une matrice de micro-lentille (2). Les fibres sont connectées aux fibres d'un V-Groove (la figure~\ref{fig:VGroove} présente des photographies de ce composant) de manière non-redondantes (4) où les sorties sont disposées sur une dimension (réarrangement de pupille). Les sorties du V-Groove sont dispersées par un spectrographe (5) et imagées sur la caméra dont une image est montrée (6). Ici les longueurs des fibres optiques ont dû être égalisées afin d'annuler les retards de phase entre les faisceaux des sous-pupilles au moment d'interférer dans le plan focal (6).

\begin{figure}[ht!]
    \centering
    \includegraphics[width=\figwidth]{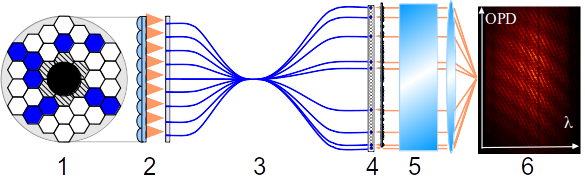}
    \caption[Schéma de principe de FIRSTv1.]{Schéma de principe de FIRSTv1. La lumière se propage de gauche à droite, sur les composants suivants : (1) le miroir segmenté représenté avec l'obstruction centrale du télescope, (2) la matrice de micro-lentilles, (3) les fibres optiques monomodes, (4) le V-Groove pour le réarrangement de pupille, (5) le spectrographe et (6) la caméra. Crédit : Elsa Huby.}
    \label{fig:FIRSTv1Scheme}
\end{figure}

L'instrument a ensuite été intégré pour sa première lumière \citep{huby2012} sur le télescope de $3 \,$m Shane de l'observatoire Lick durant la thèse d'Elsa Huby \citep{huby2013these}. La figure~\ref{fig:FIRSTv1PupilMaskingA} présente le masquage de pupille utilisé sur \ac{FIRST} (à gauche) et leur réarrangement non-redondant à $1$ dimension à l'aide d'un V-Groove (à droite). \ac{FIRST} a ainsi pu recombiner jusqu'à $18$ sous-pupilles (via deux masques sur la même pupille et deux V-Grooves). Une image des réseaux de franges mesurée lors d'une des premières observations de Véga ($\upalpha$ Lyr) est présentée sur la gauche de la figure~\ref{fig:FIRSTv1PupilMaskingB} sur laquelle l'axe vertical est l'axe des \ac{OPD}s et l'axe horizontal est l'axe des longueurs d'onde. Sur cette image on voit la superposition de $9 \times 8 / 2 = 36$ réseaux de franges donnant l'information de $36$ fréquences spatiales. Enfin, la partie droite de la figure~\ref{fig:FIRSTv1PupilMaskingB} présente la densité spectrale de puissance (obtenue par la transformée de Fourier) de l'image précédente, sur laquelle on voit les pics de fréquences spatiales échantillonnées.

\begin{figure}[ht!]
    \centering
    \begin{subfigure}[t]{0.45\textwidth}
        \centering
        \includegraphics[width=0.7\textwidth]{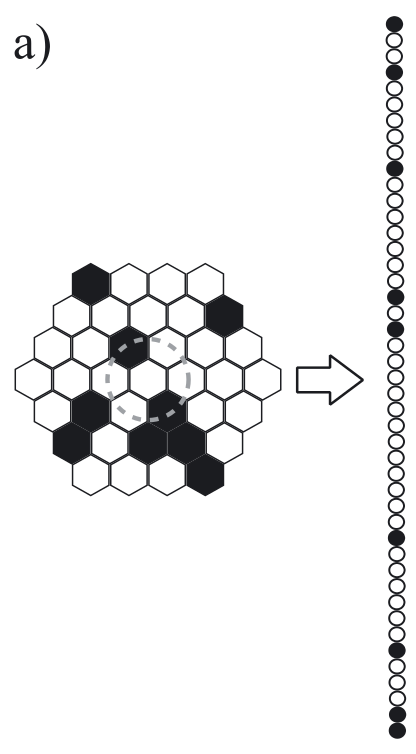}
        \caption{À gauche, la configuration des sous-pupilles choisies, à droite, la configuration 1-D choisie pour le réarrangement de pupille.}
        \label{fig:FIRSTv1PupilMaskingA}
    \end{subfigure}\hspace{0.01\textwidth}
    \begin{subfigure}[t]{0.45\textwidth}
        \centering
        \includegraphics[width=0.9\textwidth]{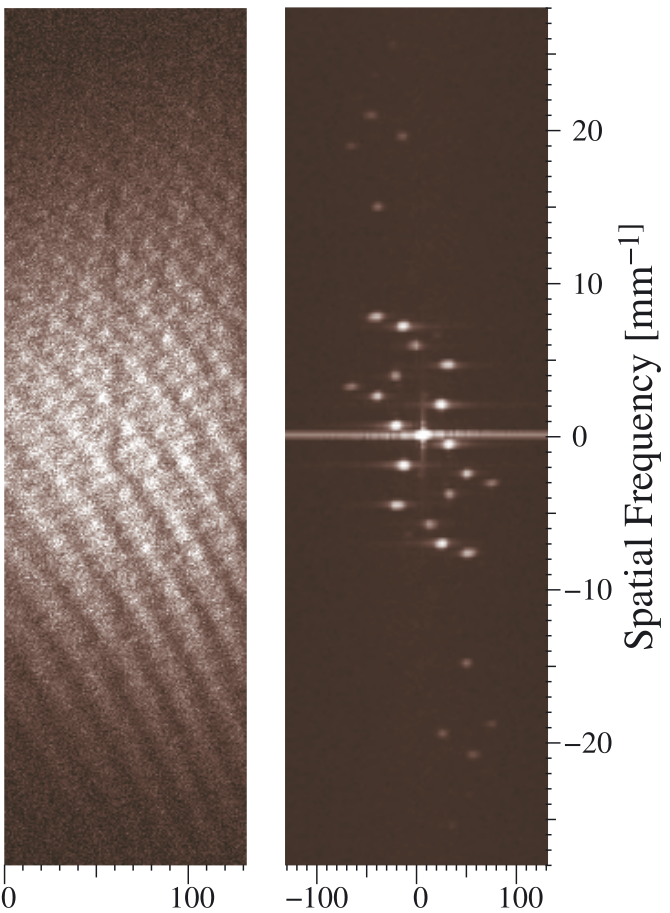}
        \caption{À gauche, une image des réseaux de franges mesurée sur la caméra lors d'une observation de Véga à l'observatoire Lick, à droite la densité spectrale de puissance de cette image, montrant la répartition de l'échantillonnage du plan UV.}
        \label{fig:FIRSTv1PupilMaskingB}
    \end{subfigure}
    \caption[Expérience de réarrangement de pupille de l'instrument FIRSTv1 sur le télescope Shane.]{Expérience de masquage et de réarrangement de pupille de l'instrument FIRSTv1 sur le télescope Shane, tiré de \cite{huby2012}.}
    \label{fig:FIRSTv1PupilMasking}
\end{figure}

De cette manière, les performances de ces techniques ont été démontrées durant la thèse d'Elsa Huby \citep{huby2013} à partir d'observations sur l'étoile multiple Capella ($\upalpha$ Aur) sur le télescope de $3 \,$m Shane de l'observatoire Lick. La séparation des deux composantes résolues par \ac{FIRST} est estimée à $50 \,$mas avec une précision de l'ordre de $1 \,$mas, ce qui est en-deçà du pouvoir de résolution du télescope ($58 \,$mas à $850 \,$nm). Les positions mesurées d'une des deux composantes (l'autre étant centrée dans le champ de vue du télescope) ont été trouvées en accord avec l'orbite connue. Cela a aussi donné lieu à la première mesure du spectre du rapport de flux des deux composantes ($\sim 1$) aux longueurs d'onde $600 - 850 \,$nm avec une résolution spectrale de $\sim 300$. Des raies d'émission ont pu être identifiées (notamment la raie \ha, les bandes TiO et CN) et analysées en regard de simulations.

À la fin de sa thèse, Elsa Huby a mené l'intégration de \ac{FIRST} sur la plateforme \ac{SCExAO} sur le télescope de $8 \,$m Subaru. Le concept est resté le même et cela a permis de bénéficier du plus grand miroir du Subaru afin d'augmenter le flux lumineux injecté dans l'instrument ainsi que son pouvoir de résolution, tout en bénéficiant de la correction extrême d'optique adaptative de la plateforme \ac{SCExAO} ainsi que la meilleure qualité de ciel (en terme de turbulence atmosphérique) à disposition sur Terre, qu'offre le site d'Hawaii. Cela permet en conséquence d'acquérir des images interférométriques avec un temps d'exposition plus long et donc d'augmenter la sensibilité de l'instrument. Les premiers résultats et la démonstration de l'instrument sur ciel ont été publiés par Sébastien Vievard dans \cite{vievard2020a} et Vievard et al. (2023, en préparation) afin de démontrer la faisabilité du concept sur un télescope faisant partie de la classe des télescopes d'une dizaine de mètres de diamètre avec un système d'optique adaptative extrême, grâce à de nouvelles observations sur le système Capella ($\upalpha$ Aur). La figure~\ref{fig:FIRSTv1PupilMaskingSubaru} présente le masquage de pupille sur \ac{FIRST}, sur le télescope Subaru \citep{vievard2020a}. Comme la figure précédente, à gauche est la configuration des sous-pupilles choisies (les deux couleurs indiquent les deux groupes de neuf sous-pupilles injectés dans les deux bras de \ac{FIRST}), au milieu est une image des sorties photométriques d'un des deux V-Grooves ce qui montre la configuration 1-D du réarrangement de pupille et à droite est une image des réseaux de frange sur la source interne.

\begin{figure}[ht!]
    \centering
    \begin{subfigure}[t]{0.49\textwidth}
        \centering
        \includegraphics[width=\textwidth]{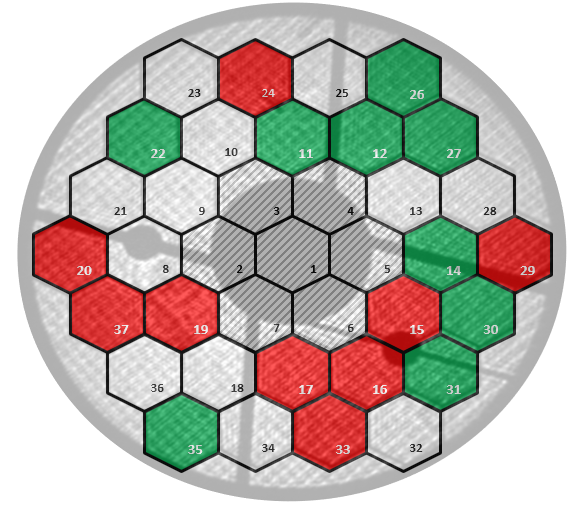}
        \caption{À gauche, la configuration des sous-pupilles choisies, superposée au plan pupille d'entrée de FIRST. Deux jeux de neuf fibres permettent l'injection et la recombinaison des deux groupes de sous-pupilles, en vert en rouge.}
        \label{fig:FIRSTv1PupilMaskingSubaruA}
    \end{subfigure}\hspace{0.01\textwidth}
    \begin{subfigure}[t]{0.49\textwidth}
        \centering
        \includegraphics[width=0.9\textwidth]{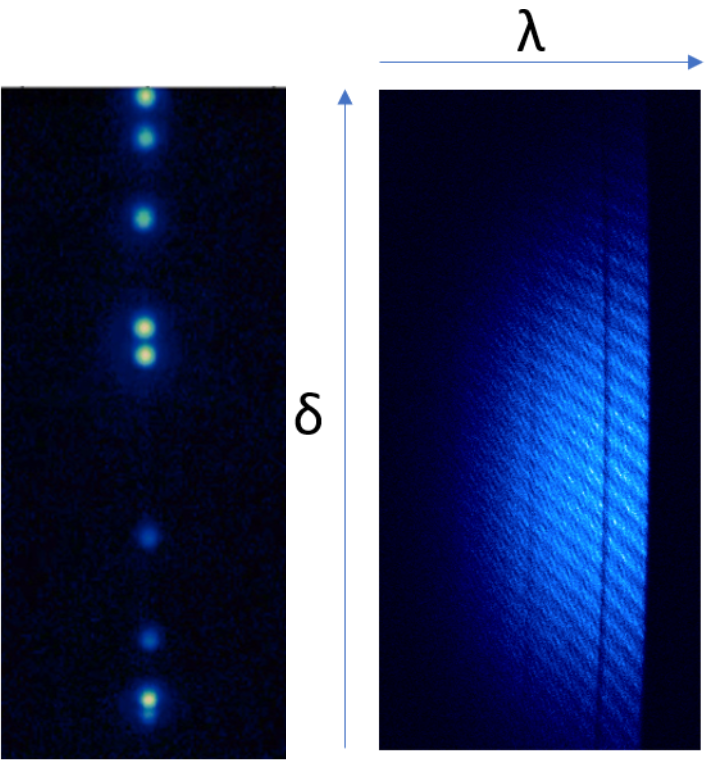}
        \caption{À gauche, une image des sorties photométriques des neuf fibres d'un des deux V-Grooves, montrant la configuration 1-D du réarrangement de pupille. À droite, une image des réseaux de franges mesurée sur la caméra sur la source interne, avec la dispersion sur l'axe horizontal et l'opd sur l'axe vertical.}
        \label{fig:FIRSTv1PupilMaskingSubaruB}
    \end{subfigure}
    \caption[Expérience de réarrangement de pupille de l'instrument FIRSTv1 sur le télescope Subaru.]{Expérience de masquage et de réarrangement de pupille de l'instrument FIRSTv1 sur le télescope Subaru, tiré de \cite{vievard2020a}.}
    \label{fig:FIRSTv1PupilMaskingSubaru}
\end{figure}

Une réplique de cet instrument a été développée par Nick Cvetojevic sur le banc de test \ac{FIRSTv2} au laboratoire \ac{LESIA} (à Meudon) dans le but de tester de nouvelles solutions de recombinaison des sous-pupilles. Ce banc de test est amplement détaillé dans la section~\ref{sec:FIRSTv2Concept} mais je résume son principe ici. Il s'agit de recombiner indépendamment les faisceaux par paire en injectant les faisceaux des sous-pupilles dans un composant d'optique intégrée, là où sur la première version les faisceaux interféraient dans le plan focal de l'instrument. Cette technologie est dite photonique car elle consiste en un bloc de verre dans lequel sont gravés des guides d'onde. Les faisceaux des n sous-pupilles y sont injectés, puis sont divisés en $\text{n} - 1$ sous-faisceaux pour être ensuite recombinés par paire. Enfin, les sorties de la puce sont connectées aux fibres d'un V-Groove, dont ses sorties (qui n'ont pas besoin d'être disposées de manière non-redondante comme précédemment) sont dispersées par un spectrographe et imagées sur un détecteur. La figure~\ref{fig:FIRSTv2FringesCameraB} présente les interférogrammes obtenus sur ce banc de test imagés par la caméra, pour la configuration de sous-pupilles montrée sur la figure~\ref{fig:FIRSTv2FringesCameraA} (pour plus de détails voir la section~\ref{sec:BaseConfig}). Chaque interférogramme est imagé sur des pixels différents sur le détecteur selon l'axe vertical, en étant dispersés selon l'axe horizontal. On note que ces interférogrammes sont pour une valeur d'\ac{OPD} et les franges sont mesurées sur l'axe des \ac{OPD}s grâce au miroir segmenté qui est contrôlable en piston. L'objectif avec cette nouvelle version est d'augmenter les performances en contraste et en sensibilité afin de permettre des mesures de systèmes exoplanétaires présentant de plus hauts contrastes.

\begin{figure}[ht!]
    \centering
    \begin{subfigure}[t]{0.4\textwidth}
        \centering
        \includegraphics[width=0.9\textwidth]{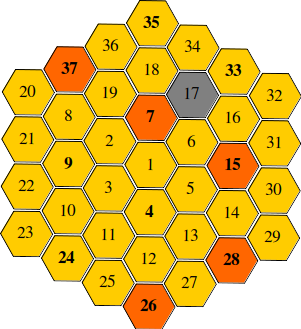}
        \caption{Configuration des sous-pupilles choisies sur FIRSTv2.}
        \label{fig:FIRSTv2FringesCameraA}
    \end{subfigure}
    \begin{subfigure}[t]{0.55\textwidth}
        \centering
        \includegraphics[width=0.9\textwidth]{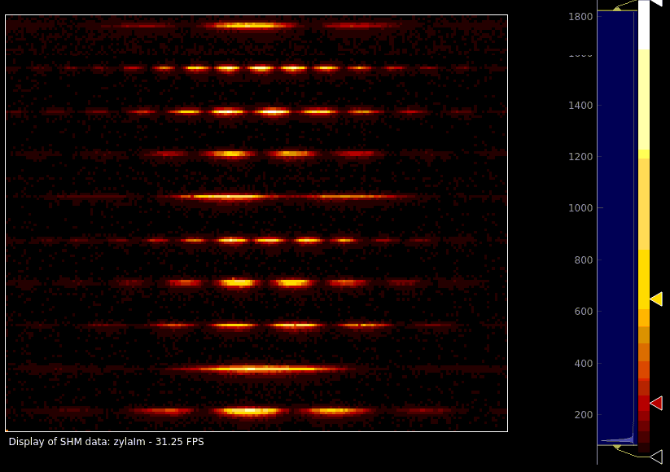}
        \caption{Image des interférogrammes sur la fenêtre de visionnage temps réel de la caméra de FIRSTv2 sur une source interne.}
        \label{fig:FIRSTv2FringesCameraB}
    \end{subfigure}
    \caption[Expérience de réarrangement de pupille de l'instrument FIRSTv2.]{Expérience de masquage et réarrangement de pupille de l'instrument FIRSTv2.}
    \label{fig:FIRSTv2PupilMasking}
\end{figure}

L'instrument \ac{GRAVITY} implémente une puce photonique \citep{perraut2018} dans le proche infrarouge qui recombine les faisceaux des quatre télescopes du \ac{VLT}. Cela a mené à la première détection directe d'une exoplanète par interférométrie \citep{lacour2019} ce qui démontre la capacité de cette technologie pour l'étude de systèmes exoplanétaires. Le développement de \ac{FIRSTv2} se fait dans les gammes de longueurs d'onde du visible afin d'améliorer les performances en résolution angulaire des mesures (celle-ci étant proportionnelle à la longueur d'onde) et de permettre la détection de la raie \ha~dans le spectre de protoplanètes, qui est probablement crucial pour la caractérisation des systèmes protoplanétaires, comme nous le verrons plus loin (section~\ref{sec:AccretionAlpha}).

Pour finir, \ac{FIRST} permet d'évaluer les futures performances de l'ensemble des techniques que nous avons vues ici sur les futurs \ac{ELT} de $> 30 \,$m de diamètre \citep{vievard2020b}. Un tel concept d'instrument sur cette classe de télescopes permettrait d'atteindre un pouvoir de résolution angulaire égal à $\sim 2 \,$mas à $700 \,$nm, sur des systèmes avec un contraste allant jusqu'à $10 ^{-6}$. Cela augmenterait le nombre de candidats d'exoplanètes au champ de l'imagerie directe mais aussi à d'autres domaines d'études comme les centres actifs des galaxies dans le visible.

%%%%%%%%%%%%%%%%
\subsection{L'apport de ma thèse au projet FIRST}

Le travail de ma thèse s'inscrit dans la continuité du développement de la réplique de \ac{FIRST} au laboratoire à Meudon. L'objectif est de tester en laboratoire la technologie d'optique intégrée pour la recombinaison interférométrique des sous-pupilles. Cette technologie dans les gammes de longueur d'onde du visible est nouvelle dans le domaine de l'astronomie. En effet, elle a été développée et optimisée ces dernières décennies dans le domaine des télécommunications en infrarouge et son adaptation dans le domaine du visible nécessite de nombreuses années de développement et de test. J'ai ainsi caractérisé deux technologies différentes de composants d'optique intégrée, présentés dans la section~\ref{sec:FIRSTv2Concept}, dans laquelle je présente toute l'expérience \ac{FIRSTv2}. 

De plus, une partie de mon travail de thèse a été de développer le logiciel de contrôle afin de le restructurer pour l'améliorer et d'ajouter de nouvelles fonctionnalités. Je présenterai ce travail dans la section~\ref{sec:ControlSoftware}. Ensuite, j'ai travaillé sur le développement du programme de traitement et d'analyse de données (section~\ref{sec:DataReduction}). Ce programme permet d'analyser des données acquises sur une source de type protoplanétaire que j'exposerai dans la section~\ref{sec:Protoplanetes}. J'ai alors démontré pendant ma thèse, la capacité de l'instrument \ac{FIRSTv2} à détecter une telle cible en intégrant au banc de test un système de sources lumineuses simulant une protoplanète compagnon d'une étoile (section~\ref{sec:SystBinaire}), en acquérant des données interférométriques sur celle-ci et en les traitant avec le programme développé pour l'occasion. Il s'agissait également d'implémenter pour la première fois sur le projet \ac{FIRST} la mesure des phases différentielles, qui est une observable particulièrement bien adaptée à la détection de systèmes protoplanétaires. Je présenterai les résultats de ces mesures dans la section~\ref{sec:BinaryCharac}. 

Enfin, la dernière partie de ma thèse a été l'intégration des puces photoniques sur la plateforme \ac{SCExAO} au télescope Subaru, pour la prise de données sur ciel lors de la première lumière de \ac{FIRSTv2}, que je présenterai dans la section~\ref{sec:FIRSTv2Subaru}. Lors de cette première lumière j'ai aussi eu l'occasion de déployer le logiciel de contrôle que j'avais développé et testé en laboratoire, à Meudon.

%%%%%%%%%%%%%%%%%%%%%%%%%%%%%%%%
\section{L'observation des systèmes protoplanétaires}
\label{sec:Protoplanetes}

%%%%%%%%%%%%%%%%
\subsection{La formation planétaire}

La variété des exoplanètes jusqu'alors détectées permet d'étudier des conditions physiques, astrométriques et chimiques différentes, constituant une source d'informations pour la compréhension des systèmes planétaires. Notamment, l'étude de leur stade évolutif retient notre attention ici et est possible en étudiant différents systèmes d'exoplanètes se trouvant à différents stades de leur formation. 

Tout d'abord, l'étoile centrale se forme par effondrement de matière au sein d'un nuage de gaz, dominé par de l'hydrogène moléculaire. Lorsqu'un corps de quelques masses de Jupiter est formé, la densité y est suffisante pour qu'il y ait un état d'équilibre hydrostatique dans lequel la force de gravité compense la force de pression de la matière. Tout en s'effondrant sur l'étoile, le nuage de gaz acquiert un moment cinétique dominant et, en conséquence, se répartit sur un disque en rotation par effet centrifuge, ce qui explique que l'on retrouve les planètes et beaucoup de corps divers du système solaire sur un même plan (l'écliptique). Ce disque protoplanétaire \citep{williams2011} se forme rapidement, en moins de $10 \,$kyr et est à des températures telles qu'il est observable dans les grandes longueurs d'ondes millimétriques et infrarouges. La figure~\ref{fig:DiskEvo} est une schématisation des étapes majeures de l'évolution de ce disque. L'étoile est totalement formée en $0.5 - 1 \,$Myr et induit sur le disque des rayonnements \ac{UV} qui ont pour effet d'évacuer tout le gaz de la partie interne du disque vers les plus lointaines orbites en l'espace de $\sim 0.1 \,$Myr (figure~\ref{fig:DiskEvo} (a), (b) et (c)). Jusqu'à environ $10 \,$Myr, le disque est encore protoplanétaire et est optiquement épais (la lumière visible ne s'échappe pas de l'intérieur du disque). Il contient encore du gaz primordial ainsi que de la poussière mais des intervalles vides se forment déjà, dus à la présence de protoplanètes. Entre $\sim 10 \,$Myr et $\sim 100 \,$Myr le disque est appelé disque de débris \citep{wyatt2008} car il ne contient plus du tout de gaz, qui a été accrété par les planètes, et il est constitué de poussières de seconde génération, provenant des nombreuses collisions d'astéroïdes et de planétésimaux (figure~\ref{fig:DiskEvo} (d)). L'âge maximal typique d'un disque de débris est de quelques $100 \,$Myr. Il est possible d'étudier ces détails des systèmes planétaires via les mesures de leur spectre électromagnétique. En effet, selon la taille des grains de poussières (qui augmente avec le temps) le rayonnement détecté évoluera vers les courtes longueurs d'ondes. De plus, les disques protoplanétaires, dans leurs premiers stades d'évolution, sont optiquement épais et l'étude de la répartition spectrale d'énergie permet d'inférer dans quelle mesure le disque modifie le spectre de l'étoile centrale.

\begin{figure}[ht!]
    \centering
    \includegraphics[width=\figwidth]{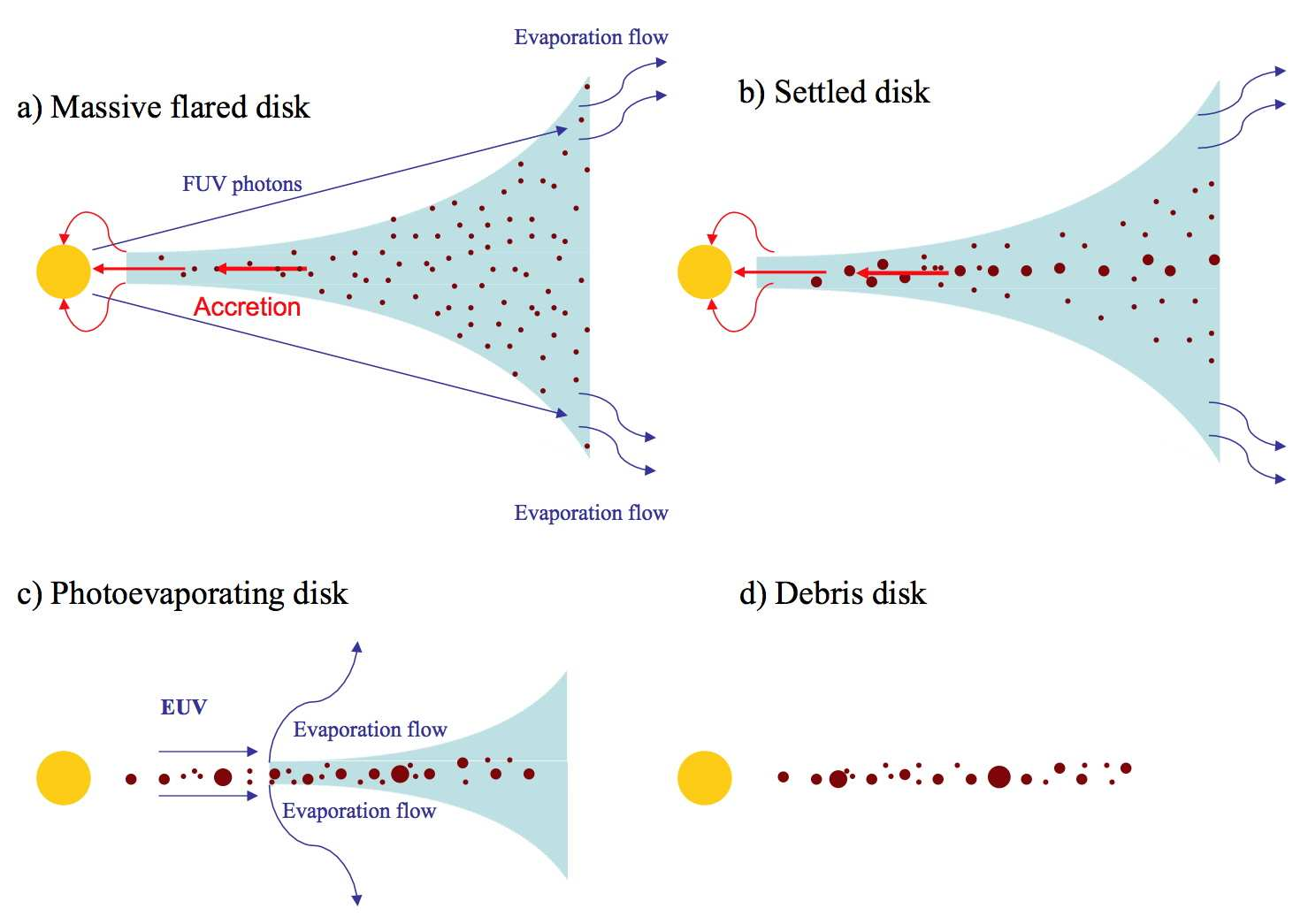}
    \caption[Évolution typique d'un disque protoplanétaire.]{Évolution typique d'un disque protoplanétaire, tiré de \cite{williams2011}. Le gaz est représenté en bleu et les poussières en marron. (a) Le disque dans son stade d'évolution primitif ($0-1\,$Myr) perd de la matière par accrétion sur l'étoile et photo-évaporation par les rayonnements \ac{UV} émis par l'étoile. (b) Les grains de poussières s'agglomèrent en débris de plus en plus gros tout en se répartissant sur un même plan. (c) L'étoile a fini de se former et le disque interne est dépourvu de gaz, devenant optiquement fin en quelques $0.1 \,$Myr. (d) Les plus petits grains de matière soit se font éjecter par pression de radiation soit sont accrétés par les planètes et il ne reste que les planétésimaux au sein du disque, qui devient de plus en plus difficile à détecter, en l'espace de quelques $100 \,$Myr.}
    \label{fig:DiskEvo}
\end{figure}

Dans le même temps, on identifie $3$ stades d'évolutions des exoplanètes observées, pendant lesquels les exoplanètes sont nommées :

\begin{enumerate}
    \item protoplanètes lorsque le système est âgé de moins de $4 \,$Myr, ce sont les exoplanètes qui m'intéressent dans le cadre de mon projet de thèse, elles sont toujours en formation et accrètent de la matière, au sein du disque protoplanétaire et ont une forte émission dans la raie \ha~(comme on le verra plus en détails dans la section~\ref{sec:AccretionAlpha});
    
    \item jeunes planètes géantes lorsque le système est âgé de $4 \,$Myr à $100 \,$Myr, elles sont au sein d'un disque de débris, elles n'accrètent plus de matière mais ont une température assez élevée ($\sim 700 - 1700 \,$K) pour émettre un rayonnement thermique visible et infrarouge et font partie du groupe d'exoplanètes les plus détectées en imagerie directe (e.g. Beta Pictoris b \citep{lagrange2010});
    
    \item planètes complètement formées lorsque le système est âgé de plus de $100 \,$Myr, constituant la grande majorité des exoplanètes connues, elles sont trop froides pour rayonner dans le visible (elles n'ont qu'un rayonnement infrarouge) comme sur les deux types d'exoplanètes précédemment cités, mais peuvent refléter suffisamment la lumière de leur étoile pour être détectées en imagerie directe (dans le visible).
\end{enumerate}

À ce jour, deux grandes théories existent pour décrire les mécanismes de formation planétaire. La première théorie est la formation par accrétion de matière (\ac{CA}, en anglais), pour la première fois énoncée dans \cite{safronov1972}. Les grains de poussières microscopiques se combinent pour former des corps de plus en plus gros : des astéroïdes, puis des planétésimaux (future planète). Enfin, si ces corps ont une masse d'au moins $10 \,$ M$_{\bigoplus}$, ils accrètent du gaz et deviennent des géantes gazeuses \citep{pollack1996}. Pour le système solaire, cela explique donc bien la présence de quatre planètes rocheuses dépourvues de gaz car il a été expulsé par le Soleil (comme expliqué dans le paragraphe précédent) sur les plus petites orbites ainsi que de planètes géantes gazeuses sur les orbites les plus lointaines et dont on soupçonne qu'elles contiennent un noyau rocheux.

Bien que cette première théorie soit la plus acceptée par la communauté, notamment pour expliquer la formation du système solaire, de récents travaux montrent qu'une deuxième théorie est plausible. Cette théorie fait appel à des processus d'instabilité gravitationnelle (\ac{GI}, en anglais) et est, par exemple, simulée dans \cite{nayakshin2017}. Ici, juste après la formation de l'étoile, des régions localisées dans le disque exoplanétaire se condensent par instabilités gravitationnelles en des corps de plusieurs fois la masse de Jupiter. Dans ces conditions, les grains de poussière sédimentent et forment un noyau rocheux central. Certaines simulations \citep{boley2010} montrent que ces amas se forment à $\gtrsim 50 \,$ua et migrent jusqu'aux basses orbites ($\sim 1 \,$ua), où ils perdent leur couche de gaz à cause des effets de marée. Cela réussit aussi à expliquer la présence de petites planètes rocheuses aux orbites les plus proches du Soleil et des géantes gazeuses dotées de noyaux rocheux sur des orbites plus éloignées, où les effets de marée ne sont pas suffisamment forts pour dissiper leur gaz.

Il est probable que les mécanismes de ces deux théories soient à l'oeuvre au cours de la formation planétaire. L'étude d'un plus grand nombre de systèmes protoplanétaires est donc indispensable pour contraindre ces théories.

%%%%%%%%%%%%%%%%
\subsection{L'accrétion de matière des protoplanètes}
\label{sec:AccretionAlpha}

Comme nous l'avons vu dans la section~\ref{sec:ImagerieDirecte}, l'imagerie directe permet l'étude de systèmes exoplanétaires encore jamais étudiés et permet une étude photométrique et spectroscopique de ceux-ci. Cette technique est adaptée à l'observation de protoplanètes car, d'une part, ce sont des systèmes jeunes et chauds donc ce sont les plus lumineux et, d'autre part, elle permet la mesure d'observables directement liées aux processus de formation planétaire.

L'une de ces observables qui nous intéresse est le spectre d'émission. En effet, l'accrétion de matière par un astre induit l'émission de raies associées à l'hydrogène dans le spectre mesuré. Lorsque le gaz, principalement composé d'hydrogène s'effondre sur une étoile ou une planète, il atteint des vitesses qui dépassent la vitesse du son locale. Le gaz génère en conséquence des ondes de choc augmentant la température de plusieurs milliers à plusieurs dizaines de milliers de Kelvins. À cette température, l'hydrogène est excité et émet les raies spectrales des bandes Lyman, Balmer, Paschen, etc. Ces émissions ont été simulées pour des proto-étoiles jeunes en accrétion à partir de modèles d'accrétion magnétosphérique \citep{muzerolle2001, natta2004, espaillat2008}, dans lesquels le gaz interagit avec le champ magnétique stellaire, permettant ainsi de déterminer les liens entre ces émissions et le taux d'accrétion. De telles émissions aux longueurs d'onde des raies de l'hydrogène ont été mesurées pour un échantillon d'étoiles jeunes du nuage Lupus \citep{alcala2014, alcala2017}, permettant d'estimer leur taux d'accrétion de matière ainsi que les liens entre la géométrie des flux d'accrétion et le taux d'accrétion. Une étude similaire sur le taux d'accrétion de l'étoile PDS 70 a aussi été conduite dans \cite{thanathibodee2020} en utilisant un modèle d'accrétion magnétosphérique. PDS 70 est dite de type T Tauri (\ac{TTS}, en anglais) \citep{appenzeller1989}, qui définit les étoiles variables jeunes en formation, de masse de moins de $2 \, \MS$, âgées de moins de $10 \,$Myr et appartenant à la pré-séquence principale (PDS 70 est de type spectral K7). Comme beaucoup d'étoiles de type \ac{TTS}, PDS 70 est dotée d'un disque de transition (stade d'évolution entre le disque protoplanétaire et le disque de débris) ainsi que d'un champ magnétique. Ces disques se caractérisent par la présence d'orbites pratiquement vides contenant parfois des protoplanètes. Les taux d'accrétion d'étoiles de type \ac{TTS} ont plusieurs fois été calculés \citep{natta2004, rigliaco2012, ingleby2013}, permettant d'éprouver les modèles et techniques de calcul de taux d'accrétion, mais de grandes incertitudes demeurent.

Ces modèles d'accrétion de matière, initialement développés pour des étoiles jeunes et peu massives, ont été adaptés pour le cas de protoplanètes \citep{aoyama2018, thanathibodee2019}. Les difficultés d'une telle adaptation sont que les conditions d'accrétion sur les protoplanètes sont différentes de celles sur les étoiles : dans le cas des protoplanètes, (1) la température du gaz en accrétion est moins élevée (d'au moins un ordre de grandeur), (2) le milieu dans le voisinage de la protoplanète (le disque circumplanétaire) n'a pas les mêmes structures (c'est parfois un bras de gaz et de poussière \citep{boccaletti2020} approvisionnant en matière l'étoile encore en formation) et (3) les champs magnétiques locaux sont différents en intensité et en structure, entraînant des écoulements de gaz différents dans l'un ou dans l'autre des cas. Plus simplement, dans le cas des étoiles de type \ac{TTS}, le gaz est excité avant le choc supersonique et est ionisé après le choc car les champs magnétiques locaux sont assez puissants pour accélérer les écoulements alors que dans le cas des protoplanètes, le gaz n'est excité seulement qu'après le choc supersonique \citep{aoyama2019}.

L'étude des raies d'émission hydrogène de Balmer, plus particulièrement la raie Balmer-$\alpha$ ($656.28 \,$nm), aussi appelée raie \ha, est un bon moyen d'estimer le taux d'accrétion de matière en cours sur les protoplanètes \citep{aoyama2019, marleau2022}. \ac{FIRSTv2} est conçu pour travailler dans cette gamme de longueur d'onde et ainsi détecter une raie \ha. C'est pour cela que, dans la suite, je vais m'intéresser essentiellement aux travaux effectués sur la détermination du taux d'accrétion à partir de la mesure de cette raie.

À partir du spectre mesuré, il est possible de contraindre le taux d'accrétion de matière de plusieurs manières différentes. La première se fait à partir de l'estimation de la luminosité d'accrétion $L_{acc}$, qui correspond au flux lumineux libéré par le mécanisme d'accrétion de matière. D'une part, celle-ci peut s'effectuer par l'utilisation de modèles magnétosphériques d'accrétion \citep{natta2004, ingleby2013, thanathibodee2019}, de modèles hydrodynamiques d'écoulement du gaz lorsqu'il s'effondre sur la protoplanète \citep{aoyama2018, aoyama2019, aoyama2020} ou de modèles d'émission par un bloc d'hydrogène à certaines conditions de température et de pression \citep{alcala2017, rigliaco2012}. Ces modèles sont ajustés aux spectres mesurés pour inférer la luminosité des raies d'émission ou directement la luminosité d'accrétion \Lacc. D'autre part, la mesure de l'intensité du pic \ha, \Lha, permet de déduire la luminosité \ha~émise par la protoplanète connaissant sa distance à la Terre. Dans le cas des étoiles de faible masse ($< 1 \, \MS$), la luminosité \Lha~est reliée à la luminosité d'accrétion $L_{acc}$ par une relation affine selon $log(\Lacc) = \text{b} + \text{a} \times log(\Lha)$. Enfin, une fois la luminosité \Lacc~estimée, il est possible d'inférer le taux d'accrétion $\dot{\text{M}}$. En effet, ce taux d'accrétion est proportionnel à la luminosité libérée à l'impact de la matière en accrétion sur la surface du corps selon une équation de chute libre provenant du modèle d'accrétion magnétosphérique \citep{gullbring1998}. \cite{wagner2018} applique cette méthode sur le système PDS 70 et \cite{rigliaco2012} l'applique sur un échantillon d'une dizaine d'étoiles de type \ac{TTS} pour plusieurs raies d'absorption de l'hydrogène. Il propose ainsi d'étalonner la relation entre l'émission \ha~et le taux d'accrétion pour ce type d'étoiles afin d'être employée à l'avenir sur le même type d'objet \citep{close2014}.

La deuxième méthode pour contraindre le taux d'accrétion se fait par la mesure de la largeur du pic \ha~émis. Plus précisément, de la largeur du pic à $10 \, \%$ de sa hauteur (\Wd) qui est élargie en fonction du taux d'accrétion. En effet, \cite{natta2004, fang2009} suggèrent un seuil indiquant que des cibles qui présentent une largeur \Wd~du pic \ha~inférieur à $4.4 \,$\AA $\,$ (ou $200 \, \text{km}.\text{s}^{-1}$) n'accrètent pas de matière, dans le cas de corps de faible masse de type stellaire (\ac{CTTS} ou \ac{WTTS}) ou géantes gazeuses. C'est aussi ce que montrent les simulations de \cite{thanathibodee2019} (figure 4b) pour les protoplanètes du système PDS 70 : il existe un seuil en dessous duquel ($\Wd = 100 \,$km.$\text{s}^{-1}$) la largeur de l'émission \ha~ne dépend pas du taux d'accrétion $\dot{\text{M}}$ pour $\dot{\text{M}} < 10^{-8} \, \MJ.\text{yr}^{-1}$. Les valeurs de seuil de taux d'accrétion diffèrent selon ces études car les modèles de simulations utilisés sont appliqués dans les premiers cas pour des corps stellaires de faible masse et dans le deuxième cas pour des protoplanètes. Mais encore, le taux d'accrétion des protoplanètes du système PDS 70 a pu être calculé à partir des largeurs \Wd~mesurées sur les spectres obtenus sur l'instrument MUSE \cite{haffert2019, hashimoto2020}.

L'intérêt d'étudier les protoplanètes dans la raie \ha~est que le contraste de luminosité planète / étoile est bien moins élevé qu'en bande H (d'un facteur $50$ à $1\,000$ selon \cite{close2014}) ce qui augmente les chances de détection par \ac{FIRSTv2}. La mesure d'une telle raie permettrait à la fois de discriminer la présence d'une protoplanète qui, dans le reste de la bande spectrale, serait en dehors de la gamme de performances de l'instrument; mais aussi d'étudier son taux d'accrétion.

Dans les cas précédemment cités où le flux de la raie d'émission est utilisé, le taux d'extinction devra, dans certains cas, être pris en compte. En effet, étant donné que les protoplanètes se trouvent dans un disque de gaz et de poussières, ce dernier atténue son flux lumineux d'un facteur $A_R$. Pour le système PDS 70, par exemple, la limite basse de ce taux a été estimé à $2.0 \,$mag et $1.1 \,$mag, respectivement pour la planète b et la planète c \citep{hashimoto2020}. Cette estimation a été faite à partir de modèles hydrodynamiques simulant la quantité d'hydrogène moléculaire qui peut être liée à l'extinction de la poussière interstellaire. Ce taux d'extinction en \ha~de l'émission mesurée d'une protoplanète se fait donc en se basant sur nos connaissances du système, notamment sur la présence de gaz et de poussières (qui diffère pour un disque de transition par rapport à un disque plus jeune ne présentant pas de transition) qui peut être inféré par l'étude des spectres du système à d'autres longueurs d'ondes.

Il est aussi intéressant de noter que ce facteur d'extinction n'affecte que la hauteur des pics d'émissions et non leur largeur. On voit donc ici l'intérêt de la mesure de la largeur du pic \ha, notamment la largeur à $10 \, \%$ de la hauteur \Wd, qui peut ainsi se faire sans se soucier de l'extinction au sein du disque. Mais pour cela, il est nécessaire de disposer d'une résolution spectrale suffisante. Par exemple, la raie spectrale mesurée sur PDS 70b \citep{haffert2019} a une largeur à mi hauteur égale à $\text{FWHM} = 0.27 \pm 0.03 \,$nm et on souhaite mesurer six points sur cette raie, correspondant à un élément de résolution de $\updelta\uplambda = 0.09 \,$nm. Cela nécessite un spectrographe avec une résolution spectrale à $656.28 \,$nm de $\text{R} = 7292$. Actuellement, la résolution spectrale du spectrographe de \ac{FIRSTv2} est de $\text{R}_{\text{FIRSTv2}} \simeq 4000$ (section~\ref{sec:InstruSpectro}) et ne permet donc pas de mesure fiable de la largeur d'une raie d'émission \ha~mais seulement sa mesure d'intensité.

Actuellement plusieurs systèmes semblent être pourvus de protoplanètes ou de candidats de protoplanètes. Mais seul les deux premières de cette liste font l'objet d'un consensus : 

\begin{itemize}
    \item les deux protoplanètes mises en évidence \citep{keppler2018, muller2018} autour de l'étoile PDS 70, déjà mentionnée plus haut, qui présentent une émission \ha~plusieurs fois mesurée et étudiée. Je les présenterai plus amplement dans la section~\ref{sec:pds70}.

    \item L'étoile AB Aurigae âgée de $1 - 3 \,$Myr semble présenter un compagnon \citep{currie2022b} à une séparation de $\sim 93 \,$ua, d'après des observations avec l'instrument \ac{CHARIS} et sur \ac{HST}. La protoplanète se situe au sein d'un disque massif avec de nombreux bras spiraux et l'ajustement de modèles spectraux montre un taux d'accrétion qui vaut $\dot{\text{M}} \simeq 1,1.10^{-6} \, \MJ.\text{yr}^{-1}$ (incertitude non fournie) mais il n'est pas certain que cela provienne de l'accrétion de matière.
    
    \item Le système LkCa 15 présentant $3$ protoplanètes potentielles, entre autres, observées par les instruments \ac{NIRC2} sur le télescope Keck-II, avec la technique \ac{SAM} \citep{kraus2012}, \ac{LMIRCam} sur le \ac{LBTI} et \ac{MagAO} sur le télescope Magellan Clay \citep{sallum2015} (détection d'émission \ha). De plus, lors d'observations avec l'instrument \ac{ISIS} sur le \ac{WHT} \citep{mendigutia2018} et avec \ac{CHARIS} sur le télescope Subaru en combinaison avec \ac{NIRC2} sur le télescope Keck-II \citep{currie2019}, il a été mis en évidence qu'il se pourrait que ce soit le disque lui-même ou un compagnon Jovien qui émettent le rayonnement \ha~et non des protoplanètes. Une étude plus approfondie est nécessaire pour conclure sur ce système.
    
    \item L'étoile HD 100546 observée par l'instrument \ac{NACO} du \ac{VLT} \citep{quanz2013a, quanz2015}, par l'instrument \ac{GPI} du \ac{GST} \citep{currie2015, follette2017} et par \ac{SPHERE} sur le \ac{VLT} \citep{mendigutia2017}. Deux protoplanètes se trouveraient à des séparations de $\sim 50 \,$ua et $\sim 13 \,$ua.
    
    \item La protoplanète Delorme 1 (AB)b \citep{eriksson2020, ringqvist2021} observée avec l'instrument \ac{MUSE} du \ac{VLT}.
    
    \item Le système MWC 758 observé avec un coronographe vortex installé sur l'instrument \ac{NIRC2} du télescope Keck-II \citep{reggiani2018} ainsi qu'avec l'instrument \ac{SPHERE} sur le \ac{VLT} \citep{cugno2019}.
    
    \item L'étoile T Cha observée par la technique de masquage de pupille avec \ac{NACO} sur le \ac{VLT} \citep{huelamo2011}. Il semblerait que le candidat compagnon ait été réfuté par la suite par \cite{olofsson2013}.

    \item L'étoile HD 169142 présente deux sources \citep{quanz2013b, biller2014, reggiani2014} dont l'une semble être un compagnon sub-stellaire (à une séparation de $\sim 23 \,$ua) et l'autre pourrait être une exoplanète (à une séparation de $\sim 50 \,$ua).
    
    \item Ainsi que l'échantillon de cibles étudié dans le cadre de la recherche de protoplanètes en accrétion \ac{GAPlanetS} avec l'instrument \ac{MagAO} \citep{follette2022}.
\end{itemize}

%%%%%%%%%%%%%%%%
\subsection{Le cas du système PDS 70}
\label{sec:pds70}
% Aoyama 2020 sec 4.1.1 gives a good review of PDS 70 system

PDS 70 est une étoile âgée de $5 - 10 \,$Myr de masse égale à $0,9 \,$\MS. Elle est au sein d'un disque de transition avec une cavité de $\sim 80 \,$ua (figure~\ref{fig:PDS70Image}, gauche) qui suggère la présence d'une exoplanète en formation. Un compagnon de masse planétaire (PDS 70 b) a d'abord été découverte \citep{keppler2018} dans les données d'observations infrarouges sur les instruments \ac{SPHERE} et \ac{NACO} du \ac{VLT} et \ac{NICI} de l'observatoire Gémini. Il est caractérisé grâce à de nouvelles données en infrarouge prises sur l'instrument \ac{SPHERE} \citep{muller2018} puis une émission \ha~venant du compagnon est détectée \citep{wagner2018} sur l'instrument \ac{MagAO}. Il est ensuite confirmé et caractérisé plus amplement dans cette raie d'émission grâce à l'instrument \ac{MUSE} sur le \ac{VLT} et un deuxième compagnon (PDS 70 c) est détecté à la même occasion \citep{haffert2019}. L'image de droite de la figure~\ref{fig:PDS70Image} présente ces deux détections dans la raie d'émission \ha.

\begin{figure}[ht!]
    \centering
    \includegraphics[width=\figwidth]{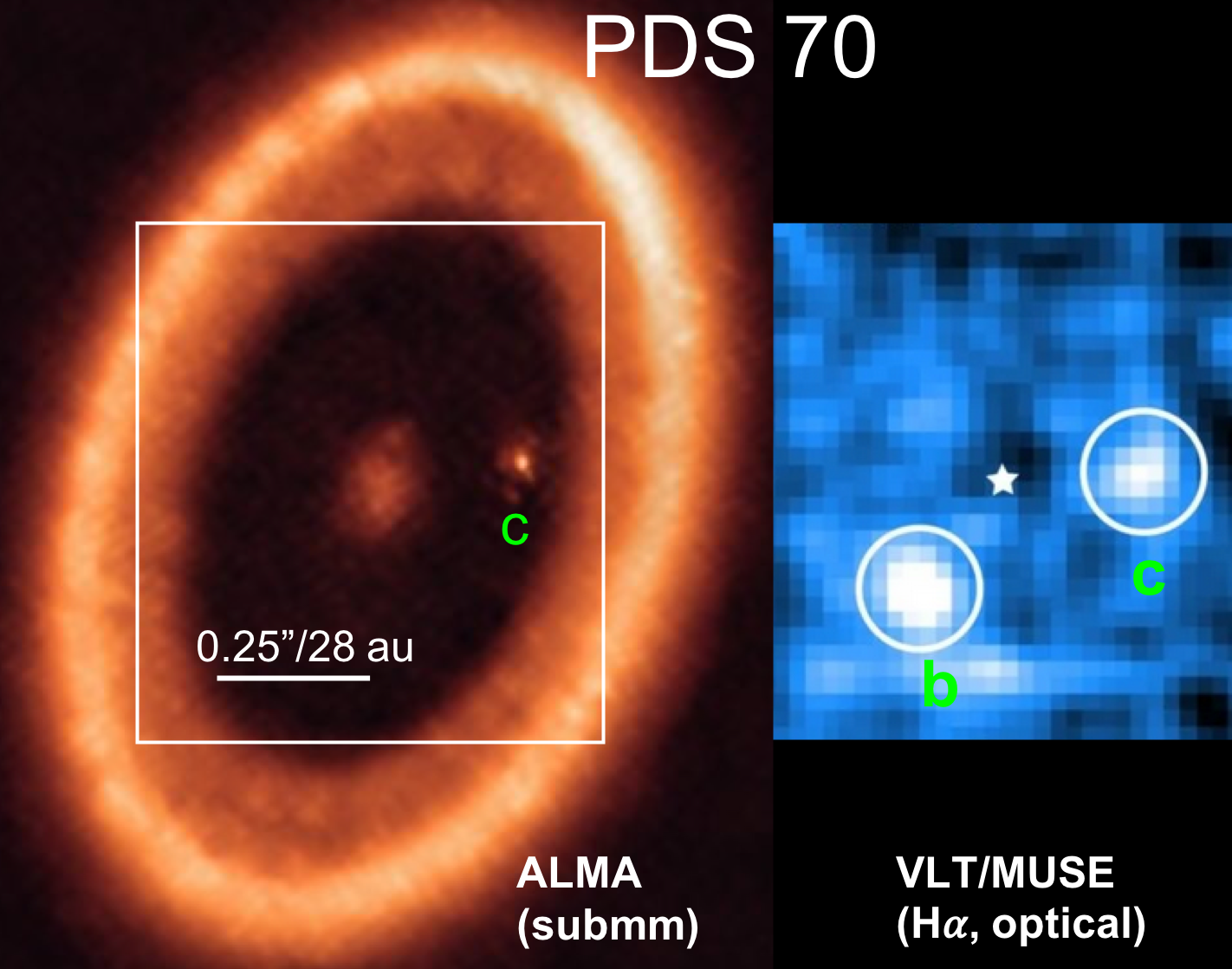}
    \caption[Images du système PDS 70.]{Image dans la gamme de longueur d'onde millimétrique obtenue sur le télescope ALMA (Atacama Large Millimeter/Submillimiter Array) \citep{benisty2021}, mettant en évidence le disque protoplanétaire et sa cavité de PDS 70 ainsi que le compagnon PDS 70 c et son disque circumplanétaire. Image dans la raie \ha~obtenue sur l'instrument MUSE du VLT \citep{haffert2019} sur laquelle les deux compagnons sont détectés. Le champ de vue de cette dernière correspond au champ de vue délimité par le rectangle blanc sur l'image de gauche. La composition d'images est tirée de \cite{currie2022a}.}
    \label{fig:PDS70Image}
\end{figure}

Les deux compagnons sont localisés dans la cavité du disque, à une séparation de $180 \,$mas (demi-grand axe de $20 \,$ua) et $240 \,$mas (demi-grand axe de $34 \,$ua) respectivement pour PDS 70 b et c \citep{wang2021}. Leur masse est estimée à $1 - 10 \, \MJ$ \citep{wang2021, haffert2019}. Les différentes méthodes exposées dans la partie précédente ont été utilisées dans plusieurs études pour estimer le taux d'accrétion. Celui-ci est trouvé dans l'intervalle $1-10 \times 10^{-8} \, \MJ.\text{yr}^{-1}$ pour les deux planètes \citep{wagner2018, haffert2019, aoyama2019, thanathibodee2019, hashimoto2020}. Au vu de ces connaissances, il s'agit à l'heure actuelle de la première détection confirmée de protoplanète.

% contraste de pds70 : $\sim 2.10^{-2}$ \citep{hashimoto2020} avec variation de $10\%$ à cause de l'étoile
Le rapport de flux (contraste) entre la planète b et l'étoile centrale est estimé à $\sim 1.10^{-3}$ \citep{wagner2018, zhou2021}. Cela induit un signal de phase valant $\sim 0,06\degree$ lors de mesures interférométriques. Ainsi, les développements de \ac{FIRST}, dont le travail de ma thèse fait partie, sont menés pour atteindre de telles sensibilités afin de détecter le signal d'une protoplanète.

Les mesures de largeur à $10\%$, \Wd, des raies d'émission \ha~de \cite{haffert2019} sont de $\sim 224 \pm 24 \, \text{km}.\text{s}^{-1}$ et $\sim 186 \pm 35 \, \text{km}.\text{s}^{-1}$ pour les planètes b et c respectivement. Cela correspond à des largeurs spectrales de $0,49 \pm 0,05 \,$nm et de $0,41 \pm 0,08 \,$nm, respectivement. Ainsi, pour obtenir $6$ points de mesure (pour être bien au-dessus du critère de Shanon) d'une raie d'émission \ha~de largeur égale à $200 \, \text{km}.\text{s}^{-1}$, un instrument doit disposer d'un pouvoir de résolution spectrale égal à $R = 9\,000$. Avec le nouveau concept de spectrographe récemment intégré à \ac{FIRSTv2} (voir la section~\ref{sec:InstruSpectro} à ce sujet), nous disposons d'un pouvoir de résolution égal à $3\,400$. Ce pouvoir de résolution permet de mesurer au moins $6$ points de raies d'émission de largeur supérieur à $\sim 529 \, \text{km}.\text{s}^{-1}$. Il est ainsi prévu que nous n'exploitions pas la mesure de la largeur \Wd~de la raie d'émission lors de futures observations, mais seulement l'intensité de celle-ci. Mais lorsque la sensibilité de l'instrument sera suffisante, la résolution spectrale de l'instrument pourra être augmenté (d'un facteur $\sim 2,6$) afin de permettre la mesure de la largeur \Wd~et contraindre le taux d'accrétion de la protoplanète observée.

% Des forts taux d'accrétion favorisent CA
% It should be noted that the community remains unsure of the formation mechanism of the PDS 70 planets. Bright and widely separated exoplanets such as the HR 8799 planets are often thought to have been formed by disk instability (Boss 2011). However, due to the presence of accretion on the PDS 70 planets, core accretion is one of the potential scenarios (as for the planets in our own solar system, e.g. Nesvorny, 2018). In the case of planets formed by core accretion, the accretion rate is expected to be at least as high as what is observed in the PDS 70 planets. The closer in time they are to the run-away accretion phase (that is when the planets acquire a critical mass to clear the gas and create a gap in the transitional disk), the more material the planets accrete (Fig. 1 top panel)

%%%%%%%%%%%%%%%%%%%%%%%%%%%%%%%%%%%%%%%%%%%%%%%%%%%%%%%%%%%%%%%%
\chapter{La réplique de FIRST en laboratoire : FIRSTv2}
\label{sec:FIRSTv2Concept}
\setcounter{figure}{0}
\setcounter{table}{0}
\setcounter{equation}{0}

\minitoc

\clearpage
\ac{FIRSTv2} est la deuxième version l'instrument \ac{FIRST} (installé sur le télescope Subaru) qui implémente une nouvelle solution de recombinaison interférométrique. Alors que la première version est un interféromètre multi-axial qui fait interférer la lumière des sous-pupilles du télescope dans un plan focal sur la caméra, dans la deuxième nous proposons de recombiner par paire ces sous-pupilles au sein d'un composant d'optique intégré.

Dans ce chapitre je présenterai le banc de test en laboratoire qui réplique \ac{FIRSTv2} dans sa globalité et je décrirai en détails les différents composants qui y sont intégrés. Je montrerai aussi la caractérisation des composants les plus cruciaux. Je décrirai également le logiciel de contrôle du banc dont j'ai poursuivi le développement et l'amélioration. Enfin, je montrerai les sous-pupilles que j'ai choisies de recombiner lors des prises de données présentées plus tard dans le manuscrit ainsi qu'une étude de stabilité des mesures de phases sur le banc.

%%%%%%%%%%%%%%%%%%%%%%%%%%%%%%%%
\section{Le banc de test à Meudon}

Le banc de test en laboratoire à Meudon (que je nommerai \ac{FIRSTv2} par simplification dans la suite) est une réplique de l'instrument \ac{FIRST}, qui est déjà installé au télescope de $8 \,$m de diamètre, le Subaru, sur le banc d'optique adaptative extrême \ac{SCExAO}. Ce banc de test a pour objectif de tester et valider les nouveaux développements expérimentaux, notamment des composants d'optique intégrée pour la recombinaison interférométrique des faisceaux pour l'imagerie haut contraste de systèmes exoplanétaires.

\begin{figure}[ht!]
    \centering
    \includegraphics[width=\figwidth]{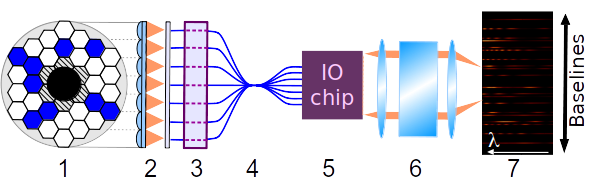}
    \caption[Schéma de principe du banc de test de FIRSTv2.]{Schéma de principe du banc de test de FIRSTv2. La lumière se propage de gauche à droite, sur les composants suivants : (1) le miroir segmenté représenté avec une obstruction centrale (non présente sur le banc de test de Meudon), (2) la matrice de micro-lentilles, (3) les lignes à retard (ODL), (4) les fibres optiques monomodes, (5) la puce d'optique intégrée, (6) le spectrographe et (7) la caméra.}
    \label{fig:FIRSTv2Scheme}
\end{figure}

La figure~\ref{fig:FIRSTv2Scheme} présente le schéma de principe de \ac{FIRSTv2}. Une source est injectée à gauche et alimente tous les composants principaux suivants :

\begin{enumerate}
    \item le miroir segmenté, fabriqué par \textit{Iris AO Inc}, composé de $37$ segments hexagonaux contrôlables en piston, tip et tilt;
    \item la matrice de micro-lentilles fabriquée par \textit{SÜSS Optics MicroOptics}, qui applique le masquage de la pupille en focalisant les faisceaux des sous-pupilles dans les fibres optiques regroupées dans un toron fabriqué par \textit{Fiberguide Industries}\footnote{\url{https://www.molex.com/molex/products/group/fiberguide}};
    \item les lignes à retard fabriquées par \textit{Oz Optics}\footnote{\url{https://www.ozoptics.com/}}, que je nommerai par la suite \ac{ODL}, qui appliquent un aller-retour en air libre aux faisceaux injectés à l'aide d'un prisme réfléchissant contrôlable en position, afin de changer la longueur de chemin optique pour compenser les longueurs différentes entre les fibres optiques;
    \item les fibres optiques monomodes fabriquées par \textit{Thorlabs}, qui sont à maintien de polarisation et qui filtrent le front d'onde des faisceaux lumineux des aberrations optiques et permettent le réarrangement de pupille;
    \item la puce d'optique intégrée qui est un bloc de verre dans lequel ont été formés des guides d'onde par diffusion ionique afin de recombiner par paires cinq faisceaux d'entrée (pour faciliter la division des faisceaux ainsi que pour des raisons de sensibilité, la solution avec neuf faisceaux étant trop peu transmissive, $< 0,1 \, \%$);
    \item le spectrographe, composé d'un réseau holographique fabriqué par \textit{Wasatch Photonics} et donnant une résolution spectrale de $\sim 3\,200$ à $650 \,$nm;
    \item la caméra sCMOS, fabriquée par \textit{Andor}, sur laquelle sont imagés les interférogrammes pour chaque base sur l'axe vertical et en étant dispersés horizontalement.
\end{enumerate}

Tous ces composants seront plus amplement détaillés par la suite. De plus, la figure~\ref{fig:FIRSTv2BenchPhoto} montre une photographie du banc en laboratoire. Tous les composants sont numérotés et la source est injectée dans un V-Groove (1) (voir la figure~\ref{fig:VGroove}) qui est un composant alignant plusieurs fibres optiques (ici $8$ fibres) avec une séparation connue et qui peut être utilisé ici pour simuler une source protoplanétaire (voir la section~\ref{sec:SystBinaire}). La lentille (2) de focale égale à $300 \,$mm focalise le faisceau à l'infini. Ce dernier est réfléchi par le miroir plan (3) puis par le demi-miroir (4) avant d'atteindre le miroir segmenté (nommé aussi \ac{MEMS} par la suite) (5) selon une incidence quasi-normale. Les lentilles (6) et (7) de focales égales à $300 \,$mm et à $125 \,$mm, respectivement, forment un doublet afocal permettant un grandissement de $125 / 300 \simeq 0,42$ entre le diamètre du cercle inscrit des segments du \ac{MEMS} égal à $606,2 \,$\um~et le diamètre des micro-lentilles (8) égal à $250 \,$\um~($250 / 606,2 \simeq 0,41$). Les micro-lentilles (8) sont disposées sous forme d'une matrice et focalisent les faisceaux des sous-pupilles dans les fibres optiques monomodes disposées dans le toron (9). Les fibres optiques sont alors branchées sur les \ac{ODL}s (10). Les fibres d'entrées (à gauche de la puce sur la photographie) de la puce photonique (11) sont branchées en sortie de ces \ac{ODL}s et les fibres de sorties (à droite de la puce sur la photographie) sont branchées à un V-Groove (12) (disposant de $36$ fibres). Les sorties des fibres de ce V-Groove sont collimatées par un objectif de microscope (13) et sont enfin dispersées par un réseau holographique (14) et imagées par un doublet de lentilles (16) de focales égales à $150 \,$mm et à $80 \,$mm, sur la caméra (17). Un prisme de Wollaston peut être placé sur la plaque de métal (15). La partie spectrographique (éléments (13) à (16)) a été conçu et installé par Manon Lallement (voir plus de détails dans la section~\ref{sec:InstruSpectro}).

\begin{figure}[ht!]
    \centering
    \begin{subfigure}[t]{0.4\textwidth}
        \centering
        \includegraphics[width=\textwidth]{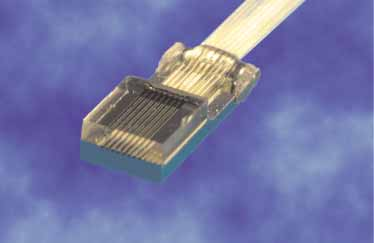}
        \caption{V-Groove.}
        \label{fig:VGrooveA}
    \end{subfigure}%
    \begin{subfigure}[t]{0.4\textwidth}
        \centering
        \includegraphics[width=0.94\textwidth]{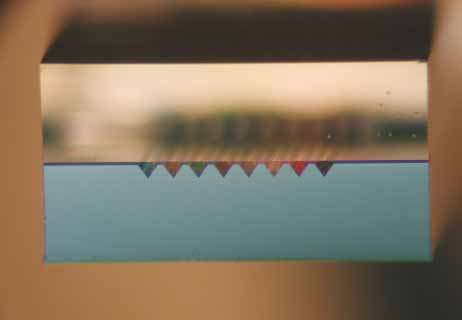}
        \caption{Face de sortie du V-Groove où l'extrémité des fibres sont placées dans des encoches en forme de V.}
        \label{fig:VGrooveB}
    \end{subfigure}
    \caption[Photographies d'un V-Groove.]{Photographies d'un V-Groove. Crédit : \textit{Oz Optics}.}
    \label{fig:VGroove}
\end{figure}

\begin{figure}[ht!]
    \centering
    \includegraphics[width=\figwidth]{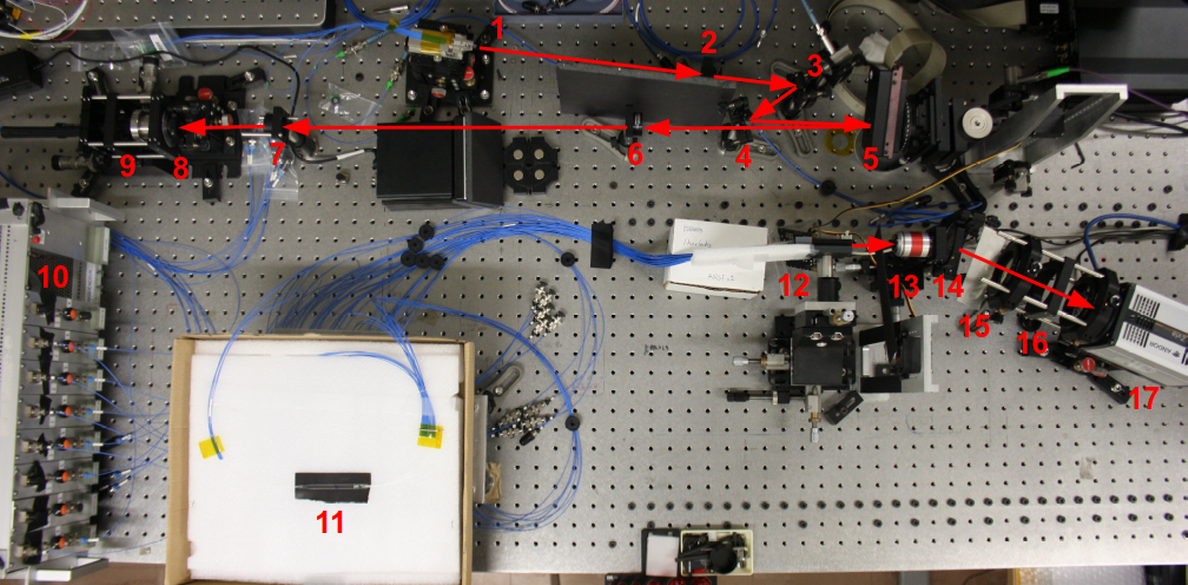}
    \caption[Photographie du banc de test FIRSTv2 à Meudon.]{Photographie du banc de test FIRSTv2 à Meudon. Tous les composants sont numérotés comme suit : (1) le V-Groove d'injection de la source, (2) une lentille, (3) un miroir plan, (4) un miroir D, (5) le MEMS, (6) et (7) le doublet de lentille, (8) la matrice de micro-lentilles, (9) le toron de fibres optiques, (10) les lignes à retard (ODL), (11) la puce d'optique intégrée, (12) un V-Groove, (13) un objectif de microscope, (14) le réseau holographique, (15) une plateforme pour disposer un prisme de Wollaston, (16) un doublet de lentille et (17) la caméra d'imagerie.}
    \label{fig:FIRSTv2BenchPhoto}
\end{figure}

%%%%%%%%%%%%%%%%
\subsection{Le miroir déformable Iris AO}

Le miroir déformable installé sur le banc de test de \ac{FIRSTv2} est fabriqué par \textit{Iris AO Inc} et je le nommerai par la suite \ac{MEMS}. La figure~\ref{fig:IrisAOMapA} est une photographie du miroir, sur laquelle on voit un des segments qui est coincé, les électrodes de lecture tout autour du miroir et la fenêtre d'ouverture noire. Le miroir dispose de $37$ segments hexagonaux identifiés sur la carte présentée sur la figure~\ref{fig:IrisAOMapB} (la carte est tournée de $+60\degree$ par rapport à la photographie) : le segment gris $17$ est le segment coincé inutilisable et les segments oranges sont ceux devant lesquels est placée une fibre optique du toron (voir plus de détails pour cela dans la section~\ref{sec:FiberInjection}). Lors du choix des cinq sous-pupilles à injecter dans la puce d'optique intégrée, les cinq segments correspondant ne peuvent donc être choisis que parmi ces $10$ segments oranges. Enfin, le cercle circonscrit de chaque segment mesure $700 \,$\um~de diamètre et celui de l'ensemble du pavage mesure $4,2 \,$mm de diamètre. Tous les segments, excepté le segment $17$, sont contrôlables en piston sur une plage de $\pm 1,5 \,$\um~et en tip-tilt sur une plage de $\pm 2\,$mrad.

\begin{figure}[ht!]
    \centering
    \begin{subfigure}[t]{0.4\textwidth}
        \centering
        \includegraphics[width=0.8\textwidth]{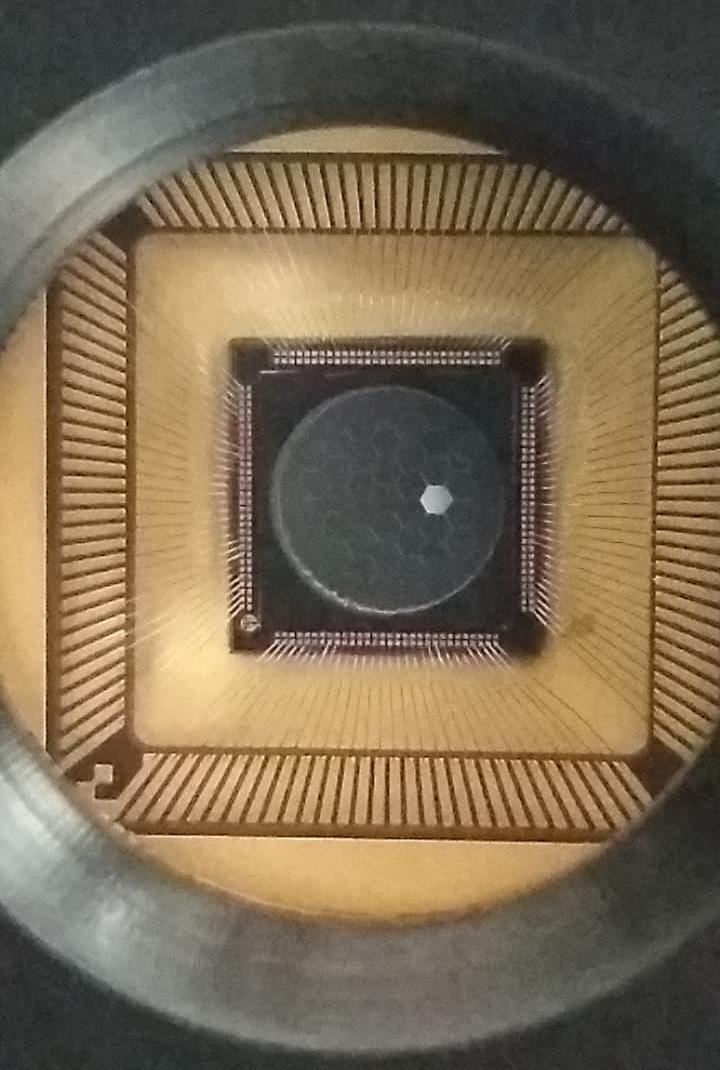}
        \caption{Photographie de la surface du miroir segmenté.}
        \label{fig:IrisAOMapA}
    \end{subfigure}\hspace{0.01\textwidth}
    \begin{subfigure}[t]{0.5\textwidth}
        \centering
        \includegraphics[width=0.8\textwidth]{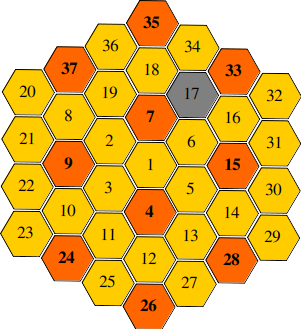}
        \caption{Carte d'identification des segments du miroir : en orange sont les segments devant lesquels est disposée une fibre optique et en gris est le segment cassé.}
        \label{fig:IrisAOMapB}
    \end{subfigure}
    \caption[Pavage du miroir segmenté \textit{Iris AO}.]{Pavage du miroir segmenté \textit{Iris AO}.}
    \label{fig:IrisAOMap}
\end{figure}

%%%%%%%%%%%%%%%%
\subsection{L'injection dans les fibres optiques}
\label{sec:FiberInjection}

%%%%%%%%
\subsubsection{Mise en oeuvre}

Les faisceaux des sous-pupilles sont injectés dans des fibres optiques monomodes à maintien de polarisation de type panda, de la gamme PM630-HP, fabriquées par \textit{Thorlabs}. Les fibres ont une ouverture numérique égale à $0,12$, un diamètre de mode égal à $4,5 \pm 0,5 \,$\um~(à $630 \,$nm) et leur cœur a un diamètre égal à $3,5 \,$\um. La propriété monomode de ces fibres permet de filtrer les aberrations du front d'onde du faisceau injecté, convertissant ainsi les variations de phases en variations d'intensité. Un ensemble de $19$ fibres est disposé dans un toron fabriqué par \textit{Fiberguide Industries}\footnote{\url{https://www.molex.com/molex/products/group/fiberguide}}, dont la photographie est montrée sur la figure~\ref{fig:InjectionCompB}, selon un motif hexagonal avec un espacement de $500 \,$\um~schématisé sur la figure~\ref{fig:InjectionCompC}. Cela explique la disposition des segments utiles (orange) sur la figure~\ref{fig:IrisAOMapB}.

\begin{figure}[ht!]
    \centering
    \begin{subfigure}{0.52\textwidth}
        \centering
        \includegraphics[width=0.9\textwidth]{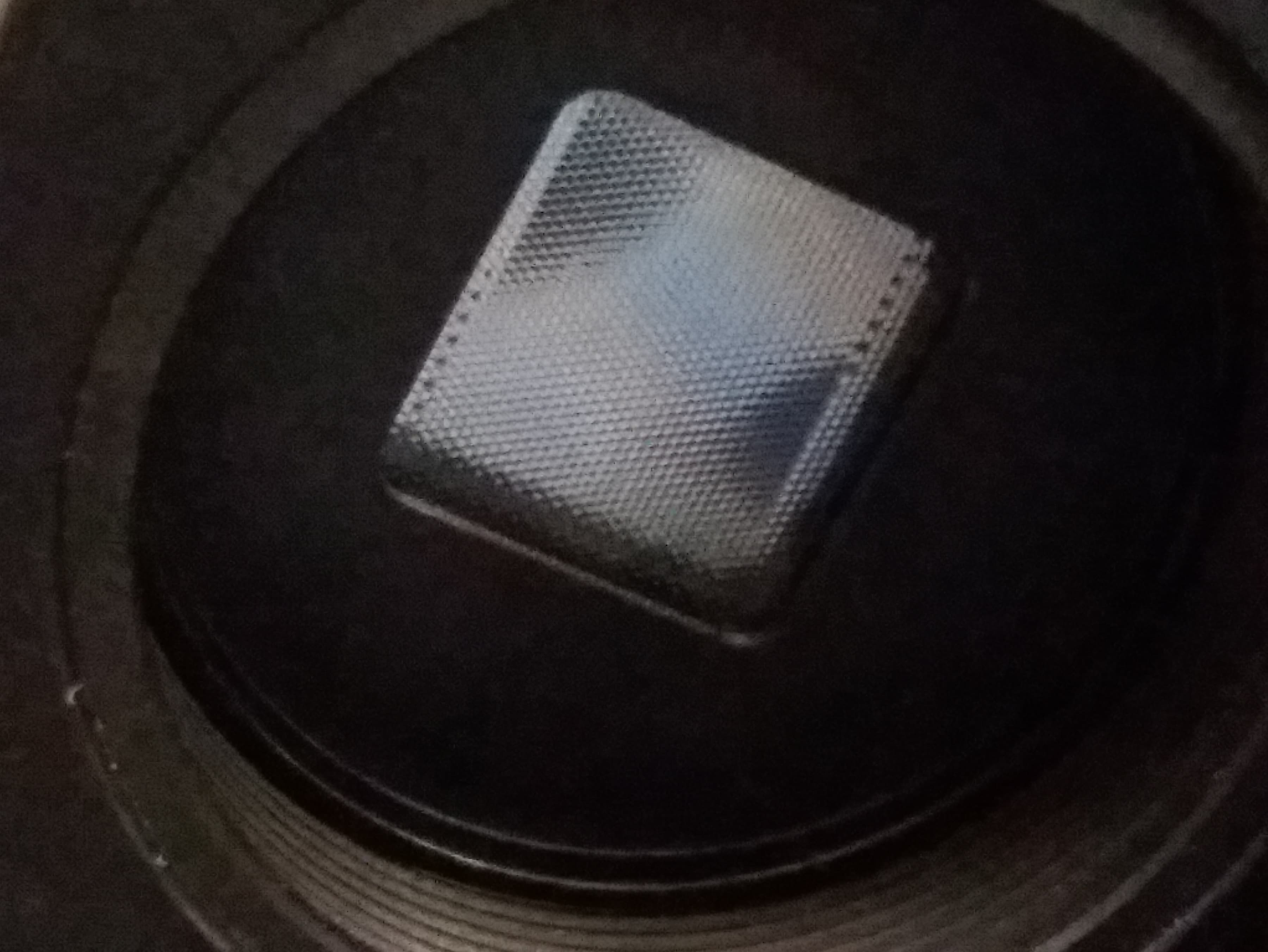}
        \caption{Photographie de la matrice de micro-lentilles utilisée pour diviser la pupille en sous-pupilles et focaliser les faisceaux dans les fibres optiques.}
        \label{fig:InjectionCompA}
    \end{subfigure}
    \begin{subfigure}[t]{0.49\textwidth}
        \centering
        \includegraphics[width=0.8\textwidth]{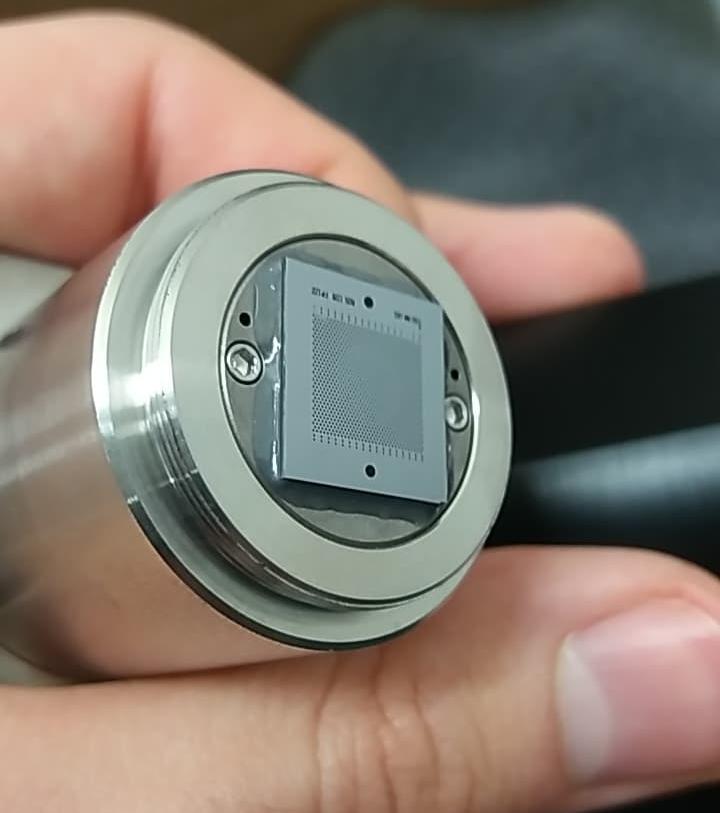}
        \caption{Photographie du toron de $19$ fibres optiques utilisé pour l'injection des faisceaux des sous-pupilles.}
        \label{fig:InjectionCompB}
    \end{subfigure}\hfill
    \begin{subfigure}[t]{0.49\textwidth}
        \centering
        \includegraphics[width=0.8\textwidth]{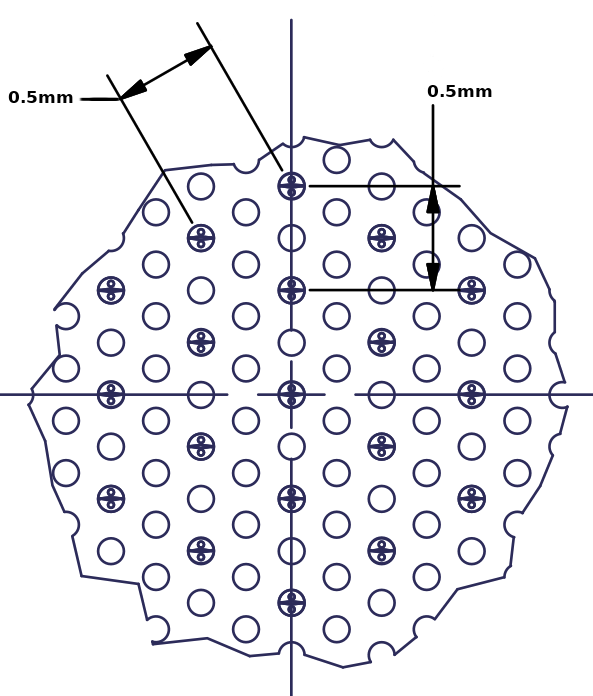}
        \caption{Schéma de la disposition des entrées des fibres optiques du toron. Crédit : \textit{Fiberguide Industries}.}
        \label{fig:InjectionCompC}
    \end{subfigure}
    \caption[Matrice de micro-lentilles et toron de fibres utilisés pour l'injection des faisceaux dans les fibres optiques.]{Matrice de micro-lentilles et toron de fibres utilisés pour l'injection des faisceaux dans les fibres optiques.}
    \label{fig:InjectionComp}
\end{figure}

Une matrice de micro-lentille, dont une photographie est présentée sur la figure~\ref{fig:InjectionCompA}, est disposé à $300 \,$\um~du toron. Les micro-lentilles ont une focale de $\sim 1 \,$mm et ont pour rôle de sous-diviser la pupille (masquage de pupille) et de focaliser les faisceaux des sous-pupilles dans les fibres optiques du toron. La matrice est placée dans le plan pupille conjugué du plan dans lequel est placé le miroir segmenté de telle sorte qu'une lentille, donc par extension une fibre optique, est alignée devant chaque segment. Les micro-lentilles ont un diamètre de $250 \,$\um~et ont donc un pas qui est la moitié du pas des fibres du toron. Un doublet de lentille de focales égales à $300 \,$mm et à $125 \,$mm adapte la taille des faisceaux provenant des segments du \ac{MEMS} de diamètre égal à $606,2 \,$\um~à la taille des micro-lentilles, par un grandissement d'un facteur $\sim 0,42$. Enfin, le faisceau de la source est en incidence quasi-normale sur le \ac{MEMS} pour déformer les sous-faisceaux le moins possible et maximiser la surface utile.

%%%%%%%%
\subsubsection{Procédure d'alignement}

Ce trio de composants nécessite d'être précisément aligné et toute la procédure qui suit est effectuée régulièrement car les dilatations mécaniques dues aux variations de température induisent un désalignement. Le toron de fibres est d'abord retiré du foyer des micro-lentilles et une caméra est installée à la place afin d'imager les segments du miroir à travers les micro-lentilles. Une telle image est montrée sur la figure~\ref{fig:MLAalignment} où l'on voit les faisceaux des sous-pupilles défocalisés à travers les micro-lentilles, elles-mêmes entourées des bords des segments hexagonaux. Le but est d'aligner les cercles au centre des hexagones et les lignes jaunes sont tracées sur l'image pour effectuer cet alignement sur toutes les micro-lentilles du champ à la fois.

\begin{figure}[ht!]
    \centering
    \includegraphics[width=0.7\textwidth]{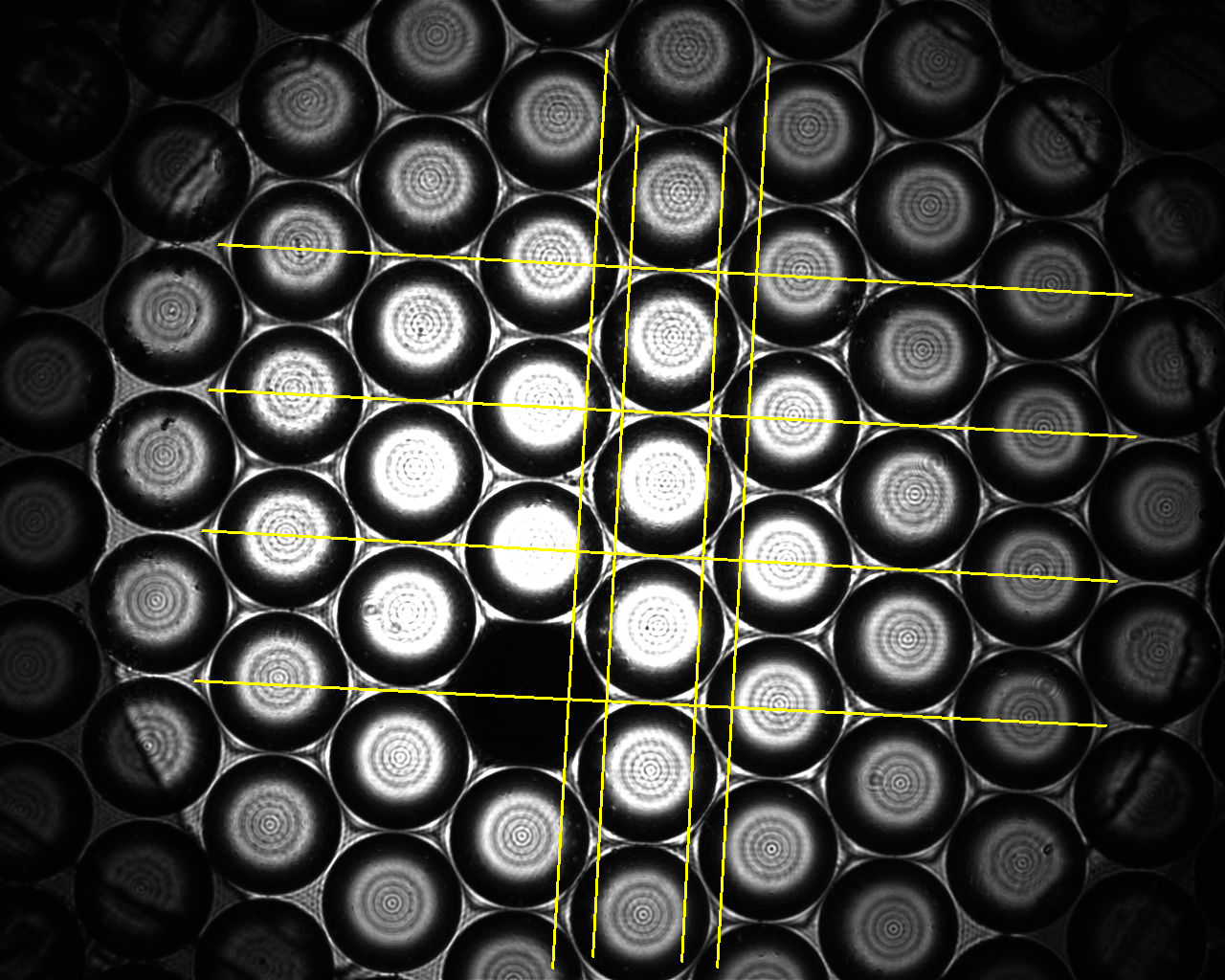}
    \caption[Image du plan pupille après la matrice de micro-lentilles, prise lors de leur alignement par rapport au MEMS.]{Image du plan pupille après la matrice de micro-lentilles où les faisceaux des sous-pupilles sont défocalisés pour aligner les micro-lentilles par rapport aux segments hexagonaux du MEMS. On peut voir les hexagones entourant les micro-lentilles et les traits jaunes sont tracés pendant l'alignement pour aligner toutes les micro-lentilles du champ en même temps.}
    \label{fig:MLAalignment}
\end{figure}

Ensuite, la caméra est retirée et le toron est remis devant la matrice de micro-lentilles. Une source lumineuse est rétro-injectée dans les fibres optiques du toron et le faisceau réfléchi par le miroir segmenté est projeté sur un écran. Les segments sont déplacés en tip-tilt les uns après les autres afin d'identifier la position de la fibre illuminée par rapport au \ac{MEMS} et le toron est déplacé sur ses axes X et Y. Pour finir, la bonne position du toron sur son axe Z est trouvée en focalisant l'image par une des micro-lentilles de la source rétro-injectée.

%%%%%%%%
\subsubsection{Optimisation de l'injection}
\label{sec:OptiInj}

Une procédure programmée dans le logiciel de contrôle du banc permet d'optimiser l'injection des faisceaux des sous-pupilles dans les fibres optiques, en contrôlant les segments du \ac{MEMS} en synchronisation avec la caméra. Pour ce faire, les segments sont un à un déplacés en tip et en tilt sur un quadrillage dont la taille et le pas sont donnés en paramètres d'entrée de la procédure. Chaque segment quadrille la zone pendant que les autres sont mis en bout de course (à $2 \,$mrad) pour ne pas injecter de lumière. Pour chaque nouvelle position du segment, une image est prise sur la caméra et le flux total sur les sorties illuminées est calculé. Une carte de ces valeurs de flux est construite en fonction des positions de tip et de tilt pour déduire les coordonnées pour lesquelles le maximum de flux est mesuré, à l'aide d'un ajustement d'une fonction gaussienne. 

La figure~\ref{fig:OptiInj} est une capture d'écran de l'ordinateur de contrôle du banc de test et montre le résultat d'une optimisation de l'injection. En haut à gauche on peut voir les cinq cartes de transmission des cinq fibres tracées côte à côte, pour un intervalle de tip et de tilt de $\pm 2 \,$mrad avec un pas de $0.25 \,$mrad. Chacune de ces cartes présente un profil quasi-gaussien, correspondant à la convolution de la fonction d'étalement du point (\ac{PSF}) du faisceau injecté par le mode de la fibre optique. Une fonction gaussienne est alors ajustée à ces cartes pour inférer les positions qui maximisent les flux et sont enregistrées afin d'être ré-utilisées par la suite pour optimiser le flux injecté avant chaque nouvelle prise de données. En haut à droite de l'écran est affichée la carte de phase du \ac{MEMS} avec les cinq segments sur les positions qui optimisent le flux. En bas à droite est montrée l'image en temps réel de la caméra, sur laquelle on peut voir des franges d'interférences sur dix bases (axe vertical), dispersées horizontalement et dont les flux sont égalisés. On peut voir en arrière plan les terminaux de commandes des différents composants (voir la section~\ref{sec:ControlSoftware} pour plus de détails).

\begin{figure}[ht!]
    \centering
    \includegraphics[width=\figwidth]{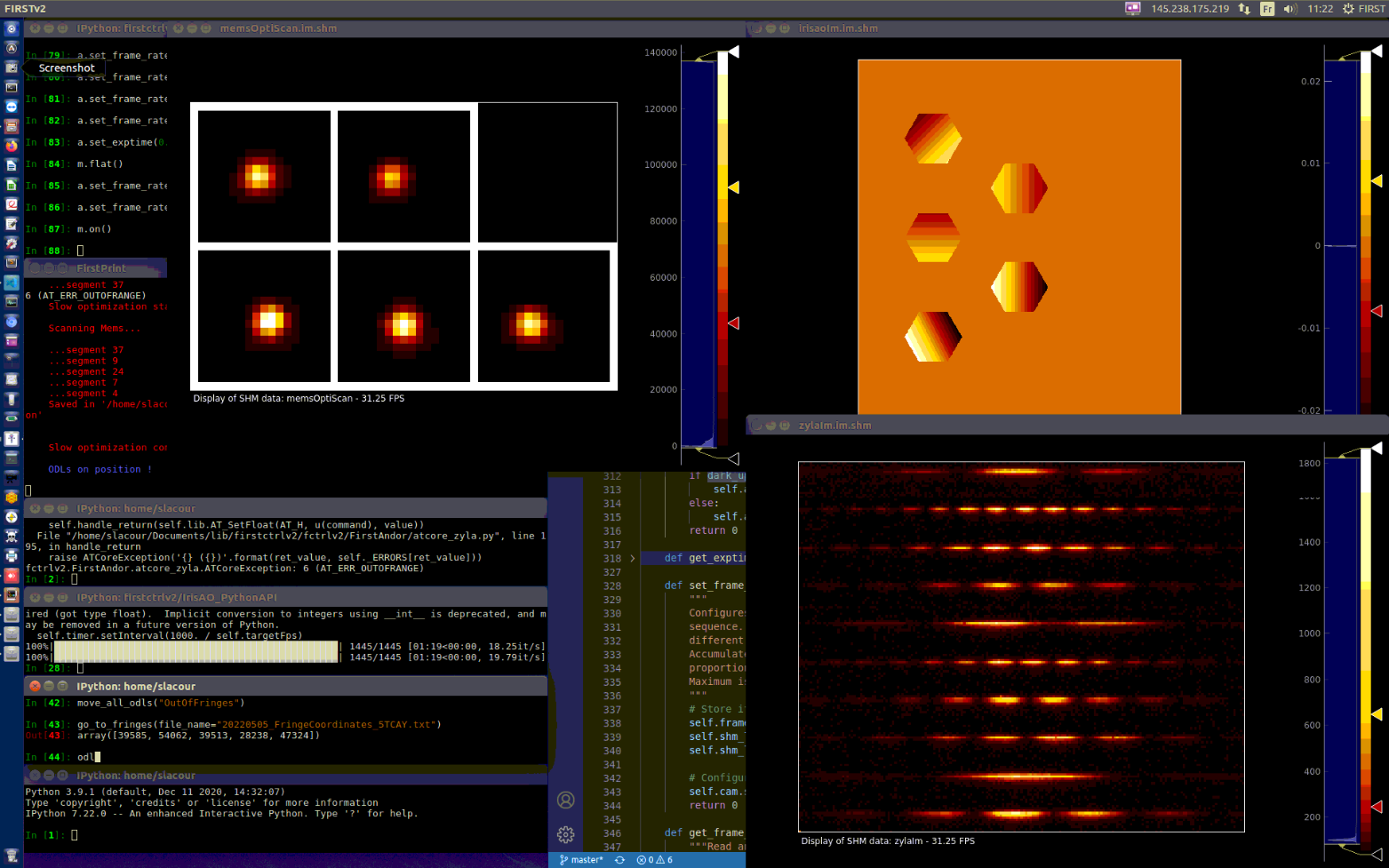}
    \caption[Capture d'écran de l'ordinateur de contrôle de FIRSTv2 montrant une optimisation de l'injection et les interférogrammes.]{Capture d'écran de l'ordinateur de contrôle de FIRSTv2. En haut à gauche : les cinq cartes de transmission des fibres, après le quadrillage par les segments du MEMS sur un intervalle de tip-tilt de $\pm 2 \,$mrad avec un pas $0.25 \,$mrad. En haut à droite : la carte de phase du MEMS avec les cinq segments en question en position d'injection du flux maximale. En bas à droite : l'image en direct de la caméra avec les dix sorties de la puce Y, présentant des franges, illuminées par une source SLED à $650 \,$nm.}
    \label{fig:OptiInj}
\end{figure}

%%%%%%%%%%%%%%%%
\subsection{Les lignes à retard}
\label{sec:InstruODL}

%%%%%%%%
\subsubsection{Concept}

Les lignes à retard (ou \acrfull{ODL}) sont cruciales pour compenser les longueurs différentes entre les fibres optiques utilisées. En effet, il est nécessaire que la différence de marche (ou \acrfull{OPD}) entre toutes les fibres optiques soit nulle pour obtenir la frange centrale des interférogrammes imagés sur la caméra. La figure~\ref{fig:ODLScheme} montre le schéma de principe d'une ligne à retard, fabriquée par \textit{Oz Optics}. Le faisceau d'une fibre d'entrée, en haut à gauche, est collimaté par une lentille (nommé \textit{lenses} sur le schéma) sur un prisme utilisé en réflexion totale (équivalent à $2$ miroirs), à droite. Ce prisme induit un aller-retour au faisceau qui est focalisé par une autre lentille dans une fibre de sortie (qui est ensuite branché à une fibre d'entrée de la puce photonique), en bas à gauche du schéma. Le prisme est monté sur un moteur contrôlable en position depuis le logiciel de contrôle du banc, sur un intervalle de $25 \,$mm, ce qui résulte en une distance parcourue de $50 \,$mm par le faisceau. Cet intervalle est subdivisé en $100\,000$ pas de $500 \,$nm.

\begin{figure}[ht!]
    \centering
    \includegraphics[width=\figwidth]{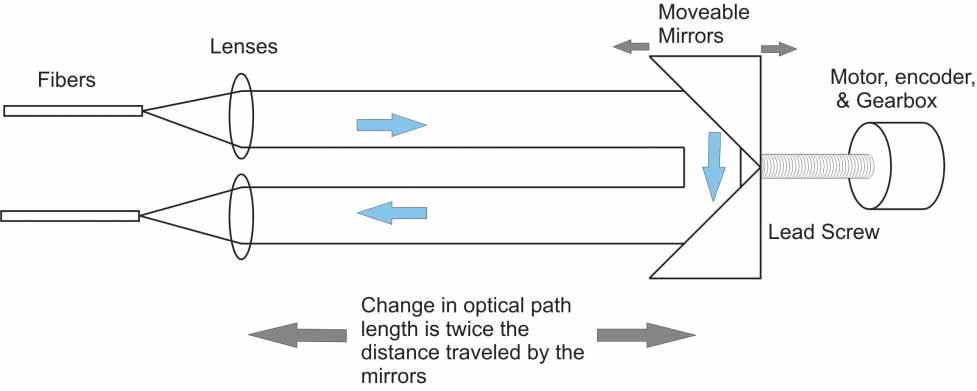}
    \caption[Schéma de fonctionnement d'une ligne à retard.]{Schéma de fonctionnement d'une ligne à retard. Le faisceau est injecté par une fibre en haut à gauche, est collimaté par une lentille (nommé \textit{lenses}) avant d'être réfléchis deux fois par un prisme, à droite, pour faire demi-tour vers la fibre de sortie en bas à gauche. Le prisme est monté sur un moteur permettant de contrôler sa position sur un intervalle de $50\,$mm avec un pas de $500 \,$nm. Crédit : \textit{Oz Optics}.}
    \label{fig:ODLScheme}
\end{figure}

%%%%%%%%
\subsubsection{Transmission}
\label{sec:OdlThroughput}

Au cours de ma thèse, on s'est rendu compte que les \ac{ODL}s transmettaient peu de flux lumineux. En effet, lors de l'intégration anticipée de cinq nouvelles \ac{ODL}s sur l'instrument \ac{FIRSTv1}, la différence d'intensité lumineuse mesurée sur la caméra a été flagrante. Les transmissions les plus basses de celles-ci ont été mesurées à $20 - 30 \%$ alors que la spécification constructeur était une perte de $1,2\,$dB maximum à $650 \,$nm, correspondant à une valeur de transmission de $75\%$. Avec Elsa Huby et Manon Lallement, on a mesuré des transmissions entre $30\%$ et $80\%$ à $675 \,$nm. La figure~\ref{fig:ODLThroughput} présente les mesures de transmission des \ac{ODL}s (exceptée l'\ac{ODL} $\#2$) en fonction de la longueur d'onde à l'aide d'un spectrographe fibré. Pour chaque \ac{ODL}, ces mesures sont effectuées en injectant la lumière soit dans la fibre d'entrée (\textit{In Out}) soit dans la fibre de sortie (\textit{Out In}) pour évaluer si on peut obtenir de meilleures transmissions en injectant dans le sens opposé. On remarque premièrement que la transmission ne dépend pas significativement du sens de propagation dans l'\ac{ODL}. Deuxièmement, la transmission est chromatique (à cause des lentilles) et atteint un maximum vers $675 \,$nm. Les transmissions les plus basses sont celles des lignes à retard $\#3$ et $\#4$. De plus, nous ne disposons pas de la mesure de transmission de l'\ac{ODL} $\#2$ car elle avait été expédiée au fabriquant pour investigation sur la très mauvaise transmission (mesurée à $12\%$ au laboratoire). Il a été diagnostiqué que les fibres d'entrée et de sortie étaient la cause de la faible transmission car elles ont été trouvées endommagées.

\begin{figure}[ht!]
    \centering
    \includegraphics[width=\figwidth]{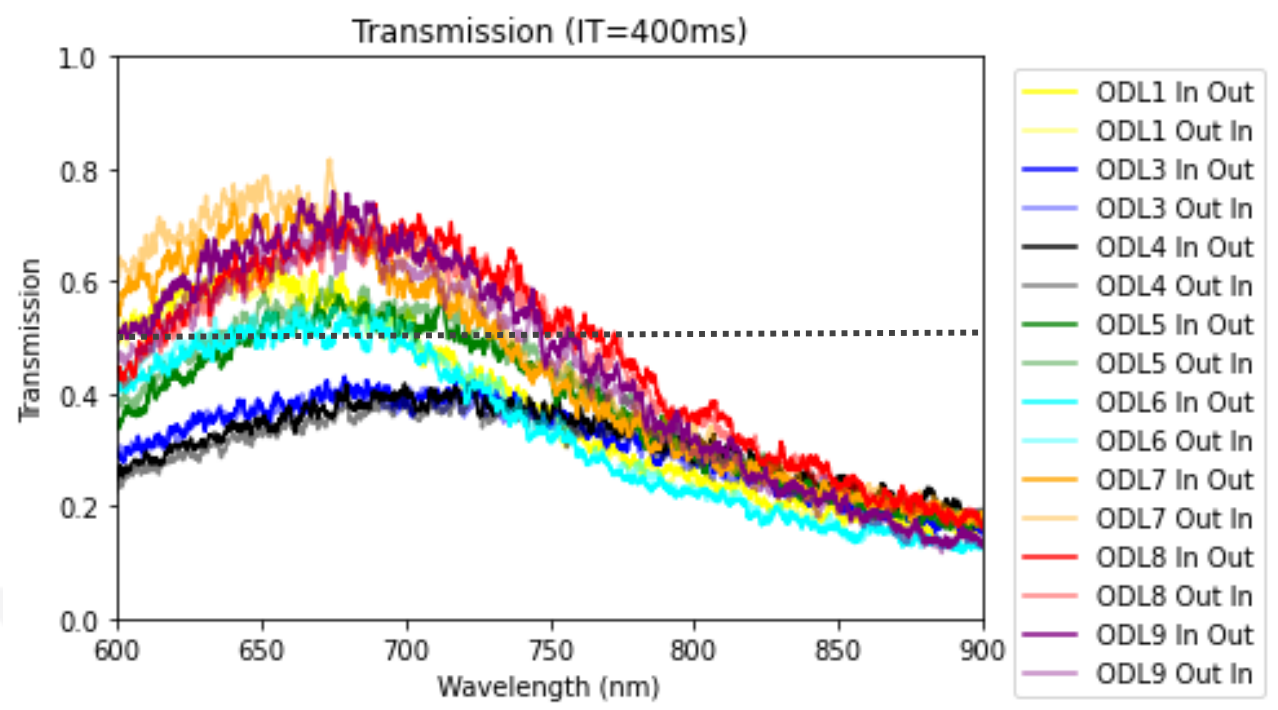}
    \caption[Transmissions des ODLs du banc de test FIRSTv2 mesurées en fonction de la longueur d'onde.]{Transmissions des ODLs du banc de test FIRSTv2 mesurées en fonction de la longueur d'onde à l'aide d'un spectrographe fibré. Pour chaque ODL, la transmission est mesurée en injectant la lumière dans la fibre d'entrée (courbe en couleur foncée) et en injectant la lumière dans la fibre de sortie (en couleur claire). Crédit : Elsa Huby et Manon Lallement.}
    \label{fig:ODLThroughput}
\end{figure}

Nous avons finalement décidé d'en ouvrir une afin d'investiguer par nous-même la cause du problème. La figure~\ref{fig:ODLOpened} présente une photographie de l'intérieur de l'\ac{ODL} $\#1$ après son ouverture au laboratoire. On peut y voir les deux fibres d'entrée (à gauche) et de sortie (à droite). Un laser rouge est injecté dans la fibre d'entrée afin de voir comment se propage le faisceau à travers le système et on peut voir la diffusion d'une partie de la lumière à travers la lentille de collimation et à travers la colle qui la maintient. Ce pourrait être à ce niveau là que la transmission est en grande partie dégradée. Enfin, on aperçoit le prisme collé sur une pièce noire vissée sur la monture motorisée, en bas de l'image et qui réfléchit le faisceau lumineux.

\begin{figure}[ht!]
    \centering
    \includegraphics[width=0.6\textwidth]{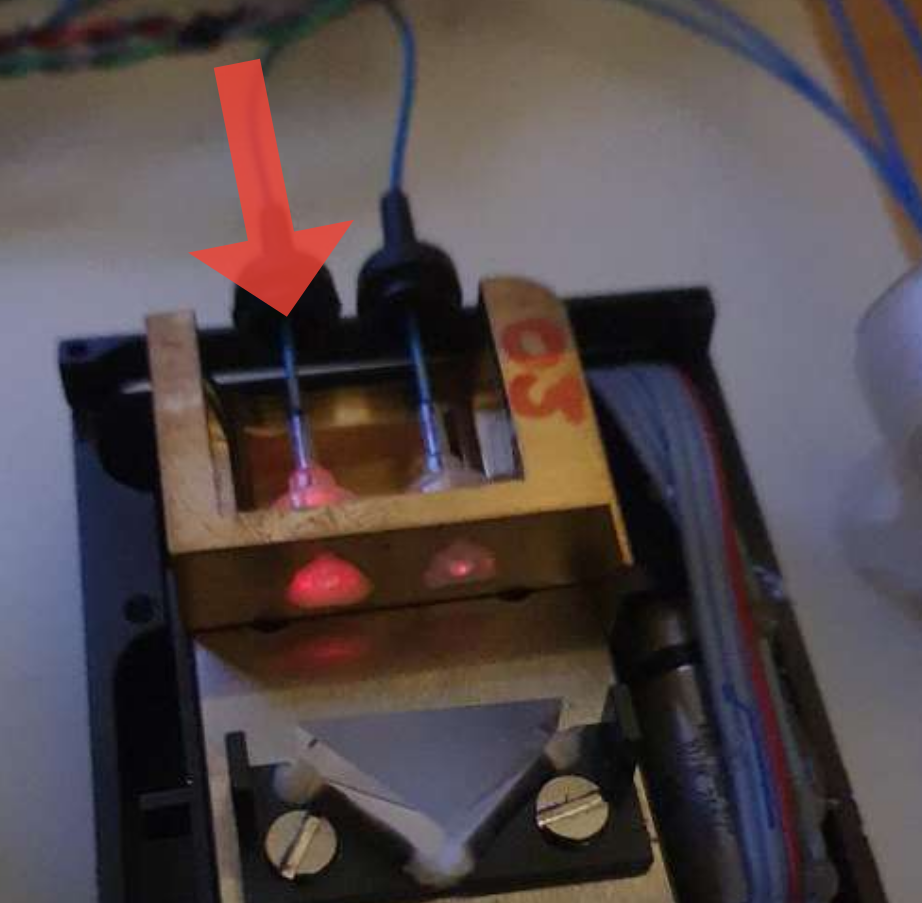}
    \caption[Photographie d'une ODL ouverte.]{Photographie de l'ODL $\#1$ ouverte. En haut, sont les fibres d'entrée (à gauche) et de sortie (à droite). La lumière d'un laser rouge est injectée dans la fibre d'entrée. Puis plus bas, on voit les lentilles de collimation des faisceaux ainsi que la colle qui les maintient. Le prisme qui réfléchit le faisceau est en bas. Crédit : Elsa Huby et Manon Lallement.}
    \label{fig:ODLOpened}
\end{figure}

%%%%%%%%
\subsubsection{La recherche des franges}

J'ai programmé une procédure de recherche de franges dans le logiciel de contrôle du banc consistant à balayer les lignes à retard sur leur intervalle de position. Plus précisément, une des lignes à retard est laissée immobile à la position la plus basse pendant que toutes les autres \ac{ODL}s sont commandées en translation. Pendant que les \ac{ODL}s se déplacent, la variation de flux est inspectée sur les quatre sorties correspondant aux quatre bases formées par le faisceau passant par l'\ac{ODL} immobile et les quatre autres faisceaux des \ac{ODL}s en mouvement. Dès que les franges sont détectées sur une base, l'\ac{ODL} correspondante est arrêtée et sa position est enregistrée. Si les franges sur les quatre bases sont trouvées, les franges sur toutes les autres bases le sont aussi, nécessairement. Si à la fin du balayage des quatre \ac{ODL}s les franges de toutes les bases ne sont pas trouvées, la ligne à retard restée immobile est déplacée en bout de course (à la position $100\,000$) et le balayage recommence. A la fin de la procédure, lorsque les franges de toutes les bases sont trouvées, on affine les positions des \ac{ODL}s à la main, en regardant l'image en temps réel de la caméra afin de trouver la frange centrale avant de les enregistrer dans un fichier local à l'ordinateur de contrôle. Dans le cas où les franges ne sont pas trouvées sur toutes les bases, cela signifie que la différence de longueur entre les fibres excède $5 \,$cm. Il y a alors la possibilité d'utiliser d'autres \ac{ODL}s dont la longueur de fibre est différente (nous disposons en tout de neuf lignes à retard car à terme il est question de recombiner neuf sous-pupilles comme sur \ac{FIRSTv1}). Les fibres internes aux lignes à retard ont une longueur égale à $\sim 1 \,$m et les incertitudes sur ces longueurs ainsi que celles sur les longueurs des fibres du toron et du V-Groove d'entrée de la puce impliquent qu'il n'est parfois pas possible d'égaliser les longueurs de toutes les fibres du banc.

%%%%%%%%
\subsubsection{Discussions}
\label{sec:ODLDiscussions}

Étant donné que la transmission est le point critique à améliorer pour augmenter la sensibilité de l'instrument, il est essentiel d'identifier les sources de pertes de transmission. L'étude présentée dans la partie~\ref{sec:OdlThroughput}, nous montre que les lignes à retard présentent une perte de transmission considérable et de nouvelles solutions sont envisagées afin de les retirer de l'instrument.

% Alternative 1 : fibres de compensation
La solution qui sera probablement implémentée prochainement est de remplacer les lignes à retard par des fibres de compensation. C'est la solution choisie sur \ac{FIRSTv1} et présentée dans la thèse d'Elsa Huby \citep{huby2013these}. Cela demanderait de mesurer les longueurs des fibres du toron et des fibres d'entrée de la puce afin de les égaliser. La longueur de cohérence étant ici égale à $\text{l}_\text{c} = 2,21 \,$mm à $650 \,$nm, pour une résolution spectrale égale à $3\,400$ (voir la section~\ref{sec:InstruSpectro}), la précision sur l'égalisation des longueurs doit donc être de $1,1 \,$mm. C'est une charge conséquente de travail en plus mais qui serait justifiée par le gain conséquent en transmission.

% Alternative 2 : puce photonique 3D
Une autre solution serait de changer tout le système d'injection et deux concept sont envisagés. Premièrement, la matrice de micro-lentille pourrait être placée directement entre le \ac{MEMS} et la puce d'optique intégrée afin d'injecter les faisceaux des sous-pupilles directement dans cette dernière. Les guides d'ondes de la puce seraient alors gravées dans les trois dimensions de la puce (et non dans un plan comme c'est le cas sur les puces utilisées lors de cette thèse) afin de les placer devant les sous-pupilles choisies pour la recombinaison interférométrique et de les placer sur une ligne en sortie de la puce. En plus d'avoir l'avantage de se passer du toron de fibres et des lignes à retard, cela permet aussi de recombiner par paire tous les faisceaux sans aucun croisement des guides d'ondes, ce qui constitue une source de diaphotie (\textit{cross-talk} en anglais) et de perte de transmission (voir plus de détails dans la section~\ref{sec:ChipCharacterization}). Ce concept est déjà exploité dans l'instrument \ac{GLINT} \citep{martinod2021} et a été développé à l'\ac{IPAG} (voir son schéma de principe sur la figure~\ref{fig:Chip5T3D}) pour le projet \ac{FIRST} et testé sur le banc de test de \ac{FIRSTv2} \citep{martin2022a}.

\begin{figure}[ht!]
    \centering
    \includegraphics[width=0.8\textwidth]{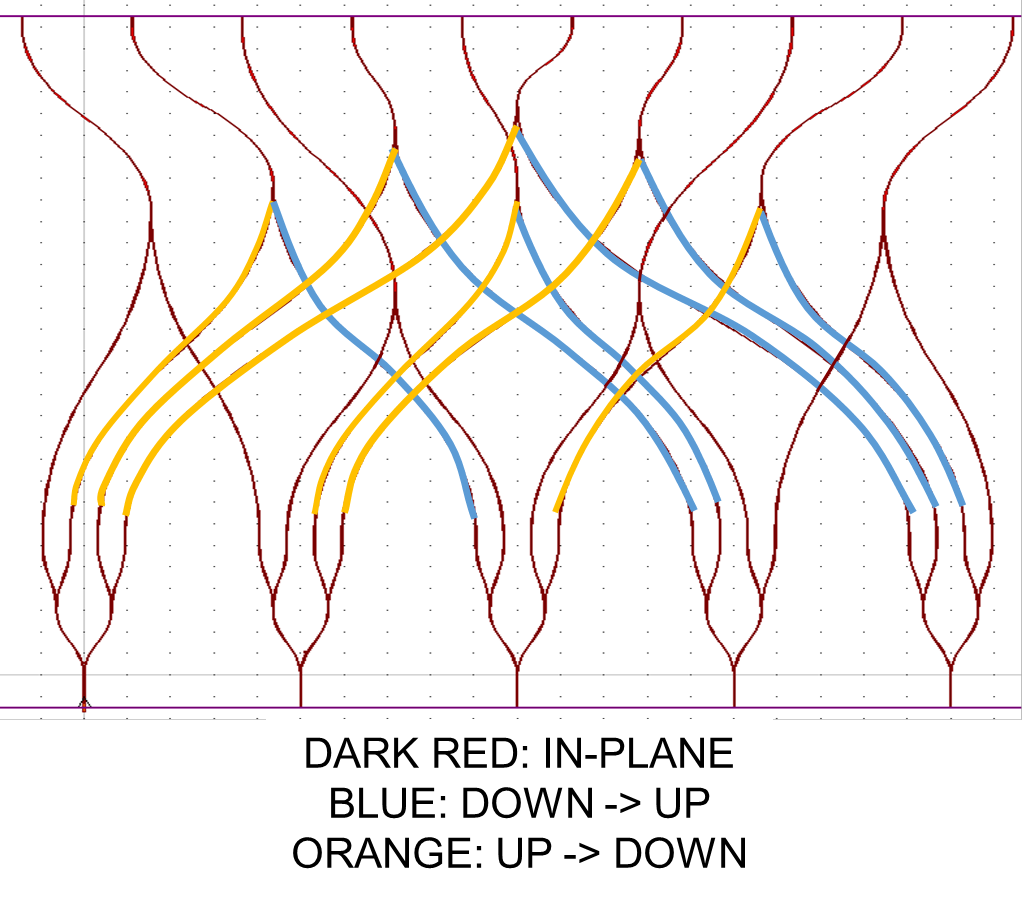}
    \caption[Schéma de principe du concept de puce photonique 3D.]{Schéma de principe du concept de puce photonique 3D. Les entrées sont en bas et les sorties en haut. Trois types de guides d'onde sont représentés : dans le plan (en rouge), qui monte puis descend (en orange) et qui descend puis monte (en bleu). Crédit : Guillermo Martin.}
    \label{fig:Chip5T3D}
\end{figure}

% Alternative 3 : photonic lantern
Deuxièmement, on envisage d'intégrer une lanterne photonique \citep{leonsaval2005} à la place des puces actuelles. Le principe de cette technologie est de convertir un faisceau injecté dans une fibre optique multi-mode en plusieurs faisceaux dans des fibres optiques mono-modes. En mesurant l'intensité en sortie des fibres optiques mono-modes, il est possible de remonter à la phase du front d'onde d'entrée. La relation entre les sorties et l'entrée d'une telle puce est non-linéaire et \cite{norris2020a} propose d'utiliser un réseau de neurones pour remonter aux aberrations du front d'onde d'entrée et utilise ce système comme senseur de front-d'onde pour l'optique adaptative. Des discussions ont lieu avec Barnaby Norris afin de dimensionner un tel composant travaillant dans le visible pour son application future dans le projet \ac{FIRST}. Cela permettrait de se passer de la matrice de micro-lentille, du toron de fibres et des lignes à retard (seule une lentille suffirait pour injecter le faisceau dans la fibre d'entrée de la lanterne photonique) et optimiserait l'utilisation de tout le flux lumineux injecté dans l'instrument puisque toute la pupille est injectée dans le composant.

%%%%%%%%%%%%%%%%
\subsection{Les composants d'optique intégrée}
\label{sec:PhotonicChip}
% Low bend radius ---> high refractive index difference
% IMEC material of Si3N4 core with SiO2 cladding but the waveguide has to be very small and difficult to use them in the visible
% The material thickness is such that at a 670nm wavelength the mode is well confined in the core. As a result the bend radius is small, at 20 microns.

% High bend radius ---> low refractive index difference
% Teem material of SiN core and SiO2 cladding
% LioniX is same material but deposited differently

% The bend radius decreases with the thickness of the waveguide

% A small bend radius allows us to use adiabatic bends which will further decrease the loss but that is the next level. An adiabatic bend (if you don't know) just means that we don't just use one bend radius but start high at the start of the bend and decrease until a bend radius minima at the middle before increasing it again. This increases the total length of the bend but removes any loss between the straight waveguides and the bend.

%%%%%%%%
\subsubsection{Concept}
\label{sec:ChipConcept}

Les puces d'optique intégrée (photonique) sont conçues à l'\ac{IPAG} et testées sur le banc \ac{FIRSTv2} \citep{martin2020, martin2022b, lallement2022} et sont fabriquées par \textit{Teem Photonics}\footnote{\url{https://www.teemphotonics.com}}. Elles consistent en un bloc de verre de quelques centimètres (voir la photographie de la puce $Y$ sur la figure~\ref{fig:ChipYPhoto}), utilisant la technologie IoNext par échange des ions $\text{K}_+ : \text{Na}_+$, dans lequel des guides d'onde sont fabriqués par des techniques classiques de photolithographie. Les fibres d'entrée et de sortie sont celles de V-Grooves collés au bloc de verre par le fabriquant et sont branchées sur les \ac{ODL}s et sur un autre V-Groove, respectivement. Cette technologie est très utilisée dans le domaine des télécommunications optiques dans l'infrarouge proche alors que les puces développées pour le projet \ac{FIRSTv2} ont été optimisées pour être transmissives dans la bande spectrales $\sim 600 - 800 \,$nm.

\begin{figure}[ht!]
    \centering
    \includegraphics[width=\figwidth]{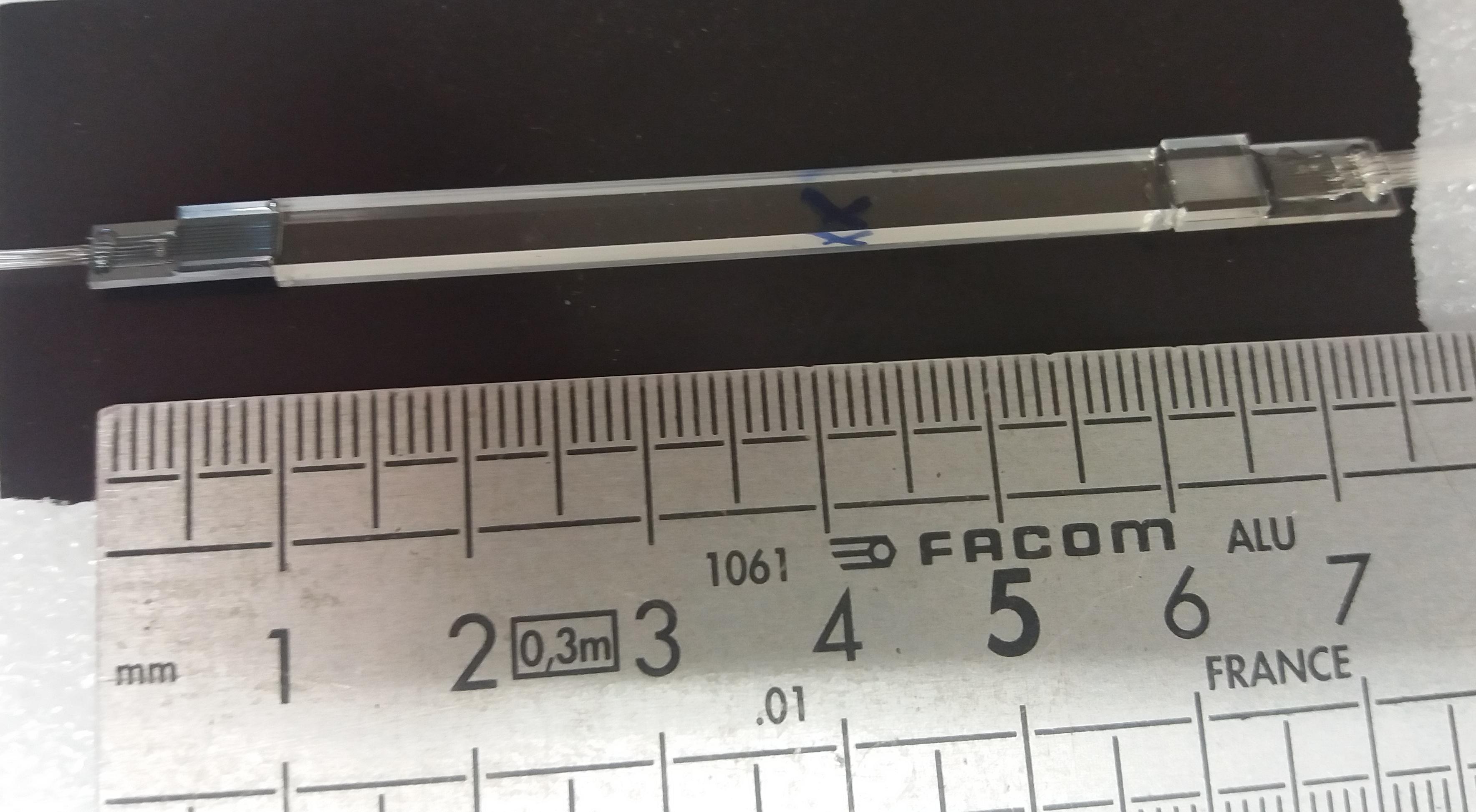}
    \caption[Photographie de la puce $Y$ testée sur le banc de test de FIRSTv2 à Meudon.]{Photographie de la puce $Y$ testée sur le banc de test de FIRSTv2 à Meudon.}
    \label{fig:ChipYPhoto}
\end{figure}

Durant ma thèse, j'ai testé \citep{barjot2020} sur \ac{FIRSTv2} deux technologies de puces photoniques nommées $X$ et $Y$, dont les schémas de principe sont montrés sur la figure~\ref{fig:ChipSchemes}, à gauche et à droite, respectivement. Les deux puces ont cinq guides d'onde en entrée (à gauche des schémas) qui sont chacun divisé en quatre (zone nommée \textit{splitting}). Chaque faisceau est recombiné par paires avec les quatre autres (zone nommée \textit{recombination}). Les faisceaux des cinq sous-pupilles du banc sont injectés dans ces entrées. La puce $X$ contient des coupleurs dit directionnels qui recombinent les faisceaux par couplage évanescent : il y a deux guides d'entrée et deux guides de sortie. La puce $Y$ recombine les faisceaux à l'aide de coupleurs $Y$ qui fusionne deux guides en un seul : il y a deux guides d'entrée et un guide de sortie. En sortie, dix bases ($5 \times 4 / 2$) sont ainsi formées. Dans le cas de la puce $X$ on dispose de deux sorties par base, du fait de la technique de couplage directionnel. On mesure alors le double de flux avec cette puce par rapport à la puce $Y$. Pour chaque base, les franges d'interférences obtenues sur les deux sorties ont un déphasage théorique de $\pi \,$rad.

\begin{figure}[ht!]
    \centering
    \begin{subfigure}{0.5\textwidth}
        \centering
        \includegraphics[width=\textwidth]{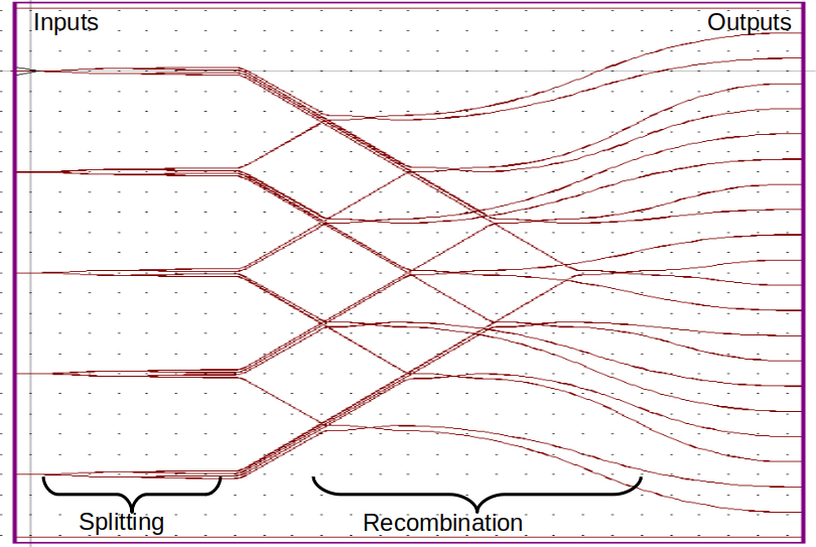}
    \end{subfigure}%
    \begin{subfigure}{0.5\textwidth}
        \centering
        \includegraphics[width=\textwidth]{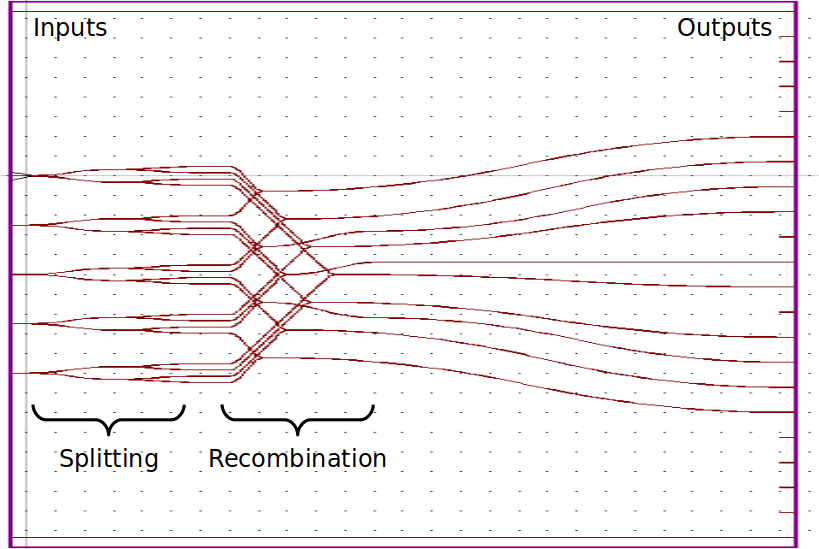}
    \end{subfigure}
    \caption[Schémas de principe des puces photoniques $X$ et $Y$.]{Schémas de principe des puces photoniques $X$ à gauche et $Y$ à droite. Les lignes représentent les guides d'onde dont les entrées sont à gauche et les sorties sont à droite. De la gauche vers la droite, les deux puces ont cinq guides d'onde en entrée qui sont d'abord divisés en quatre (zone nommée \textit{splitting}), afin d'être (plus loin) recombinés par paires avec les quatre autres (zone nommée \textit{recombination}). La puce $X$ a deux fois plus de fibres de sortie que la puce $Y$ du fait de la technique de couplage directionnel. Crédit : Guillermo Martin.}
    \label{fig:ChipSchemes}
\end{figure}

%%%%%%%%
\subsubsection{Caractérisation}
\label{sec:ChipCharacterization}

Pour caractériser les deux puces photoniques, j'ai utilisé une source large bande halogène de la gamme \textit{HL-2000-FHSA-HP}, fabriquée par \textit{Ocean Insight}\footnote{\url{https://www.oceaninsight.com/}}, lumineuse dans la gamme spectrale $360 - 2400 \,$nm. Cette caractérisation est publiée dans \cite{barjot2020} (voir la section~\ref{sec:SPIEproceeding}) et je la rappelle ici. Trois études sont conduites sur les deux puces :

\begin{enumerate}
    \item la diaphotie (\textit{cross-talk} en anglais), qui caractérise la quantité de fuite du signal entre les différents guides d'onde
    \item la transmission
    \item le contraste interférométrique
\end{enumerate}

% Caractérisation du cross-talk
\threesubsection{Diaphotie (\textit{cross-talk})}

Pour estimer la diaphotie (\textit{cross-talk} en anglais), cinq images de la caméra sont acquises successivement en illuminant les cinq entrées des puces photoniques, sans passer par le reste du banc de test en amont. La figure~\ref{fig:ChipCrossTalk} présente les cinq histogrammes correspondant à l'intensité lumineuse mesurée sur toutes les sorties de la puce $X$ en haut et de la puce $Y$ en bas. On s'attend à mesurer une intensité lumineuse sur $4$ ou $8$ sorties (représentées par les barres de couleur bleue) pour les puces $Y$ et $X$, respectivement et une intensité nulle sur les sorties restantes (représentées par les barres de couleur rouge). Ainsi, le flux mesuré sur les sorties en rouge est l'estimation de la diaphotie. On estime ainsi que celle-ci est en moyenne égal à $\sim 1\%$ pour les deux puces et est dans le pire des cas égal à $10 \%$ et à $20 \%$ en moyenne pour la puce $X$ et pour la puce $Y$, respectivement. Notre objectif est d'abaisser ce niveau à moins de $1\%$ (dans le pire des cas) et de nouvelles puces utilisant d'autre technologies et d'autres matériaux sont en cours de développement et de test par l'équipe travaillant à Grenoble, dans ce but.

La diaphotie intervient ici de deux façons différentes. Premièrement, la lumière fuit au niveau des croisements entre les guides et la fuite est d'autant plus importante que l'angle au croisement est petit \citep{labeye2008}. En effet, les niveaux de diaphotie les plus élevés sur la figure~\ref{fig:ChipCrossTalk} correspondent aux guides d'onde qui croisent les guides dans lesquels la lumière est injectée. Par exemple, sur le graphique du haut (de la puce $X$), dans l'histogramme du cas où la lumière est injectée dans l'entrée $1$, ce sont les sorties $5$, $6$ et $7$ qui sont concernées par ce phénomène. La difficulté dans la conception des puces réside dans le fait que la courbure des guides d'onde est limitée (la fuite de la lumière en dehors des guides augmente avec la courbure des guides) car ces puces sont basées sur des guides à faible contraste d'indice ($10^{-3} - 10^{-4}$). Cela ne permet donc pas de croiser les guides avec des angles plus grand que $5\degree$, non atteignables avec le faible confinement obtenu (c'est l'ordre de grandeur des angles des croisements utilisés dans les puces testées sur \ac{FIRSTv2}). Il est possible d'utiliser des matériaux différents dans le but de diminuer les rayons de courbure des guides d'onde. Il s'agit pour cela de choisir les matériaux du coeur et de la gaine de sorte que la différence de leurs indices de réfraction soit plus élevée. Les inconvénients de cette solution sont que ces puces sont moins transmissives dans le visible et que les coeurs doivent être plus petits : typiquement de l'ordre de quelques centaines de nanomètres pour des guides avec des contrastes d'indice de $0,1 - 0,5$. Cela augmente la difficulté de couplage entre les modes de petite taille des guides de la puce avec les modes des fibres optiques standard. Deuxièmement, une partie du flux est diffusée dans toute la puce à l'extérieur des guides, au niveau de l'injection de la lumière en entrée de la puce, due aux erreurs de couplage entre les modes d'injection et les modes guidés. Ce phénomène peut se mesurer sur les sorties qui ne sont pas en aval de croisements avec les guides dont la lumière est injectée. Sur les mesures de caractérisation, le niveau de diaphotie sur ces sorties est mesuré $10 \times$ inférieur que les autres sorties (en aval de croisement). Par exemple sur le graphique du haut de la figure~\ref{fig:ChipCrossTalk}, lorsque l'entrée $1$ est illuminée, les sorties dont le niveau de diaphotie est concerné par ce phénomène sont celles numérotées de $12$ à $20$. De plus, on remarque que ce niveau décroît avec la distance entre la sortie et l'entrée dans laquelle la lumière est injectée. On retrouve cet effet sur tous les histogrammes des deux graphiques.

\begin{figure}[ht!]
    \centering
    \begin{subfigure}{0.9\textwidth}
        \centering
        \includegraphics[width=\textwidth]{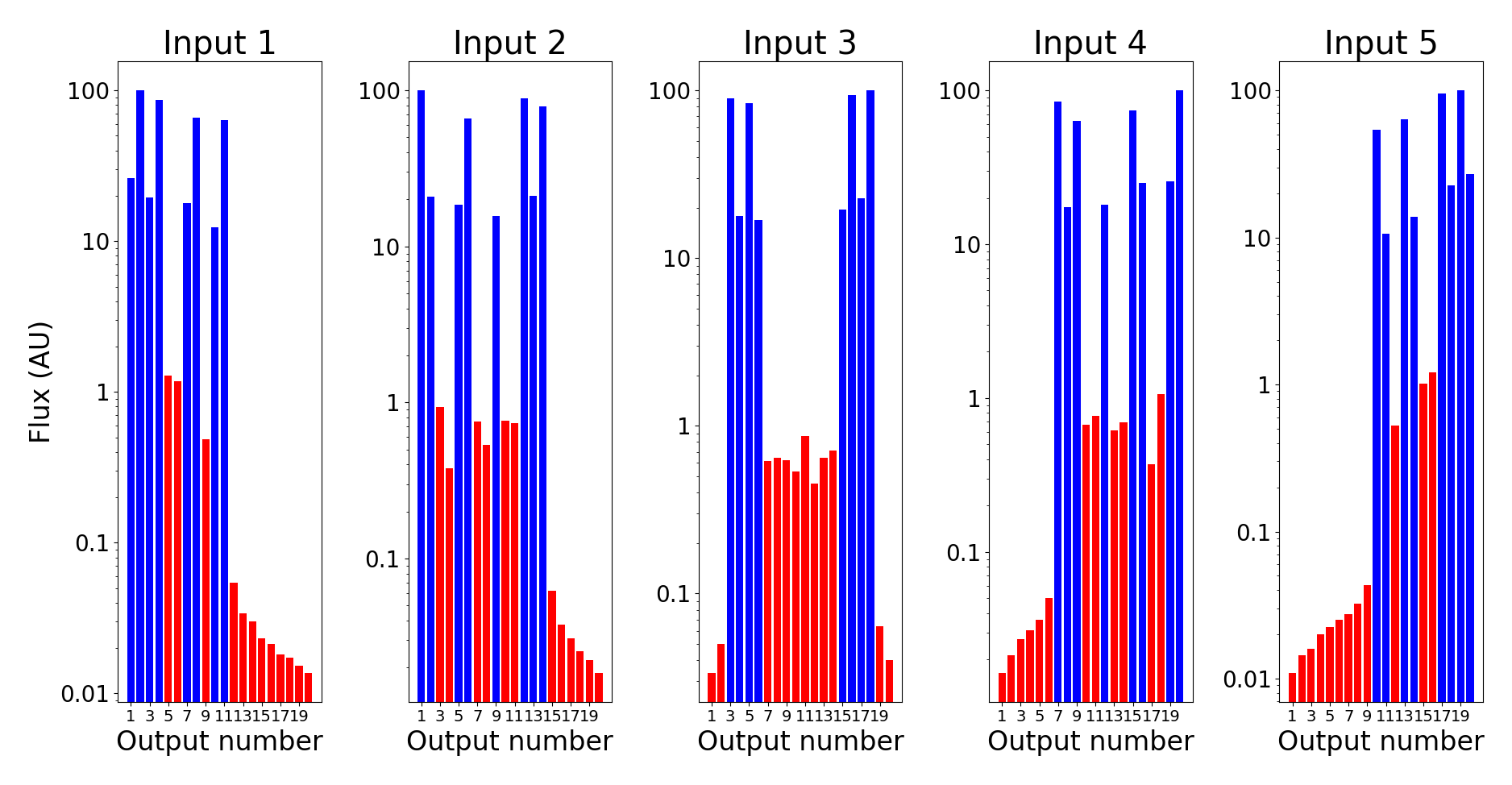}
    \end{subfigure}
    \begin{subfigure}{0.9\textwidth}
        \centering
        \includegraphics[width=\textwidth]{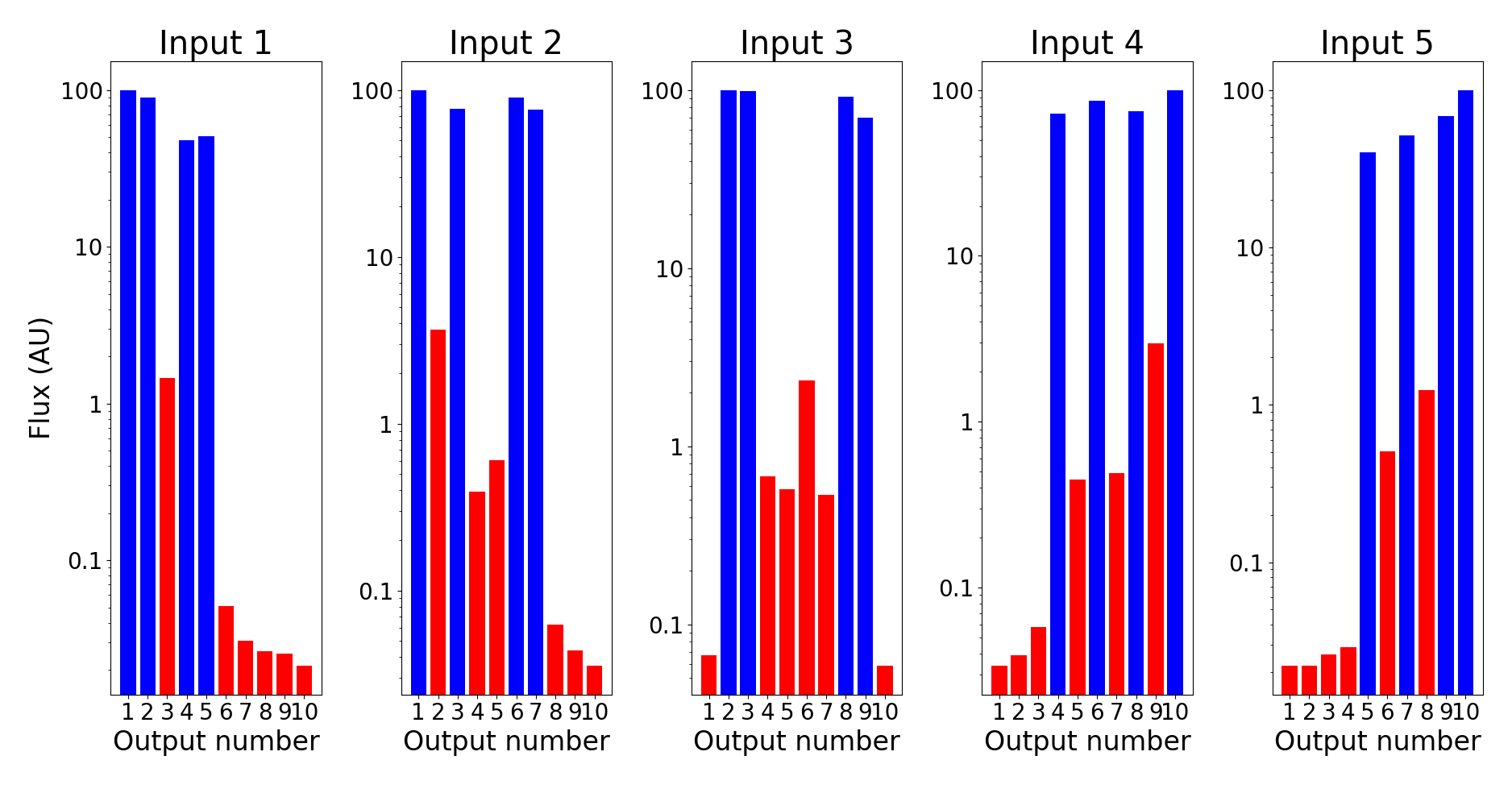}
    \end{subfigure}
    \caption[Histogrammes de l'estimation de la diaphotie des puces photoniques $X$ et $Y$.]{Histogrammes en échelle logarithmique de l'estimation de la diaphotie des puces photoniques $X$ (en haut) et $Y$ (en bas). La source lumineuse est successivement injectée dans les cinq entrées des puces (nommé \textit{input \#} dans les sous-titres) et l'intensité lumineuse de toutes les sorties est à chaque fois estimée sur l'image de caméra. La mesure de l'intensité lumineuse des sorties illuminées par l'entrée ($4$ et $8$ pour les puces $Y$ et $X$) est représentée par des barres bleues et la mesure de l'intensité lumineuse des sorties non-illuminées par l'entrée ($6$ et $12$ pour les puces $Y$ et $X$) est représentée par des barres rouges sur les histogrammes.}
    \label{fig:ChipCrossTalk}
\end{figure}

% Caractérisation de la transmission
\threesubsection{Transmission}
\label{sec:IOChipThroughput}

Le flux total transmis par la puce est calculé par la somme des flux mesurés sur les cinq images acquises pour la caractérisation de la diaphotie. On note que le flux provenant de la diaphotie n'a pas été retiré dans la somme du flux total, ainsi les valeurs de transmission présentées ci-après sont surestimées d'environ $5\%$. Ensuite on obtient la transmission par la normalisation de ce flux par le flux mesuré en injectant la lumière juste après la puce, dans une des fibres du V-Groove connecté aux sorties de la puce. La figure~\ref{fig:ChipThroughput} présente les courbes de transmission calculées de cette façon, en fonction de la longueur d'onde, pour la puce $X$ en trait continu et pour la puce $Y$ en pointillés. Dans la bande spectrale $600 - 800 \,$nm, la transmission de la puce $X$ et de la puce $Y$ est mesurée à $\sim 30\%$ et à $\sim 13\%$, respectivement. Ces valeurs de transmission sont bien plus élevées que celles des puces précédentes testées avant ma thèse (moins de $1\%$) et sont mêmes suffisantes pour qu'on puisse intégrer les puces sur \ac{SCExAO} afin de les tester sur des cibles astrophysiques (pour plus de détails voir la section~\ref{sec:FIRSTv2Subaru}). Le facteur $2$ entre les deux transmissions mesurées est attendu, étant donné que le flux injecté dans le coupleur $Y$ est à moitié transmis dans la sortie et à moitié diffusé en dehors des guides. Le couplage directionnel transmet cette moitié diffusée dans le deuxième guide de sortie (voir plus de détails dans la section II-B-6 de \cite{labeye2008}). On pourrait avoir une préférence pour l'utilisation de la puce $X$ sur \ac{FIRSTv2} pour sa meilleure transmission, mais comme on le verra par la suite, elle ne permet pas d'estimation satisfaisante des observables interférométriques et la puce $Y$ se révèle bien meilleure de ce point de vue là.

\begin{figure}[ht!]
    \centering
    \includegraphics[width=\figwidth]{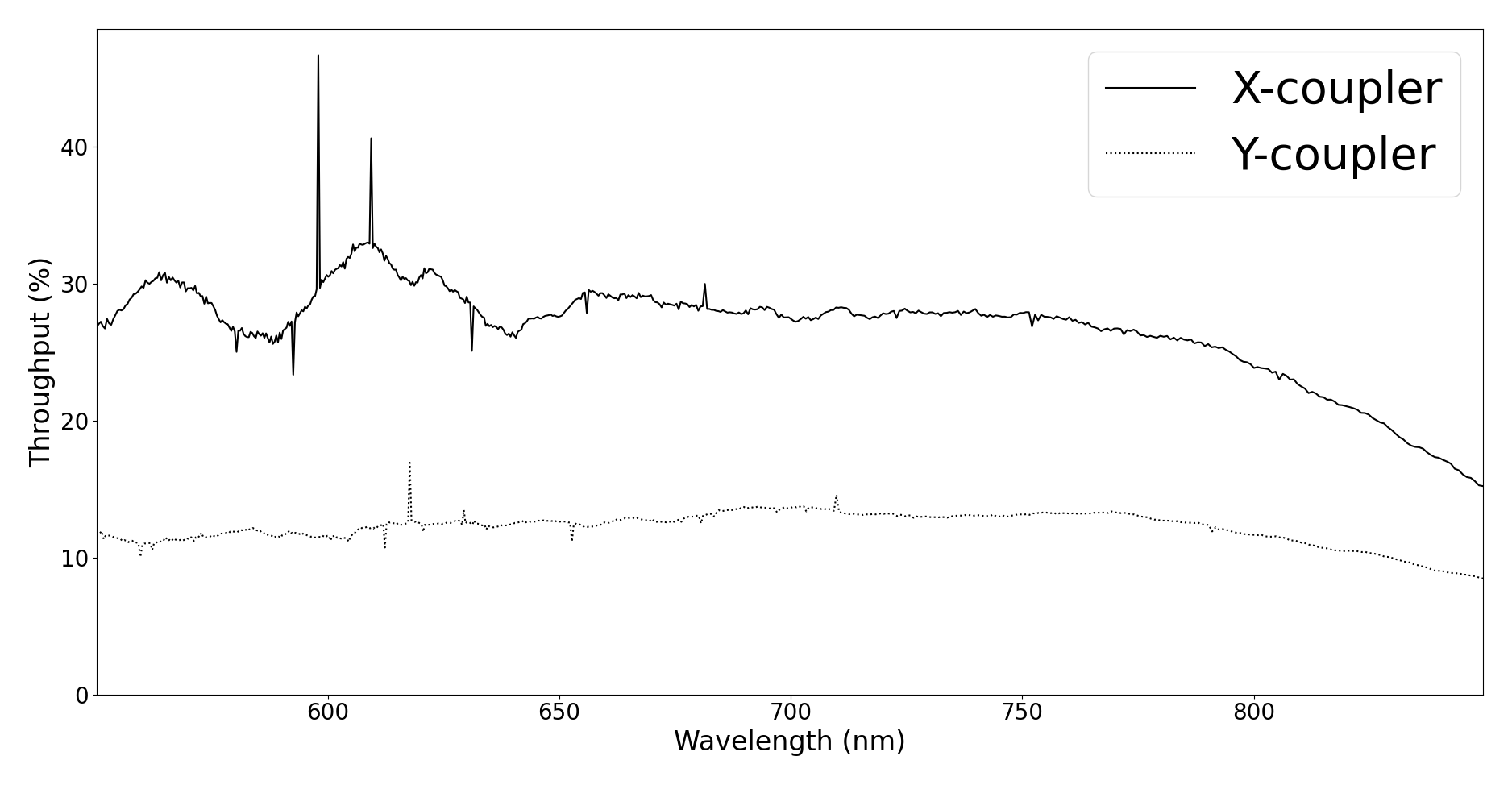}
    \caption[Transmission spectrale mesurée des puces $X$ et $Y$.]{Transmission spectrale mesurée des puces $X$ en trait continu et $Y$ en pointillés.}
    \label{fig:ChipThroughput}
\end{figure}

% Caractérisation du contraste interférométrique
\threesubsection{Contraste interférométrique}

Cette partie présente la limite haute du contraste interférométrique des bases des deux puces calculé par la différence de flux entre les paires de faisceaux qui interfèrent. Pour une base donnée résultante de la combinaison des faisceaux $n$ et $n'$, je calcule ce contraste C à partir de la mesure de l'intensité des faisceaux $\text{I}_{n}$ et $\text{I}_{n'}$ selon l'équation :

\begin{equation}
    \text{C} = \frac{2 \sqrt{\text{I}_{n} \text{I}_{n'}}}{\text{I}_{n} + \text{I}_{n'}}
\end{equation}

J'utilise les cinq images décrites précédemment pour estimer les valeurs d'intensité de la sortie illuminée par les deux entrées $n$ et $n'$. Autrement dit, les valeurs $\text{I}_{n}$ et $\text{I}_{n'}$ sont obtenues sur la sortie qui est commune aux deux images obtenues en illuminant les entrées $n$ et $n'$ et qui est représentée par une barre bleue sur les histogrammes de la figure~\ref{fig:ChipCrossTalk}. Par exemple, le contraste de la base $5$, formée par les entrées $2$ et $3$, est estimé à partir des intensités mesurées sur la sortie $3$ des histogrammes nommées \textit{Input 2} et \textit{Input 3}, pour la puce $Y$. La figure~\ref{fig:ChipContrast} présente les contrastes interférométriques estimés de cette façon pour toutes les sorties de la puce $X$ (points) et de la puce $Y$ (croix). Les contrastes obtenus avec les puces $X$ et $Y$ sont en moyenne égaux à $0,78$ et $0,994$, respectivement, avec un écart-type de $0,04$ et $0,004$, respectivement. On remarque que le contraste est pratiquement maximal pour la puce $Y$ et est dégradé de $\sim 20\%$ pour la puce $X$ ce qui peut s'expliquer par le fait que les coupleurs directionnels ne sont pas parfaitement équilibrés en transmission de flux sur les deux sorties (visible sur les histogrammes du haut de la figure~\ref{fig:ChipCrossTalk}). Ce contraste correspond à une transmission dans les deux bras de sortie d'environ $80\% / 20\%$ alors que dans l'idéal on s'attend à une transmission de $50\% / 50\%$.

\begin{figure}[ht!]
    \centering
    \includegraphics[width=\figwidth]{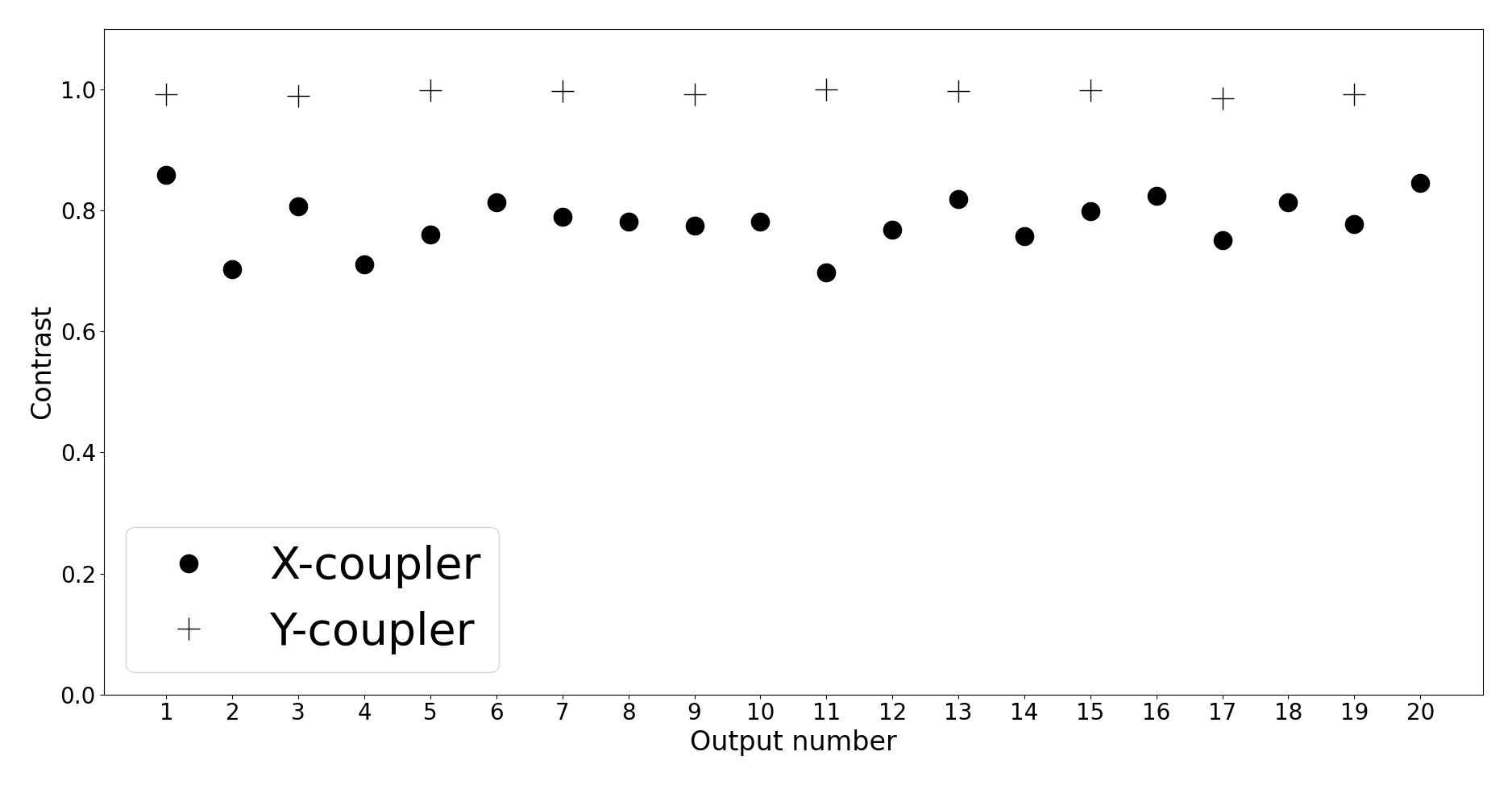}
    \caption[Contraste interférométrique estimé pour toutes les bases avec les puces $X$ et $Y$.]{Contraste interférométrique estimé pour toutes les bases avec les puces $X$ (représenté par les points) et $Y$ (représenté par les croix).}
    \label{fig:ChipContrast}
\end{figure}

% Conclusion
\threesubsection{Conclusion sur la caractérisation}
\label{sec:ChipCharacDiscu}

En résumé, j'ai mesuré un niveau de diaphotie équivalent pour les deux puces, mais un contraste interférométrique dégradé de $\sim 20\%$ sur les mesures de la puce $X$. De plus, les transmissions des puces $X$ et $Y$ sont estimées à $30\%$ et $13\%$, respectivement. Le facteur $2$ observé entre les transmissions des deux puces est attendu du fait des natures transmissives différentes des coupleurs directionnel et des jonctions $Y$. La puce $X$ offre ainsi le double avantage, premièrement, d'être plus transmissive et, deuxièmement, de disposer de deux points de mesures des franges déphasés de $\pi \,$rad sur chaque image acquise, permettant de diviser par 2 la quantité de données nécessaire (plus de détails dans la section~\ref{sec:Modulation}).

On peut alors conclure qu'il sera préférable d'utiliser la puce $X$ lors de futures prises de données, mais comme nous le verrons dans la section~\ref{sec:BinaryCharac}, je ne suis pas parvenu à estimer les observables interférométriques avec cette puce et la puce $Y$ produit de bien meilleurs résultats.

Aussi, les valeurs de transmission mesurées sont suffisantes pour permettre leur intégration et de les tester sur le banc \ac{SCExAO} en conditions d'observations du ciel, ce que j'ai pu faire durant ma thèse et que j'exposerai dans la section~\ref{sec:FIRSTv2Subaru}. En revanche, elles ne sont pas suffisantes pour une exploitation de l'instrument \ac{FIRSTv2} par la communauté scientifique. Guillermo Martin et Manon Lallement travaillent à Grenoble sur le développement de nouvelles puces avec de meilleures performances. Par exemple, des puces fabriquées avec un dopage aux ions Ag+, augmentant le contraste d'indice des guides ce qui permet des rayons de courbure plus petits. Le but est de faire les croisements avec des angles plus grands afin de diminuer la diaphotie. Mais aussi, des puces 3D \citep{martin2022a} dont les guides d'onde ont été gravés par laser non pas dans un même plan mais dans le volume de la puce, permettent d'éviter les croisements qui engendrent la diaphotie et les pertes en transmission. De même, des puces avec une modulation électro-optique fabriquées par \textit{FEMTO-ST}\footnote{\url{https://www.femto-st.fr/en}} dans un matériau différent (Niobate de Lithium LiNi) avec une meilleure transmission ont été caractérisées sur le banc de test \ac{FIRSTv2} durant ma thèse \citep{martin2022b}. Enfin, des puces utilisant différents types de coupleurs (ABCD, directionnel asymétrique) dont les paramètres physiques sont explorés ont été fabriquées et caractérisées dans \cite{lallement2022}.

%%%%%%%%
\subsubsection{La polarisation}
\label{sec:ChipPolar}

Les deux polarisations ne sont pas transmises de la même façon à travers les composants d'optique intégrée. En effet, en plaçant un prisme de Wollaston en sortie de la puce (juste après le réseau holographique) pour imager les deux polarisations et un polariseur en entrée de la puce (juste avant la matrice de micro-lentilles) que l'on place successivement sur les deux polarisations, on peut estimer la façon dont les deux polarisations se propagent dans la puce. On note que le prisme de Wollaston ne peut pas être réglé en rotation autour de son axe $Z$, contrairement au polariseur en entrée. La figure~\ref{fig:PolaComparison} présente le flux des $20$ sorties de la puce $Y$ ($10$ pour chaque polarisation) lorsque le polariseur sélectionne la polarisation verticale (V) (dénommée ici \og Pola X \fg) sur l'image de gauche, lorsque le polariseur est enlevé du chemin optique sur l'image du milieu et lorsque le polariseur sélectionne la polarisation horizontale (H) (dénommée ici \og Pola Y \fg) sur l'image de droite. Les flèches bleues et rouges identifient les polarisations V et H séparées par le prisme de Wollaston, respectivement. La source utilisée ici est la source SLED 650 fabriquée par \textit{Exalos Inc.}\footnote{\url{https://www.exalos.com/}} de largeur spectrale mesurée de $\sim 10 \,$nm. Ce qu'on peut déduire de ces mesures c'est que premièrement, lorsque le polariseur sélectionne l'une ou l'autre des polarisations à l'injection de la puce, on mesure une intensité lumineuse sur les sorties dans les deux polarisations discriminées par le prisme de Wollaston (et aucune position du polariseur ne permet d'annuler totalement le flux d'une des deux moitiés des sorties sur la caméra). Il semblerait qu'une partie de la polarisation sélectionnée est projetée dans l'autre polarisation au cours de la transmission dans la puce. Deuxièmement, on remarque qu'il y a moins de flux converti dans l'autre polarisation lorsque c'est la polarisation V qui est sélectionnée par le polariseur en entrée de la puce.

\begin{figure}[ht!]
    \centering
    \includegraphics[width=0.7\textwidth]{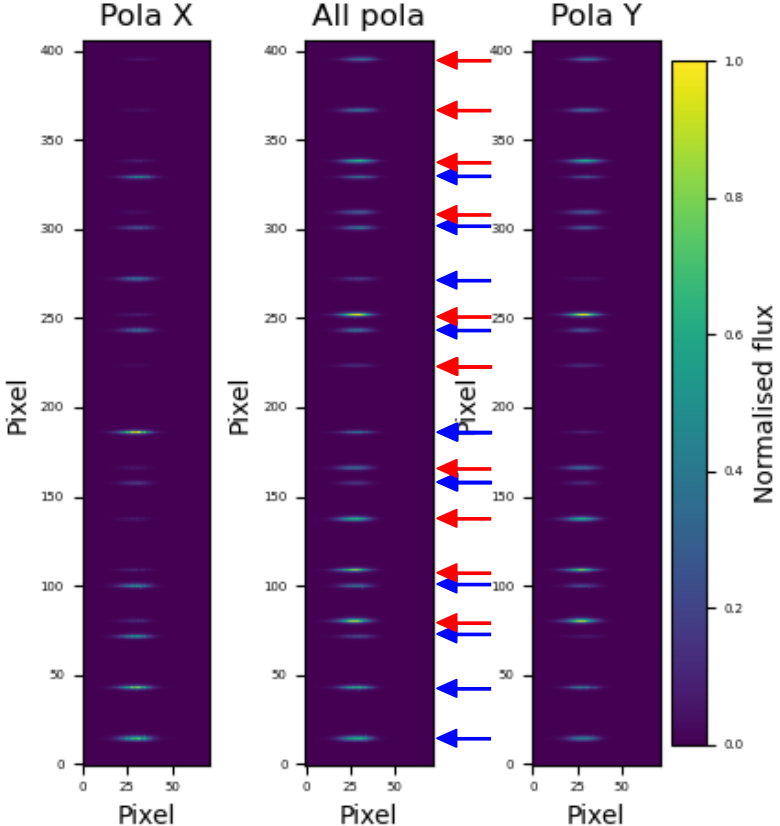}
    \caption[Images de la caméra montrant le flux sur les dix sorties de la puce $Y$ dans les deux polarisations.]{Images de la caméra montrant le flux normalisé par le maximum de l'image sur les dix sorties de la puce $Y$ (multiplié par deux par le prisme de Wollaston), lorsque le polariseur en amont de la puce sélectionne la polarisation V (à gauche), est retiré (au milieu) et sélectionne la polarisation H (à droite). Dans les titres, les polarisations V et H sont nommées X et Y, respectivement et l'axe horizontal est l'axe de dispersion. Les flèches bleues et rouges identifient la polarisation V et H, respectivement.}
    \label{fig:PolaComparison}
\end{figure}

Ainsi, lors de l'intégration du polariseur sur le banc de test, l'angle de celui-ci est réglé sur la polarisation V et affiné de façon à minimiser le flux observé sur les sorties dans la polarisation H sur l'image de la caméra. La figure~\ref{fig:PolaRotation} montre l'intensité lumineuse moyenne de toutes les sorties de la puce $X$, en polarisation V en trait continu et en polarisation H en trait-point, en fonction de l'angle du polariseur. La polarisation V est sélectionnée lorsque l'angle du polariseur est égal à $\sim -15\degree$. Ce graphique permet de quantifier l'échange de polarisation dans la puce photonique et on voit que la polarisation H (angle du polariseur égal à $\sim 75\degree$) est celle dont le flux est le plus converti dans la polarisation opposée (V). On souhaite donc placer le polariseur à l'angle égal à $\sim 0\degree$ qui sélectionne la polarisation V. Autrement dit, on souhaite régler l'angle du polariseur qui minimise le flux dans la polarisation opposée à celle qu'il sélectionne.

\begin{figure}[ht!]
    \centering
    \includegraphics[width=\figwidth]{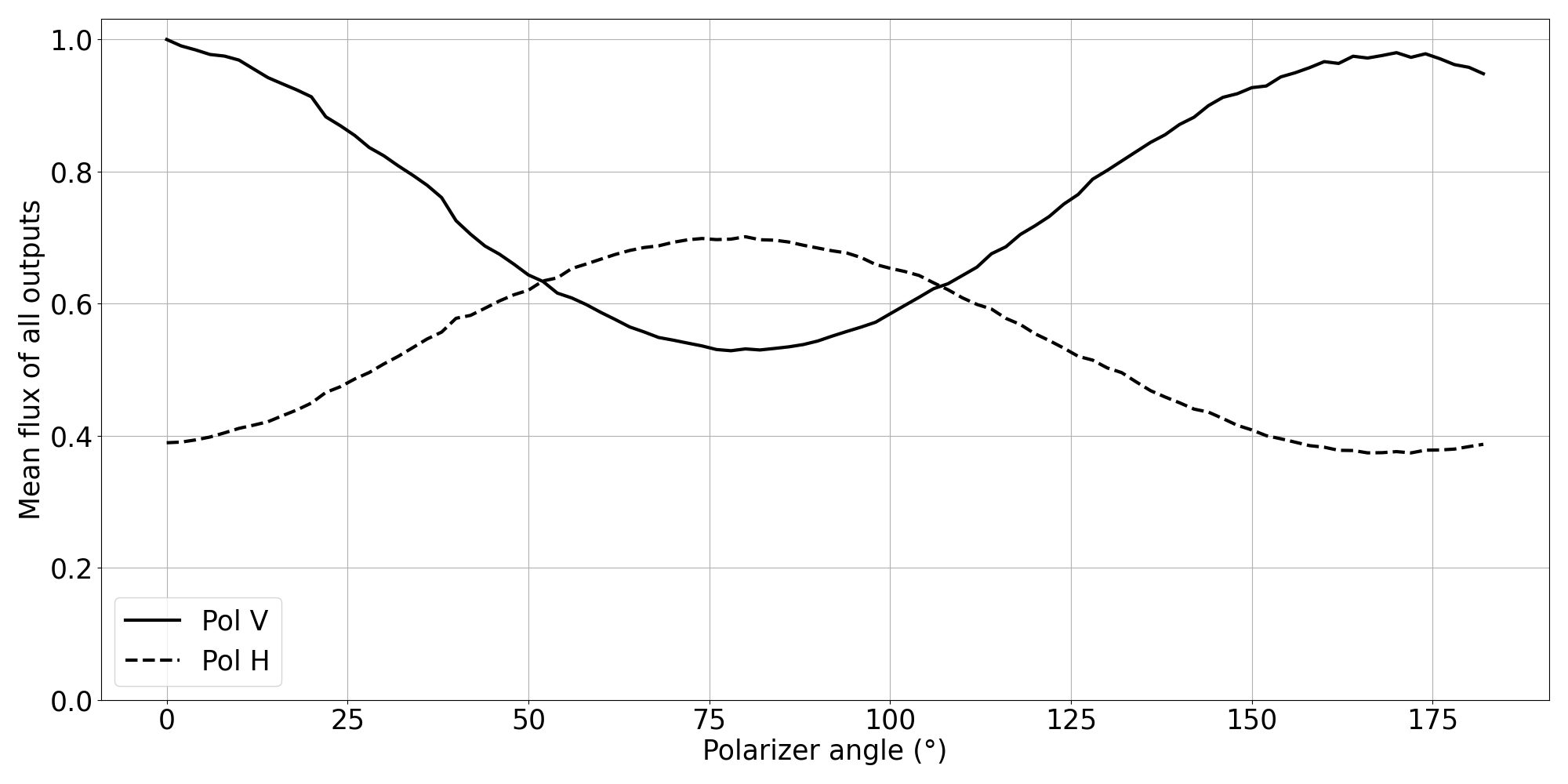}
    \caption[Flux intégré moyenné sur toutes les sorties de la puce $X$ en fonction de l'angle du polariseur, dans les deux polarisations.]{Flux intégré moyenné sur toutes les sorties de la puce $X$, sur toute la gamme spectrale transmise, en polarisation V (en trait continu) et en polarisation H (en trait discontinu) sélectionnées par le prisme de Wollaston, en fonction de l'angle du polariseur en amont de la puce. Les courbes sont normalisées par le maximum des deux (le premier point de la courbe en polarisation V).}
    \label{fig:PolaRotation}
\end{figure}

Des travaux sont actuellement en cours par Harry-Dean Kenchington Goldsmith et Manon Lallement, afin d'améliorer notre compréhension du comportement de la polarisation dans les puces photoniques. À ce jour nous pensons que les axes principaux des guides de la puce ne sont pas alignés avec les axes de toutes les fibres optiques ainsi que du polariseur en entrée et du prisme de Wollaston. Cela a pour conséquence de rendre la polarisation elliptique et expliquerait les mesures montrées sur le graphique de la figure~\ref{fig:PolaRotation}. Ce sujet est lié aux problèmes de \wiggles~que nous discuterons par la suite dans la section~\ref{sec:wiggles}. De plus, de récentes analyses montreraient que la biréfringence des fibres optiques ait un effet sur la polarisation des faisceaux injectés. Cela constitue une source de perturbations à prendre en compte dans l'analyse présentée ici.

%%%%%%%%%%%%%%%%
\subsection{Le spectro-imageur}
\label{sec:InstruSpectro}

Le principe optique du spectro-imageur du banc de test a été changé au cours de ma thèse dans le but d'augmenter sa résolution spectrale. À l'origine, il était composé d'un objectif de microscope, d'un prisme de Wollaston, d'un prisme équilatéral en SF2 et d'une lentille d'imagerie et fournissait une résolution spectrale égale à $\sim 600$. Durant son stage en 2021, Manon Lallement a spécifié, optimisé et intégré au banc le nouveau concept de spectro-imageur. Les caractéristiques requises (récapitulées dans la deuxième colonne du tableau~\ref{tab:SpectroSpec}) étaient (CR1) d'imager une bande spectrale égale à $\Delta \lambda = 140 \,$nm (entre $633 \,$nm et $773 \,$nm) sur moins que la largeur de la caméra; (CR2) que la résolution spectrale $\lambda / \delta \lambda$ soit égale à $2\,200$ pour $\lambda = 656,3 \,$nm; (CR3) que les sorties dans les deux polarisations ne se superposent pas lorsqu'elles sont imagées sur la caméra.

\begin{table}[ht!]
    \centering
    \renewcommand*{\arraystretch}{1}
    \begin{tabular}{|c|c|c|}
        \hline
        Composant & Spécification & Solution \\
        \hline
        Source & N/A & V-Groove de séparation de $127 \,$\um\\
        \hline
        \multirow{5}{*}{Collimateur} & \multirow{5}{*}{N/A} & objectif de microscope Thorlabs TL2X SAP \\
         & & NA $= 0,1$ \\
         & & distance de travail d$= 56,3 \,$mm \\
         & & focale $\text{f'} = 100\,$mm \\
         & & grossissement $\times 2$ \\
        \hline
        \multirow{2}{*}{Wollaston} & \multirow{2}{*}{N/A} & dimensions $50 \times 50 \times 11,4\,$mm \\
         & & séparation $10,8' = 0,18\degree$ \\
        \hline
        \multirow{2}{*}{Réseau} & \multirow{2}{*}{(CR1), (CR2)} & holographique en transmission \\
         & & fréquence spatiale $600 \,\text{l}.\text{mm}^{-1}$\\
        \hline
        \multirow{3}{*}{Imageur} & \multirow{3}{*}{(CR1), (CR2), (CR3)} & objectif de Lister de focale $83 \,$mm (deux lentilles) \\
         & & $\text{f'}_1 = 150 \,$mm \\
         & & $\text{f'}_2 = 80 \,$mm \\
        \hline
        \multirow{3}{*}{Caméra} & \multirow{3}{*}{N/A} & Andor Zyla 5.5 USB 3.0 \\
         & & taille des pixels $6,5 \,$\um \\
         & & $2\,560 \,$px en largeur \\
        \hline
    \end{tabular}
    \caption[Spécifications de conception et caractéristiques des composants du nouveau spectro-imageur.]{Spécifications de conception et caractéristiques des composants du nouveau spectro-imageur. Le type de composant est indiqué dans la colonne de gauche, la caractéristique requise est rappelée dans la colonne du milieu (N/A est indiqué lorsque le composant existait déjà au laboratoire) et les caractéristiques finales du composant sont présentées dans la colonne de droite. CR1 : imager une bande spectrale égale à $\Delta \lambda = 160 \,$nm (entre $623 \,$nm et $781 \,$nm) sur moins que la largeur de la caméra ($1430 \,$px). CR2 : la résolution spectrale $\lambda / \Delta \lambda$ doit être égale à $2\,200$ pour $\lambda = 656,3 \,$nm. CR3 : les sorties dans les deux polarisations ne doivent pas se superposer lorsqu'elles sont imagées sur la caméra.}
    \label{tab:SpectroSpec}
\end{table}

Les caractéristiques des composants choisis pour la solution du nouveau spectro-imageur sont récapitulées dans la colonne de droite du tableau~\ref{tab:SpectroSpec}. Les composants dont la spécification n'est pas renseignée sont ceux qui étaient déjà disponibles au laboratoire, constituant les paramètres fixés de l'optimisation de la conception du spectro-imageur.

La figure~\ref{fig:SpectroPhoto} montre une photographie du nouveau spectro-imageur après intégration sur le banc de test. La source est le V-Groove (1) sur lequel sont branchées les fibres optiques de sortie de la puce photonique. Les faisceaux des fibres sont ensuite collimatés par un objectif de microscope (2) fabriqué par \textit{Thorlabs}\footnote{\url{https://www.thorlabs.com/}} qui a un grossissement $\times 2$. L'élément de dispersion (3) choisi est un réseau \ac{VPH} fabriqué par \textit{Wasatch Photonics}\footnote{\url{https://wasatchphotonics.com/}}, que j'appelle réseau holographique dans tout le manuscrit, de fréquence spatiale égale à $600 \,\text{l}.\text{mm}^{-1}$. Celui-ci travaille en transmission et est constitué d'une couche de gélatine photosensible dans laquelle le réseau de diffraction a été gravé au laser. Cette couche est confinée entre deux parois de verre en BK7 traitées en surface avec un revêtement anti-réflection pour les longueurs d'onde du visible. Le prisme de Wollaston (4) est fabriqué par \textit{Optique Fichou}\footnote{\url{https://optique-fichou.com/}} et les nouvelles optiques ont été dimensionnées en s'assurant que les sorties de la puce photonique imagées sur la caméra, après la séparation des polarisations par le prisme, soient séparées d'environ $5 - 8 \,$px sur l'axe vertical pour éviter une superposition du flux de deux sorties adjacentes. Le système imageur juste avant la caméra doit avoir une focale de $83 \,$mm pour répondre aux performances attendues et il a été choisi d'installer un objectif de Lister\footnote{\url{https://www.degruyter.com/document/doi/10.1515/aot-2019-0002/html}} afin de limiter les aberrations optiques, composé de deux lentilles (5) et (6) de focales égales à $150 \,$mm et $80 \,$mm, respectivement, disposées à $80 \,$mm l'une de l'autre. Enfin, la caméra (7) est de la gamme Andor Zyla $5.5$ disposant de $2\,560 \,$px de taille égale à $6,5 \,$\um. Le nombre de pixels choisis sur lesquels la bande spectrale est imagée est de $1\,430 \,$px ce qui permet d'intégrer le spectro-imageur sur le banc \ac{SCExAO} avec des caméras de tailles différentes, sans changer son concept et la largeur de la bande spectrale imagée.

\begin{figure}[ht!]
    \centering
    \includegraphics[width=\figwidth]{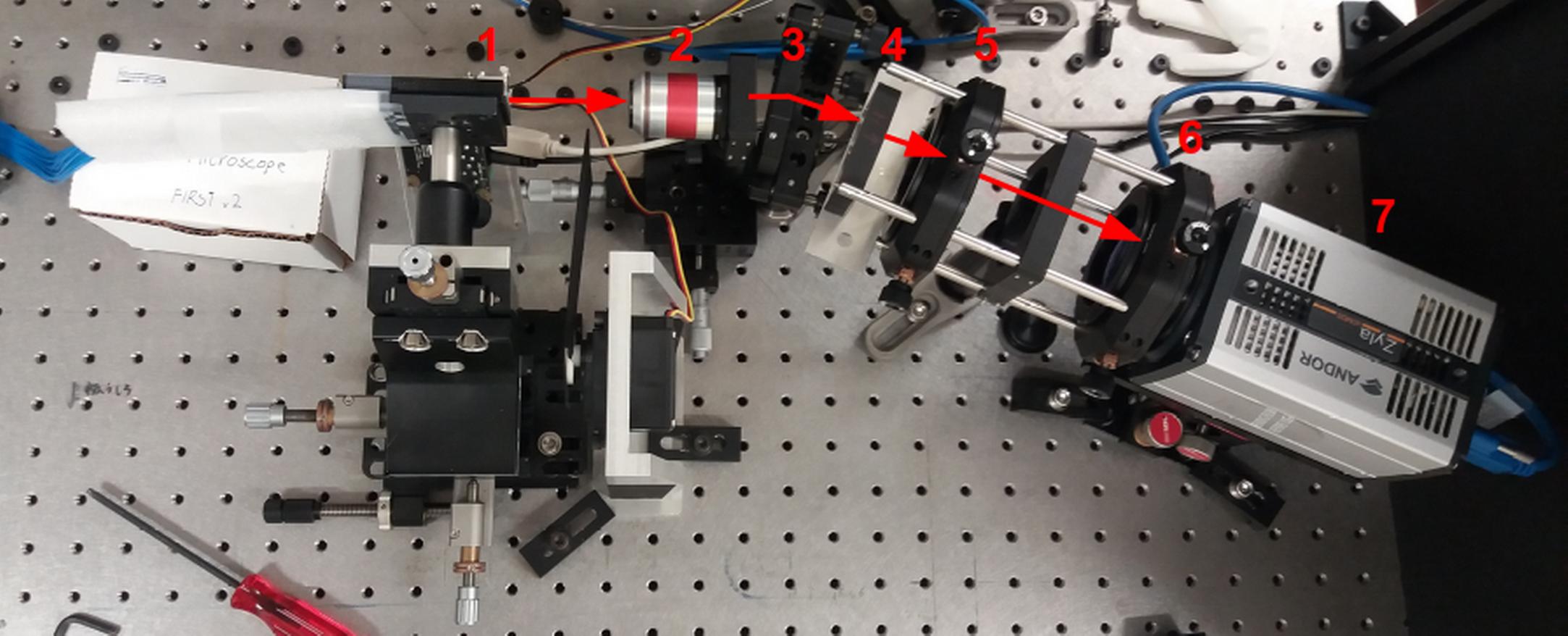}
    \caption[Photographie du spectro-imageur du banc de test de FIRSTv2.]{Photographie du spectro-imageur du banc de test de FIRSTv2. Les composants sont numérotés comme suit : (1) le V-Groove branché aux fibres de sorties de la puce photonique; (2) l'objectif de microscope de collimation (grossissement $\times 2$); (3) le réseau holographique; (4) le prisme de Wollaston; (5) la première lentille d'imagerie de focale égale $150 \,$mm; (6) la deuxième lentille d'imagerie de focale égale à $80 \,$mm; (7) la caméra.}
    \label{fig:SpectroPhoto}
\end{figure}

La caméra de la gamme ORCA-Quest qCMOS fabriquée par \textit{Hamamatsu}\footnote{\url{https://www.hamamatsu.com/}} a récemment été acquise et est intégrée dans \ac{FIRSTv1} au télescope Subaru. Elle dispose de meilleures performances que la caméra utilisant la technologie \ac{EMCCD} jusqu'ici utilisée : entre autre, l'écart-type sur le bruit de courant d'obscurité est égal à $0,006 \,\text{e}^-.px^{-1}.s^{-1}$ et l'efficacité quantique est de $65 - 80\%$ dans la gamme $600 - 700\,$nm. Le capteur est composé de $4\,000 \times 2\,300 \,$px et les pixels ont une taille égale à $4,6 \,$\um. Cela nous permettrait d'augmenter, à terme, la résolution spectrale, lorsque la transmission de l'instrument sera plus élevée, ce qui est un atout non négligeable pour l'étude de protoplanètes, comme nous l'avons vu dans la section~\ref{sec:Protoplanetes}.

L'analyse des mesures effectuées pour l'étalonnage spectral de l'instrument, présentés dans la section~\ref{sec:EtalonnageSpectral}, permet d'estimer la résolution spectrale du nouveau spectro-imageur à $\sim 3\,400$. Celle-ci est plus élevée que prévu car les composants ont été alignés à l'aide du visionnage de l'image de la caméra ce qui ne permet pas de s'assurer que les spots lumineux aient une taille de $2 \,$px a minima (pour respecter le critère de Shanon). Comme nous le verrons par la suite, cela n'empêche pas de caractériser la source protoplanétaire simulée présentée plus loin dans la section~\ref{sec:SystBinaire} car la source utilisée pour le compagnon a une largeur spectrale s'étendant sur plus de quatre pixels.

%%%%%%%%%%%%%%%%
\subsection{La caméra}
\label{sec:InstruCamera}

% Brève description
La caméra utilisée sur le banc \ac{FIRSTv2} est de la gamme Andor Zyla $5.5$ fabriquée par \textit{Oxford Instruments}\footnote{\url{https://www.oxinst.com/}}. Elle utilise la technologie de capteur \ac{CMOS}, d'une taille de $2\,160 \,\text{px} \times 2\,560 \,\text{px}$ de taille de pixel égale à $6,5 \,$\um. Ses caractéristiques sont résumées dans le tableau~\ref{tab:CameraSpec}. Elle est refroidie par air à l'aide d'un ventilateur. Le spectro-imageur et la caméra sont isolés des lumières parasites de la pièce grâce à un coffrage en carton amovible (l'ordinateur de contrôle étant dans la même pièce). La caméra est opérée avec le mode obturateur déroulant (\textit{rolling shutter}) de lecture des pixels à une fréquence de $280 \,$MHz. Les charges des pixels de chaque image sont donc lues en $3,6 \,$ns, garantissant que les franges ne sont pas brouillées pendant l'acquisition d'une image.

\begin{table}[ht!]
    \centering
    \renewcommand*{\arraystretch}{1}
    \begin{tabular}{cc}
        \hline
        \hline
        \multicolumn{2}{c}{Caractéristiques du détecteur} \\
        \hline
        \hline
        Technologie & CMOS \\
        Connexion & USB $3.0$ \\
        Nombre de pixels & $2\,160 \times 2\,560$ \\
        Taille des pixels & $6,5 \,$\um \\
        Efficacité quantique max & $60$\% \\
        Efficacité quantique $> 30 \%$ & $400 - 800 \,$nm \\
        Mode de lecture & \textit{Rolling shutter} \\
        Fréquence de lecture & $280 \,$MHz \\
        Sensibilité & $0,49 \, \text{e}^- \,$/ADU \\
        Bruit de lecture & $1,11 \, \text{e}^- \,$RMS \\
        Courant d'obscurité & $0,1453 \, \text{e}^{-}.\text{px}^{-1}.\text{s}^{-1}$ \\
        Dynamique & $65\,536:1$\\
        \hline
    \end{tabular}
    \caption[Caractéristiques de la caméra Andor Zyla 5.5 USB 3.0 du banc de test de FIRSTv2.]{Caractéristiques de la caméra Andor Zyla 5.5 USB 3.0 du banc de test de FIRSTv2.}
    \label{tab:CameraSpec}
\end{table}

% Image de Dark
La figure~\ref{fig:CameraDark} présente la médiane de cent images de la caméra sans flux, à un temps d'exposition de $100 \,$ms. Cette image permet d'étalonner le bruit sur le courant d'obscurité et le bruit de lecture de la caméra sur toutes les données acquises (pour plus de détails voir la section~\ref{sec:CameraDark}). On remarque des motifs par colonne visibles aussi sur la visualisation en temps réel du capteur de la caméra. Ce sont des motifs fixes qui résultent de la structure de l'électronique d'amplification des pixels inhérente aux capteurs \ac{CMOS}\footnote{\url{https://www.mdpi.com/2076-3417/10/11/3694/htm}}. Ces structures étant constantes en fonction du temps, elles se corrigent très bien lors du traitement de données.

\begin{figure}[ht!]
    \centering
    \includegraphics[width=\figwidth]{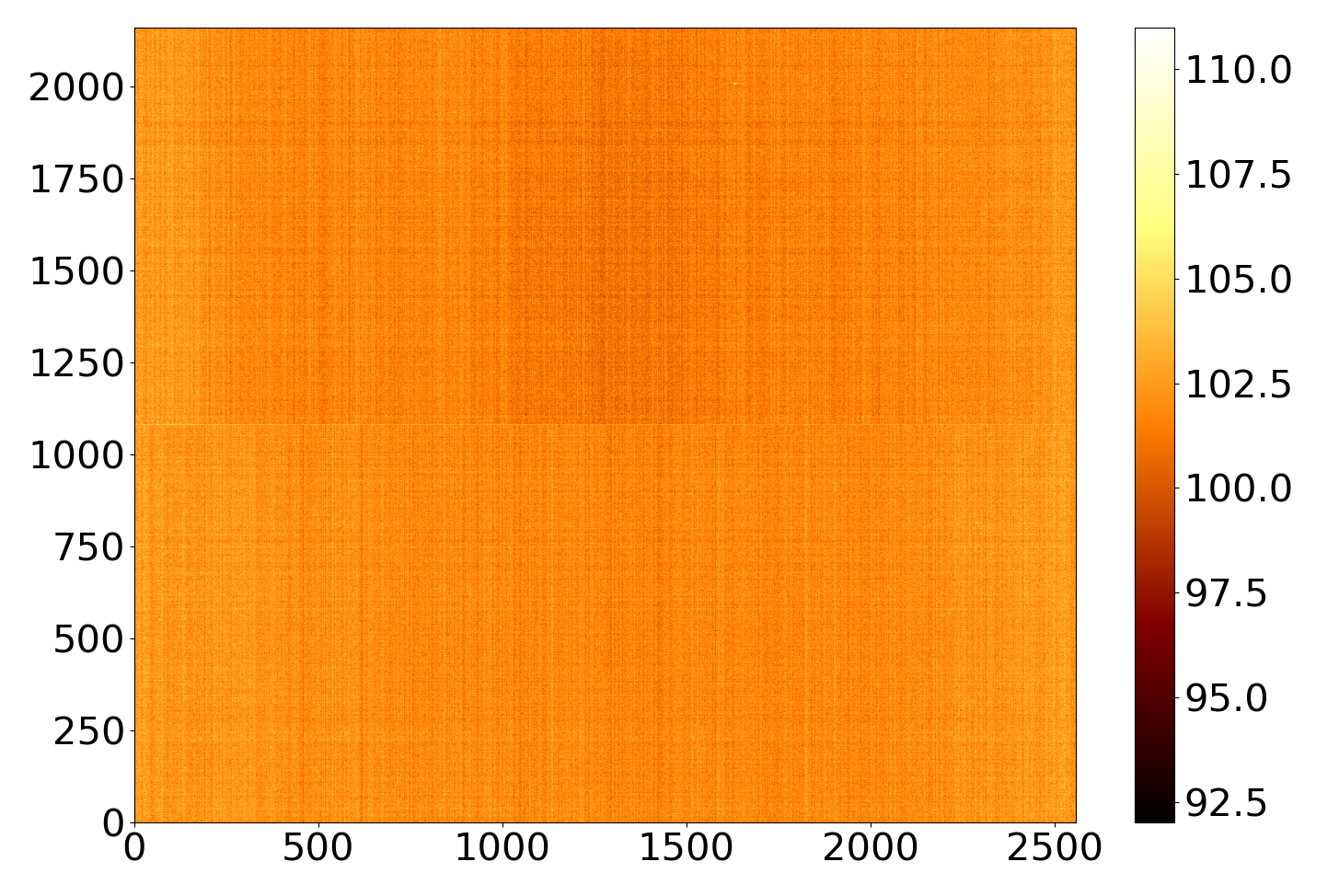}
    \caption[Image sans flux de la caméra Andor Zyla de FIRSTv2.]{Médiane de cent images de temps d'exposition égal à $100 \,$ms de la caméra Andor Zyla, sans flux. Les valeurs des pixels données sur les échelles sont en \ac{ADU}.}
    \label{fig:CameraDark}
\end{figure}

% Calcul du SNR
Le rapport signal sur bruit obtenu ou \ac{SNR} sur les images de la caméra avec un temps d'exposition t et pour un flux de photons P, s'écrit :

\begin{equation}
    \text{SNR} = \frac{\eta \times \text{P} \times \text{t}}{\sqrt{\sigma_{lec}^{2} + \sigma_{obsc}^{2} + \sigma_{p}^{2}}}
\end{equation}

\noindent avec $\eta$ l'efficacité quantique, $\sigma_{lec} = 1,11 \,\text{e}^-$ l'écart-type du bruit de lecture, $\sigma_{obsc} = \sqrt{\text{t} \times 0,1453 \,\text{e}^-.px^{-1}.s^{-1}}$ l'écart-type du bruit sur le courant d'obscurité et l'écart-type du bruit de photons se calculant par $\sigma_{p} = \sqrt{\eta \times P \times t}$ . Les capteurs de type \ac{CMOS} actuellement fabriqués remplacent petit à petit les capteurs \ac{CCD} qui étaient incontournables ces dernières décennies. En effet, les capteurs \ac{CMOS} consomment moins d'énergie, sont plus rapides lors de la lecture des charges des pixels du capteur et présentent un bruit de courant d'obscurité et de lecture bien plus faibles. La caméra de \ac{FIRSTv1} utilise la technologie \ac{EMCCD} qui dispose d'un gain d'amplification du signal mais, comparée à la technologie \ac{CMOS}, cela nécessite un fort gain induisant une faible dynamique et un bruit d'amplification est ajouté (voir plus de détails dans la section 2.1.1.3 de la thèse d'Elsa Huby \cite{huby2013these}).

% Par exemple, l'équipe travaillant sur \ac{SCExAO} a pu montré que leur caméra fabriquée par \textit{Hamamatsu}\footnote{\url{https://www.hamamatsu.com/eu/en.html}} comptait les photons incident au capteur (de type \ac{CMOS}). Comme le montre le graphique de la figure~\ref{fig:HamamatsuSCExAO}, qui est le nombre d'occurrence (axe des ordonnées) des valeurs d'\ac{ADU} (axe des abscisses) mesurées à faible intensité lumineuse, le bas bruit permet d'atteindre une résolution sur la mesure de l'intensité lumineuse pour mesurer.
% First, we've tried photon counting in the lab in Hilo with one of the Hamamatsu cameras. Attached is a histogram of pixel counts in the near-dark and another in the slightly less dark. #of occurence (y-axis) vs. px value in ADU (x-axis). Blue is the standard readout mode (fast) and orange is the "ultraquiet readout". As you can see, we're counting photons with probably just a few percent of errors due to the feet of each lobe.
% \begin{figure}[ht!]
%     \centering
%     \includegraphics[width=\figwidth]{Figure_Chap3/}
%     \caption[]{}
%     \label{fig:HamamatsuSCExAO}
% \end{figure}

Ainsi, les bruits de lecture et sur le courant d'obscurité sont négligeables par rapport au bruit de photons (dans le régime de flux élevé), le \ac{SNR} peut se ré-écrire comme suit :

\begin{equation}
    \text{SNR} = \sqrt{\eta \times \text{P} \times \text{t}}
\end{equation}

Les détecteurs permettent aujourd'hui l'acquisition de données très peu affectées par le bruit instrumental et le \ac{SNR} est donc uniquement limité par le bruit de photons. La technique d'interférométrie \textit{nulling} qui permet de s'affranchir du bruit de photon de l'étoile, se révèle alors très pertinente et constitue l'avenir du projet \ac{FIRST}.

%%%%%%%%%%%%%%%%%%%%%%%%%%%%%%%%
\section{Le logiciel de contrôle}
\label{sec:ControlSoftware}

%%%%%%%%%%%%%%%%
\subsection{L'architecture}

Une partie du travail de ma thèse a consisté en l'amélioration du logiciel de contrôle du banc de test \ac{FIRSTv2}, écrit en Python\footnote{\url{https://www.python.org/}}. Il s'agissait d'une part de faire la mise à jour des programmes depuis la version $2.7$ vers la version $3.9$ de Python (avec l'aide précieuse de Pierre Fedou, Franck Marchis et Clément Chalumeau), ce qui n'est pas évident surtout vis-à-vis des \ac{SDK} associées aux composants, fournis par les constructeurs. Le \ac{SDK} du miroir déformable \textit{Iris AO} que nous avons adapté à la version $3+$ de Python est disponible sur GitHub\footnote{\url{https://github.com/scexao-org/Iris-AO_MEMS_Software-control.git}} et celui de la caméra est fourni directement par \textit{Andor}. D'autre part, il s'agissait de changer toute l'architecture du logiciel car un unique script (dont un schéma est présenté sur la figure~\ref{fig:SoftwareArchitecture} du haut) était exécuté pour établir la connexion et contrôler tous les composants (\ac{MEMS}, \ac{ODL} et caméra), ce qui induisait une lenteur dans l'utilisation du terminal de commande par l'opérateur. Sur le schéma, chaque rectangle jaune correspond à un script et décrit son contenu. Certaines parties du logiciel étaient écrites dans d'autres scripts comme \texttt{memsCtrl.py} et \texttt{odlCtrl.py} mais étaient exécutés dans le script principal. Les rectangles bleus clairs représentent des connexions entre processus/processus ou processus/composants et les rectangles bleus foncés représentent les composants.

\begin{figure}[ht!]
    \centering
    \begin{subfigure}{1\textwidth}
        \centering
        \includegraphics[width=\textwidth]{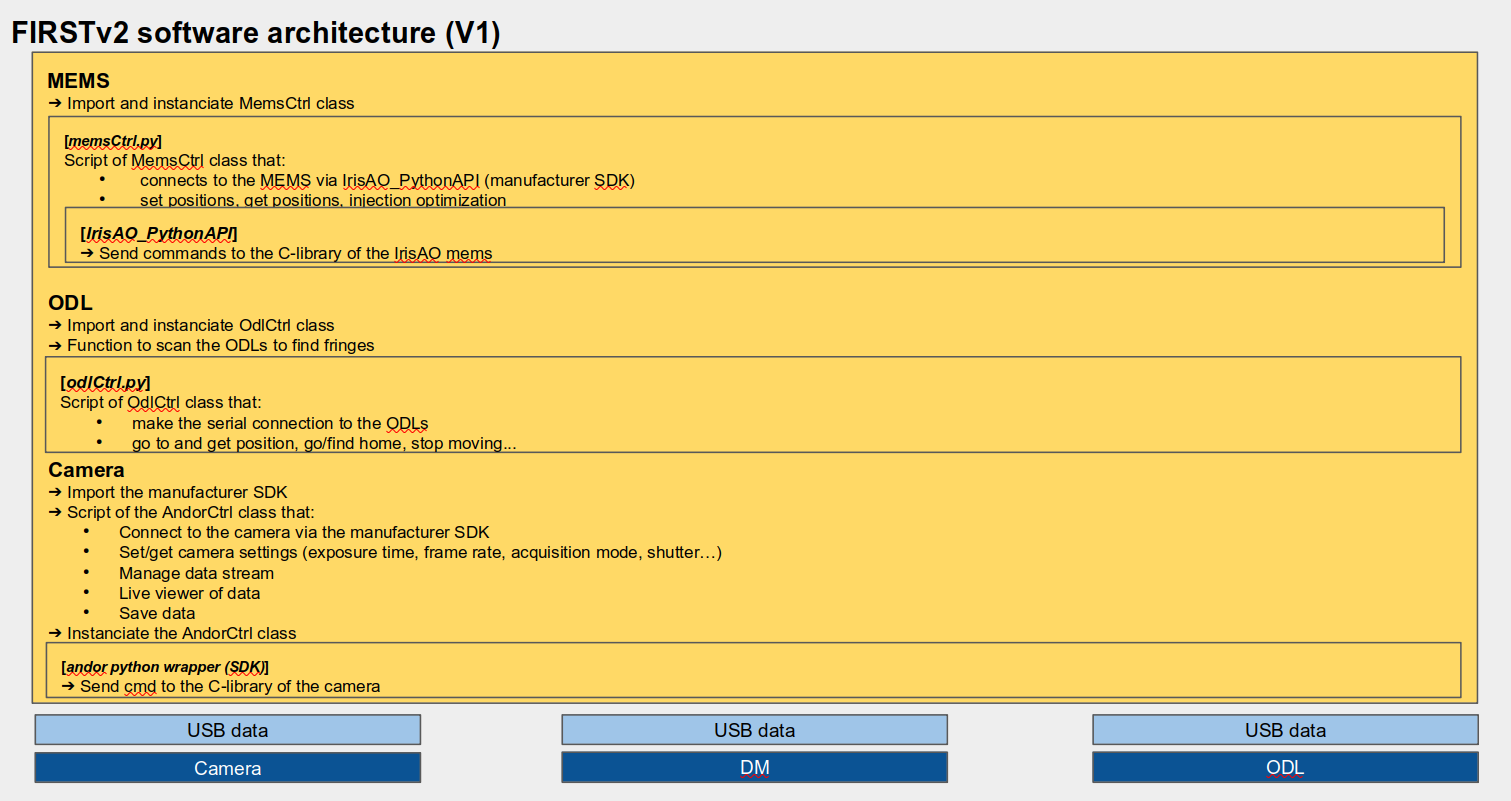}
    \end{subfigure}
    \line(1,0){430}\\
    \begin{subfigure}{1\textwidth}
        \centering
        \includegraphics[width=\textwidth]{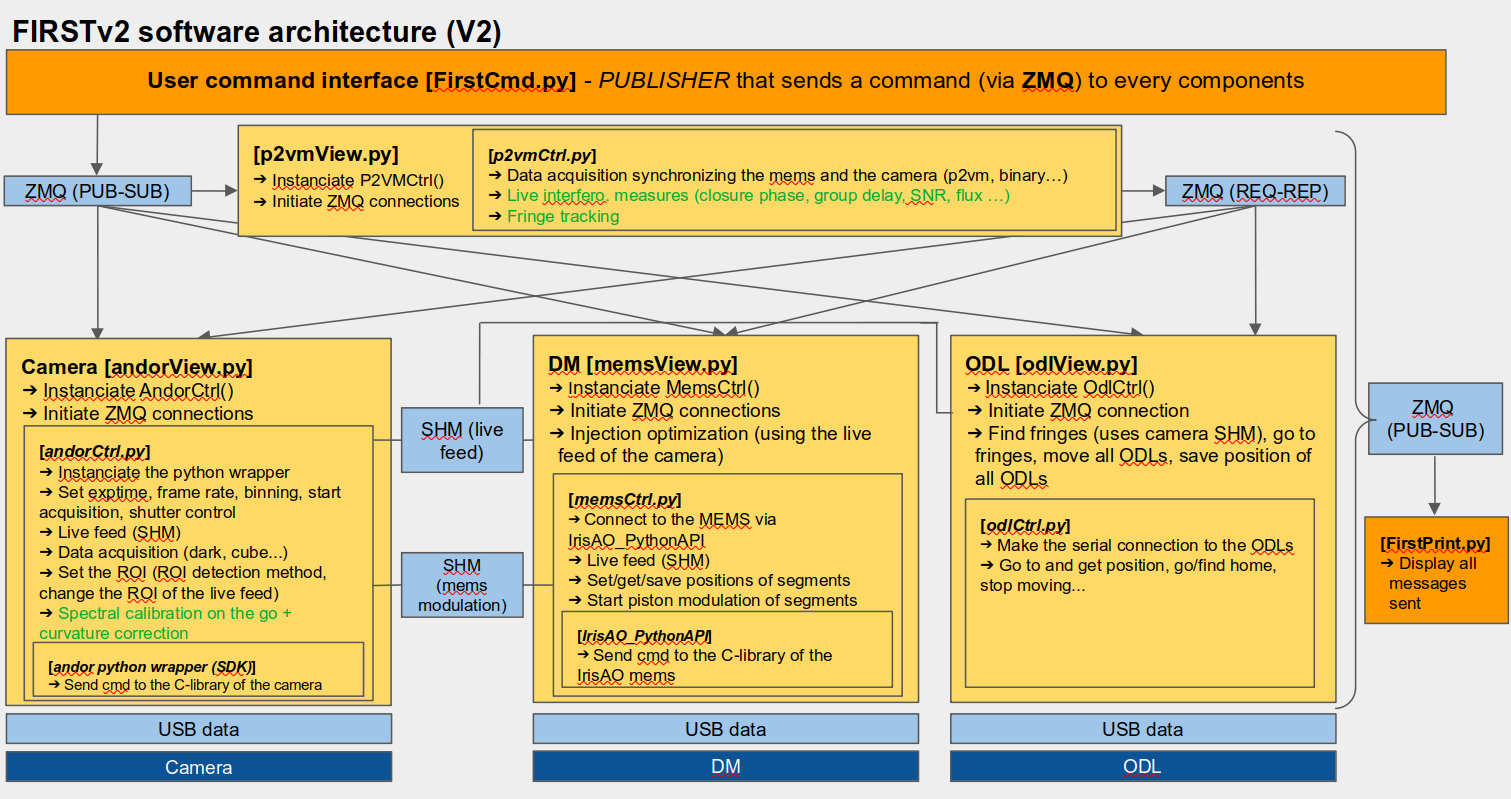}
    \end{subfigure}
    \caption[Schémas de l'architecture du logiciel de contrôle de FIRSTv2 avant et après ma thèse.]{Schémas de l'architecture du logiciel de contrôle de FIRSTv2 avant (en haut) et après (en bas) ma thèse. Les rectangles jaunes sont des scripts, les rectangles oranges sont les scripts qui implémentent une interface utilisateur, les rectangles bleus clairs sont des interfaces et les rectangles bleus foncés sont les composants. Les flèches montrent les liens établis entre les différents processus et ce qui est écrit en vert est ce qui n'est pas encore implémenté.}
    \label{fig:SoftwareArchitecture}
\end{figure}

La figure~\ref{fig:SoftwareArchitecture} du bas présente le schéma de cette nouvelle architecture, avec les mêmes codes couleurs que celui du haut, précédemment décrit et où les flèches indiquent une interaction entre processus. J'ai ainsi amélioré l'architecture du logiciel en créant un script pour chaque composant afin de les exécuter en parallèle. Les rectangles oranges sont des terminaux d'interface logiciel/utilisateur. Le terminal nommé \textit{User command interface} permet de centraliser l'envoi de toutes les commandes vers les différents processus par l'opérateur, via des protocoles de communication serveur/client (voir plus de détails dans la section~\ref{sec:ZMQ}) et le terminal nommé \texttt{FirstPrint.py} centralise l'affichage de tous les messages de tous les processus. Toutes les fonctions de bas niveau permettant d'établir la connexion et de contrôler les composants sont dans un script nommé \texttt{XCtrl.py} et sont importés dans un script nommé \texttt{XView.py} (\texttt{X} étant le nom du composant) qui les exécute et implémente des fonctions de plus haut niveau, comme l'optimisation de l'injection par le \ac{MEMS}. Ce sont ces derniers scripts qui sont exécutés lors du démarrage du logiciel de contrôle.

De plus, par rapport à l'ancienne version du logiciel, un processus, écrit dans un fichier nommé \texttt{p2vmView.py}, qui n'est pas directement lié à un composant a été ajouté. Il fonctionne avec la même architecture que les autres programmes liés aux composants et permet l'envoi successif de commandes à la caméra, au \ac{MEMS} et aux \ac{ODL}s, via également des protocoles de communication serveur/client (démarrer une acquisition d'images, changer les positions des segments du \ac{MEMS}, envoyer les \ac{ODL}s en position de zéro \ac{OPD}, etc...). C'est ce processus qui fait l'acquisition automatique des données (une soixantaine de fichiers d'images) permettant de calculer les \ac{V2PM} et \ac{P2VM} ainsi que les données interférométriques (moins d'une dizaine de fichiers d'images) sur la source protoplanétaire simulée (voir plus de détails dans la section~\ref{sec:SystBinaire}). Les fonctionnalités écrites en vert sont celles qui ne sont pas encore implémentées.

Enfin, pour la visualisation en temps réel des données (les images de la caméra et la carte de phase de la surface du miroir segmenté) j'utilise un système de mémoire partagée sur l'ordinateur, nommé \ac{SHM} sur le schéma (voir plus de détails dans la section~\ref{sec:SHM}). Par exemple, le processus associé à la caméra envoie les images dans une mémoire partagée allouée et un autre processus récupère les images de cet espace mémoire en la lisant et les affiche. Cela permet aussi de transmettre des données entre les différents programmes en cours d'exécution.

La figure~\ref{fig:SoftwareScreenShot} montre une capture de l'écran de l'ordinateur de contrôle du banc de test lorsque le logiciel de contrôle est ouvert. Le premier terminal en haut à gauche permet la saisie des commandes par l'utilisateur. Trois commandes y ont été exécutées : (1) \texttt{m.on()} commandant les cinq segments du \ac{MEMS} dans leur position d'optimisation de l'injection du flux dans les fibres optiques préalablement enregistrées; (2) \texttt{a.get\_exptime()} interrogeant la caméra sur son temps d'exposition actuel; (3) \texttt{odl3.get\_position()} interrogeant la troisième ligne à retard sur sa position actuelle. Le terminal du dessous est pour l'affichage de tous les messages et affiche ceux envoyés lors de l'ouverture du logiciel : initialisation du \ac{MEMS}, le processus associé au script \texttt{p2vmView.py} est prêt, les informations générales sur la caméra sont rappelées et elle est prête à l'emploi, ainsi que les messages de couleur blanche rendant compte de la mise en place des liens de communication. Les derniers messages sont ceux envoyés par le processus de la caméra renseignant la valeur du temps d'exposition actuel et par le processus de la troisième ligne à retard renseignant sa position actuelle, en réponse aux commandes envoyées sur le premier terminal. Une couleur de message est associée à chaque processus afin de faciliter leur lecture. Ensuite, ce sont les terminaux qui s'ouvrent à l'exécution des programmes de la caméra, du \ac{MEMS}, des \ac{ODL}s et de \texttt{p2vmView.py}. Ces quatre derniers sont ouverts pour des raisons de développement et de débogage, mais ils pourraient ne pas être affichés. Enfin, en haut à droite est la fenêtre d'affichage en temps réel du flux d'images de la caméra (actuellement sans flux lumineux) et en bas à droite est la fenêtre d'affichage en temps réel de la carte de phase de la surface du \ac{MEMS}. Les cinq segments, dont les faisceaux sont recombinés, sont sur leur position qui optimise l'injection du flux dans les fibres et sont ceux choisis dans la section~\ref{sec:BaseConfig}. Ils ont été commandés sur ces positions à la suite de l'exécution de la commande \texttt{m.on()} dans le premier terminal.

\begin{figure}[ht!]
    \centering
    \includegraphics[angle=90,height=0.78\textheight]{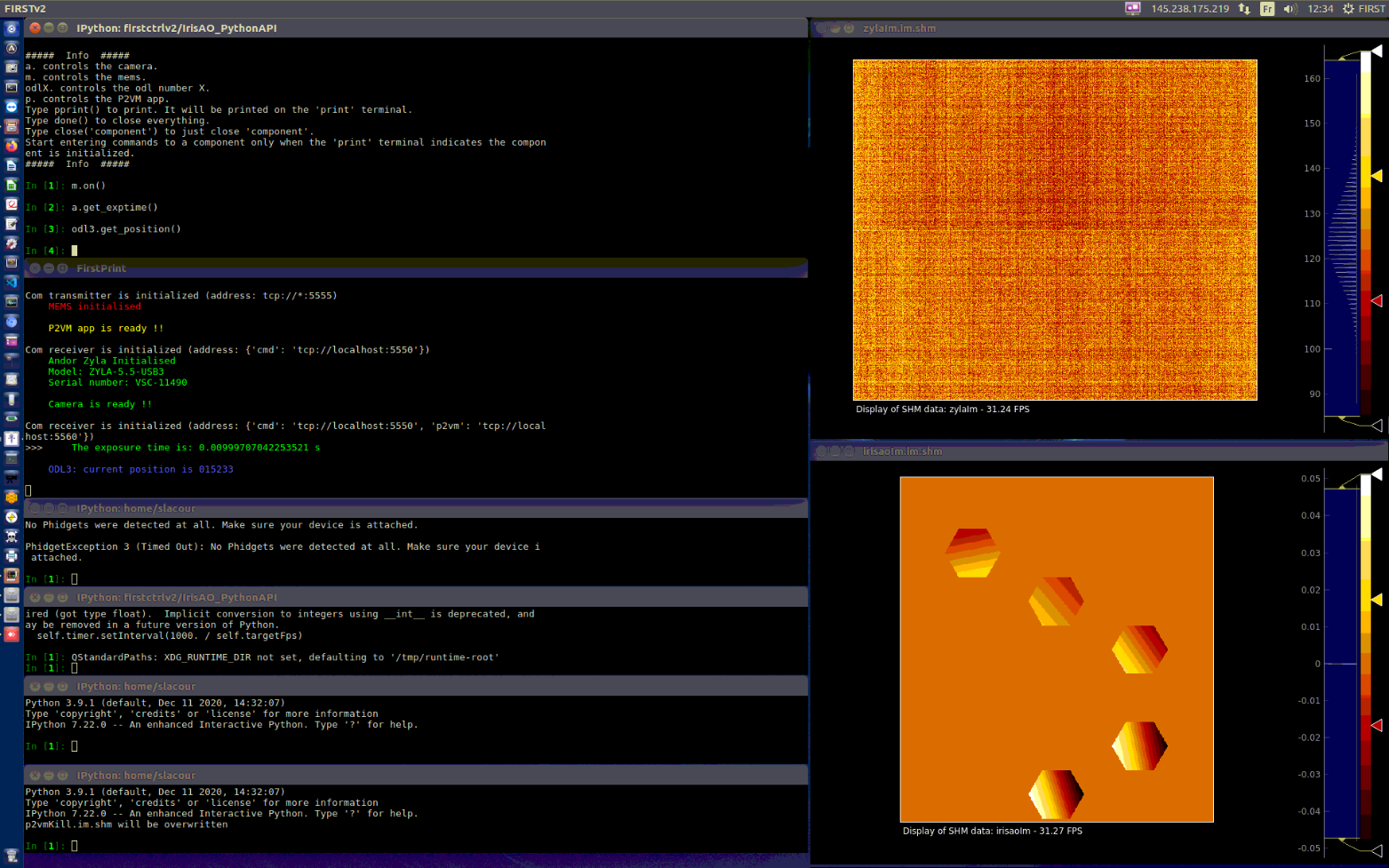}
    \caption[Capture d'écran du logiciel de contrôle du banc de test de FIRSTv2, à Meudon.]{Capture d'écran du logiciel de contrôle du banc de test de FIRSTv2, à Meudon. À gauche, sont affichés les terminaux, de haut en bas, pour l'exécution des commandes par l'utilisateur, de l'affichage de tous les messages des différents processus et ceux qui s'ouvrent à l'exécution des programmes de la caméra, du MEMS, des ODLs et de \texttt{p2vmView.py}. En haut à droite est la fenêtre d'affichage en temps réel du flux d'images de la caméra. En bas à droite est la fenêtre d'affichage en temps réel de la carte de phase de la surface du MEMS, dont cinq segments sont commandés sur leur position qui optimise l'injection du flux dans les fibres.}
    \label{fig:SoftwareScreenShot}
\end{figure}

%%%%%%%%%%%%%%%%
\subsection{La communication serveur/client avec la librairie ZeroMQ}
\label{sec:ZMQ}

La version écrite en Python de la librairie \textit{ZeroMQ}\footnote{\url{https://github.com/zeromq/pyzmq.git}} permet de créer en parallèle des processus se comportant comme des serveurs et des clients se connectant ensemble afin de s'échanger des chaînes de caractères ou des dictionnaires (objet simple du langage Python). Cette librairie rend l'implémentation de ces processus très facile (une vingtaine de lignes de code suffisent). Il existe plusieurs types de schéma de communication et les deux que j'ai utilisés sont schématisés sur la figure~\ref{fig:ZMQProtocols}. Le premier (schématisé à gauche) est le schéma \textit{request/reply} nommé \textit{REQ/REP}. Un script est écrit pour le processus serveur et un autre est écrit pour le processus client. Le principe est que le processus associé au client envoie une requête sous forme d'une chaîne de caractères à une adresse définie et se met en attente d'une réponse. Le serveur est implémenté en étant associé à cette adresse et se met dans un premier temps en attente d'une requête et dans un second temps envoie une réponse (sous forme d'une chaîne de caractères) lorsqu'il reçoit la requête, avant de se remettre en attente. Enfin, le client reçoit à son tour la réponse du serveur. Cela permet, par exemple, de mettre en place un serveur hébergeant un site internet sur lequel des utilisateurs font des requêtes de connexion. La librairie \textit{ZeroMQ} se charge d'implémenter les ports de connexion identifiés par une adresse, l'envoi des informations, la mise en attente du serveur et du client, etc de manière très simple. J'ai ainsi utilisé ce schéma de communication pour que le processus nommé \texttt{p2vmView.py} puisse se connecter successivement à chaque composant, en s'assurant que les composants sont disponibles, afin d'envoyer des commandes de contrôle et d'attendre que l'exécution de ces commandes se termine avant d'en envoyer une autre.

\begin{figure}[ht!]
    \centering
    \begin{subfigure}{0.45\textwidth}
        \centering
        \includegraphics[width=0.5\textwidth]{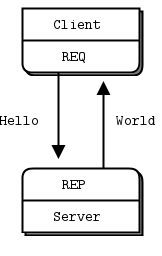}
    \end{subfigure}%
    \begin{subfigure}{0.45\textwidth}
        \centering
        \includegraphics[width=\textwidth]{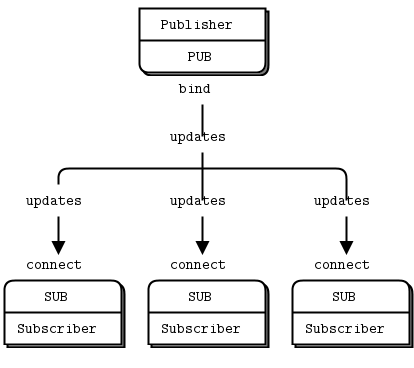}
    \end{subfigure}
    \caption[Schémas des processus de communication request/reply et publisher/subscriber de la librairie ZeroMQ.]{Schémas des processus de communication request/reply (à gauche) et publisher/subscriber (à droite) implémenté par la librairie ZeroMQ. Crédit : \textit{ZeroMQ}.}
    \label{fig:ZMQProtocols}
\end{figure}

Le deuxième schéma de communication (à droite sur la figure~\ref{fig:ZMQProtocols}) que j'ai utilisé est le \textit{publisher/subscriber} nommé \textit{PUB/SUB}. Il s'agit d'implémenter un publieur qui peut envoyer à n'importe quel moment une chaîne de caractère et des abonnés (\textit{subscriber}) qui se connectent au publieur et reçoivent la chaîne de caractère. Je me suis servi de ce schéma de communication pour l'envoi des commandes par l'utilisateur (le terminal de commande utilisateur est le publieur) aux processus associés aux composants (qui sont les abonnés). Une partie du message envoyé contient un caractère d'identification lu par tous les abonnés afin de savoir quel composant doit exécuter la commande en question. J'ai aussi implémenté ce système pour afficher tous les messages sur un unique terminal (\texttt{FirstPrint.py}) dont le processus s'abonne à tous les processus des composants qui sont des publieurs. Chaque script peut en fait implémenter des processus publieur, abonné, serveur et client en parallèle (via la technique de \textit{Threading}\footnote{\url{https://fr.wikipedia.org/wiki/Thread_(informatique)}}).

%%%%%%%%%%%%%%%%
\subsection{La gestion des flux de données avec la librairie pyMilk}
\label{sec:SHM}

La librairie écrite en Python \ac{pyMILK}\footnote{\url{https://github.com/milk-org/pyMilk.git}} permet un interfaçage simple (moins d'une dizaine de lignes de codes suffisent) entre les scripts du logiciel écrits en Python et la librairie \ac{MILK}\footnote{\url{https://github.com/milk-org/milk-package.git}} écrite en langage C et utilisée dans le contrôle du banc \ac{SCExAO} \citep{guyon2020}. Elle permet d'allouer un espace mémoire partagée sur la \ac{RAM} de l'ordinateur (nommé \ac{SHM}) pour le stockage, la lecture, la réduction et l'analyse de données, en temps réel et à haute fréquence.

Ce flux de données se compose :
\begin{itemize}
    \item des données stockées dans l'espace de mémoire partagée \ac{SHM}, qui sont accessibles par n'importe quel processus sur l'ordinateur;
    \item des méta-données composées de compteurs du nombre de données, du temps d'accès et d'écriture des données, des numéros d'identification des processus qui consultent la \ac{SHM} et sont aussi accessibles par les autres processus;
    \item des sémaphores\footnote{\url{https://fr.wikipedia.org/wiki/S\%C3\%A9maphore_(informatique)}} qui permettent la synchronisation entre le processus qui écrit et celui qui lit.
\end{itemize}

Ces sémaphores permettent aux processus qui lisent les \ac{SHM}s de se mettre en attente de manière passive jusqu'à ce qu'une nouvelle donnée soit écrite et, lorsqu'elle est écrite, de reprendre le cours de l'exécution. Cette pratique permet ainsi d'économiser des ressources et du temps de calcul. De plus, en les utilisant en combinaison avec la connaissance de la durée (qui est en fait quasi-déterministe) que prend une action sur le système d'exploitation de l'ordinateur et avec l'étalonnage de la latence de ce système, il est possible de synchroniser plusieurs éléments logiciels ou matériels (comme un miroir déformable) à l'horloge interne d'une caméra à des fréquences de plus de $1 \,$kHz et avec une précision de quelques $\upmu$s rms. C'est ce qui est implémenté pour faire fonctionner l'optique adaptative sur \ac{SCExAO} et il a été mesuré des fréquences de fonctionnement de la caméra et du miroir déformable en synchronisation jusqu'à $\sim 10 \,$kHz et plus. Un ordinateur doté d'une puissance de calcul suffisante pour faire une telle synchronisation a récemment été acheté pour le projet \ac{FIRST} sur \ac{SCExAO}. Lors de futurs développements, il s'agira de mettre en place la modulation des franges à une fréquence de l'ordre du kHz, permettant la mesure des interférogrammes sans que les franges subissent les perturbations de phase de l'environnement du banc et des résidus de l'optique adaptative.

Dans le logiciel de contrôle du banc \ac{FIRSTv2}, cette librairie est cruciale pour échanger des données entre différents processus de manière fluide et rapide. En effet, par exemple, pour la modulation des franges, une \ac{SHM} (nommée \og mems modulation \fg~sur la figure~\ref{fig:SoftwareArchitecture} du bas) est créée par le processus de la caméra. Celle-ci contient un nombre qui est incrémenté par la caméra après l'acquisition de chaque image et le processus associé au \ac{MEMS} change les segments de positions en conséquence. Cela permet ainsi de parfaitement synchroniser la caméra et le \ac{MEMS} pour la modulation rapide des franges (typiquement à $20 \,$Hz mais pouvant aller jusqu'à la centaine de Hz). Mais encore, ce principe est utilisé pour l'optimisation de l'injection du flux dans les fibres par le logiciel. En effet, le processus associé au \ac{MEMS} se connecte à la mémoire partagée des images de la caméra (nommée \og live feed \fg) et à chaque fois qu'un segment est déplacé en tip-tilt, une image de la caméra est récupérée pour estimer l'intensité du flux des sorties afin de construire les cartes de transmission des fibres optiques (pour plus de détails voir la section~\ref{sec:OptiInj}). La même chose est implémentée sur le programme associé aux \ac{ODL}s pour la recherche des franges.

La librairie dispose aussi d'un module de visualisation de données qui passent par les \ac{SHM}s, qui est utilisé par le logiciel de \ac{FIRSTv2} pour visualiser en temps réel le flux d'images de la caméra ainsi que la carte de phase instantanée de la surface du \ac{MEMS}. Le module ouvre une fenêtre sur l'écran de l'ordinateur, dans laquelle la dernière image est affichée. Ce sont les deux fenêtres sur la droite de la capture d'écran présentée sur la figure~\ref{fig:SoftwareScreenShot}.

%%%%%%%%%%%%%%%%%%%%%%%%%%%%%%%%
\section{La configuration des bases}
\label{sec:BaseConfig}

Les sous-pupilles choisies sur le banc de test lors de l'acquisition de toutes les données, qui sont présentées et analysées par la suite, sont représentées sur la figure~\ref{fig:SegUVSimuleA} à l'aide de la carte des segments du \ac{MEMS}. Les segments considérés pour former les bases sont en orange et le plan UV correspondant est tracé sur la figure~\ref{fig:SegUVSimuleB}. J'ai choisi ces sous-pupilles pour échantillonner au mieux le plan UV dans la direction horizontale, qui correspond à la direction du système simulé. En effet, comme l'indique l'équation~\ref{eq:PhaseBinaireCentree} les signaux de phases dépendent de la projection des bases sur l'orientation du système protoplanétaire observé et étant orienté horizontalement par rapport à la carte des segments (dans la même direction que la base $37-33$ par exemple), il n'y a bien que trois signaux d'intensités différentes qui peuvent être mesurés. Ces trois types de signaux proviennent des groupes de bases suivants :
\begin{itemize}
    \item les bases qui sont orthogonales à la source et qui donnent un signal nul e.g. les bases $7-26$ et $15-28$;
    \item les bases projetées sur l'orientation de la source qui s'étendent sur la longueur d'un segment et demi e.g. $37-7$, $37-26$, $7-15$, $7-28$, $26-15$ et $26-28$, de longueur égale à $1,05 \,$mm 2,1;
    \item et les bases projetées sur l'orientation de la source qui s'étendent sur la longueur de trois segments e.g. $37-15$ et $37-28$, de longueur égale à $2,10 \,$mm, équivalentes aux plus grandes bases sur le télescope Subaru de longueur $6,86 \,$m.
\end{itemize}

\begin{figure}[ht!]
    \centering
    \begin{subfigure}{0.39\textwidth}
        \centering
        \includegraphics[width=\textwidth]{Figure_Chap2/BaselineMap_Meudon_37_7_26_15_28.png}
        \caption{Configuration des sous-pupilles choisies (en orange) dans le plan pupille sur la carte des segments du MEMS.}
        \label{fig:SegUVSimuleA}
    \end{subfigure}\hfill
    \begin{subfigure}{0.59\textwidth}
        \centering
        \includegraphics[width=\textwidth]{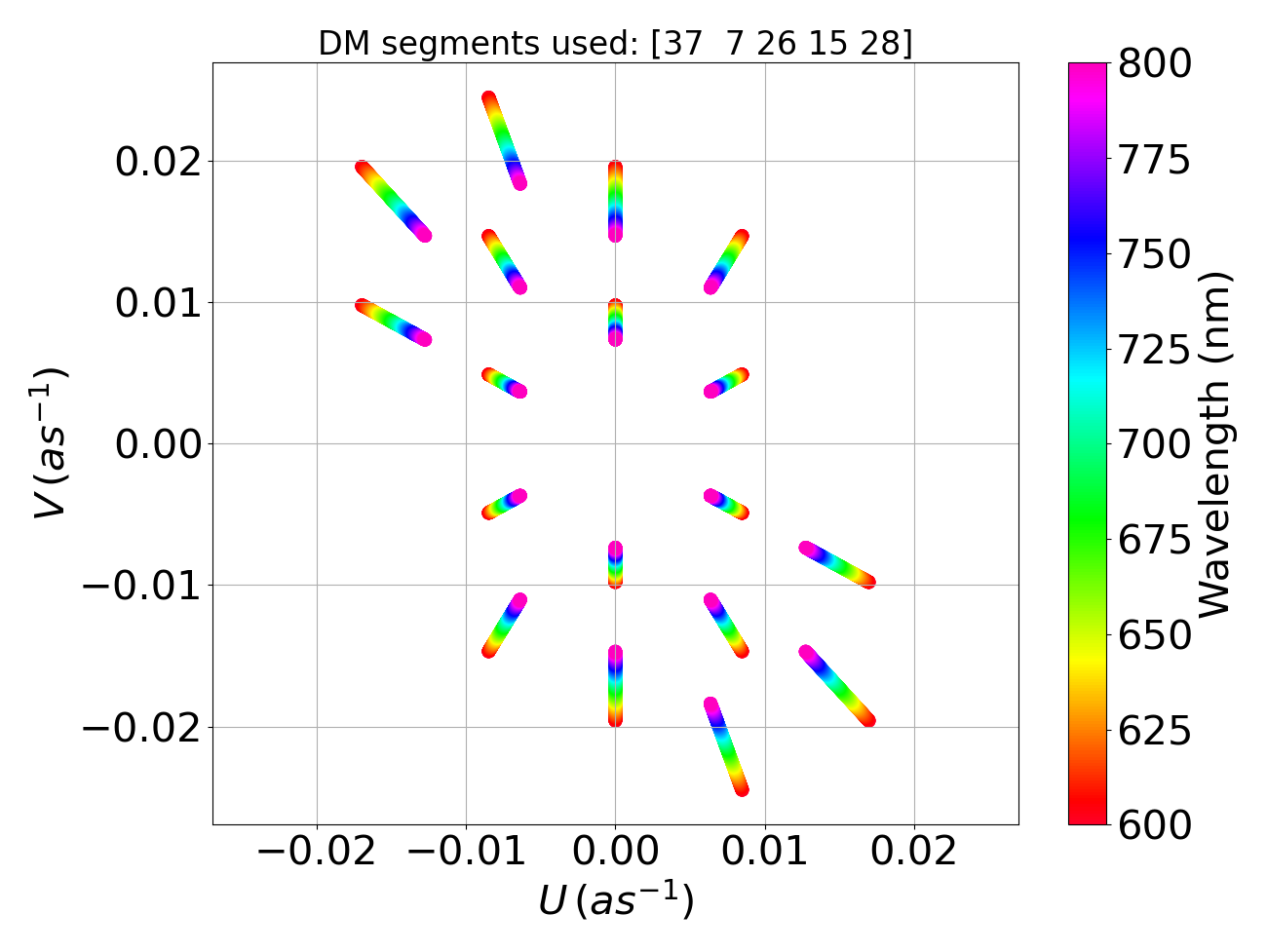}
        \caption{La répartition des bases choisies représentée dans l'espace de Fourier, appelée aussi la couverture du plan UV des fréquences spatiales. Les couleurs représentent la longueur d'onde.}
        \label{fig:SegUVSimuleB}
    \end{subfigure}
    \caption[Configuration des sous-pupilles et couverture du plan UV du banc de test de FIRSTv2.]{Configuration des sous-pupilles et couverture du plan UV du banc de test de FIRSTv2.}
    \label{fig:SegUVSimule}
\end{figure}

Historiquement, sur les instruments implémentant la technique de masquage de pupille, les bases sont choisies non-redondantes pour éviter le brouillage des franges (voir la section~\ref{sec:PupilMasking}). Ici, les bases peuvent être choisies redondantes car on utilise le principe de réarrangement de pupille ce qui empêche les interférogrammes provenant de chaque base de se superposer sur le détecteur. Il n'y a donc pas de risque que les franges se brouillent. Cependant, lorsque l'instrument sera utilisé sur une cible astrophysique, les bases seront choisies de manière la plus non-redondante possible pour que la couverture du plan UV soit maximisée et homogène dans le but de sonder le plus possible de fréquences spatiales différentes. C'est ce que j'ai fait ici et les bases $37-7$ et $7-15$ sont les seules à être redondantes. Finalement, seulement deux fréquences spatiales sont sondées (correspondant à deux longueurs de base projetée différentes) car le montage du banc n'en permet pas plus du fait que seul les segments $37 - 9 - 24 - 35 - 7 - 4 - 26 - 33 - 15 - 28$ peuvent être utilisés (pour plus de détails, voir la section~\ref{sec:FiberInjection}). Enfin, on note que la couverture du plan UV présentée sur la figure~\ref{fig:SegUVSimuleB} est calculée pour toute la bande spectrale transmise par l'instrument, mais que lors de la détection de protoplanètes grâce à leur raie d'émission \ha, la couverture de ce plan UV effective est alors réduite aux canaux spectraux correspondant.

%%%%%%%%%%%%%%%%%%%%%%%%%%%%%%%%
\section{Conclusion}

Ce chapitre nous a permis de présenter les avancées de l'intégration de la technologie d'optique intégrée sur le banc de test \ac{FIRSTv2} tout en le présentant dans sa globalité. Une partie de mon travail de thèse, qui est présentée dans ce chapitre, est la caractérisation en transmission, diaphotie, contraste interférométrique et polarisation de deux puces photoniques utilisant deux concepts de recombinaison différentes : couplage directionnel et jonction $Y$. La transmission de la première est estimée à $30\%$ et celle de la deuxième à $13\%$, ce qui nous a conduit à décider de les intégrer sur le banc \ac{SCExAO} pour la première lumière de \ac{FIRSTv2} (présentée dans le chapitre~\ref{sec:FIRSTv2Subaru}). Ainsi, de nouvelles puces sont actuellement en cours de développement à l'\ac{IPAG} pour améliorer ces performances, dans l'objectif d'atteindre une transmission d'au moins $75\%$.

Dans le même but d'augmenter la transmission de l'instrument, il est envisagé de retirer les lignes à retard car leur transmission s'avère être trop faible. Pour cela, on pourra concevoir des fibres de compensation qui égaliseraient tous les chemins optiques ou opter pour une nouvelle technologie de puce photonique telle qu'un composant 3D ou une lanterne photonique qui permettraient aussi de se passer du toron de fibres en plus des \ac{ODL}s.

Une autre partie de mon travail est la restructuration et la continuation du développement du logiciel de contrôle du banc de test. Cela a permis d'améliorer sa rapidité d'exécution, sa modularité ainsi que d'ajouter des processus d'acquisitions automatiques de données interférométriques. Je l'ai ainsi déployé sur l'ordinateur de contrôle de \ac{FIRSTv1} au télescope Subaru et il a été utilisé lors de quelques nuits d'observations avec succès.

%%%%%%%%%%%%%%%%%%%%%%%%%%%%%%%%
\clearpage
\section*{Conférence SPIE}
\label{sec:SPIEproceeding}
\phantomsection
\addcontentsline{toc}{section}{Conférence SPIE}

\clearpage
\includepdf[pages=-, pagecommand={}, offset=-10 -20, templatesize={0.8\textwidth}{0.8\textheight}]{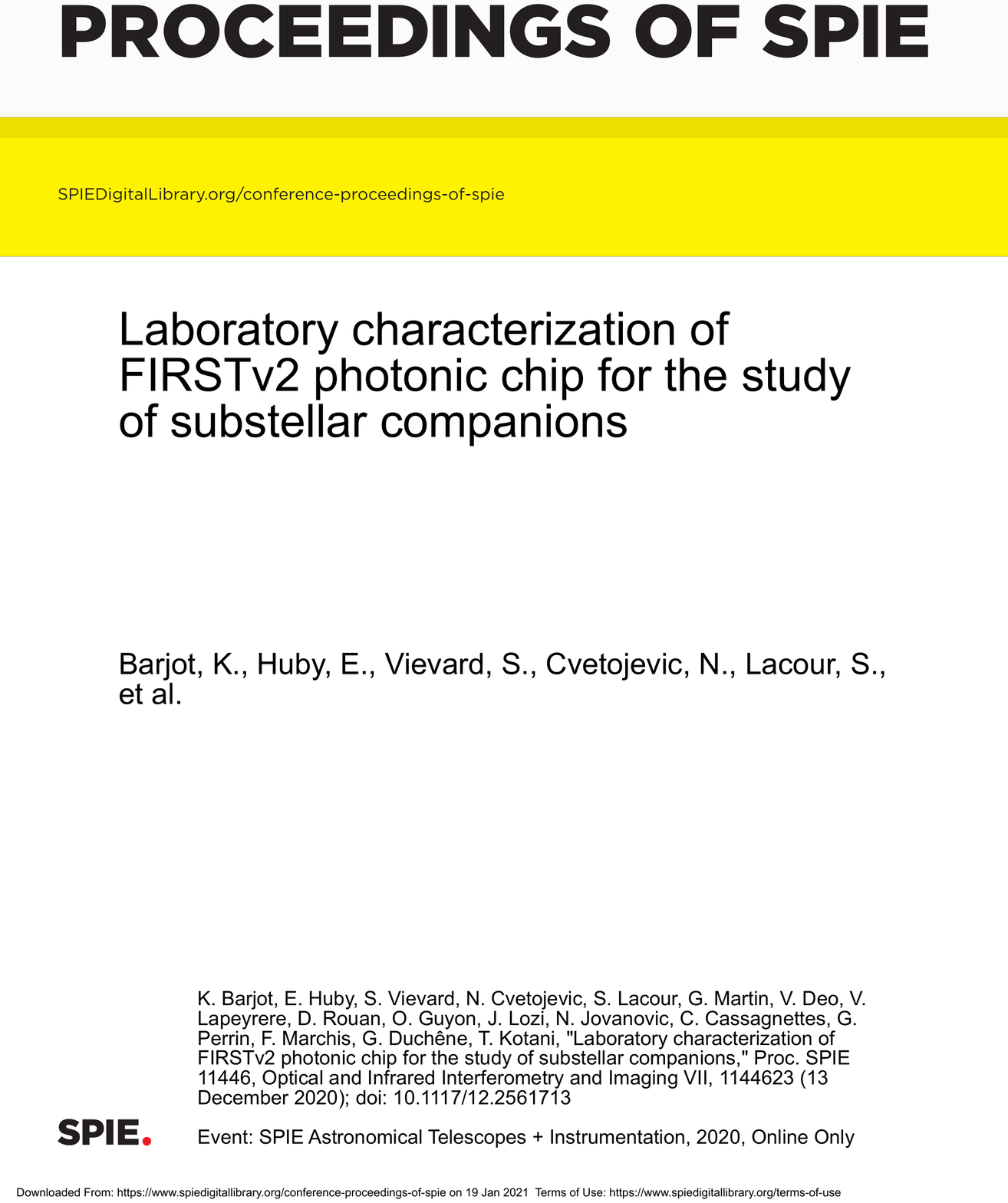}

%%%%%%%%%%%%%%%%%%%%%%%%%%%%%%%%%%%%%%%%%%%%%%%%%%%%%%%%%%%%%%%%
\chapter{Le traitement des données FIRSTv2}
\label{sec:DataReduction}
\setcounter{figure}{0}
\setcounter{table}{0}
\setcounter{equation}{0}

\minitoc

\clearpage
L'objectif de cette section est de présenter les données de \ac{FIRSTv2}, depuis leur acquisition et les procédures mises en place dans ce cadre jusqu'à leur traitement. Il s'agit aussi de présenter le formalisme mathématique de ce traitement de données afin d'établir les observables que l'on infère des données de \ac{FIRSTv2}.

% On the sensitivity of closure phases to faint companions in optical long baseline interferometry https://ui.adsabs.harvard.edu/abs/2012A%26A...541A..89L/abstract

%%%%%%%%%%%%%%%%%%%%%%%%%%%%%%%%
\section{L'acquisition des données}

%%%%%%%%%%%%%%%%
\subsection{Morphologie des images}

%%%%%%%%
\subsubsection{Étalonnage du détecteur}
\label{sec:CameraDark}

Comme décrit dans la section~\ref{sec:InstruCamera}, les images sont d'une taille de $2\,160 \,$px de hauteur par $2\,559 \,$px de largeur. La figure~\ref{fig:DarkFull} montre, à gauche, une image sans frange prise avec un temps d'exposition de $100 \,$ms et à droite la médiane d'un cube de $100$ images sans frange au même temps d'exposition. Ces images permettent de caractériser le courant d'obscurité et le bruit du détecteur de la caméra. 

Un tel cube sans frange est systématiquement acquis à chaque prise de données interférométriques. Le début du traitement des images consiste au calcul de la médiane de ce cube sans frange qui est ensuite soustraite aux images interférométriques diminuant ainsi le biais du détecteur et les structures fixes mentionnées dans le paragraphe précédent, présents sur toutes les données. Afin d'effectuer cette soustraction sans se soucier d'une possible dépendance non linéaire du courant d'obscurité du détecteur en fonction du temps d'exposition, ce dernier est choisi identique pour les deux types de données (sans et avec franges).

\begin{figure}[ht!]
    \begin{subfigure}{.5\textwidth}
        \centering
        \includegraphics[width=\textwidth]{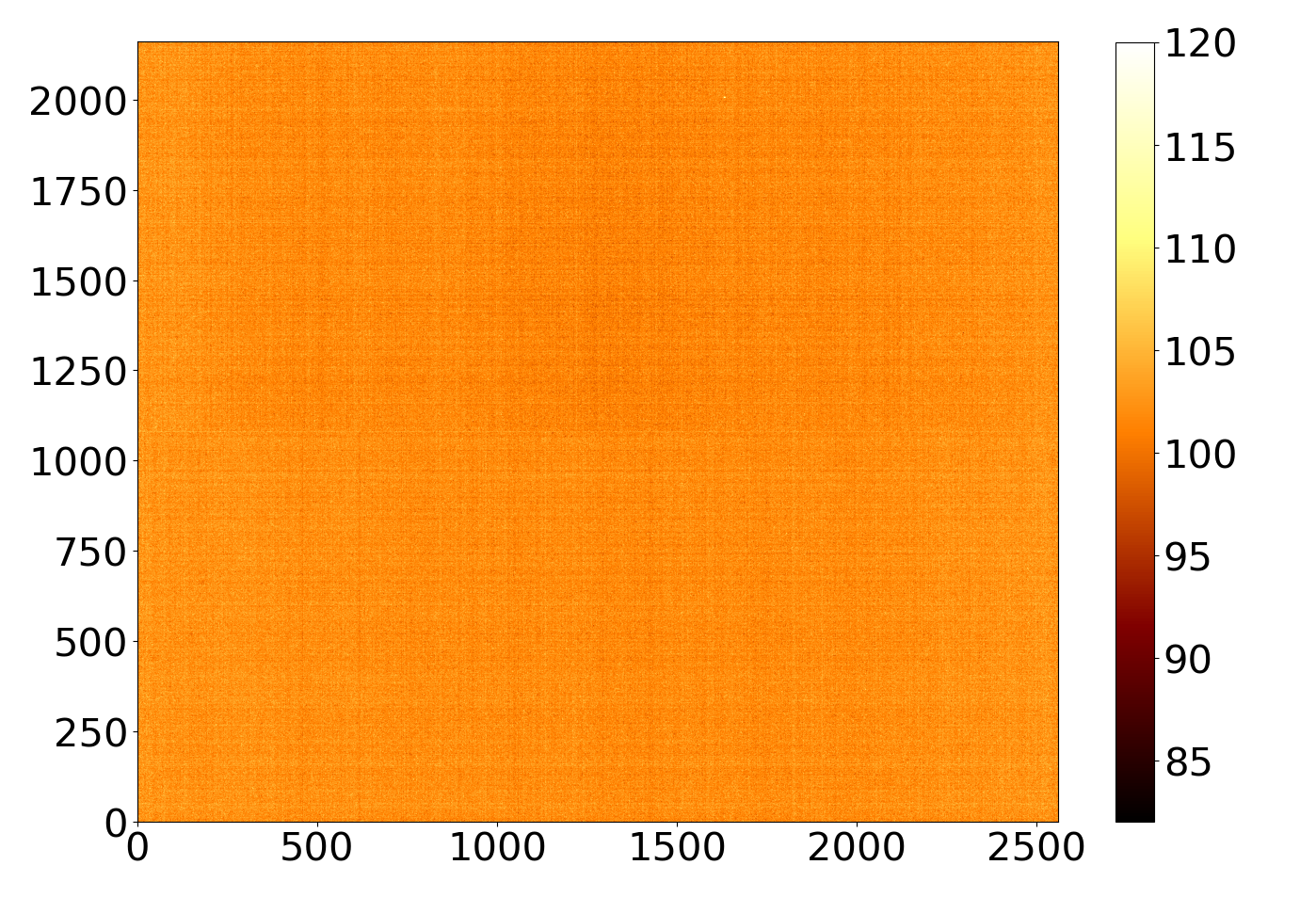}
    \end{subfigure}%
    \begin{subfigure}{.5\textwidth}
        \centering
        \includegraphics[width=\textwidth]{Figure_Chap3/20220705_DarkFullImage_100ms_24C_median_clip.png}
    \end{subfigure}
    \caption[Images du courant d'obscurité de la caméra de FIRSTv2.]{A gauche une image de la caméra non illuminée et à droite la médiane de $100$ de ces images, pour le même temps d'exposition de $100 \,$ms. Les valeurs des pixels données sur les échelles sont en \ac{ADU}.}
    \label{fig:DarkFull}
\end{figure}

%%%%%%%%
\subsubsection{Les images interférométriques}

Les images acquises avec la caméra sont rognées pour cadrer les sorties de la puce d'optique intégrée. La puce $Y$ a $10$ sorties et la puce $X$ a $20$ sorties, qui peuvent être dédoublées lorsqu'elles sont imagées par la caméra quand le prisme de wollaston est installé sur le chemin optique. Les différentes configurations possibles peuvent donc amener à avoir $10$, $20$ ou $40$ sorties à cadrer. Ce rognage dépend aussi de la bande spectrale de la source utilisée (changeant la largeur de l'image) et est choisi de façon à encadrer celle-ci le plus largement possible. Pour chaque montage sur le banc toutes les images (sans franges, \textit{flats} ou interférométriques) sont acquises avec le même rognage. La figure~\ref{fig:FringeCrop} présente deux images obtenues avec la caméra avec les sorties de la puce $X$ imagées sans que le prisme de wollaston soit installé ($20$ sorties sont donc visibles). La \sk est la source utilisée et les images ont alors une taille de $320 \,$px par $1\,800 \,$px ($1\,800$ canaux spectraux). En haut est montrée la médiane de $300$ images sans frange, qui est soustraite à une image avec des franges dont le résultat est montré en bas, pour un temps d'exposition de $50 \,$ms.

\begin{figure}[ht!]
    \centering
    \begin{subfigure}{\textwidth}
        \includegraphics[width=\textwidth]{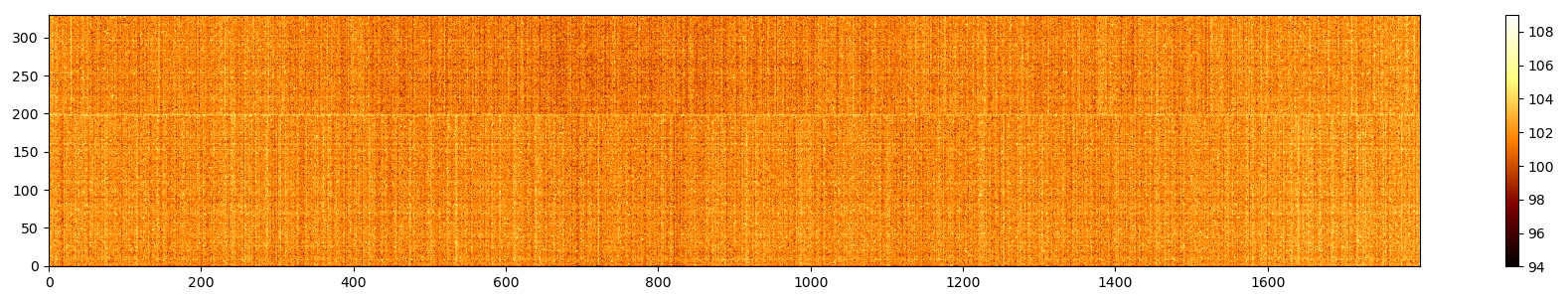}
    \end{subfigure}
    \begin{subfigure}{\textwidth}
        \includegraphics[width=\textwidth]{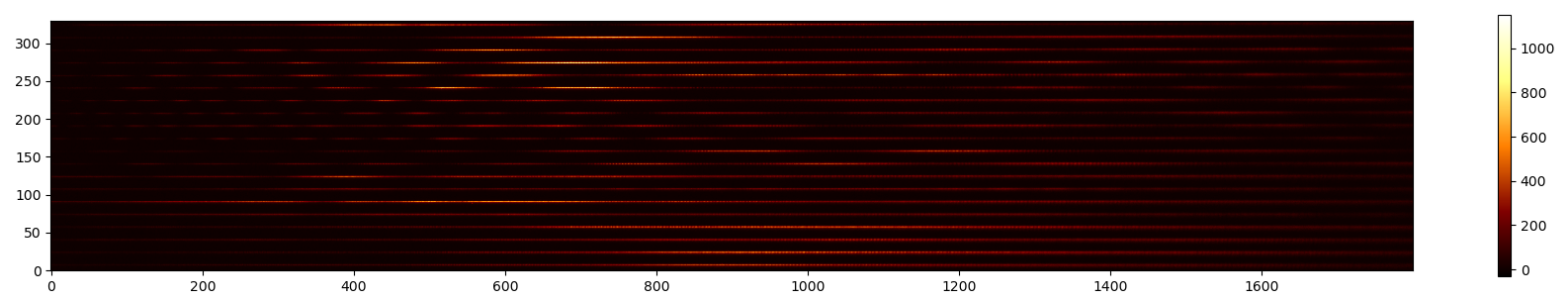}
    \end{subfigure}
    \caption[Image interférométrique obtenues sur FIRSTv2.]{En haut la médiane de $300$ images de la caméra sans franges et en bas une image avec des franges, avec la source \sk. Les deux images ont été acquises avec un temps d'exposition de $50 \,$ms et le même rognage induisant une taille d'image de $320 \, px$ de hauteur par $1\,800 \, px$ de largeur.}
    \label{fig:FringeCrop}
\end{figure}

Les lignes à retard sont au voisinage de leur position d'\ac{OPD} nulle, ce qui se traduit par la présence de franges sur l'image du bas. Il y a donc pour chacun des $1\,800$ canaux spectraux, un point de mesure du flux interférométrique pour chaque sortie, correspondant à une valeur d'\ac{OPD}. L'objectif est alors de reconstruire toutes les figures d'interférences par la mesure du flux des sorties pour au moins $4$ valeurs d'\ac{OPD}s par base. A cette fin, nous avons besoin de contrôler en piston les cinq segments du \ac{MEMS} qui illuminent les cinq entrées de la puce selon la méthode décrite dans la section~\ref{sec:Modulation}.

%%%%%%%%%%%%%%%%
\subsection{Étalonnage spectral}
\label{sec:EtalonnageSpectral}

Ici on souhaite associer à chaque pixel une valeur de longueur d'onde. Pour cela, j'utilise une lampe à vapeur de néon (Avalight-CAL-NEON par Avantes, avec sortie fibrée FC-PC). Le spectre d'émission de la vapeur de Néon est connu et présente de nombreuses raies sur la bande spectrale détectée par \ac{FIRSTv2}. La figure~\ref{fig:NeonReference} montre deux spectres de référence de l'émission de la vapeur de Néon, que j'ai utilisés pour les étalonnages spectraux durant ma thèse, celui de gauche étant un spectre que j'ai trouvé sur le site internet Astrosurf\footnote{\url{http://www.astrosurf.com/buil/us/spe2/calib2/neon1.gif}} et le deuxième étant celui mesuré par \ac{FIRSTv1} pendant la thèse d'Elsa Huby. Les deux spectres présentent des pics d'émission à des longueurs d'onde différentes et permettent donc d'identifier toutes les raies mesurées par \ac{FIRSTv2}.

\begin{figure}[ht!]
    \centering
    \begin{subfigure}{0.9\textwidth}
        \includegraphics[width=\textwidth]{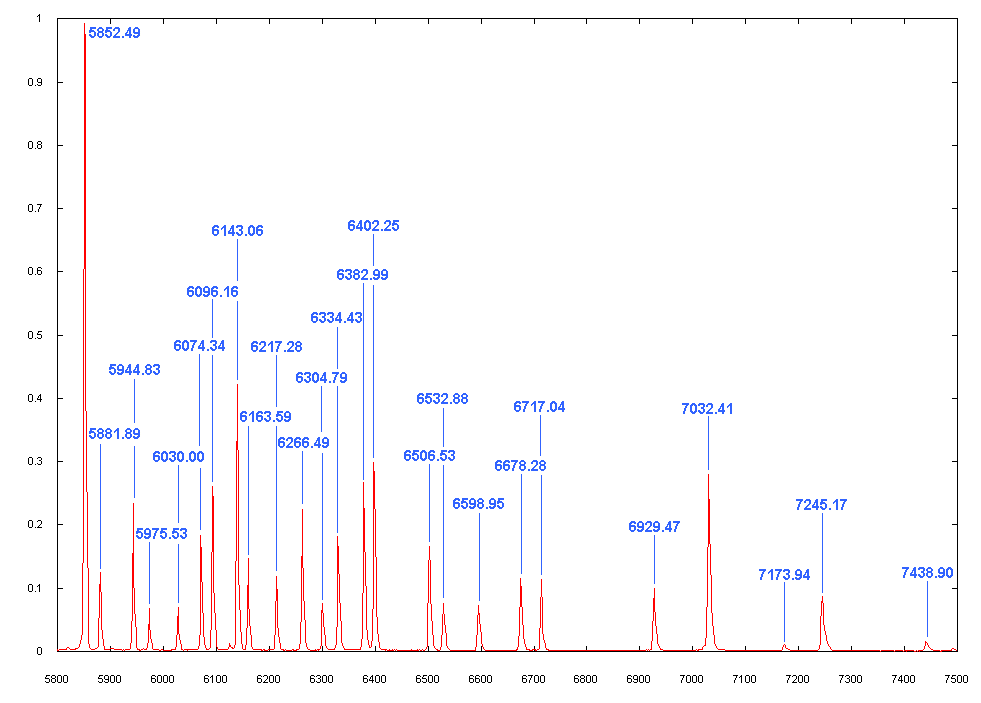}
    \end{subfigure}
    \begin{subfigure}{0.9\textwidth}
        \includegraphics[width=\textwidth]{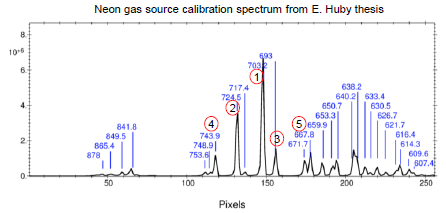}
    \end{subfigure}
    \caption[Spectres de référence d'émission du Néon utilisés pour l'étalonnage spectral de FIRSTv2.]{Spectres de références d'émission du Néon utilisés pour l'étalonnage spectral de FIRSTv2. À gauche est un spectre trouvé sur le site Astrosurf, présentant l'intensité des raies d'émissions du Néon (axe des ordonnées) en fonction de la longueur d'onde en Angstrom (axe des abscisses). À droite est le spectre tel que mesuré par FIRSTv1 pendant la thèse d'Elsa Huby traçant l'intensité des raies d'émission du Néon (axe des ordonnées) en fonction des pixels (axe des abscisses) de la caméra utilisée. Sur les deux spectres la longueur d'onde des pics est notée sur le graphique au-dessus de chaque pic.}
    \label{fig:NeonReference}
\end{figure}

La méthodologie utilisée pour conduire l'étalonnage spectral est la suivante :

\begin{enumerate}
    \item cinq séquences d'images sont acquises en illuminant successivement les cinq entrées de la puce avec la lampe néon;
    \item la somme des cinq moyennes de ces séquences est calculée pour donner une unique image avec toutes les sorties de la puce;
    \item une première étape d'identification des raies est effectuée avec une fonction de détection de pics pour chaque sortie;
    \item ensuite ces positions des pics sont passées en paramètres d'une fonction d'ajustement gaussien, afin de calculer finement la position centrale des pics et permettre leur identification en longueur d'onde;
    \item enfin, les longueurs d'ondes des pics sont ajustées aux positions des pics à l'aide d'une fonction polynomiale de degré 4.
\end{enumerate}

La figure~\ref{fig:NeonSpecCal} du haut montre l'image résultante de l'étape \textit{1.} en échelle logarithmique, la courbe du milieu montre l'identification précise des pics à l'issue de l'étape \textit{4.} sur le spectre d'émission de la lampe néon mesurée sur la sortie 1 de \ac{FIRSTv2} et la courbe du bas présente l'ajustement polynomial de l'étape \textit{5.}, pour la sortie 1 également. L'équation du polynôme trouvée par ajustement est montrée dans la légende et on remarque que la répartition des longueurs d'ondes sur les pixels du détecteur suit une loi affine. Cet ajustement a l'avantage de permettre l'extrapolation du modèle sur l'entièreté du détecteur, ce qui est utile lorsqu'on utilise la source \sk qui a une plus large bande que la lampe néon. Finalement, nous disposons d'autant d'équations de cet ajustement qu'il y a de sorties imagées sur la caméra et elles seront utilisées dans la suite du traitement de données, à chaque calcul en fonction de la longueur d'onde (notamment sur les phases) ou à chaque fois que des courbes seront tracées en fonction de la longueur d'onde. Enfin, il est indiqué dans le titre du graphique du milieu la résolution spectrale mesurée lors de l'étalonnage, qui est ici de $3\,184$ à $651 \,$nm. Elle a été mesurée sur toutes les sorties en moyenne à $3\,254$ avec un écart-type de $127$.

\begin{figure}[ht!]
    \centering
    \begin{subfigure}{0.8\textwidth}
        \includegraphics[width=\textwidth]{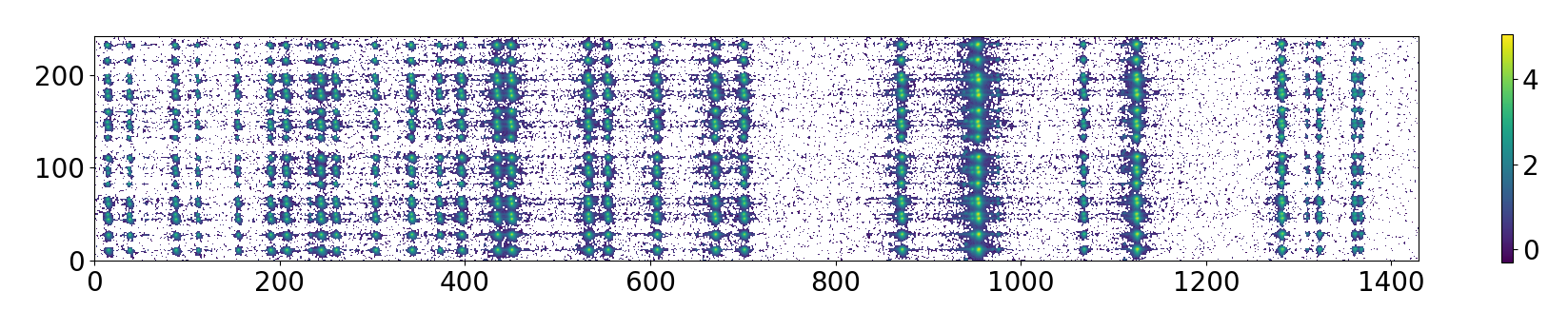}
    \end{subfigure}
    \begin{subfigure}{0.8\textwidth}
        \includegraphics[width=\textwidth]{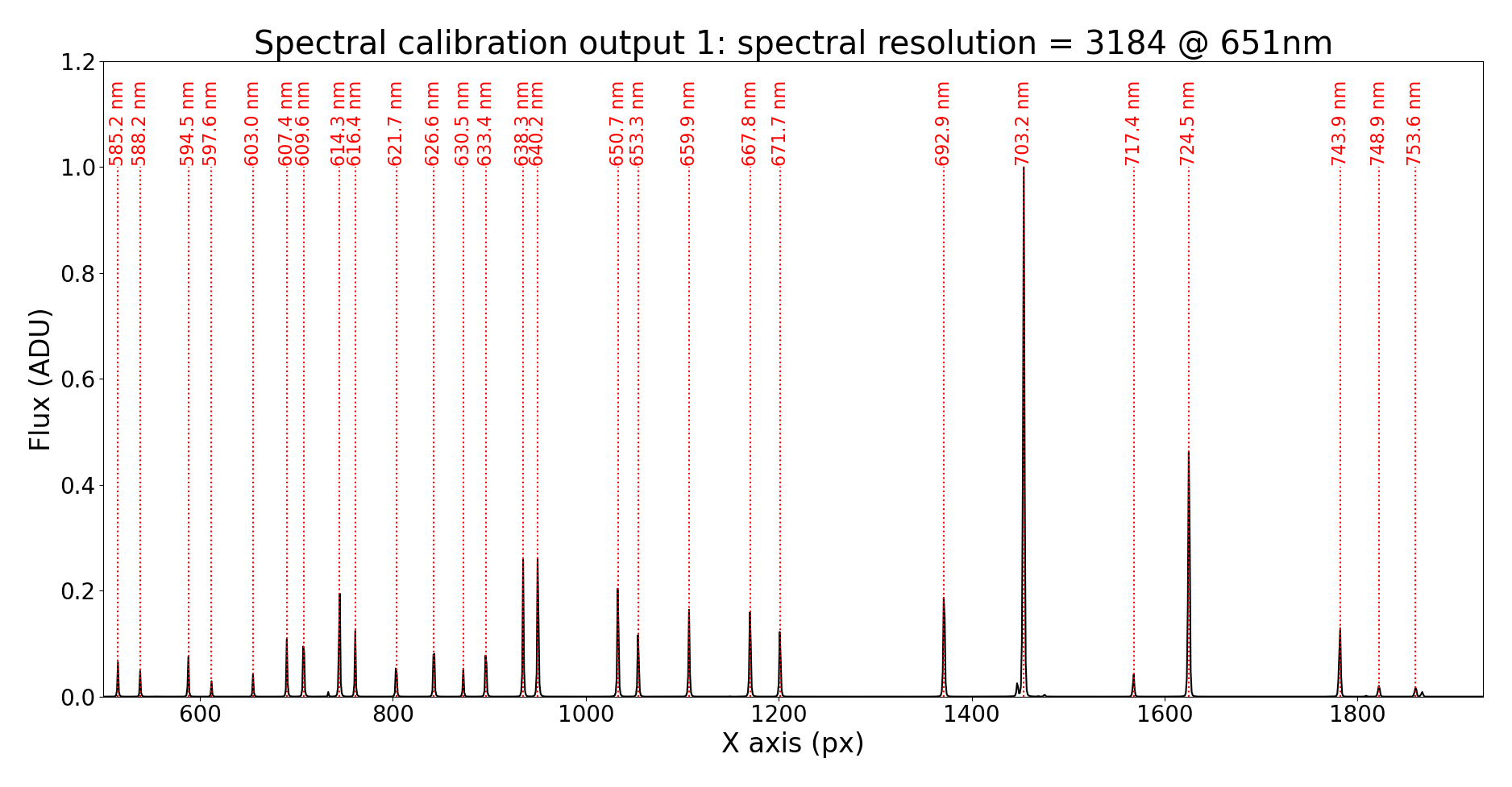}
    \end{subfigure}
    \begin{subfigure}{0.8\textwidth}
        \includegraphics[width=\textwidth]{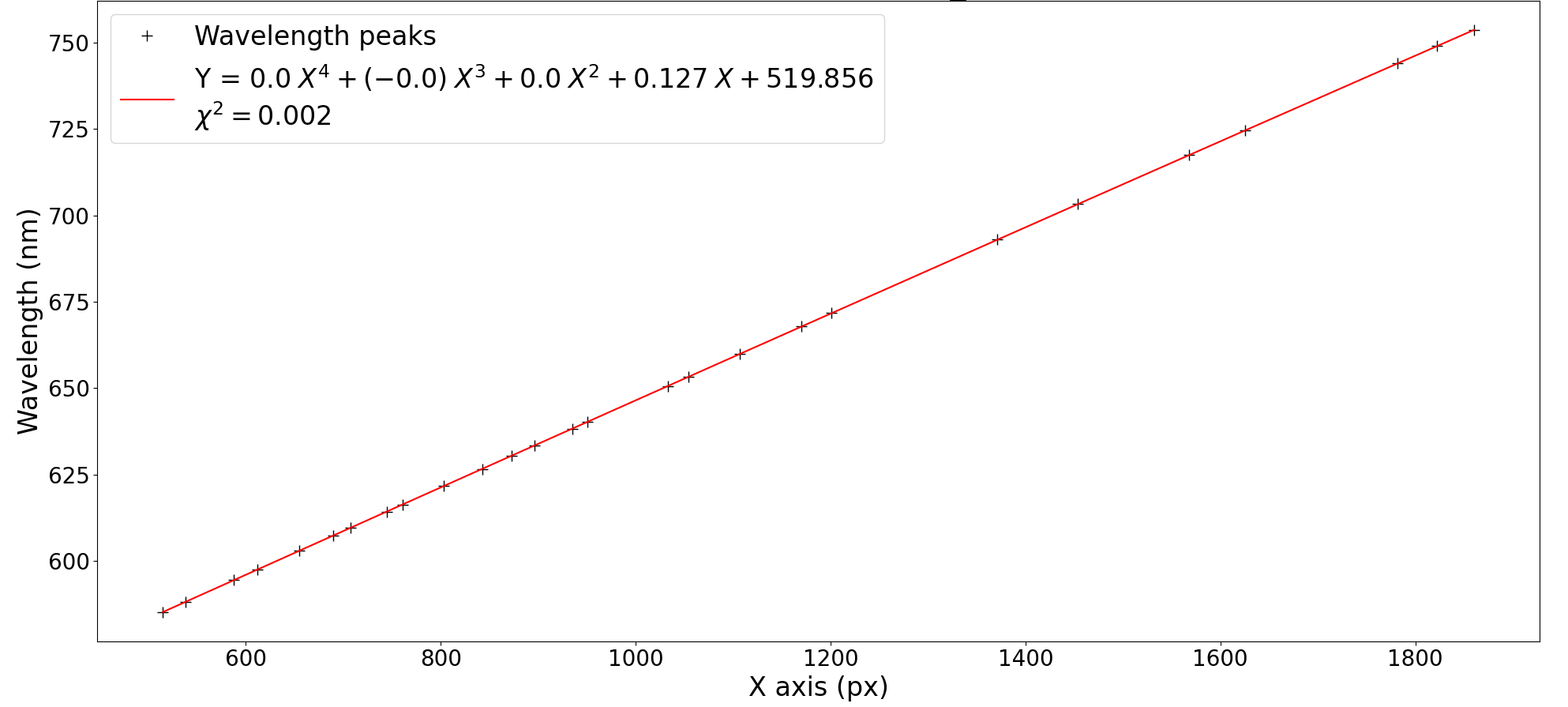}
    \end{subfigure}
    \caption[Résultat de l'étalonnage spectrale de FIRSTv2 avec une lampe spectrale au néon.]{Étalonnage spectral de FIRSTv2 à l'aide d'une lampe spectrale au néon. En haut : image de la caméra avec le spectre de la lampe néon sur chaque sortie. Au milieu : identification des pics du spectre du néon sur les pixels de la caméra pour la sortie 1. En bas : les longueurs d'onde du néon en fonction des positions des pics sur la caméra ainsi que l'ajustement polynomial associé, de la sortie 1.}
    \label{fig:NeonSpecCal}
\end{figure}

%%%%%%%%%%%%%%%%
\subsection{La modulation des franges}
\label{sec:Modulation}

L'acquisition des données interférométriques s'effectue en modulant les franges à l'aide des \ac{MEMS}. Ces derniers sont synchronisés avec la caméra de telle sorte que chaque image d'une séquence contienne les franges pour une valeur différente d'\ac{OPD}. La figure~\ref{fig:FringeSamplingABCD} montre la méthode ABCD utilisée pour mesurer une frange, représentée sur le graphique par une période de la fonction sinusoïdale ($2 \pi$). Cette période est divisée en $8$ et un déphasage de $\frac{3}{8} 2 \pi \,$rad, $\frac{-1}{8} 2 \pi \,$rad, $\frac{1}{8} 2 \pi \,$rad et $\frac{-3}{8} 2 \pi \,$rad est appliqué sur chaque base pour mesurer quatre points de la frange. Pour cela on applique quatre valeurs de piston sur deux segments du \ac{MEMS}, définis comme multiples de la longueur d'onde, prise égale à $750 \,$nm : $\frac{3}{8} \frac{0.750}{4} \,$\um, $\frac{-1}{8} \frac{0.750}{4} \,$\um, $\frac{1}{8} \frac{0.750}{4} \,$\um~et $\frac{-3}{8} \frac{0.750}{4} \,$\um. Comme ce sont les valeurs mises en consignes des segments, deux facteurs $1/2$ sont présents pour prendre en compte que (1) l'\ac{OPD} effective est la somme des déplacements de $2$ segments ainsi que (2) la lumière parcourt deux fois la même distance en étant réfléchie sur les segments.

\begin{figure}[htp!]
    \centering
    \includegraphics[width=\figwidth]{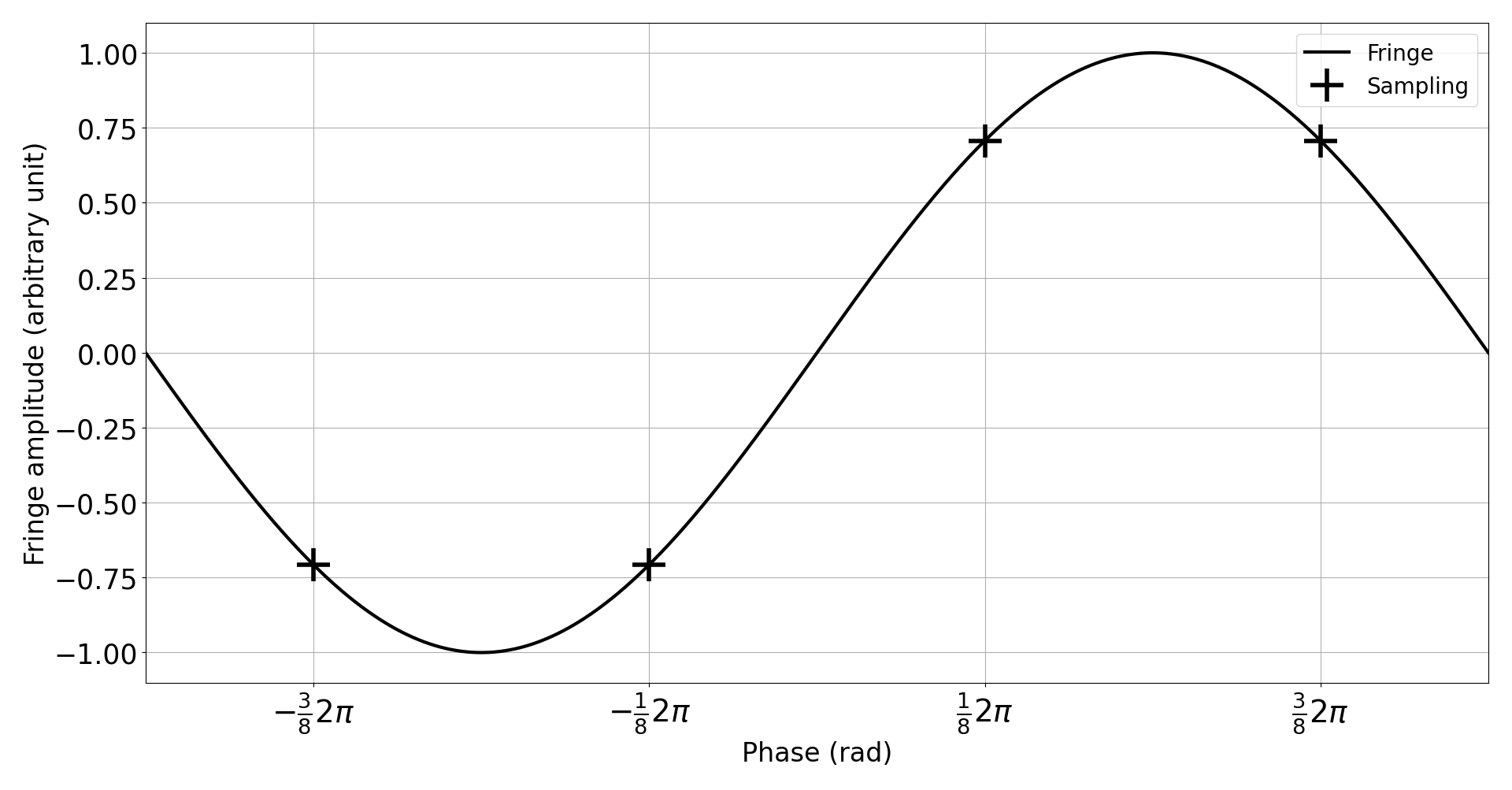}
    \caption[Méthode ABCD d'échantillonnage d'une frange.]{Méthode ABCD d'échantillonnage d'une frange. Ici la frange est représentée par une période de la fonction sinusoïdale ($2 \pi$) et pour l'échantillonner, les quatre valeurs de déphasage égales à $\frac{3}{8} 2 \pi \,$rad, $\frac{-1}{8} 2 \pi \,$rad, $\frac{1}{8} 2 \pi \,$rad et $\frac{-3}{8} 2 \pi \,$rad sont appliquées par les segments du MEMS.}
    \label{fig:FringeSamplingABCD}
\end{figure}

Une première approche est d'appliquer quatre valeurs de pistons, successivement, par paire de segments, pour effectuer les quatre points de mesures par base, ce qui constitue une séquence de modulation de $20$ points de mesure. La figure~\ref{fig:ModSeq20} du haut montre les valeurs de piston (en couleur) appliquées aux segments (selon l'axe vertical) en fonction des étapes de modulation (selon l'axe horizontal). La figure du bas montre les valeurs effectives d'\ac{OPD}s tout au long de la séquence de modulation pour les dix bases. On remarque qu'il y a deux types de quadruplets de valeurs d'\ac{OPD}s dans cette séquence. Le premier type (aux valeurs les plus élevées sur la figure) contient les valeurs précédemment définies ($\frac{3}{8} \frac{0.750}{4} \,$\um, $\frac{-1}{8} \frac{0.750}{4} \,$\um, $\frac{1}{8} \frac{0.750}{4} \,$\um~et $\frac{-3}{8} \frac{0.750}{4} \,$\um) car elles sont obtenues en déplaçant deux segments et sont appliquées sur les bases $1-2$, $2-3$, $3-4$, $4-5$ et $1-5$. Le deuxième type contient quatre valeurs dont les amplitudes sont moitié moindres que celles du premier type ($\frac{3}{8} \frac{0.750}{8} \,$\um, $\frac{-1}{8} \frac{0.750}{8} \,$\um, $\frac{1}{8} \frac{0.750}{8} \,$\um~et $\frac{-3}{8} \frac{0.750}{8} \,$\um) car elles sont obtenues avec un des deux segments immobile et sont appliquées sur les bases $1-3$, $1-4$, $2-4$, $2-5$ et $3-5$. Par conséquent, la moitié des bases sont modulées sur une OPD deux fois trop petite ce qui ne permet pas un bon échantillonnage des franges et empêchant le code de traitement de données de converger.

\begin{figure}[ht!]
    \centering
    \begin{subfigure}{0.8\textwidth}
        \centering
        \includegraphics[width=\textwidth]{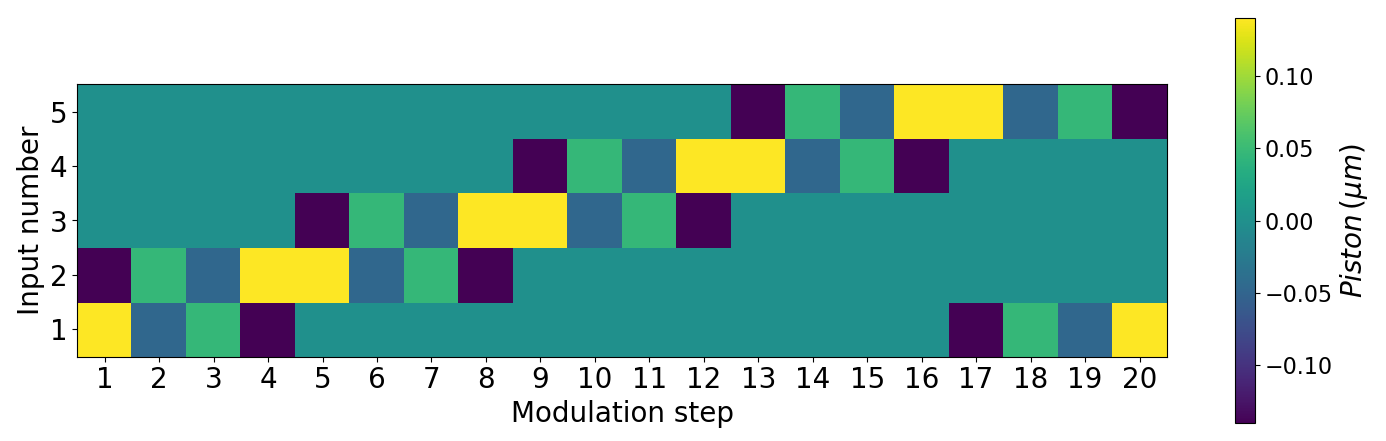}
    \end{subfigure}
    \begin{subfigure}{0.8\textwidth}
        \centering
        \includegraphics[width=\textwidth]{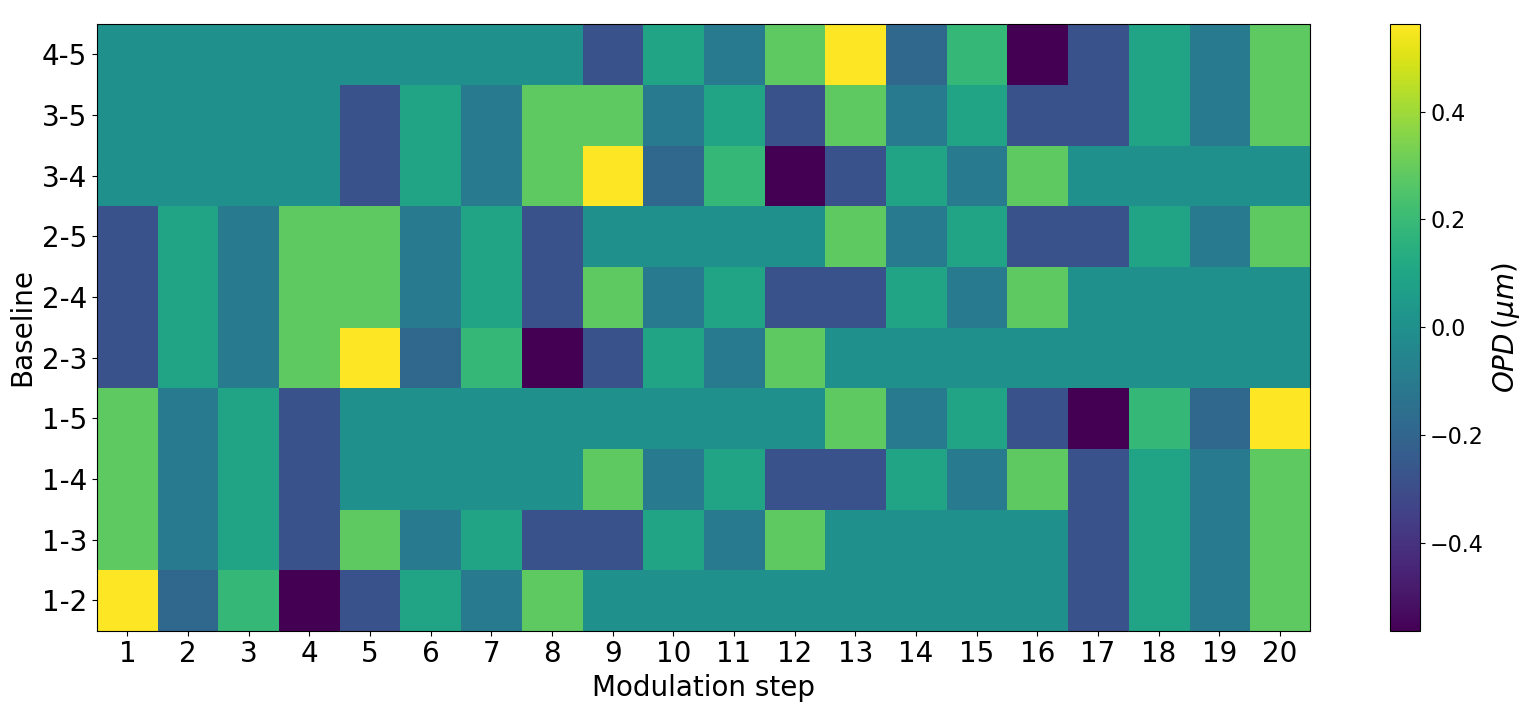}
    \end{subfigure}
    \caption[Séquence basique de modulation des MEMS à 20 pas pour échantillonner les franges sur FIRSTv2.]{En haut la séquence des pistons (en $\upmu$m) appliqués aux cinq segments du MEMS qui illuminent les entrées de la puce. En bas les valeurs d'OPDs (en $\upmu$m) induites sur les dix bases en application de cette séquence de modulation. Les couleurs sont les valeurs appliquées, les axes horizontaux sont les étapes de la séquence et les axes verticaux sont les segments et les bases, respectivement, pour la figure du haut et du bas.}
    \label{fig:ModSeq20}
\end{figure}

Une deuxième approche est de moduler plusieurs bases en même temps et non une seule à la fois comme précédemment. Il faut alors s'assurer que chaque base soit modulée au moins une fois par le bon quadruplet de pistons durant la séquence de modulation. La figure~\ref{fig:ModSeq12} montre un exemple d'une telle séquence. De même que sur la figure~\ref{fig:ModSeq20}, en haut est présentée la séquence des pistons appliqués aux segments. Sur les quatre premiers pas, tous les segments sont modulés, la condition étant que la moitié des segments soient modulés avec des valeurs opposées à l'autre moitié (ici trois segments sont modulés de la même façon et deux sont modulés avec la valeur opposée). A partir de la figure du bas on peut voir les valeurs d'\ac{OPD}s induites sur les bases pour ces quatre premières étapes, nous permettant de définir les bases sur lesquelles appliquer le quadruplet pour les étapes suivantes. Ici seules les bases $1-3$, $1-5$, $2-4$ et $3-5$ ne sont pas modulées et doivent donc l'être par la suite. Ces quatre bases ne pouvant pas être modulées sur le bon quadruplet d'\ac{OPD}s en même temps, il faut alors les moduler sur huit étapes de plus. La séquence résultante est ainsi composée de $12$ étapes. Cette nouvelle séquence a les deux atouts suivants :

\begin{enumerate}
    \item elle permet la modulation des franges sur les quatre points d'échantillonnage, à la bonne amplitude;
    \item elle a aussi le bon goût d'être plus courte ($12$ étapes contre $20$) ce qui permet de réduire l'influence des perturbations atmosphériques (lorsque l'instrument fait des mesures sur ciel) et/ou présents sur le banc (mouvements d'air, vibrations, etc.).
\end{enumerate}

On note que la longueur de cette séquence peut être divisée par deux lorsqu'on utilise la puce $X$. En effet, avec cette puce, chaque image contient deux mesures de frange, théoriquement déphasées de $\pi \,$rad (voir la section~\ref{sec:ChipConcept}). Mais au cours de cette thèse, cette propriété de la puce $X$ n'a pas été utilisée et les interférogrammes mesurés avec cette puce ont été exploités séparément car il serait nécessaire d'écrire une partie en plus dans le programme de réduction de données et de faire de nouveaux tests pour vérifier la valeur du déphasage entre les couples de sorties. La même séquence de modulation a ainsi été utilisée pour les deux puces.

\begin{figure}[ht!]
    \centering
    \begin{subfigure}{0.8\textwidth}
        \centering
        \includegraphics[width=\textwidth]{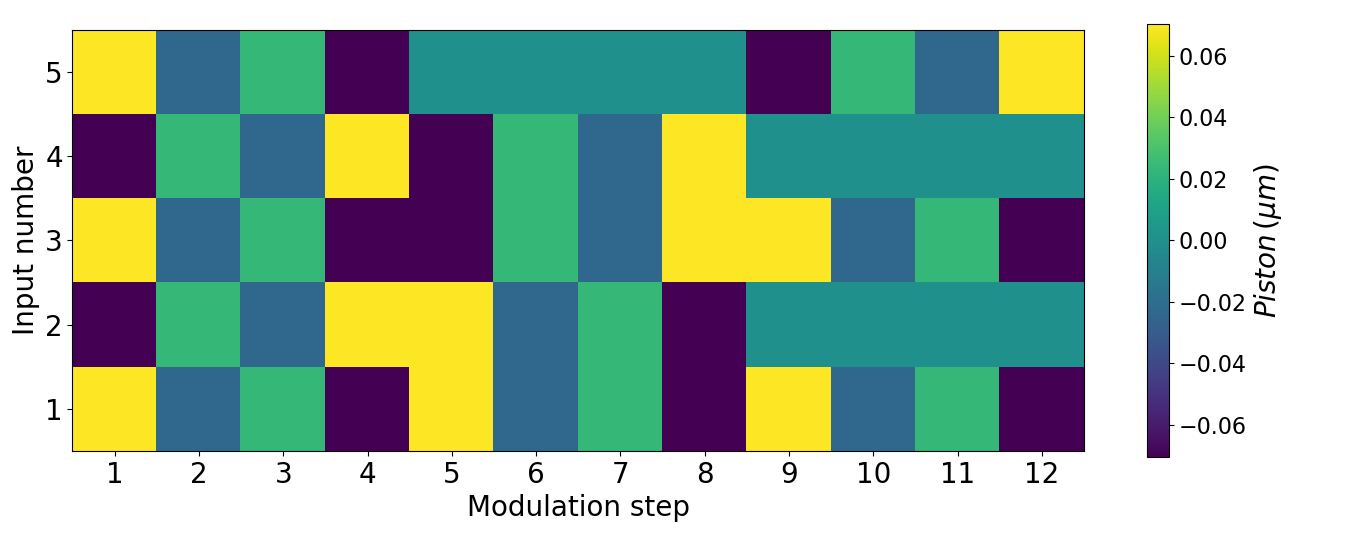}
    \end{subfigure}
    \begin{subfigure}{0.8\textwidth}
        \centering
        \includegraphics[width=\textwidth]{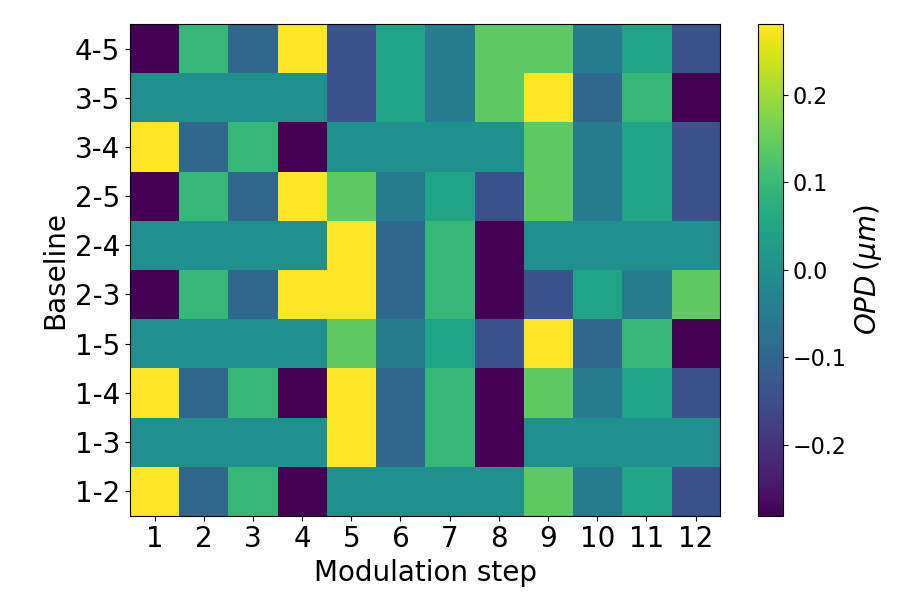}
    \end{subfigure}
    \caption[Séquence optimisée de modulation des MEMS à 12 pas pour échantillonner les franges sur FIRSTv2.]{En haut la séquence optimisée des pistons (en $\upmu$m) appliqués aux cinq segments du MEMS qui illuminent les entrées de la puce. En bas les valeurs d'OPDs (en $\upmu$m) induites sur les dix bases en application de cette séquence de modulation. Les couleurs sont les valeurs appliquées, les axes horizontaux sont les étapes de la séquence et les axes verticaux sont les segments et les bases, respectivement, pour la figure du haut et du bas.}
    \label{fig:ModSeq12}
\end{figure}

%%%%%%%%%%%%%%%%%%%%%%%%%%%%%%%%
\section{Étalonnage de l'instrument}

La méthode pour calculer les termes de visibilité complexe qui est utilisée dans le traitement des données de \ac{FIRSTv2} a été adaptée de celle déjà utilisée sur \ac{FIRSTv1} \citep{huby2012, huby2013these} et Vievard et al. (2023, en préparation).

%%%%%%%%%%%%%%%%
\subsection{L'interférométrie : une mesure de la cohérence des ondes lumineuses}

Pour écrire cette partie, je me suis beaucoup aidé du fabuleux cours de Guy Perrin, que j'ai pu suivre pendant ma deuxième année de master, sur l'interférométrie astronomique et qui fait référence dans le domaine.

Commençons par considérer une source astrophysique étendue incohérente à l'infini dans la direction $\vv{Z}$, émettant une onde lumineuse qui a pour champ électrique, au point $\vv{P}$ et à l'instant $t$ : $E(P,\vv{Z},t) = |E(P,\vv{Z},t)| e^{i(\omega t - \vv{k}.\vv{P})}$, avec le vecteur d'onde $\vv{k} = 2\pi \frac{\vv{Z}}{\lambda}$ et la pulsation $\omega = 2\pi \frac{c}{\lambda}$. Alors, en un point $P$ du foyer image d'un interféromètre qui combine deux faisceaux séparés par la base $\vv{B}$, est mesurée l'intensité $I_{tot}(\vv{B})$ de la superposition de deux champs électriques, intégré sur la surface de la source, selon l'équation :

\begin{align}
	I_{tot}(\vv{B}) &= \int_{source} \langle | E(P,\vv{Z},t) + E(P+\vv{B},\vv{Z},t) |^2 \rangle d^2\vv{Z} \label{eq:Interferogramme1} \\
	&= 2 \int_{source} I(\vv{Z},\lambda) d^2\vv{Z} + 2 \Re \left[\int_{source} I(\vv{Z},\lambda) e^{-2i\pi \vv{Z}.\frac{\vv{B}}{\lambda}} d^2\vv{Z} \right] \\
	&= 2 \int_{source} I(\vv{Z},\lambda) d^2\vv{Z} + 2 \left(\int_{source} I(\vv{Z},\lambda) d^2\vv{Z} \right) \Re [\mu_{nn'}(\vv{B})] \label{eq:Interferogramme3}
\end{align}

Où l'intensité lumineuse des deux faisceaux est supposée égale et où le facteur de cohérence complexe de la base $n-n'$ est défini selon :

\begin{equation}
    \mu_{nn'}(\vv{S}) = \frac{\int_{\alpha, \beta} I_{\lambda}(\alpha,\beta) e^{-2i\pi (\alpha u + \beta v)} d\alpha d\beta}{\int_{\alpha, \beta} I_{\lambda}(\alpha,\beta) d\alpha d\beta} \label{eq:ZVC}
\end{equation}

Avec $(\alpha,\beta)$ les coordonnées du vecteur $\vv{Z}$, $(u, v)$ les coordonnées du vecteur fréquence spatiale $\vv{S} = \vv{B}/\lambda$ projeté selon $\vv{Z}$. L'équation~\ref{eq:ZVC} est le résultat du théorème de Zernike - Van Cittert, qui énonce que la cohérence complexe est la transformée de Fourier de la distribution spatiale d'intensité de la source normalisée par le flux de cette dernière. C'est donc ce terme là que l'on cherche à mesurer pour obtenir des informations sur la cible observée. On l'appelle aussi communément la visibilité complexe que l'on note donc comme suit :

\begin{equation}
	\mu_{nn'} = |V_{nn'}| e^{i\varphi_{nn'}} A_n A_{n'} e^{i(\psi_{nn'} + \Delta\Phi_{nn'})} \label{eq:mu}
\end{equation}

Les termes $|V_{nn'}|$ et $\varphi_{nn'}$ sont, respectivement, le module et la phase de la visibilité complexe. $A_n$ et $A_{n'}$ sont, respectivement, le flux des sous-pupilles $n$ et $n'$. $\psi_{nn'}$ est le piston différentiel intrinsèque à l'instrument (résultant des \ac{OPD}s résiduelles entre les différentes fibres et autres aberrations optiques) dont nous verrons l'étalonnage dans la section~\ref{sec:P2VM}. Enfin, $\Delta\Phi_{nn'}$ est le piston atmosphérique correspondant aux aberrations non corrigées par l'optique adaptative (dans le cas où l'instrument est sur ciel) qui sera étalonné par la méthode d'analyse de données présentée dans la section~\ref{sec:PhaseSpecDiff}.

A partir d'une séquence d'images sur laquelle les franges sont modulées selon la séquence présentée dans la section~\ref{sec:Modulation}, l'interférogramme en fonction de l'\ac{OPD} est reconstruit et un modèle est ajusté pour en déduire les termes de visibilités complexes, comme décrit dans la section suivante.

%%%%%%%%%%%%%%%%
\subsection{Pixel-To-Visibility Matrix (la pitouvièm)}
\label{sec:P2VM}

Dans cette section, je vais exposer la partie du traitement de données consistant à retrouver les termes de visibilités complexes dans les interférogrammes mesurés. Il s'agit d'une méthode adaptée aux instruments possédant une pupille de sortie peu diluée (la dilution étant définie par le rapport entre les longueurs de base et le rayon des sous-pupilles), ce qui est le cas sur FIRSTv2. Elle a été pour la première fois utilisée pour les données de l'instrument \ac{AMBER} \citep{millour2004, tatulli2007} et consiste en la modélisation des franges dans le but de les ajuster et d'étalonner le terme complexe dû à l'instrument (le terme $e^{i(\psi_{nn'})}$ de l'équation~\ref{eq:mu}).

On commence par exprimer l'intensité de la base $n-n'$ de l'équation~\ref{eq:Interferogramme1}, avec un retard temporel $\updelta \text{t} = x / c$ (avec $x$ la variable spatiale discrète, selon l'axe des \ac{OPD}s) :

\begin{align}
    I_{nn'}(x) &= \int_{source} \langle | E(P,\vv{Z},t) + E(P+\vv{B},\vv{Z},t + \updelta \text{t}) |^2 \rangle d^2\vv{Z}
\end{align}

Et d'après l'équation~\ref{eq:Interferogramme3} combinée avec l'équation~\ref{eq:mu}, on écrit : 

\begin{align}
    I_{nn'}(x) &= \sum_{i \in \{n,n'\}} A_i^2 E_i^2 + 2 \Re \left[ A_n E_n A_{n'} E_{n'} e^{i(\frac{2 \pi}{\lambda}x + \varphi_{nn'} + \psi_{nn'} + \Delta\Phi_{nn'})}  \right] \\
    &= \sum_{i \in \{n,n'\}} A_{i}^{2} E_{i}^{2} + 2 A_n E_n A_{n'} E_{n'} cos \left(\frac{2 \pi}{\lambda}x + \varphi_{nn'} + \psi_{nn'} + \Delta\Phi_{nn'} \right)
\end{align}

Avec $A_i$ le flux et $E_i$ l'enveloppe de la cohérence temporelle, de la sous-pupille $i$. On considère les termes $A_i$ et $E_i$ indépendants de $x$. Les termes $A_i^2 E_i^2$ caractérisent notamment la transmission de la puce. En effet, chaque sortie imagée sur le détecteur est la combinaison interférométrique de faisceaux qui traversent des guides d'onde différents dans la puce.

En reformulant cette équation et en utilisant la relation~\ref{eq:mu}, on trouve une relation linéaire entre l'interférogramme $I_{nn'}(x)$ et le terme de cohérence complexe $\mu_{nn'}$, s'écrivant :

\begin{equation}
    I_{nn'}(x) = \sum_{i \in \{n,n'\}} A_{i}^{2} E_{i}^{2}(x) + \Re [\mu_{nn'}] C_{nn'}(x) + \Im [\mu_{nn'}] S_{nn'}(x) \label{eq:InterferogrammeLineaire}
\end{equation}

Avec
\begin{align}
    \left\{
    \begin{array}{r@{}l}\displaystyle
    &C_{nn'}(x) = 2E_n E_{n'}\cos \left( \frac{2 \pi}{\lambda}x \right) \\
    &S_{nn'}(x) = -2E_n E_{n'}\sin \left( \frac{2 \pi}{\lambda}x \right)
    \end{array}\right. \label{eq:V2PMtermesComplexes}
\end{align}

Ainsi, $\{ F_{nn'} \text{; } C_{nn'}(x) \text{; } S_{nn'}(x)  \}$ forme une base sur laquelle sont projetés les interférogrammes (avec $F_{nn'} = \sum_{i \in \{n,n'\}} A_{i}^{2} E_{i}^{2}$). Ce sont les termes qui contiennent les erreurs instrumentales de l'instrument et ils devront être étalonnés, comme cela sera montré plus loin.

Enfin, on peut ré-écrire l'équation~\ref{eq:InterferogrammeLineaire} pour toutes les bases sous forme matricielle, ce qui est plus adapté au calcul numérique. On note $n_B$ le nombre de bases correspondant à toutes les combinaisons par paire des faisceaux $n$ et $n'$ ($n_B = n_I(n_I-1)/2$ pour $n_I$ sous-pupilles), le nombre de pixels $n_p$ sur une séquence de modulation ($n_p = 12$ pour la séquence montrée dans la section~\ref{sec:Modulation}). La matrice est alors la concaténation en ligne des équations des interférogrammes de toutes les bases $b_i$. Pour $5$ sous-pupilles d'entrée, cette matrice s'écrit donc :

\begin{align}
    \text{\textbf{I}}=
    \begin{bmatrix}
    	I_{b_1}(x_1) \\
    	\vdots \\
    	I_{b_1}(x_{n_p}) \\
    	\vdots \\
    	I_{b_{n_B}}(x_1) \\
    	\vdots \\
    	I_{b_{n_B}}(x_{n_p})
    \end{bmatrix}
    = V2PM \cdot
    \begin{bmatrix}
        A_{1}^2 \\
        \vdots \\
        A_{n_I}^2 \\
        \Re [\mu_{b_1}] \\
        \vdots \\
        \Re [\mu_{b_{n_B}}] \\
        \Im [\mu_{b_1}] \\
        \vdots \\
        \Im [\mu_{b_{n_B}}]
    \end{bmatrix}
    =V2PM\cdot\text{\textbf{P}} \label{eq:IV2PMP}
\end{align}

% \begin{align}\tiny
%     \begin{bmatrix}
%         E_{1}^2(x_1)     & E_{2}^2(x_1)     &        & \vdots               & \vdots             & C_{b_1}(x_1)     &        & 0                    & S_{b_1}(x_1)     &        & 0 \\
%         \vdots           & \vdots           &        & \vdots               & \vdots             & \vdots           &        & \vdots               & \vdots           &        & \vdots \\
%         E_{1}^2(x_{n_p}) & E_{2}^2(x_{n_p}) &        & \vdots               & \vdots             & C_{b_1}(x_{n_p}) &        & 0                    & S_{b_1}(x_{n_p}) &        & 0 \\
%         \vdots           & \vdots           & \ddots & \vdots               & \vdots             & 0                & \ddots & 0                    & 0                & \ddots & 0 \\
%         \vdots           & \vdots           &        & E_{n_I-1}^2(x_1)     & E_{n_I}^2(x_1)     & 0                &        & C_{b_{n_B}}(x_1)     & 0                &        & S_{b_{n_B}}(x_1) \\
%         \vdots           & \vdots           &        & \vdots               & \vdots             & \vdots           &        & \vdots               & \vdots           &        & \vdots \\
%         \vdots           & \vdots           &        & E_{n_I-1}^2(x_{n_p}) & E_{n_I}^2(x_{n_p}) & 0                &        & C_{b_{n_B}}(x_{n_p}) & 0                &        & S_{b_{n_B}}(x_{n_p})
%     \end{bmatrix}
% \end{align}

avec V2PM =
\begin{align}\tiny
    \hspace{-1ex}\bordermatrix{
        & \tikzmark{harrowleft1} & & & & \tikzmark{harrowright1} & \tikzmark{harrowleft2} & & \tikzmark{harrowright2} & \tikzmark{harrowleft3} & & \tikzmark{harrowright3} \cr
        & E_{1}^2(x_1)     & E_{2}^2(x_1)     &        & \vdots               & \vdots             & C_{b_1}(x_1)     &        & 0                    & S_{b_1}(x_1)     &        & \hphantom{mmi} 0 \hphantom{mmii} \tikzmark{varrowtop} \cr
        & \vdots           & \vdots           &        & \vdots               & \vdots             & \vdots           &        & \vdots               & \vdots           &        & \vdots \cr
        & E_{1}^2(x_{n_p}) & E_{2}^2(x_{n_p}) &        & \vdots               & \vdots             & C_{b_1}(x_{n_p}) &        & 0                    & S_{b_1}(x_{n_p}) &        & 0 \cr
        & \vdots           & \vdots           & \ddots & \vdots               & \vdots             & 0                & \ddots & 0                    & 0                & \ddots & 0 \cr
        & \vdots           & \vdots           &        & E_{n_I-1}^2(x_1)     & E_{n_I}^2(x_1)     & 0                &        & C_{b_{n_B}}(x_1)     & 0                &        & S_{b_{n_B}}(x_1) \cr
        & \vdots           & \vdots           &        & \vdots               & \vdots             & \vdots           &        & \vdots               & \vdots           &        & \vdots \cr
        & \vdots           & \vdots           &        & E_{n_I-1}^2(x_{n_p}) & E_{n_I}^2(x_{n_p}) & 0                &        & C_{b_{n_B}}(x_{n_p}) & 0                &        & S_{b_{n_B}}(x_{n_p}) \tikzmark{varrowbottom} \cr
        }
\end{align}
% Set and draw arrows on the sides of the matrix
\tikz[overlay,remember picture] {
    \draw[<->] ([yshift=0.1ex,xshift=-2ex]harrowleft1) -- ([yshift=0.1ex,xshift=3ex]harrowright1) node[midway,above] {\scriptsize $n_I$};
    \draw[<->] ([yshift=0.1ex,xshift=-3ex]harrowleft2) -- ([yshift=0.1ex,xshift=3ex]harrowright2) node[midway,above] {\scriptsize $n_B$};
    \draw[<->] ([yshift=0.1ex,xshift=-3ex]harrowleft3) -- ([yshift=0.1ex,xshift=2ex]harrowright3) node[midway,above] {\scriptsize $n_B$};
    \draw[<->] ([yshift=1.5ex,xshift=3ex]varrowtop) -- ([xshift=3ex]varrowbottom) node[midway,right] {\scriptsize \rotatebox{90}{$n_B \times n_p$}};
}

La \acrfull{V2PM}, relie le vecteur $\textbf{P}$ des termes de cohérence complexe aux interférogrammes $\textbf{I}$. Elle contient les termes qui caractérisent les erreurs (e.g. diaphotie, désalignement, mauvais pixels du détecteur, ...) et les caractéristiques instrumentales (e.g. utilisation d'une puce photonique, les transmissions différentes dans les fibres, ...). Cette matrice constitue un modèle de l'instrument et est calculée pour chaque canal spectral. Elle permet alors d'étalonner la réponse de l'instrument à un signal d'entrée modulé selon la séquence de modulation.

Enfin, à partir de l'équation~\ref{eq:IV2PMP} nous cherchons à déduire le vecteur $\textbf{P}$, or la \ac{V2PM} est une matrice rectangulaire de dimensions $(n_B \times n_p) \times (n_I + 2 n_B)$ et ne peut donc être simplement inversée. Nous calculons donc la matrice pseudo-inverse (de Moore-Penrose) de la \ac{V2PM} afin d'estimer les paramètres :

\begin{equation}
    \text{\textbf{P}} = P2VM \cdot \text{\textbf{I}} \label{eq:VisibiliteMes}
\end{equation}

avec $P2VM = V2PM^\dag$ (\textit{Pixel To Visibility Matrix}) la matrice pseudo-inverse de la \ac{V2PM}. La \ac{P2VM} associe donc le flux mesuré à une visibilité en faisant la transformation entre les deux.

%%%%%%%%%%%%%%%%
\subsection{Étalonnage de la V2PM}
\label{sec:V2PMEtalonnage}

Dans cette partie sont présentées les procédures de mesure de la \ac{V2PM}. Une nouvelle \ac{V2PM} est mesurée sur une source interne avant chaque nouvelle session de mesures : si les conditions instrumentales ont changé (e.g. changement de la séquence de modulation) ou avant chaque nuit d'observation. La figure~\ref{fig:V2PMP2VMmesure}, à gauche, est une représentation graphique de \ac{V2PM} mesurées sur \ac{FIRSTv2} pour le canal spectral $\sim 706 \,$nm et à droite la \ac{P2VM} associée, pour la puce $Y$ (ligne du haut) et la puce $X$ (ligne du bas). Pour $5$ sous-pupilles, la \ac{V2PM} de \ac{FIRSTv2} a donc $25$ colonnes : les cinq premières contiennent les termes $E_{n}^2$, les dix suivantes contiennent les termes $C_{b_i}(x)$ et les dix dernières contiennent les termes $S_{b_i}(x)$. Enfin, il y a $n_B$ paquets de $n_p$ (nombre de pixels sur l'axe des \ac{OPD}s) lignes, ce qui correspond ici à $10 \times 12 = 120$ lignes et $10 \times 20 = 200$ lignes pour la puce $X$, respectivement. Cette différence du nombre de lignes pour l'une et l'autre puce est due au fait que les données sur la puce $X$ ont été prises antérieurement aux données prises sur la puce $Y$, à une période où la séquence de modulation n'avait pas encore été optimisée (de $20$ à $12$ pas) de la façon expliquée dans la section~\ref{sec:Modulation}. De plus, on note que les paires de sorties par base des données de la puce $X$ ont été traitées indépendamment durant ma thèse et seront traitées ensemble lors de futurs développement. Enfin, chaque colonne contient en majorité des zéros et la matrice agit comme un masque sur les pixels des bases lors de la multiplication matricielle de la \ac{V2PM} par les données.

\begin{figure}[ht!]
    \begin{subfigure}{0.5\textwidth}
        \centering
        \includegraphics[width=0.9\textwidth]{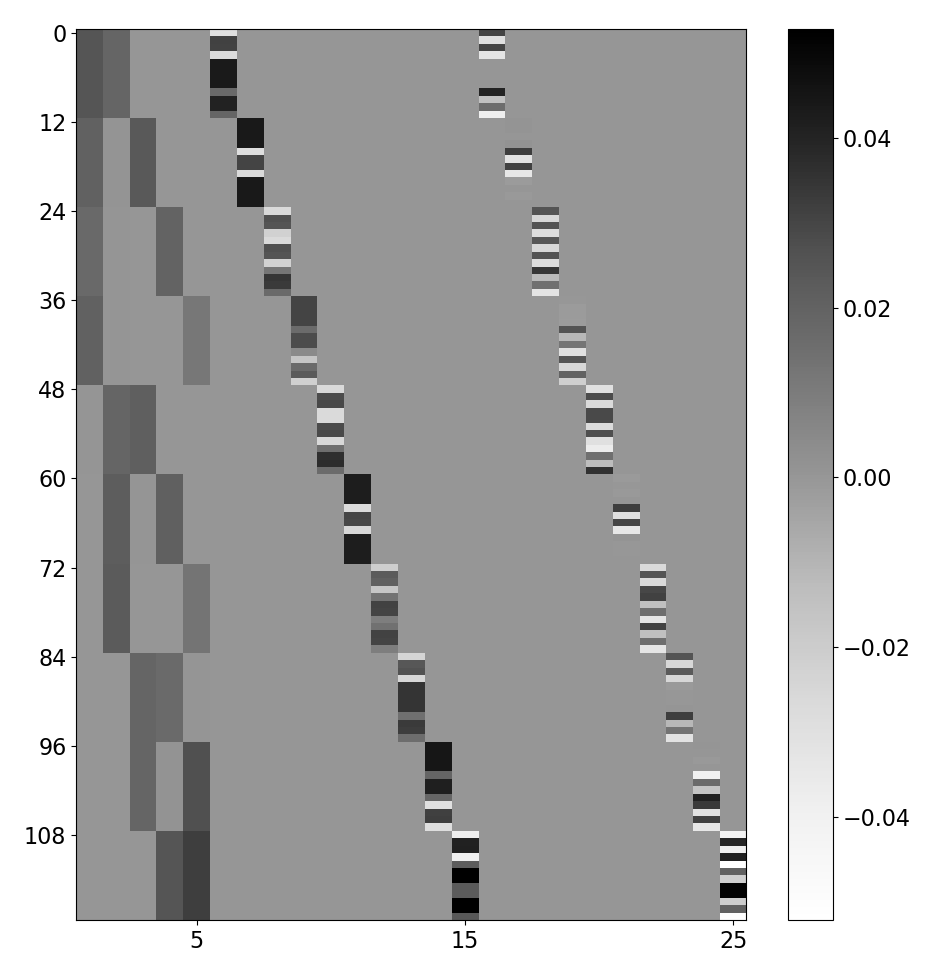}
    \end{subfigure}%
    \begin{subfigure}{0.5\textwidth}
        \centering
        \includegraphics[width=0.9\textwidth]{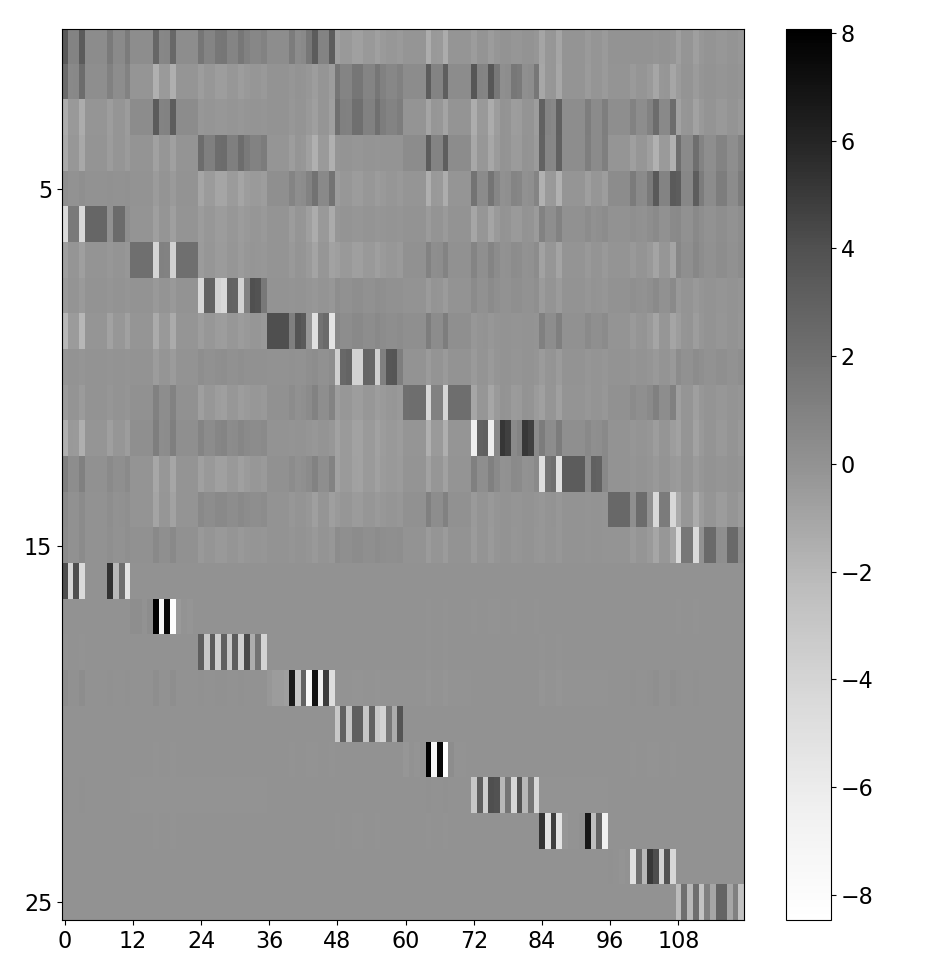}
    \end{subfigure}
    \begin{subfigure}{0.5\textwidth}
        \centering
        \includegraphics[width=0.9\textwidth]{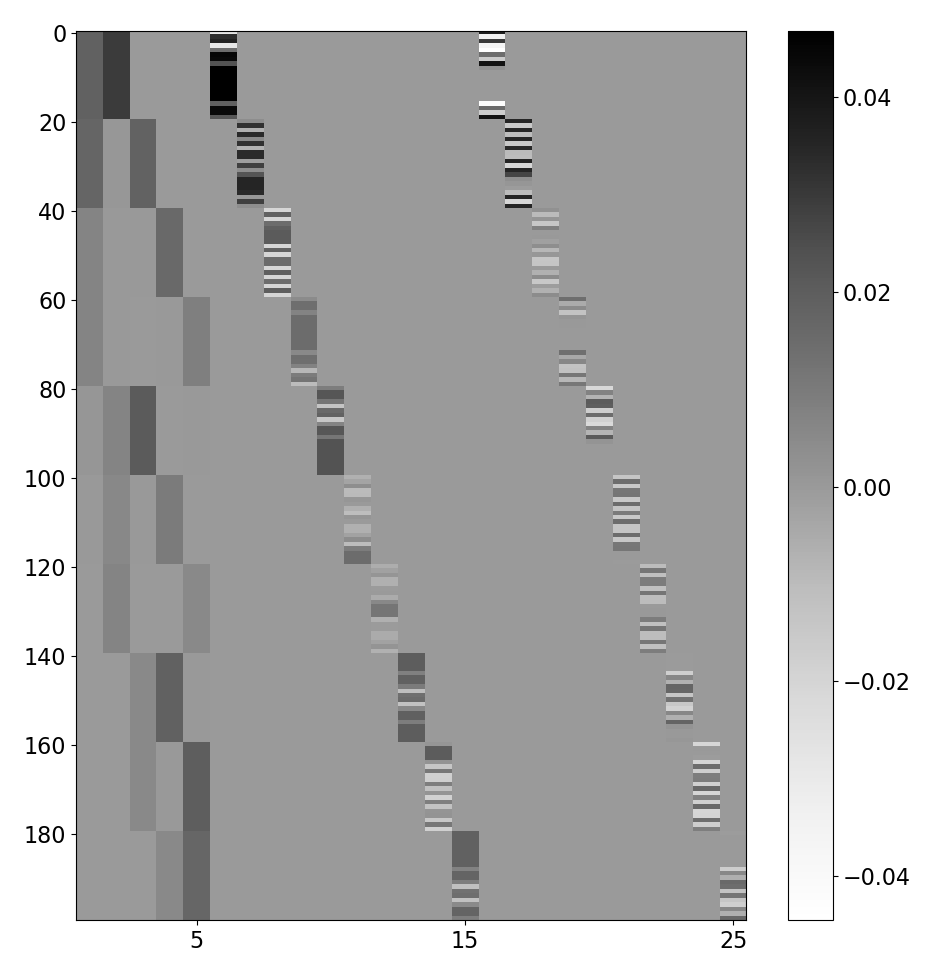}
    \end{subfigure}%
    \begin{subfigure}{0.5\textwidth}
        \centering
        \includegraphics[width=0.9\textwidth]{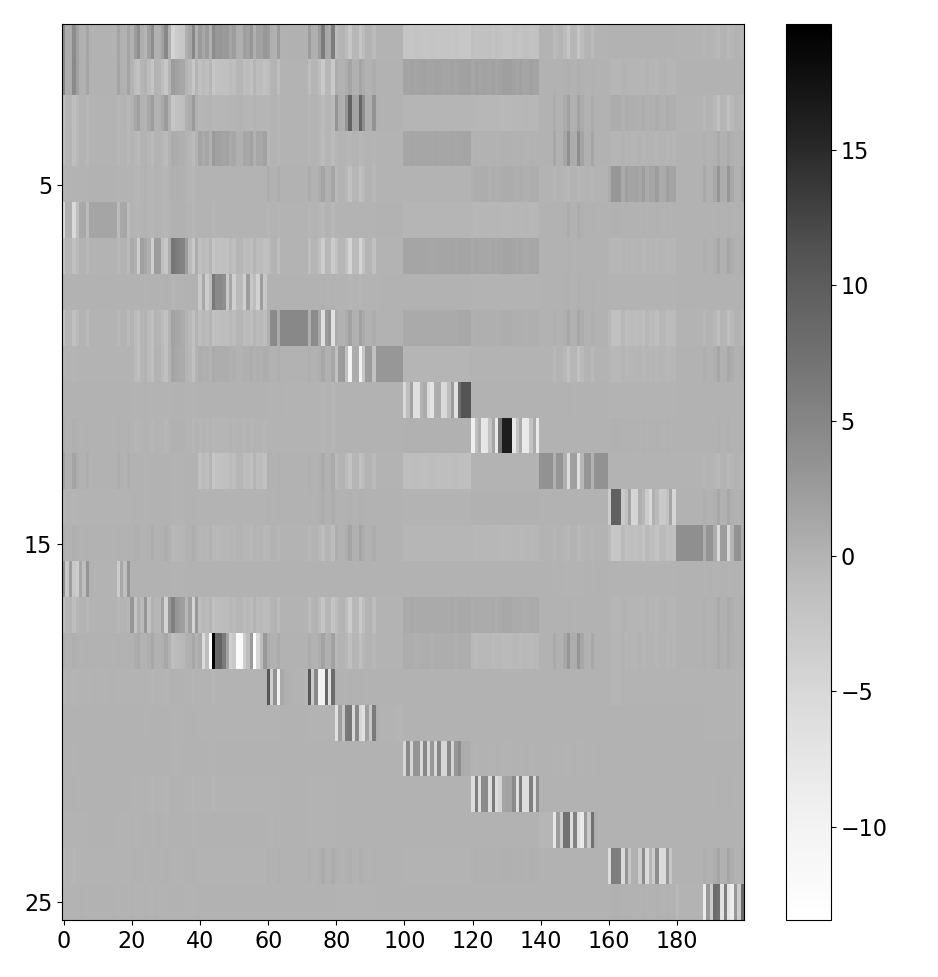}
    \end{subfigure}
    \caption[Représentation graphique de la V2PM et de sa P2VM mesurée sur FIRSTv2, pour les deux puces.]{Représentation graphique de la V2PM (colonne de gauche) et de la P2VM associée (colonne de droite) mesurée sur la puce $Y$ (ligne du haut) et la puce $X$ (ligne du bas). Ce sont les V2PM (P2VM) du canal spectral $\sim 706 \,$nm, mesurées sur FIRSTv2 avec une source \sk. L'axe vertical (respectivement l'axe horizontal) de la V2PM (respectivement de la P2VM) est le nombre de points de la séquence de modulation (OPD) multiplié par le nombre de bases ($12 \times 10 = 120$ pour la $Y$ et $20 \times 10 = 200$ pour la $X$) et l'axe horizontal (respectivement l'axe vertical) contient $5$ termes de transmission en plus de $2 \times 10$ termes interférométriques.}
    \label{fig:V2PMP2VMmesure}
\end{figure}

%%%%%%%%
\subparagraph{Mesure des termes de transmission \\}

Les termes de transmission $E_{n}^2$, arrangés par couple de sous-pupilles,  remplissent les cinq premières colonnes. Pour un paquet de $n_p$ lignes donné (correspondant à une base) les lignes se composent des deux termes $E_{n}^2$ et $E_{n'}^2$ et de zéros. Les bases sont ordonnées, de haut en bas, dans l'ordre des bases $1-2$, $1-3$, $1-4$, $1-5$, $2-3$, $2-4$, $2-5$, $3-4$, $3-5$ et $4-5$, les termes non nuls étant placés dans les colonnes $n$ et $n'$ correspondantes à la base.

Afin d'effectuer leur mesure, cinq séries d'images sont prises en illuminant tour à tour les sous-pupilles. Pour cela, tous les segments sont positionnés en bout de course du tip-tilt sauf le segment de la sous-pupille d'intérêt qui est envoyé sur sa position d'optimisation de l'injection (voir section~\ref{sec:OptiInj}). Ainsi, $4$ ($8$ pour la puce $X$) sorties sont illuminées sur la caméra et une série de quelques centaines d'images est acquise pendant que les segments sont modulés en piston selon la séquence de modulation des franges pour prendre en compte les potentielles variations du flux dans les fibres. La figure~\ref{fig:flats} montre les cinq images obtenues après la moyenne de ces cinq séries, où les sorties sont réparties sur l'axe vertical et la dispersion spectrale sur l'axe horizontal. Les bases sont identifiées sur la caméra en détectant les sorties qui sont illuminées.

\begin{figure}[ht!]
    \begin{subfigure}{\textwidth}
        \centering
        \includegraphics[width=0.8\textwidth]{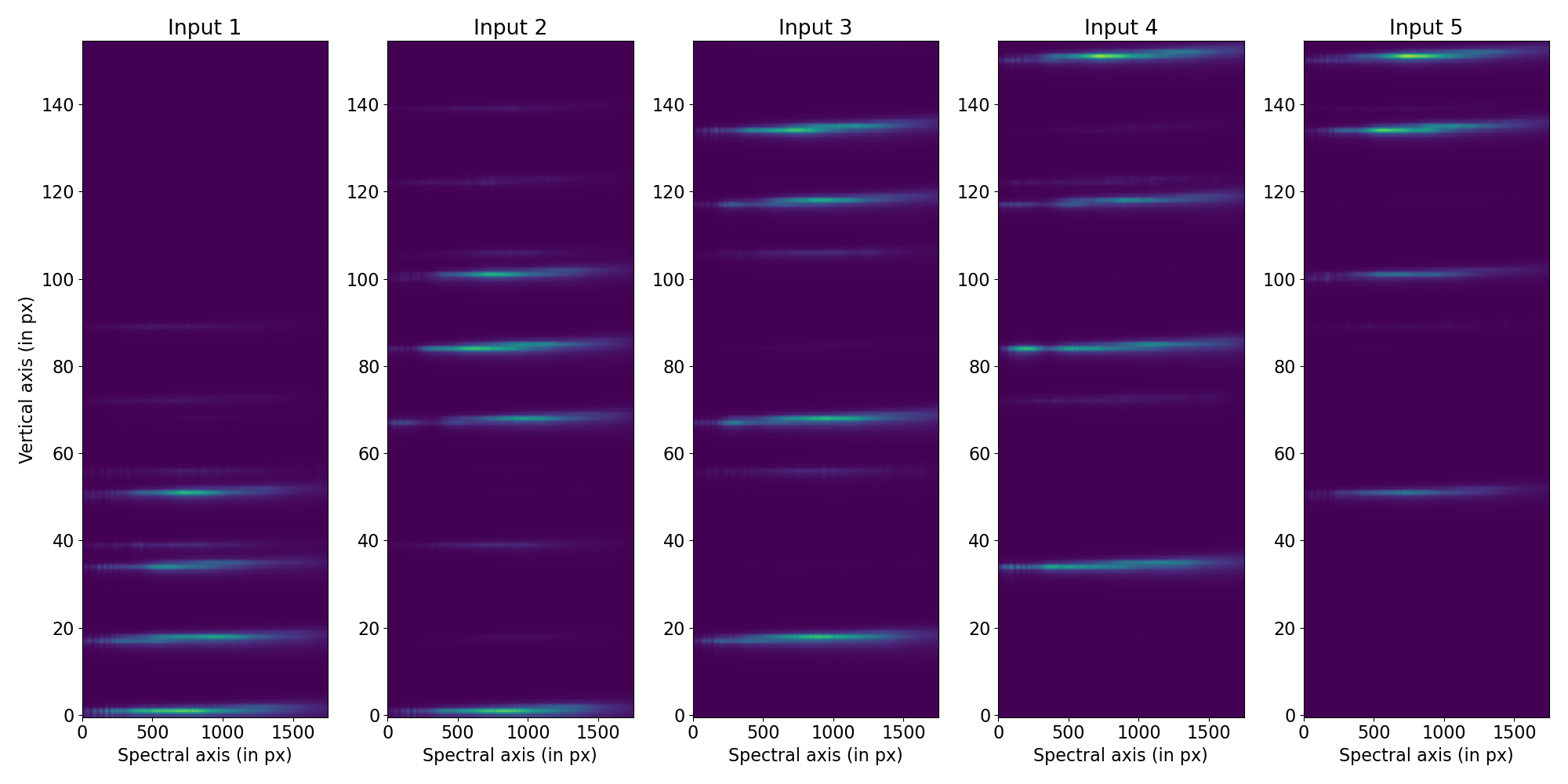}
    \end{subfigure}
    \begin{subfigure}{\textwidth}
        \centering
        \includegraphics[width=0.8\textwidth]{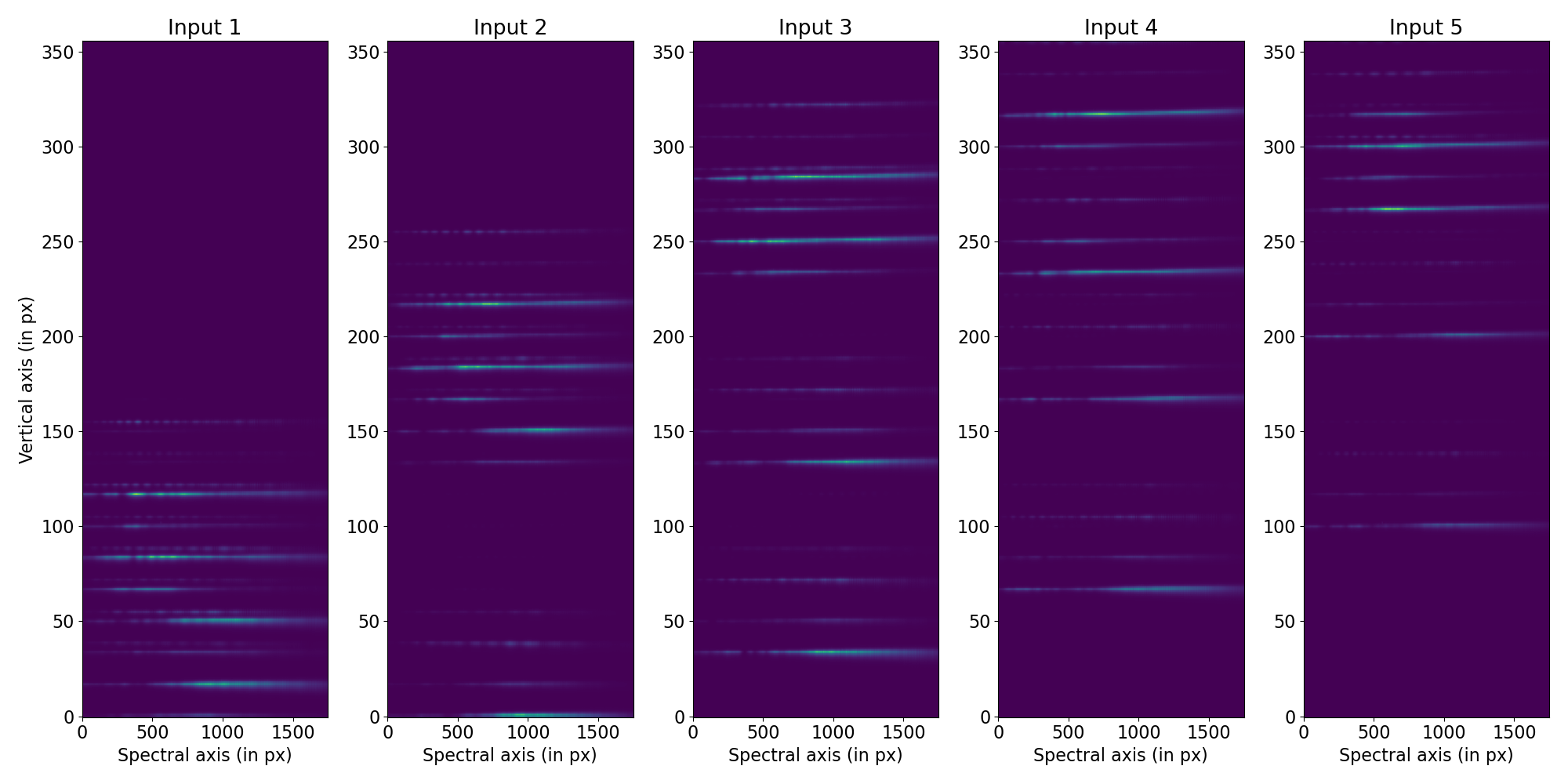}
    \end{subfigure}
    \caption[Les flats mesurés sur les deux puces, sur FIRSTv2.]{Les cinq images de la caméra lorsque les entrées de la puce $Y$ (en haut) et de la puce $X$ (en bas) sont individuellement éclairées. Sur chaque image, on peut identifier les $4$ ($8$ pour la puce $X$) sorties selon l'axe vertical et la dispersion spectrale selon l'axe horizontal. On note que sur ces images, le nombre de sorties visibles est en réalité le double à cause de la présence d'un prisme de wollaston sur le chemin optique.}
    \label{fig:flats}
\end{figure}

Pour chaque entrée illuminée, le flux de chaque sortie est normalisé par la somme des flux des $4$ ($8$ pour la puce $X$) sorties illuminées par cette entrée. Et cela est fait pour chaque pas de la séquence de modulation et pour chaque canal spectral. La figure~\ref{fig:V2PMtransmission} présente les spectres de transmission à la suite de cette opération pour les quatre sorties des cinq entrées de la puce $Y$, pour le premier pas de la séquence de modulation. On remarque que les deux courbes, pour une sortie donnée, ne sont pas nécessairement les mêmes selon que la sortie est illuminée par l'une ou l'autre des deux entrées. Cela est dû au fait que chaque sortie est illuminée par deux chemins différents. De plus, la partie inférieure à $640 \,$nm présente de grandes variations par rapport au reste du spectre, car la puce a été spécifiée monomode dans la bande spectrale $650 - 800 \,$nm et ne l'est pas en dehors. Les prochaines puces seront conçues pour être monomodes au minimum dans la gamme spectrale $600 - 700 \,$nm pour être capable de faire des mesures de raie d'émission \ha~à $656.28 \,$nm, ce qui correspond à notre cas scientifique (voir la section~\ref{sec:AccretionAlpha}).

\begin{figure}[ht!]
    \begin{subfigure}{\textwidth}
        \centering
        \includegraphics[width=0.8\textwidth]{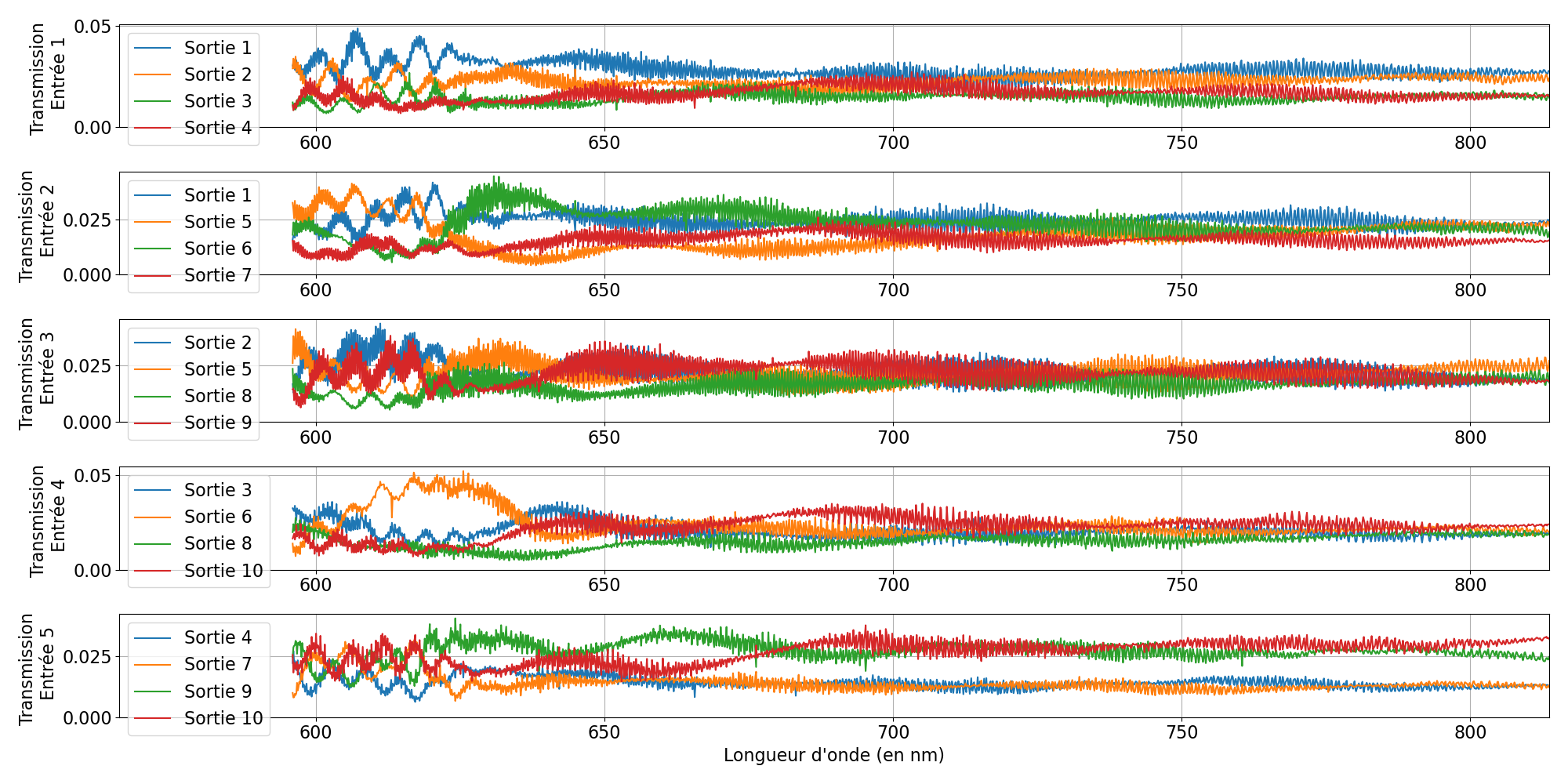}
    \end{subfigure}
    \begin{subfigure}{\textwidth}
        \centering
        \includegraphics[width=0.8\textwidth]{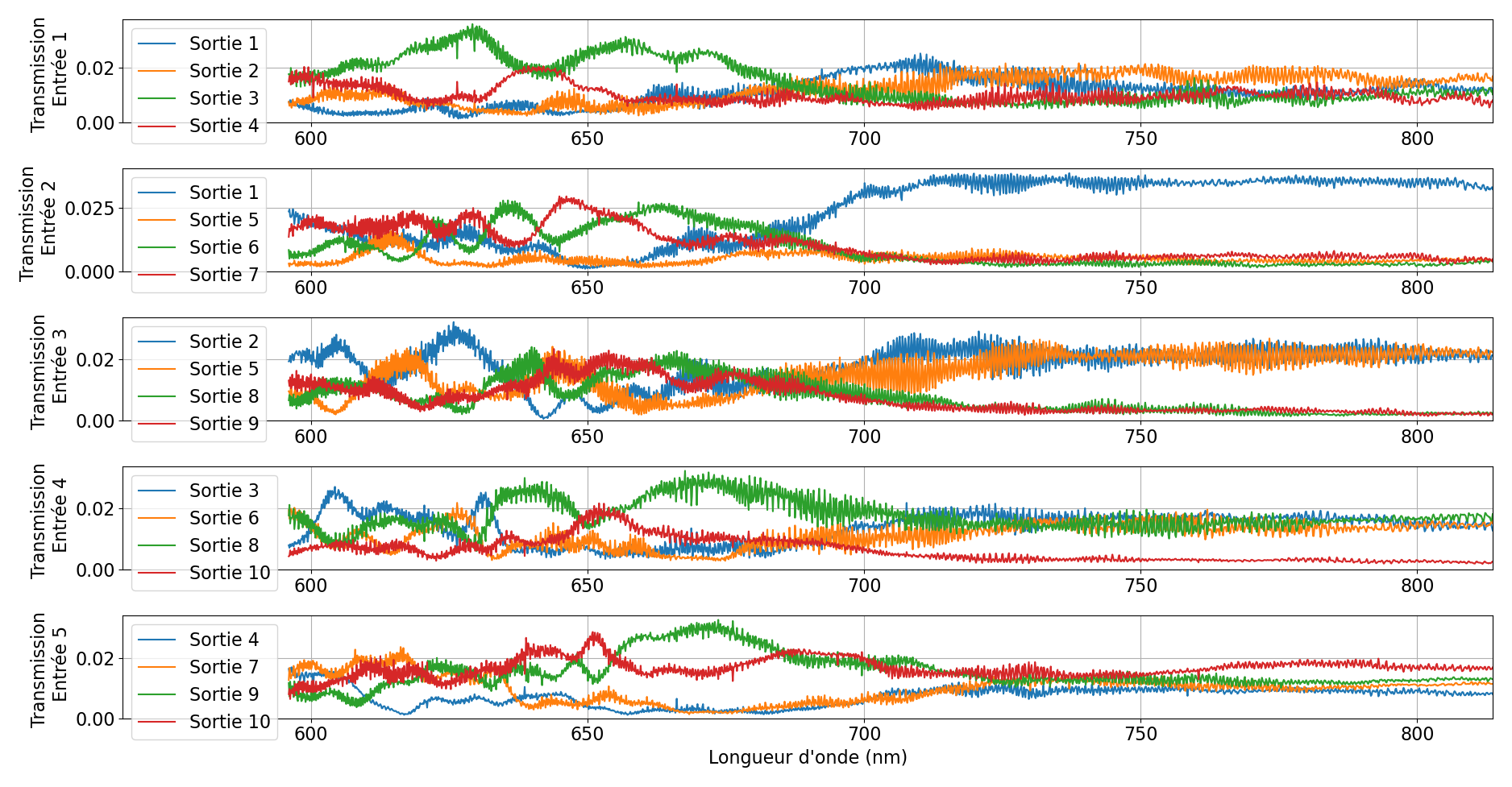}
    \end{subfigure}
    \caption[Transmissions relatives mesurées sur chaque entrée des puces $Y$ et $X$, constituant les 5 premières colonnes de la V2PM.]{Les spectres de transmissions relatives pour le premier pas de la séquence de modulation mesurés sur les quatre sorties (voir les légendes) en illuminant une à une les cinq entrées de la puce (de haut en bas). Les cinq graphiques du haut sont les transmissions mesurées avec la puce $Y$ et les cinq du bas avec la puce $X$. Ces courbes remplissent les cinq premières colonnes des V2PMs pour les canaux spectraux correspondant.}
    \label{fig:V2PMtransmission}
\end{figure}

%%%%%%%%
\subparagraph{Mesure des termes interférométriques \\}

Dans cette partie, on souhaite mesurer expérimentalement la séquence de modulation des franges. Pour cela, les interférogrammes sont modulés en phase les uns après les autres et la phase des franges est estimée pour chaque pas de la séquence et remplit les dernières colonnes de la \ac{V2PM}, selon la procédure décrite ci-après.

En effet, on souhaite estimer les termes $C_{b_i}(x)$ et $S_{b_i}(x)$ qui remplissent les vingt dernières colonnes de la \ac{V2PM}. La procédure de l'instrument \ac{AMBER} préconise de mesurer les franges d'interférences indépendamment sur chaque base. Pour cela, seules les entrées $n$ et $n'$ de la puce formant la base $b_i$ sont illuminées et une série d'images est acquise en appliquant la séquence de modulation des franges. Cette acquisition est répétée cinq fois en translatant de manière aléatoire de quelques micromètres les cinq lignes à retard (\acrfull{ODL}) afin de mesurer les interférogrammes sur une plus grande diversité de phase. Il résulte ainsi $5$ séries d'images avec les franges modulées pour chacune des $10$ bases. 

Par la suite, une image est formée à partir des $5$ séries, pour chaque base successivement, en considérant une partie du spectre seulement. Dans l'idéal on ne souhaiterait utiliser qu'un seul canal spectral afin d'estimer la phase de la séquence évaluée à une longueur d'onde donnée. Mais en pratique on choisit de l'ordre de la dizaine de canaux sur la partie du spectre avec le plus de flux (ce qui correspond à une largeur spectrale de $\sim 6\,$nm), afin d'améliorer le \ac{SNR} de l'estimation des phases dans la suite. Cette image est formée en disposant les interférogrammes mesurés ($12$ points) par colonne, donnant une taille d'image de $12$ lignes par $5 \times n_{sm} \times \Delta\lambda_{px}$, $5$ étant le nombre de série d'images acquises, $n_{sm}$ le nombre de séquence de modulation effectuées lors de cette série et $\Delta\lambda_{px} = 50$ le nombre de canaux spectraux utilisés. Pour une série de $400$ images ($20$ acquisitions répétées $5$ fois), la taille de l'image finale est donc de $12$ lignes par $5 \times 33 \times 50 = 8\,250$ colonnes. Une décomposition en valeurs singulières, ou \ac{SVD}, est alors opérée sur cette image afin d'en déduire les vecteurs propres sur lesquels l'image se projette. Comme elle est composée des interférogrammes mesurés, on s'attend à ce que seulement deux vecteurs propres aient des valeurs singulières qui dominent la décomposition. Ces deux vecteurs propres sont orthonormaux et peuvent donc être assimilés aux fonctions cosinus et sinus de l'équation~\ref{eq:InterferogrammeLineaire}. La phase de l'interférogramme est ainsi déduite de l'angle formé par ceux-ci.

La figure~\ref{fig:SVD} illustre le résultat de la \ac{SVD} appliquée aux données qui ont permis d'estimer la \ac{V2PM} de la figure~\ref{fig:V2PMP2VMmesure}. À gauche sont représentées les valeurs singulières pour chaque base (selon les couleurs de la légende), normalisées par rapport à la première et on voit effectivement que les deux premières dominent les autres. Sur la figure de droite sont tracés les deux vecteurs propres associés aux deux premières valeurs singulières, dans le plan complexe et pour chaque base. Dans cette représentation, une frange idéalement échantillonnée formerait un cercle, constitué de points régulièrement espacés. La séquence de modulation a pour but d'échantillonner au mieux ces cercles, avec au minimum $4$ points déphasés de $\uppi / 2 \,$rad (il s'agit de la méthode ABCD, voir la figure~\ref{fig:FringeSamplingABCD}). Or, la séquence de modulation des franges utilisée ne mesure pas seulement ces quatres points, mais elle en mesure quatre autres pour quatre valeurs d'\ac{OPD} différentes (voir la section~\ref{sec:Modulation}). Cela se remarque en comptant le nombre de points tracés sur les graphiques (au nombre de 8 ou de 9 suivant si le point à zéro \ac{OPD} est mesuré ou non). Aussi, les bases 2 et 6 ne présentent que cinq points de mesure, ce qui se comprend en regardant les valeurs d'\ac{OPD} appliquées par la séquence de modulation présentée sur le bas de la figure~\ref{fig:ModSeq12}.

Ces graphes nous ont servi de point de vérification que la \ac{SVD} a bien fonctionné, nous permettant de faire des corrections ou des améliorations du code lui-même ou nous indiquant que le traitement peut poursuivre. Dans le cas où cela ne fonctionne pas et que les deux premières valeurs singulières ne dominent pas les autres ou que les vecteurs propres ne forment pas des cercles (cela peut apparaître comme des ellipses ou des formes non régulières ou non équitablement échantillonnés), cela peut vouloir dire qu'il y a un problème dans la prise de données (e.g. il est arrivé que les segments du \ac{MEMS} soient restés immobiles), dans la séquence de modulation choisie (plusieurs séquences ont été testées) ou dans le processus de réduction de données en amont.

\begin{figure}[ht!]
    \begin{subfigure}{0.5\textwidth}
        \centering
        \includegraphics[width=\textwidth]{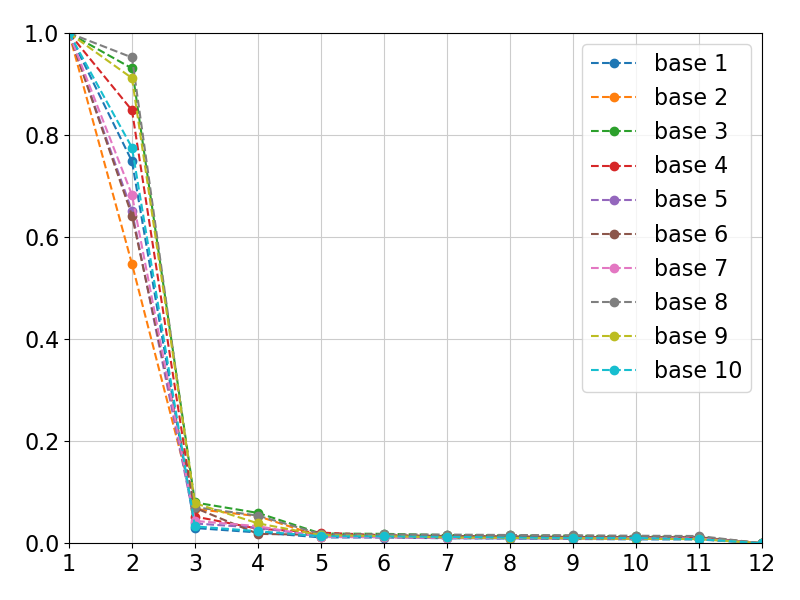}
    \end{subfigure}%
    \begin{subfigure}{0.5\textwidth}
        \centering
        \includegraphics[width=\textwidth]{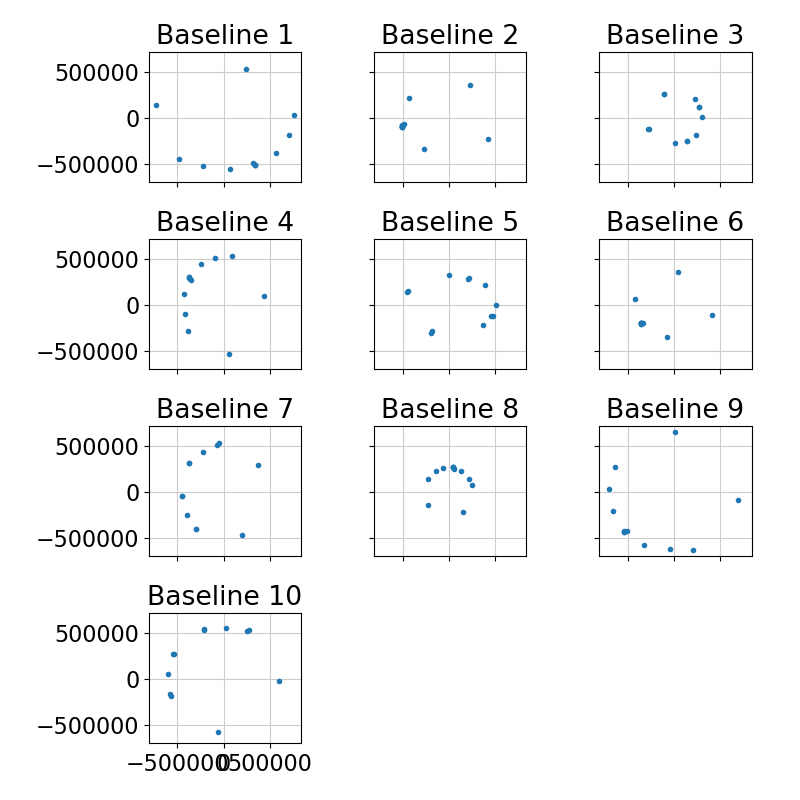}
    \end{subfigure}
    \caption[Produits finaux de la SVD appliquée aux données de FIRSTv2 mesurées avec la puce $Y$.]{Produits finaux de la SVD appliquée aux données de FIRSTv2 mesurées avec la puce $Y$. A gauche sont tracées les valeurs singulières trouvées pour chaque base (selon les couleurs de la légende), normalisées par rapport à la première et pour chaque étape de la séquence de modulation en abscisse. A droite sont représentées les vecteurs propres dans le plan complexe.}
    \label{fig:SVD}
\end{figure}

Ensuite, sont déduites les phases pour chaque point de la séquence de modulation à partir de ces deux vecteurs propres. Les \ac{OPD}s $\delta_0$ sont déduites des phases et sont comparées aux \ac{OPD}s théoriques appliquées par la séquence de modulation. Pour cela, la longueur d'onde $\lambda_0$ à laquelle sont mesurées ces phases $\psi_0$ est calculée comme la moyenne des longueurs d'ondes sur les $50$ canaux spectraux choisis. Enfin, les phases $\psi_i$ sur toutes les longueurs d'ondes $\lambda_i$ de la bande spectrale sont obtenues à partir de $\psi_i = 2 \pi \delta_0 / \lambda_i$. Le résultat est tracé sur la figure~\ref{fig:PhaseEstimees}, pour chaque base, où la phase attendue est tracée en trait continu noir et la phase estimée est tracée en trait discontinue bleue, ici pour la longueur d'onde $\sim 706 \,$nm. On voit que les phases sont correctement estimées par rapport à celles attendues. Finalement, ces phases forment les termes de visibilité complexe dont les parties réelles et imaginaires remplissent les $20$ dernières colonnes d'autant de \ac{V2PM} qu'il y a de canaux spectraux, selon le système d'équations~\ref{eq:V2PMtermesComplexes}.

\begin{figure}[ht!]
    \centering
    \includegraphics[width=\figwidth]{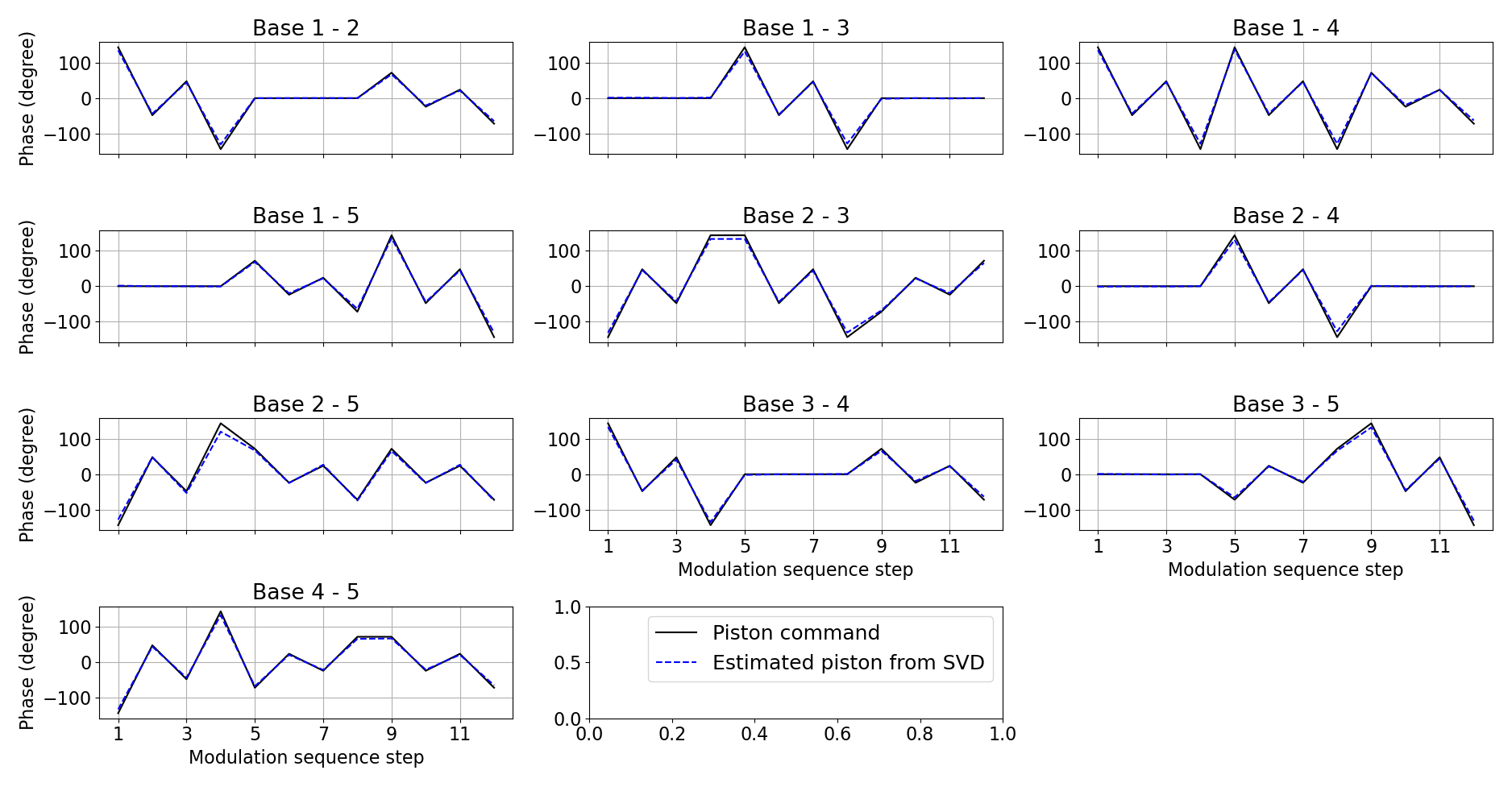}
    \caption[Phases estimées par la SVD comparées aux phases attendues en fonction des pas de la séquence de modulation.]{Phases estimées par la SVD comparées aux phases attendues en fonction des pas de la séquence de modulation, sur la puce $Y$. Pour chaque base, les phases attendues, induites par les pistons appliqués sur le MEMS, sont tracées en trait continu noir et les phases estimées sont tracées par dessus en trait discontinu bleu. Ces graphiques sont tracés pour la longueur d'onde $\sim 706 \,$nm.}
    \label{fig:PhaseEstimees}
\end{figure}

%%%%%%%%%%%%%%%%
% \subsection{Estimation du contraste interférométrique}
% les premiers termes de la V2PM
% comparaison pour les 2 puces
% comparaison avec le contraste obtenu dans la caractérisation des puces (section précédente)

%%%%%%%%%%%%%%%%%%%%%%%%%%%%%%%%
\section{Ajustement des franges}
\label{sec:FullOnFit}

Maintenant que la \ac{V2PM} est estimée, il s'agit de faire l'acquisition d'images contenant toutes les bases modulées en phase simultanément. La figure~\ref{fig:FullOnData} présente les interférogrammes mesurés pour la puce $Y$ et la puce $X$ sur les lignes d'images $1$ et $3$, respectivement (pour le canal spectral $\sim 706 \,$nm). Chaque colonne d'images est associée à une base et chaque image est une matrice de dimension $n_p \times n_S$ avec $n_p$ le nombre de points échantillonnés par base en fonction de l'\ac{OPD} (correspondant à une séquence de modulation) et $n_S$ le nombre de fois que la séquence de modulation a été acquise (les interférogrammes sont en fait empilés verticalement). On dispose ainsi de ces matrices pour chaque canal spectral et elles sont ensuite chacune multipliées par la \ac{P2VM} du canal spectral correspondant selon l'équation~\ref{eq:VisibiliteMes}. Le résultat sont les matrices des visibilités $\text{\textbf{P}}$, de dimensions $25 \times n_S$ pour cinq sous-pupilles considérées (et dix bases). On a alors pour chaque base (et chaque canal spectral) le terme de visibilité complexe selon l'équation~\ref{eq:mu}, dans lequel est inclus le terme de piston différentiel intrinsèque à l'instrument, $\psi_{nn'}$. Nous verrons dans la section~\ref{sec:PhaseSpecDiff} comment extraire le terme de phase associée à l'objet observé $\varphi_{nn'}$.

\begin{figure}[ht!]
    \begin{subfigure}{\textwidth}
        \centering
        \includegraphics[width=0.9\textwidth]{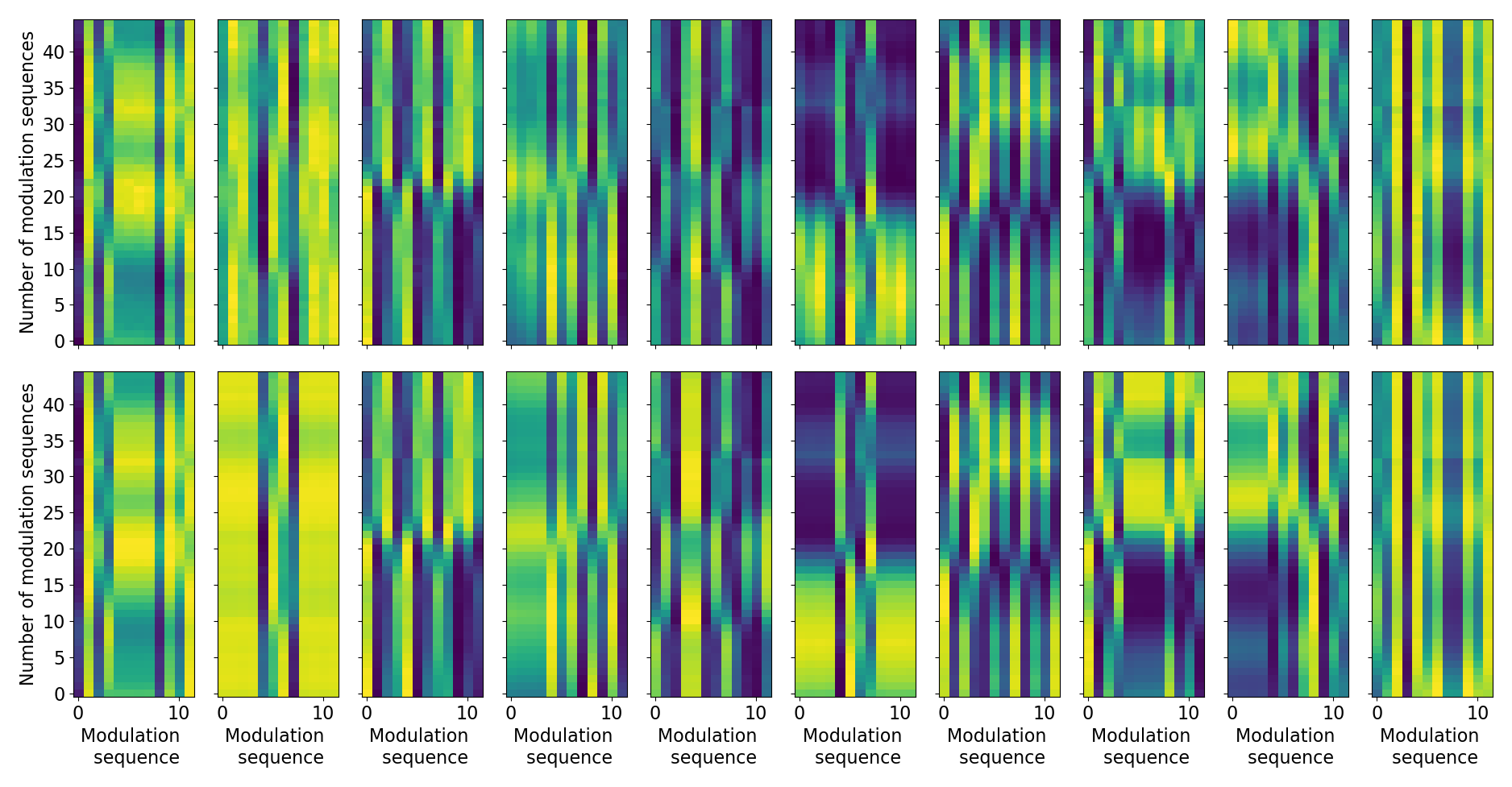}
    \end{subfigure}
    \begin{subfigure}{\textwidth}
        \centering
        \includegraphics[width=0.9\textwidth]{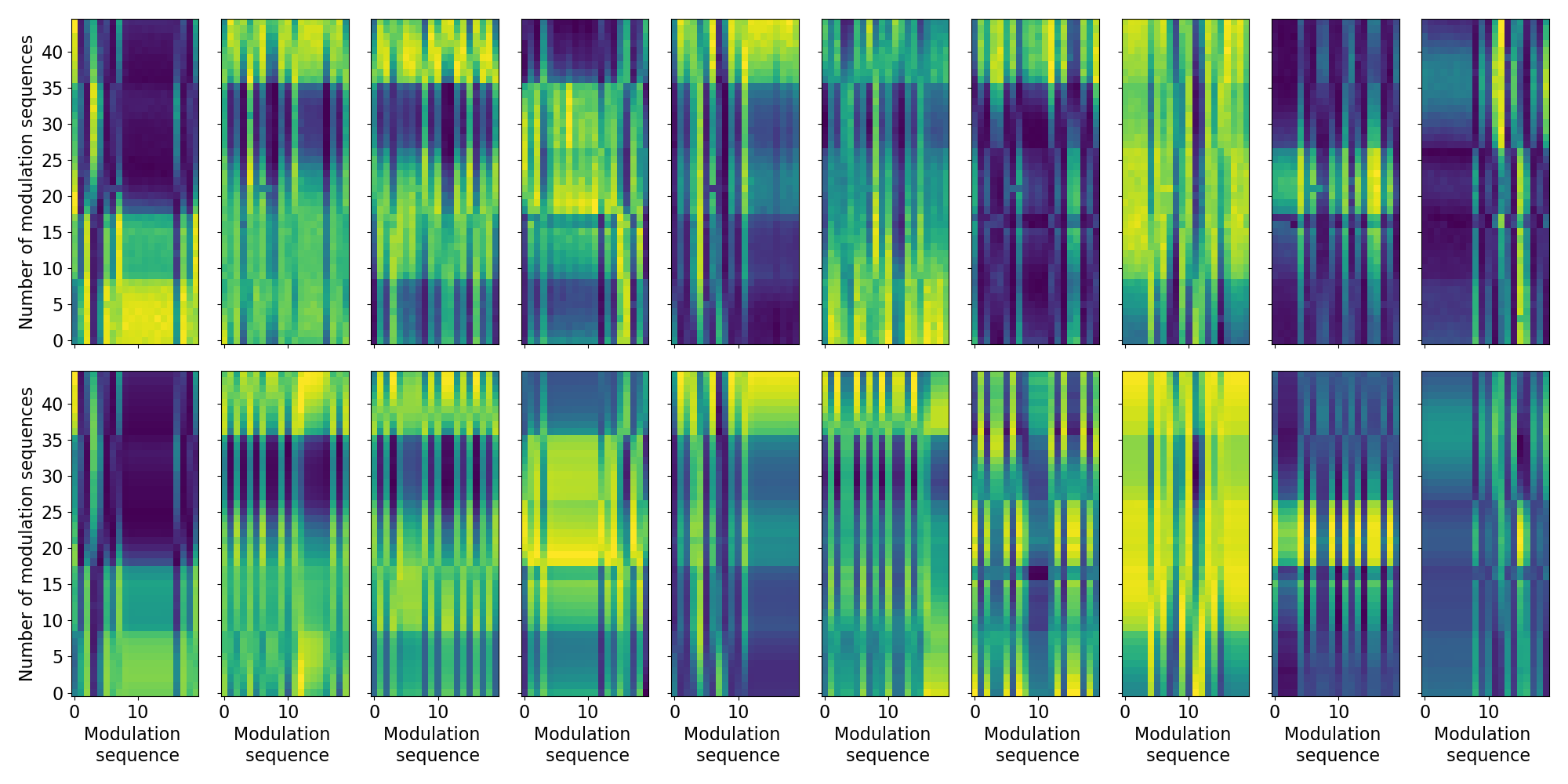}
    \end{subfigure}
    \caption[Interférogrammes mesurés et ajustés par la P2VM des puces $Y$ et $X$ mesurés sur FIRSTv2.]{Images des interférogrammes mesurés (lignes du haut) et ajustés (lignes du bas) sur les $10$ bases (en colonne), pour la puce $Y$ (les deux premières lignes) et la puce $X$ (les deux dernières lignes). L'axe horizontal est composé des $12$ et $20$ pas des séquences de modulation pour la puce $Y$ et la puce $X$, respectivement. Les axes verticaux sont le nombre de fois que la séquence de modulation a été acquise. Ces images sont normalisées et sont pour le canal spectral $\sim 706 \,$nm.}
    \label{fig:FullOnData}
\end{figure}

En multipliant la matrice des visibilités $\text{\textbf{P}}$ par la \ac{V2PM} selon l'équation~\ref{eq:FullOnFit}, on reconstruit l'ajustement des interférogrammes mesurés $I_r$. Les lignes d'image $2$ et $4$ de la figure~\ref{fig:FullOnData} sont les ajustements des interférogrammes des lignes $1$ et $3$, respectivement pour la puce $Y$ et la puce $X$.

\begin{equation}
    I_r = V2PM \cdot \text{\textbf{P}} \label{eq:FullOnFit}
\end{equation}

On note que sur tous ces interférogrammes on peut reconnaître la séquence de modulation en s'attardant sur les variations d'intensité selon l'axe horizontal des interférogrammes mesurés de la figure~\ref{fig:FullOnData}, en comparant chaque image de celle-ci à chaque ligne des séquences appliquées des figure~\ref{fig:ModSeq12} et figure~\ref{fig:ModSeq20}.

%%%%%%%%%%%%%%%%%%%%%%%%%%%%%%%%
\section{La phase différentielle spectrale : une observable auto-étalonnée}
\label{sec:PhaseSpecDiff}
% (gravity \citep{amorim2020}) mesure de la masse du trou noir central de la galaxie active IRAS 09149-6206 avec la phase différentielle

%%%%%%%%%%%%%%%%
\subsection{Estimation de la phase différentielle}

La phase différentielle spectrale est une observable auto-étalonnée des erreurs et biais de phase instrumentaux qui s'ajoutent à la phase inhérente à la source astrophysique lors de la mesure. Pour l'obtenir, on utilise le fait que la phase est mesurée en fonction de la longueur d'onde. Le concept de base \citep{buscher2015} est de soustraire la moyenne de la phase sur deux canaux spectraux $\lambda_0 - \Delta\lambda$ et $\lambda_0 + \Delta\lambda$ à la phase du canal spectral central $\lambda_0$. Trois mesures sont donc requises a minima. Dans notre cas, nous disposons de bien plus de canaux spectraux, ce qui permet de modéliser plus finement le continuum.

En pratique, on explicite l'expression de la phase mesurée en fonction du nombre d'onde $\sigma = 1 / \lambda$ pour mettre en évidence les erreurs de phase instrumentales, comme cela a été fait pour le traitement des données d'\ac{AMBER} \citep{millour2008} mais aussi de l'instrument \ac{GRAVITY} \citep{lapeyrere2014}. Premièrement, la phase de l'objet astrophysique peut être développée selon :

\begin{equation}
    \varphi^{ob}_{nn'}(\sigma) = a^{ob}_0 + a^{ob}_1 \sigma + \delta\varphi^{ob}_{nn'}(\sigma) \label{eq:PhaseSource}
\end{equation}

\noindent où $\delta\varphi^{ob}_{nn'}(\sigma)$ est la phase de l'objet qui ne peut pas être exprimée par une pente de phase et qu'on cherche à mesurer : la raie d'émission d'une protoplanète ou une composante en rotation comme cela a été mesuré par l'instrument \ac{GRAVITY} sur le quasar 3C 273 \citep{sturm2018} et sur le trou noir central de la galaxie active IRAS 09149-6206 \citep{amorim2020}.

De plus, l'atmosphère ajoute un terme de phase $\varphi^{atmos}_{nn'}(t, \sigma)$ via une différence de marche achromatique (appelée piston) $p_{nn'}(t)$ s'exprimant :

\begin{equation}
    \varphi^{atmos}_{nn'}(t, \sigma) = 2\pi p_{nn'}(t) \sigma
\end{equation}

Les lignes à retard compensent des différences de longueur de chemin optique dans le verre des fibres optiques en ajoutant une longueur de chemin optique dans l'air. La dépendance spectrale de l'indice de réfraction implique que cette compensation ne peut se faire que pour une longueur d'onde et induit un terme de dispersion chromatique (voir la section~\ref{sec:PhaseDiffFIRSTv2}) noté $\varphi^{disp}_{nn'}(\sigma^2)$.
% \kevinco{pour la dispersion quadratique, peut-on l'exprimer comme dans lacour2014 (section 5.2) ? $D(l1, l2) = delta_air * (n^{l1}_{air} - n^{l2}_{air}) + delta_fibre * (n^{l1}_{fibre} - n^{l2}_{fibre})$ d'où dans nos phases mesurées le terme $(2 pi D(1/l)) / l, en sigma**2$ ?}

Aussi, un terme de phase inhérent à l'instrument est mesuré, de différence de marche $\delta_{nn'}$ et la phase mesurée $\varphi^{m}_{nn'}(t, \sigma)$ s'écrit finalement :

\begin{align}
    \varphi^{m}_{nn'}(t, \sigma) &= \varphi^{ob}_{nn'}(\sigma) + 2\pi\delta_{nn'}\sigma + \varphi^{atmos}_{nn'}(t, \sigma) + \varphi^{disp}_{nn'}(\sigma)\\
    &= a^{ob}_0 + 2\pi(a^{ob}_1 + \delta_{nn'} + p_{nn'}(t)) \sigma + \varphi^{disp}_{nn'}(\sigma^2) + \delta\varphi^{ob}_{nn'}(\sigma)\label{eq:phasefit}
\end{align}

La phase différentielle $\delta\varphi^{m}_{nn'}(\sigma)$ s'obtient par la soustraction de la phase mesurée à une longueur d'onde de travail $\lambda_{work} = 1 / \sigma_{work}$ par la phase mesurée à une longueur d'onde de référence $\lambda_{ref} = 1 / \sigma_{ref}$. Pour cela, on calcule l'argument du produit inter-spectral de la cohérence complexe dans ces deux canaux spectraux :

\begin{equation}
	\delta\varphi^{m}_{nn'}(\sigma) = arg \left ( \langle \mu_{nn'}(t, \sigma_{work}) \mu_{nn'}(t, \sigma_{ref})^* \rangle_t \right ) = \delta\varphi^{ob}_{nn'}(\sigma)
\end{equation}

%%%%%%%%%%%%%%%%
\subsection{En pratique sur FIRSTv2}
\label{sec:PhaseDiffFIRSTv2}

Les canaux spectraux de référence sont définis par toute la bande spectrale en excluant les canaux spectraux de travail. Pour un signal mesuré sur une protoplanète en accrétion qui présente une forte raie d'émission \ha~(voir la section~\ref{sec:Protoplanetes}), les canaux spectraux de travail sont ceux englobant la longueur d'onde \ha. Le signal de phase sur les canaux spectraux de référence correspond au signal du continuum non résolu par l'instrument (l'étoile centrale) et il est nul. Par conséquent, le signal qu'on mesure effectivement sur les canaux spectraux de référence est égal à la phase provenant des pistons atmosphériques et instrumentaux et son ajustement permet de le corriger par interpolation sur les canaux de travail.

On ajuste donc une fonction polynomiale du second degré aux mesures de phases sur les canaux spectraux de référence, image par image (selon la variable $t$), afin de déterminer la fonction de l'équation~\ref{eq:phasefit}. La soustraction de cette fonction ajustée aux phases mesurées sur toute la bande spectrale permet de mettre en évidence le signal provenant uniquement de la protoplanète résolue, correspondant au terme de hauts ordres $\delta\varphi^{*}_{nn'}(\sigma)$ dans l'équation~\ref{eq:PhaseSource} de la phase de la source.

La figure~\ref{fig:FitPhaseVis} présente une telle procédure sur des données prises sur une source binaire (comme expliqué dans la section~\ref{sec:SystBinaire}). Pour chaque base (identifiée dans les sous-titres par \textit{B\#}), un graphique présente les phases des visibilités mesurées en vert et son ajustement en trait discontinu rouge en fonction du nombre d'onde $\sigma = 1/ \lambda$. Le signal du compagnon est entre $0.0015$ et $0.0016 \, \text{nm}^{-1}$ sur les courbes vertes, mais est faible et difficile à voir. La fonction polynomiale trouvée par ajustement est écrite dans le sous-titre de chaque graphique. On note que les phases sont quadratiquement très dépendantes du nombre d'onde.

\begin{figure}[ht!]
    \centering
    \includegraphics[width=\figwidth]{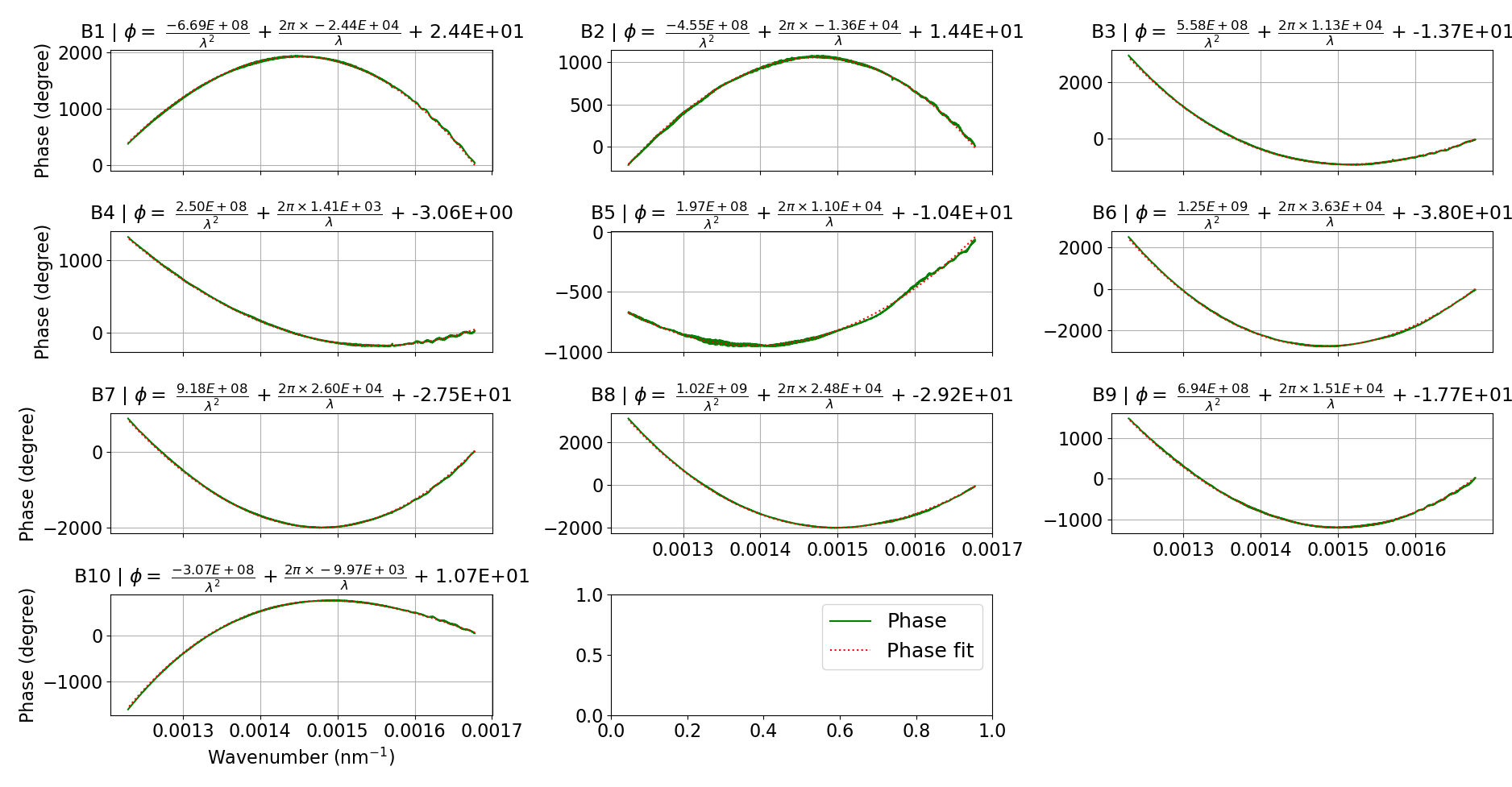}
    \caption[Phases mesurées et ajustées en fonction du nombre d'onde sur la puce $Y$.]{Phases mesurées déroulées (en vert) et leur ajustement (en rouge) en fonction du nombre d'onde pour la puce $Y$. Les numéros des bases sont indiqués dans le titre de chaque graphique par la dénomination \textit{B\#} et suivis par les termes du polynôme de second degré ajusté aux phases.}
    \label{fig:FitPhaseVis}
\end{figure}

Les phases différentielles résultantes de la soustraction entre les fonctions ajustées et les phases mesurées sont tracées sur la figure~\ref{fig:PhaseDiffEx} et seront analysées plus loin dans la section~\ref{sec:PhaseDiffAnalyse}.

\begin{figure}[ht!]
    \centering
    \includegraphics[width=\figwidth]{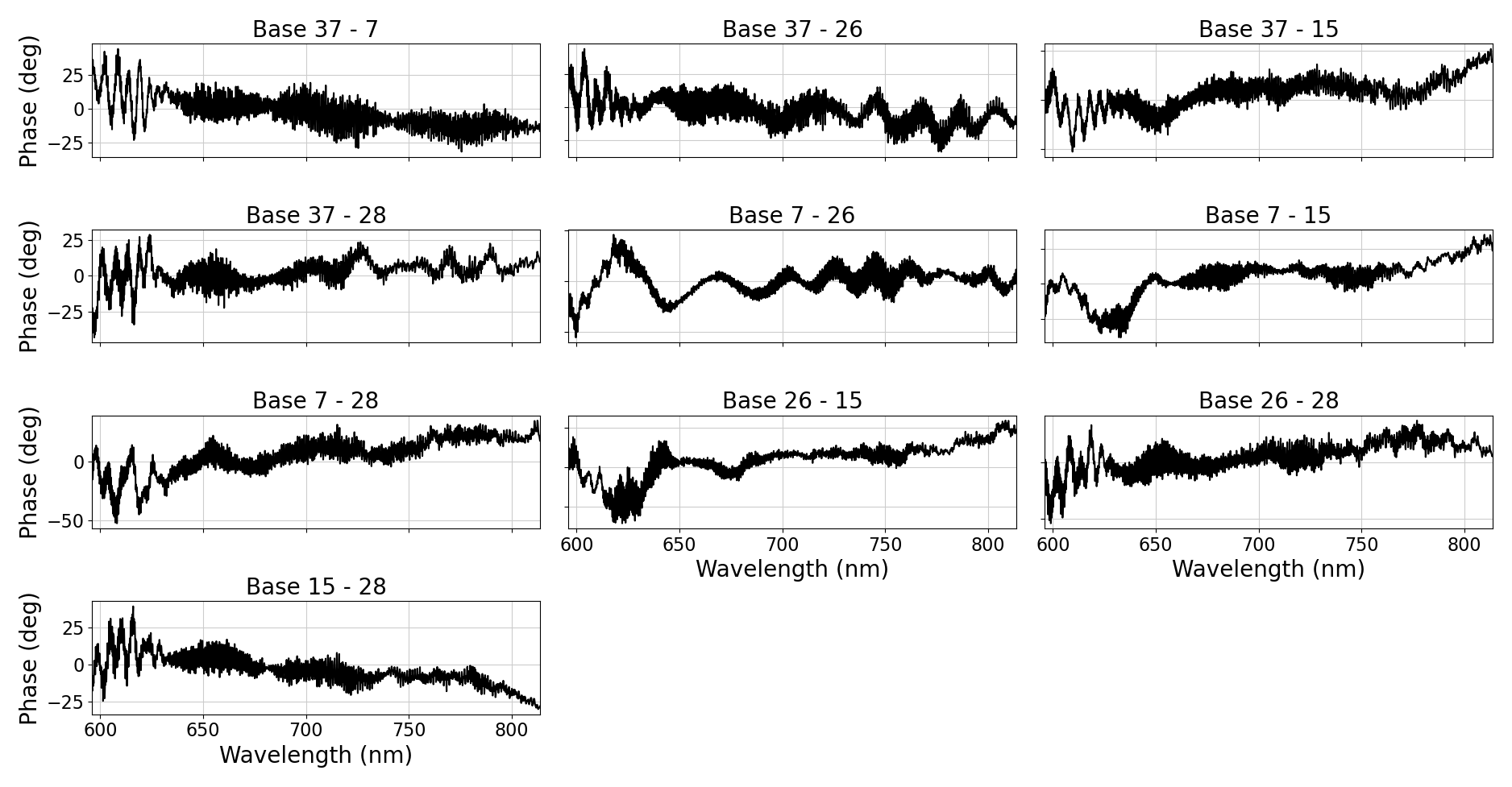}
    \caption[Phases différentielles mesurées sur la source interne du banc de test.]{Phases différentielles mesurées sur la source interne du banc de test (non résolue), pour les dix bases de FIRSTv2, avec la puce $Y$. Les mesures sont présentées en fonction de la longueur d'onde entre $620 \,$nm et $660 \,$nm.}
    \label{fig:PhaseDiffEx}
\end{figure}

%%%%%%%%%%%%%%%%
\subsection{Les intérêts de la phase différentielle}

Comme nous l'avons vu, la phase différentielle est une observable auto-étalonnée des pistons atmosphériques et des biais instrumentaux. Du fait qu'elle s'obtienne par la soustraction du signal de la cible observée sur quelques canaux spectraux de travail par le signal du continuum sur le reste de la bande spectrale, il se trouve qu'elle est très bien adaptée au cas scientifique qui m'intéresse dans le cadre de ma thèse. En effet, comme je l'ai présenté dans la section~\ref{sec:Protoplanetes}, les protoplanètes en accrétion de matière émettent une forte raie d'émission \ha. Le signal de phase mesuré par \ac{FIRSTv2} sur un tel système présenterait alors un fort pic sur quelques canaux spectraux (canaux de travail), ce qui se prête bien au calcul de la phase différentielle.

De plus, il n'est pas nécessaire d'observer une cible non-résolue (de référence) pour étalonner les phases différentielles, contrairement aux clôtures de phase. C'est un avantage considérable car cela nécessite moins de temps d'observations sur télescope (rare et précieux).

Enfin, la phase différentielle est une observable au premier ordre et à la différence des clôtures de phase qui sont au troisième ordre. Pour N sous-pupilles combinées, on obtient $N(N-1)/2$ phases différentielles indépendantes et $(N-1)(N-2)/2$ clôtures de phase indépendantes. Ainsi, pour $5$ sous-pupilles on obtient $10$ phases différentielles indépendantes et $6$ clôtures de phase indépendantes \citep{millour2006}.

%%%%%%%%%%%%%%%%%%%%%%%%%%%%%%%%
\section{La stabilité des mesures sur le banc FIRSTv2}
\label{sec:CPStabilityMeudon}

Pour caractériser la stabilité des mesures de phases sur le banc de test de \ac{FIRSTv2}, j'acquiers des données sur la source interne non résolue. Ainsi, on s'attend à ce que le module des visibilités complexes estimées sur une telle source soit égale à $|V_{nn'}| = 1$ et sa phase à $\varphi_{nn'} = 0$. Par conséquent, l'équation~\ref{eq:mu} du terme de cohérence complexe de la base formée par les sous-pupilles $n$ et $n'$ devient $\mu_{nn'} = A_n A_{n'} e^{i(\Delta\Phi_{nn'})}$. La mesure de la phase de la cohérence complexe sur la source interne revient donc à mesurer le piston différentiel des perturbations sur le banc $\Delta\Phi_{nn'} = 2 \pi \sigma \delta_{nn'} [2 \pi]$. Pour ce faire, une fonction polynomiale d'ordre $2$ est ajustée à chaque mesure de phase déroulée selon la méthode présentée dans la section~\ref{sec:PhaseDiffFIRSTv2}, ce qui permet d'extraire le terme d'\ac{OPD} $\delta_{nn'}$, qui est le coefficient du terme polynomial d'ordre $1$.

Ainsi, la figure~\ref{fig:OPDfitVStime} est un graphique qui trace en fonction du temps, l'\ac{OPD} des bases $37-7$, $37-15$ et $7-15$ (en haut) et la clôture de phase \citep{weigelt1977, lohmann1983} associée est tracée en rouge (en bas). Les courbes sont tracées pour $1\,980$ images de temps d'exposition égal à $100 \,$ms. La moyenne de chaque courbe est tracée en trait discontinu noir et celle de la clôture de phase vaut $\upmu = 90 \,$nm avec comme écart-type $\sigma = 50 \,$nm. L'écart-type des \ac{OPD}s sont estimées à $90 \,$nm, $130 \,$nm et $80 \,$nm, respectivement. Cela montre que le calcul de la clôture de phase corrige les perturbations présentes sur le banc.

\begin{figure}[ht!]
    \centering
    \includegraphics[width=\figwidth]{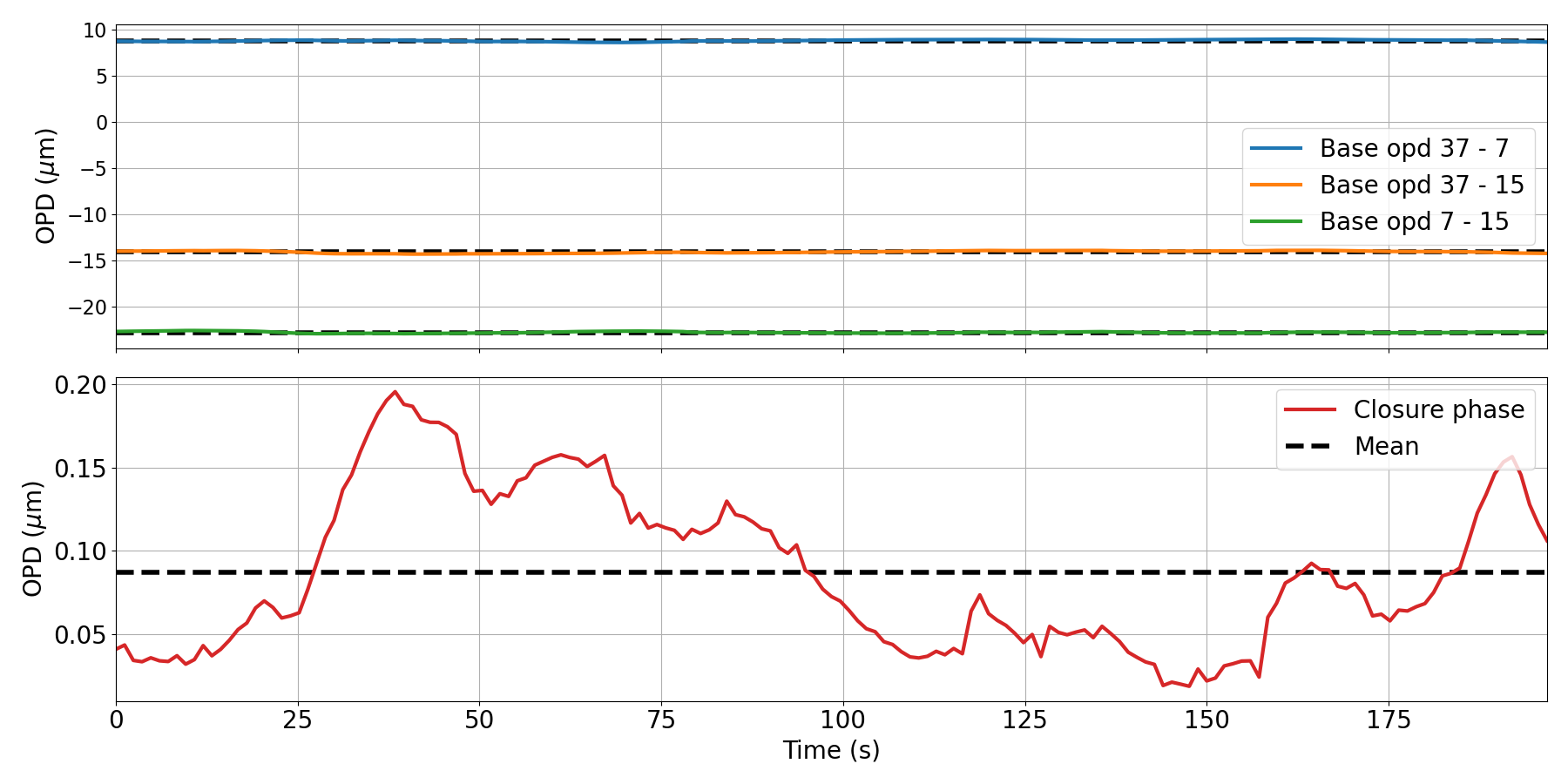}
    \caption[Graphique de l'OPD de trois bases et de la clôture de phase en fonction du temps, mesurée sur FIRSTv2 avec la puce $Y$.]{Graphique de l'OPD des bases $37-7$, $37-15$ et $7-15$ (en haut), en fonction du temps, mesurée sur FIRSTv2 avec la puce $Y$. La clôture de phase formée par ces trois bases est tracée en rouge (en bas) et les moyennes des quatre courbes sont tracées en trait noir discontinue. La moyenne de la clôture de phase est égale à $90 \,$nm et l'écart-type est de $50 \,$nm. Les courbes sont tracées pour $1\,980$ images de temps d'exposition égal à $100 \,$ms.}
    \label{fig:OPDfitVStime}
\end{figure}

Sous l'hypothèse que les variations temporelles des perturbations instrumentales se moyennent à zéro, le terme d'\ac{OPD} $\delta_{nn'}$ mesuré est la différence de longueur de chemin optique entre les faisceaux. La figure~\ref{fig:FiberPiston} montre les longueurs relatives des cinq bras de \ac{FIRSTv2}, induites par résolution du système d'équation linéaire reliant les \ac{OPD}s précédemment présentées aux longueurs de chemin relatives des bras. Chaque faisceau est identifié par le numéro du segment du \ac{MEMS} qu'il illumine (axe des abscisses). Les segments utilisés sont ceux présentés dans la section~\ref{sec:BaseConfig}. Il faut noter que ce sont les valeurs de longueurs relatives des faisceaux au moment de la mesure et celles-ci peuvent être minimisées en changeant les positions des \ac{ODL}s, rapprochant les interférogrammes de la frange centrale d'\ac{OPD} nulle.

\begin{figure}[ht!]
    \centering
    \includegraphics[width=\figwidth]{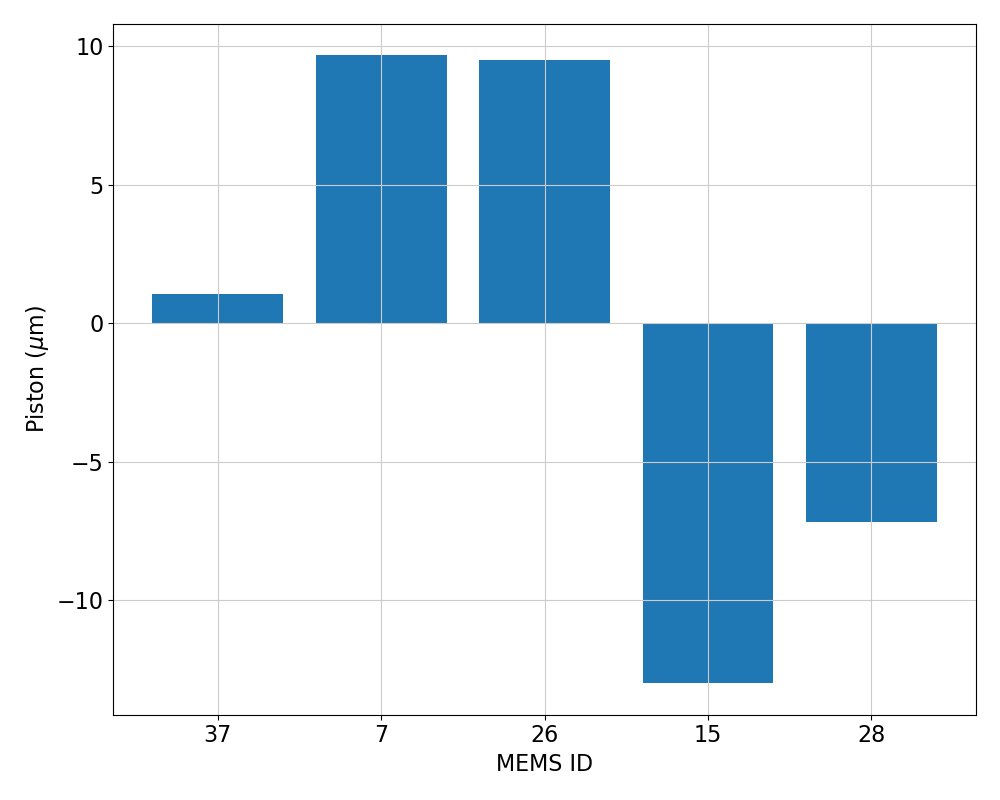}
    \caption[Longueurs relatives des $5$ bras de l'interféromètre de FIRSTv2, mesurées avec la puce $Y$.]{Longueurs relatives des $5$ bras de l'interféromètre de FIRSTv2, mesurées avec la puce $Y$. Chaque faisceau est identifié par le numéro du segment du MEMS qu'il illumine (axe des abscisses).}
    \label{fig:FiberPiston}
\end{figure}

%%%%%%%%%%%%%%%%%%%%%%%%%%%%%%%%
\section{Conclusion}

Dans ce chapitre j'ai présenté le processus du traitement des données de \ac{FIRSTv2} que j'ai développé pendant ma thèse. Celui-ci se base sur la méthode de calcul de la \ac{P2VM} qui est implémentée pour \ac{FIRSTv1} et précédemment pour \ac{AMBER}.

On a vu les différents étalonnages de l'instrument qui sont primordiaux dans le processus du traitement de données. Notamment l'étalonnage spectral sur toutes les sorties imagées sur le détecteur permettant l'identification des canal spectraux sur les pixels de la caméra. Mais encore, j'ai présenté tout le processus d'étalonnage de la \ac{P2VM} consistant à estimer la phase des interférogrammes pour chaque pas d'une séquence de modulation des franges connue et appliquée sur les segments du \ac{MEMS}.

Enfin, j'ai présenté la phase différentielle qui est une observable permettant de s'affranchir des perturbations instrumentales de bas ordres et que j'ai implémenté dans le programme de traitement de données pendant ma thèse. La phase différentielle est adaptée à la mesure interférométrique sur des sources lumineuses telles que les systèmes protoplanétaires dont le compagnon présente un faible contraste dans une raie d'émission. Elle sera exploitée dans le chapitre suivant pour l'analyse des données acquises sur un système protoplanétaire simulé sur le banc de test.

%%%%%%%%%%%%%%%%%%%%%%%%%%%%%%%%%%%%%%%%%%%%%%%%%%%%%%%%%%%%%%%%
\chapter{Caractérisation de systèmes protoplanétaires}
\label{sec:BinaryCharac}
\setcounter{figure}{0}
\setcounter{table}{0}
\setcounter{equation}{0}

\minitoc

\clearpage
Il est crucial pour montrer les performances d'un interféromètre de caractériser un système présentant plusieurs composantes lumineuses (système binaire ou protoplanétaire). Il s'agit d'estimer à la fois le contraste entre les deux composantes ainsi que leur séparation. Dans cette section, je montrerai d'abord comment j'ai pu simuler un système protoplanétaire sur le banc de test de \ac{FIRSTv2} afin de montrer sa détection. Dans un second temps, j'exposerai le modèle de phases différentielles pour un tel système, leur mesure et analyse sur la source protoplanétaire simulée du banc de test. Nous verrons la détection du compagnon d'un tel système à partir de l'analyse des phases différentielles. Enfin, j'exposerai et analyserai les limitations rencontrées sur ces mesures.

Dans toute cette étude, je ne montrerai aucune mesure de visibilité car il n'est pas possible de les normaliser dans la configuration de \ac{FIRSTv2}. En effet, pour ce faire nous avons besoin de la mesure indépendante de la photométrie sur chacune des sous-pupilles (comme expliqué dans le cadre du traitement de données de l'instrument \ac{AMBER} \citep{tatulli2007}), ce que nous n'avons pas à disposition. Il est possible de se passer de cette normalisation en ajoutant de la redondance lors du choix des bases, ce qui augmenterait le nombre de mesures par rapport au nombre d'inconnues \citep{lacour2007}. Mais cette méthode est mathématiquement plus complexe à mettre en oeuvre que de se passer d'exploiter la norme des visibilités et d'utiliser les observables auto-étalonnées que sont les phases différentielles.

% Dans ces deux instruments, une partie du ﬂux de chaque faisceau est prélevée et mesurée (Che et al., 2010, pour MIRC), aﬁn de pouvoir étalonner les observables interférométriques, c’est-à-dire étalonner les mesures de cohérences pour aboutir aux mesures de visibilité.

%%%%%%%%%%%%%%%%%%%%%%%%%%%%%%%%
\section{Le simulateur de source protoplanétaire sur le banc optique}
\label{sec:SystBinaire}

Afin d'évaluer les capacités de l'instrument à détecter et étudier un système exoplanétaire, j'ai mis en place un système optique simulant une telle source. Dans le but de simuler une protoplanète en accrétion (voir section~\ref{sec:AccretionAlpha}) j'utilise une source à bande spectrale étroite (un laser) et pour simuler une étoile j'utilise une source à bande spectrale large. Les deux sources doivent être injectées sur le banc, vues à l'infini avec une séparation angulaire définie. À cette fin, j'injecte ces deux sources dans les fibres d'un V-Groove (voir la figure~\ref{fig:VGroove}) qui est un composant alignant plusieurs fibres optiques côte à côte avec une séparation de $127 \,$\um. Une lentille convergente de $30 \,$cm de focale collimate ensuite les faisceaux de sorties de ce V-Groove selon le schéma présenté sur la figure~\ref{fig:BinarySystA}. Ainsi, j'obtiens une source centrale (tracés bleus) qui simule l'étoile sur l'axe optique de l'instrument et une source décentrée (tracés rouges) qui simule un compagnon hors axe avec un angle $\Theta$. Cet angle se calcule simplement à partir de la séparation $r$ des fibres du V-Groove et de la focale $f'$ de la lentille comme suit : $\Theta_m = r / f' = 4,23 . 10^{-4} \,$rad (ou $87,3 \,$as), soit une taille de $0,68 \lambda / B$ et de $1,37 \lambda / B$ pour les bases de longueur $1,05 \,$mm et $2,10 \,$mm, respectivement. L'axe du système simulé est en effet horizontal, ce qui permet de le sonder avec deux bases de longueur différentes avec la configuration des sous-pupilles choisies (voir la figure~\ref{fig:SegUVSimuleRappel}). À titre de comparaison, la base la plus grande à laquelle on a accès sur le télescope Subaru a une longueur de $\text{B} = 6,86 \,$m, ce qui correspond à l'observation d'un système exoplanétaire avec une séparation de $13 \,$mas ($0,68 \lambda / B$).

\begin{figure}[ht!]
    \centering
    \begin{subfigure}{\textwidth}
        \centering
        \includegraphics[width=0.95\textwidth]{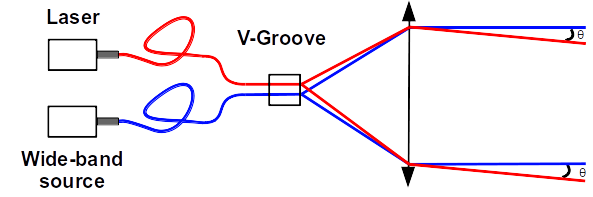}
        \caption{Schéma du système de source exoplanétaire avec, de la gauche vers la droite, la source large bande simulant le continuum et le laser simulant un compagnon, le V-Groove et une lentille convergente de $30 \,$cm de focale.}
        \label{fig:BinarySystA}
    \end{subfigure}
    \begin{subfigure}[t]{0.95\textwidth}
        \centering
        \includegraphics[width=\textwidth]{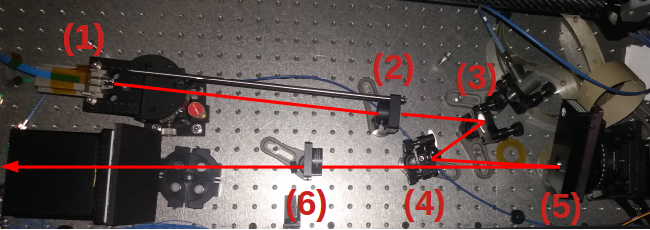}
        \caption{Photo du simulateur de source exoplanétaire sur le banc de test de FIRSTv2. On y voit le V-Groove d'injection des sources (1) la lentille de collimation (2), le miroir plan de renvoi (3), le miroir en forme de D (4), le miroir MEMS (5) et la première lentille du système afocal (6).}
        \label{fig:BinarySystB}
    \end{subfigure}
    \caption[Simulateur du système protoplanétaire sur le banc de test de FIRSTv2 à Meudon.]{Simulateur du système protoplanétaire sur le banc de test de FIRSTv2 à Meudon.}
    \label{fig:BinarySyst}
\end{figure}

La figure~\ref{fig:BinarySystB} est une photo de ce montage installé sur le banc de test. On peut voir le V-Groove (1) et ses fibres bleues sur la gauche qui illuminent une lentille (2) au centre de l'image. Un premier miroir plan (3) renvoie le faisceau sur le miroir en forme de D (4) qui sert à illuminer le miroir segmenté (5) avec une incidence quasi-normale. La lentille (6) est la première lentille du doublet afocal d'agrandissement du faisceau (voir la section~\ref{sec:FiberInjection}) avant l'injection des faisceaux dans les fibres optiques vers la gauche de la photo.

\begin{figure}[ht!]
    \centering
    \begin{subfigure}{0.39\textwidth}
        \centering
        \includegraphics[width=\textwidth]{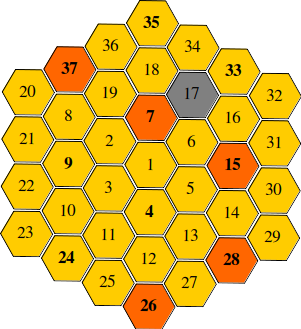}
        \caption{Configuration des sous-pupilles choisies (en orange) dans le plan pupille sur la carte des segments du MEMS.}
        \label{fig:SegUVSimuleRappelA}
    \end{subfigure}\hfill
    \begin{subfigure}{0.59\textwidth}
        \centering
        \includegraphics[width=\textwidth]{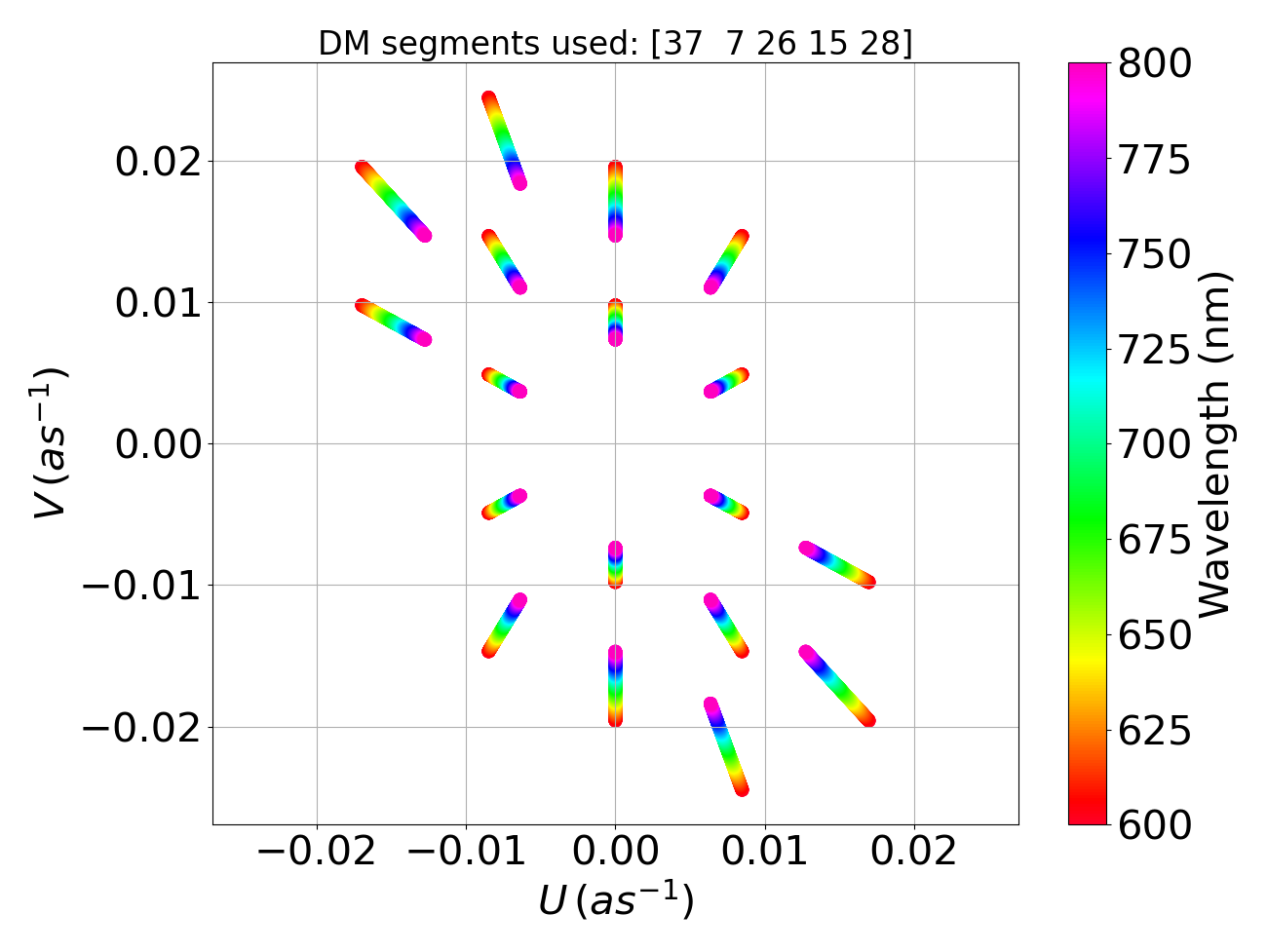}
        \caption{La répartition des bases choisies représentée dans l'espace de Fourier, appelée aussi la couverture du plan UV des fréquences spatiales. Les couleurs représentent la longueur d'onde.}
        \label{fig:SegUVSimuleRappelB}
    \end{subfigure}
    \caption[Configuration des sous-pupilles et couverture du plan UV du banc de test de FIRSTv2.]{Configuration des sous-pupilles et couverture du plan UV du banc de test de FIRSTv2. Plus de détails sont présentés dans la section~\ref{sec:BaseConfig}.}
    \label{fig:SegUVSimuleRappel}
\end{figure}

%%%%%%%%%%%%%%%%%%%%%%%%%%%%%%%%
\section{L'analyse des phases différentielles sur une source protoplanétaire simulée}
\label{sec:PhaseDiffAnalyse}
%\kevinco{préciser sur quelle puce les phases sont mesurées, notamment si je ne présente les phases que d'une puce}

%%%%%%%%%%%%%%%%
\subsection{Les observables interférométriques sur un système protoplanétaire}

Pour comprendre et analyser les phases différentielles on souhaite exprimer la forme d'un tel signal de phase pour un système protoplanétaire tel que le système PDS 70. Pour commencer, la distribution d'intensité d'une telle source s'exprime comme :

\begin{equation}
    I = \frac{1}{1 + \rho(\lambda)} (\delta(\vv{r} - \vv{r_1}) + \rho(\lambda) \delta(\vv{r} - \vv{r_2}))
\end{equation}

avec $\vv{r} = (\alpha, \beta)$ le vecteur position angulaire, $\vv{r_1}$, $\vv{r_2}$ les positions de l'étoile et du compagnon respectivement, $\delta(\vv{r} - \vv{r_i})$ le pic de Dirac modélisant la source $i$ à la position $r_i$. Pour modéliser le contraste du système, j'utilise une fonction de Gauss centrée sur la longueur d'onde $\lambda_c$ du pic d'émission du compagnon et d'écart-type $\sigma_c$ de la façon suivante : 

\begin{equation}
    \rho(\lambda) = \rho_{\lambda_c} e^{\frac{-(\lambda - \lambda_c)^2}{2 \sigma_{c}^2}}
\end{equation}

où $\rho_{\lambda_c}$ est le contraste du système à la longueur d'onde $\lambda_c$. La visibilité complexe associée à une telle source s'écrit alors :

\begin{equation}
    V = \frac{e^{2i\pi \vv{r_1} \cdot \vv{f}}}{1 + \rho(\lambda)} \left( 1 + \rho(\lambda) e^{2i\pi \vv{\Theta} \cdot \vv{f}} \right)
\end{equation}

avec $\vv{\Theta} = \vv{r_2} - \vv{r_1}$ la séparation angulaire des deux composantes du système et $\vv{f} = \vv{B} / \lambda$ le vecteur fréquence spatiale. $\vv{B}$ est le vecteur de base formé par deux sous-pupilles. On en déduit ainsi la phase comme suit :

\begin{equation}
    \varphi = atan \left( \frac{sin(2\pi \vv{r_1} \cdot \vv{f}) + \rho(\lambda) sin(2\pi \vv{r_2} \cdot \vv{f})}{cos(2\pi \vv{r_1} \cdot \vv{f}) + \rho(\lambda) cos(2\pi \vv{r_2} \cdot \vv{f})} \right) \label{eq:PhaseBinaire}
\end{equation}

La phase dépend donc de la position de la source par rapport au point centré par le télescope. Nous considérons donc que le télescope pointe sur l'étoile centrale, i.e. $\vv{r_1} = (0, 0)$ nous permettant de simplifier l'équation~\ref{eq:PhaseBinaire} comme suit :

\begin{equation}
    \varphi = atan \left( \frac{\rho(\lambda) sin(2\pi \vv{\Theta} \cdot \vv{f})}{1 + \rho(\lambda) cos(2\pi \vv{\Theta} \cdot \vv{f})} \right) \label{eq:PhaseBinaireCentree}
\end{equation}

De plus, pour un système avec un haut contraste telle que les systèmes exoplanétaires, i.e. $\rho(\lambda) \ll 1$, l'équation~\ref{eq:PhaseBinaireCentree} s'approxime de la manière suivante :

\begin{equation}
    \varphi \approx \rho(\lambda) sin(2\pi \vv{\Theta} \cdot \vv{f})
    \label{eq:PhaseBinaireHautContraste}
\end{equation}

%%%%%%%%%%%%%%%%
\subsection{Le modèle des signaux de phase différentielle pour un système protoplanétaire}

Le signal de phase différentielle mesuré par un interféromètre dépend des bases (de leur longueur et de leur orientation) mais aussi de la séparation entre les deux composantes de la source observée. Pour illustrer ce propos, le signal de phase attendu, d'après l'équation~\ref{eq:PhaseBinaireCentree}, pour les bases présentées précédemment, est tracé en fonction de la séparation entre l'étoile et le compagnon pour un contraste de luminosité égal à $1$ sur la figure~\ref{fig:PhaseDiffBinSizeSimuleA}, à $0,1$ sur la figure~\ref{fig:PhaseDiffBinSizeSimuleB} et à $0,01$ sur la figure~\ref{fig:PhaseDiffBinSizeSimuleC}. On retrouve les trois amplitudes de phase correspondant aux trois groupes de bases identifiés dans la section~\ref{sec:BaseConfig} avec celle qui est nulle quelle que soit la séparation (correspondant aux bases orthogonales au système protoplanétaire). De plus, comme l'indique l'équation~\ref{eq:PhaseBinaireHautContraste}, la phase est sinusoïdale pour des hauts contrastes ($< 0,1$), avec une amplitude égale à ce contraste, ici $0,1 \, \text{rad} = 5,7\degree$ et $0,01 \, \text{rad} = 0,57\degree$, comme attendu, pour les figures~\ref{fig:PhaseDiffBinSizeSimuleB} et \ref{fig:PhaseDiffBinSizeSimuleC}, respectivement. Par conséquent, plus le système protoplanétaire présente un contraste élevé, plus le signal de phase attendu est faible et plus le compagnon est difficile à détecter. Les systèmes protoplanétaires qui présentent un contraste plus faible dans la raie d'émission \ha~sont donc particulièrement intéressants car le signal de phase à détecter est de plus forte amplitude.

\begin{figure}[ht!]
    \centering
    \begin{subfigure}{0.5\textwidth}
        \centering
        \includegraphics[width=\textwidth]{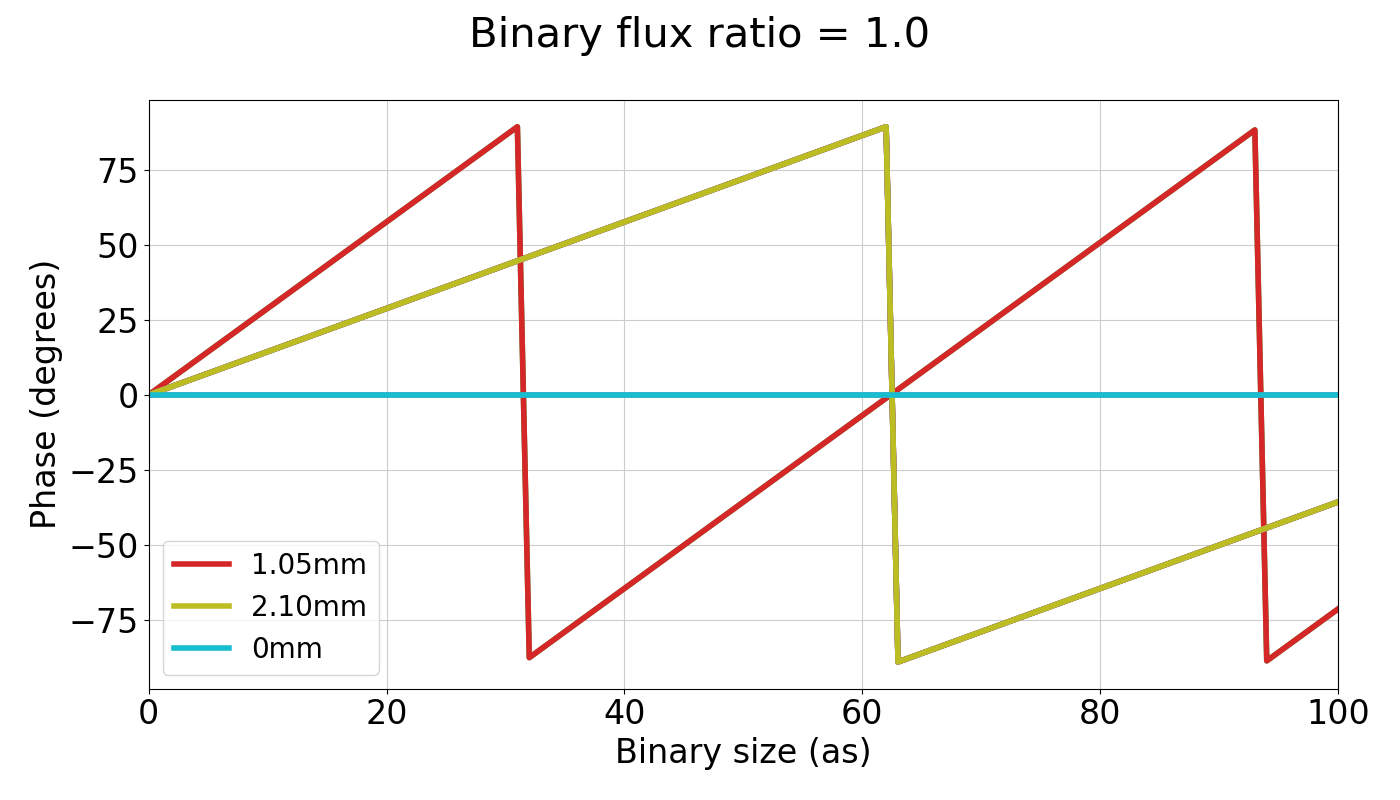}
        \caption{}
        \label{fig:PhaseDiffBinSizeSimuleA}
    \end{subfigure}
    \begin{subfigure}{0.5\textwidth}
        \centering
        \includegraphics[width=\textwidth]{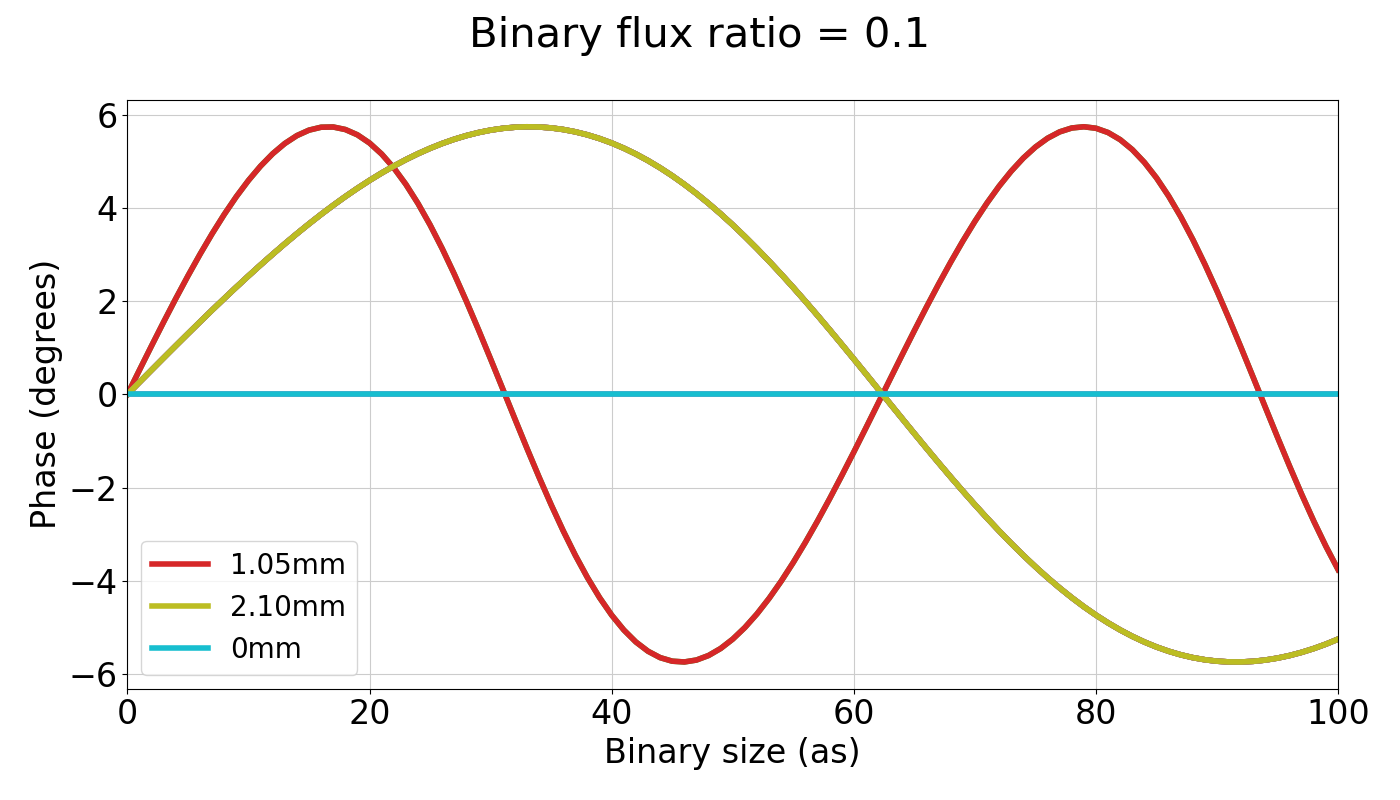}
        \caption{}
        \label{fig:PhaseDiffBinSizeSimuleB}
    \end{subfigure}%
    \begin{subfigure}{0.5\textwidth}
        \centering
        \includegraphics[width=\textwidth]{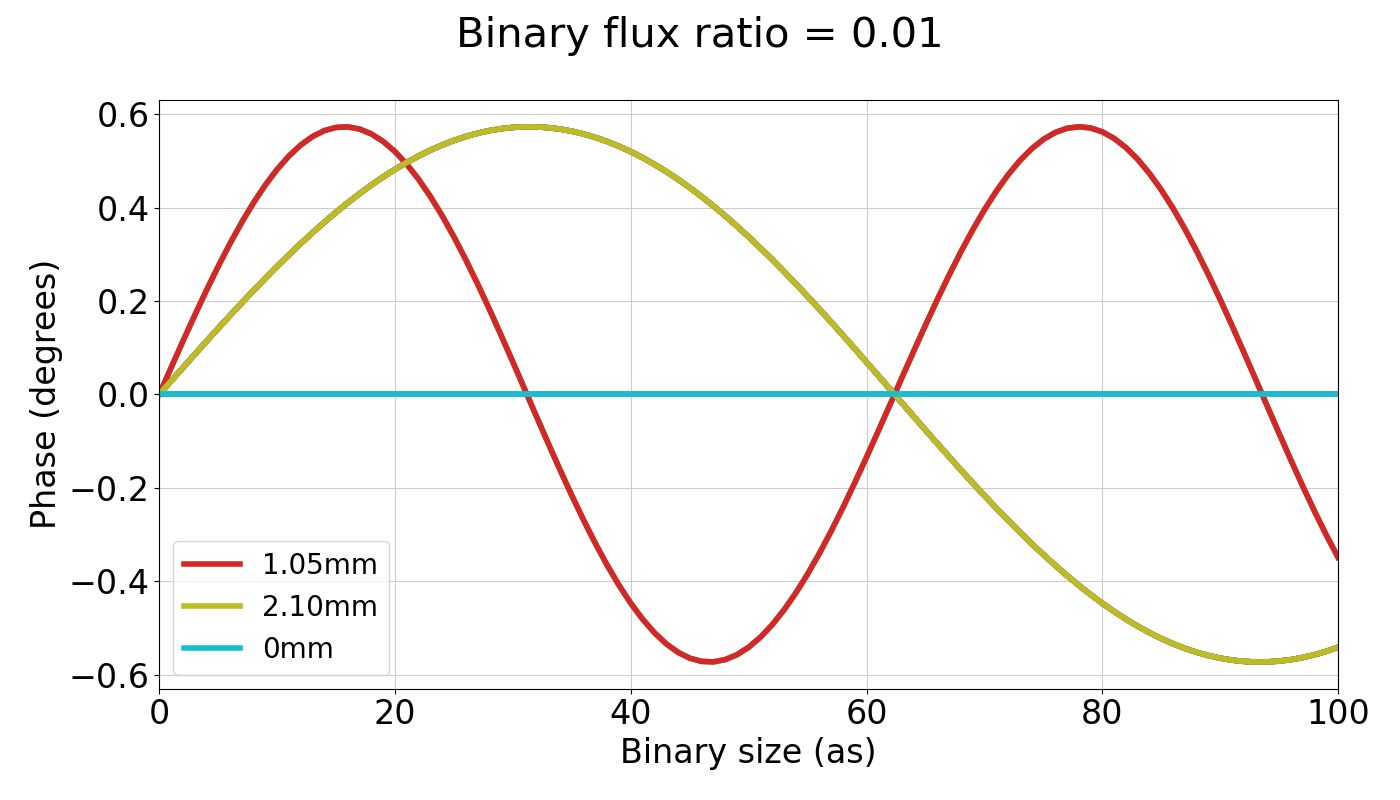}
        \caption{}
        \label{fig:PhaseDiffBinSizeSimuleC}
    \end{subfigure}
    \caption[Courbes de phases théoriques en fonction de la séparation du système observé.]{Courbes de phases théoriques en fonction de la séparation du système observé pour des contrastes de $1$ \textit{(a)}, $0,1$ \textit{(b)} et $0,01$ \textit{(c)} et pour la longueur d'onde $635 \,$nm.}
    \label{fig:PhaseDiffBinSizeSimule}
\end{figure}

Mais encore, l'équation~\ref{eq:PhaseBinaireHautContraste} nous renseigne sur les séparations pour lesquelles les signaux de toutes les bases sont nuls (possible ici car la longueur de la plus grande base est le double de la longueur de la plus petite) : i.e. pour $2\pi \vv{\Theta} \cdot \vv{f} = \pi [\pi]$ soit une séparation valant $\text{k} \lambda / 2B$ avec k un entier relatif. Ce qui donne pour B, la longueur de la base la plus grande, pour la longueur d'onde du laser simulant la protoplanète sur le banc de test $\uplambda = 635 \,$nm, ainsi que pour $\text{k} \in \llbracket 0; 3 \rrbracket$, les séparations $0 \,$as, $31,2 \,$as, $62,4 \,$as et $93,6 \,$as. En revanche, le signal de phase est maximal lorsque $2\pi \vv{\Theta} \cdot \vv{f} = \pi / 2 [\pi]$, i.e. pour une séparation valant : $(2\text{k}+1) \lambda / 4B$. De même, pour la plus grande longueur de base B et pour $\text{k} \in \llbracket 0; 3 \rrbracket$, les séparations qui induisent un signal de phase maximal sont $16,1 \,$as, $48,3 \,$as, $80,5 \,$as et $113 \,$as. Pour l'autre longueur de base la séparation égale à $96,6 \,$as maximise le signal. La séparation choisie sur le banc de test, égale à $87,3 \,$as (voir la section~\ref{sec:SystBinaire}) est un bon compromis entre les séparations $80,5 \,$as et $96,6 \,$as, pour maximiser le signal de phase sur toutes les bases et tout en utilisant le matériel qui était à disposition pour simuler la source protoplanétaire (le V-Groove et la lentille).

A présent, pour la séparation choisie, les phases différentielles sont simulées en fonction de la longueur d'onde et sont présentées pour un contraste de la source égal à $1$ sur la figure~\ref{fig:PhaseDiffWaveSimuleA}, à $0,1$ sur la figure~\ref{fig:PhaseDiffWaveSimuleB} et à $0,01$ sur la figure~\ref{fig:PhaseDiffWaveSimuleC}. Dans ces simulations, le compagnon présente une raie d'émission à $656 \,$nm (raie \ha), avec une largeur à mi-hauteur égale à $0,4 \,$nm, donnant à cette longueur d'onde les contrastes cités. Cette largeur est prise pour correspondre à la largeur de bande du laser qu'on utilise sur le banc de test, comme mesuré dans la section~\ref{sec:PhaseDiffMesure}. La résolution spectrale a été fixée à $3\,200$ ce qui correspond à la résolution spectrale de \ac{FIRSTv2} (mesurée et présentée dans la section~\ref{sec:EtalonnageSpectral}).

\begin{figure}[ht!]
    \centering
    \begin{subfigure}{0.5\textwidth}
        \centering
        \includegraphics[width=\textwidth]{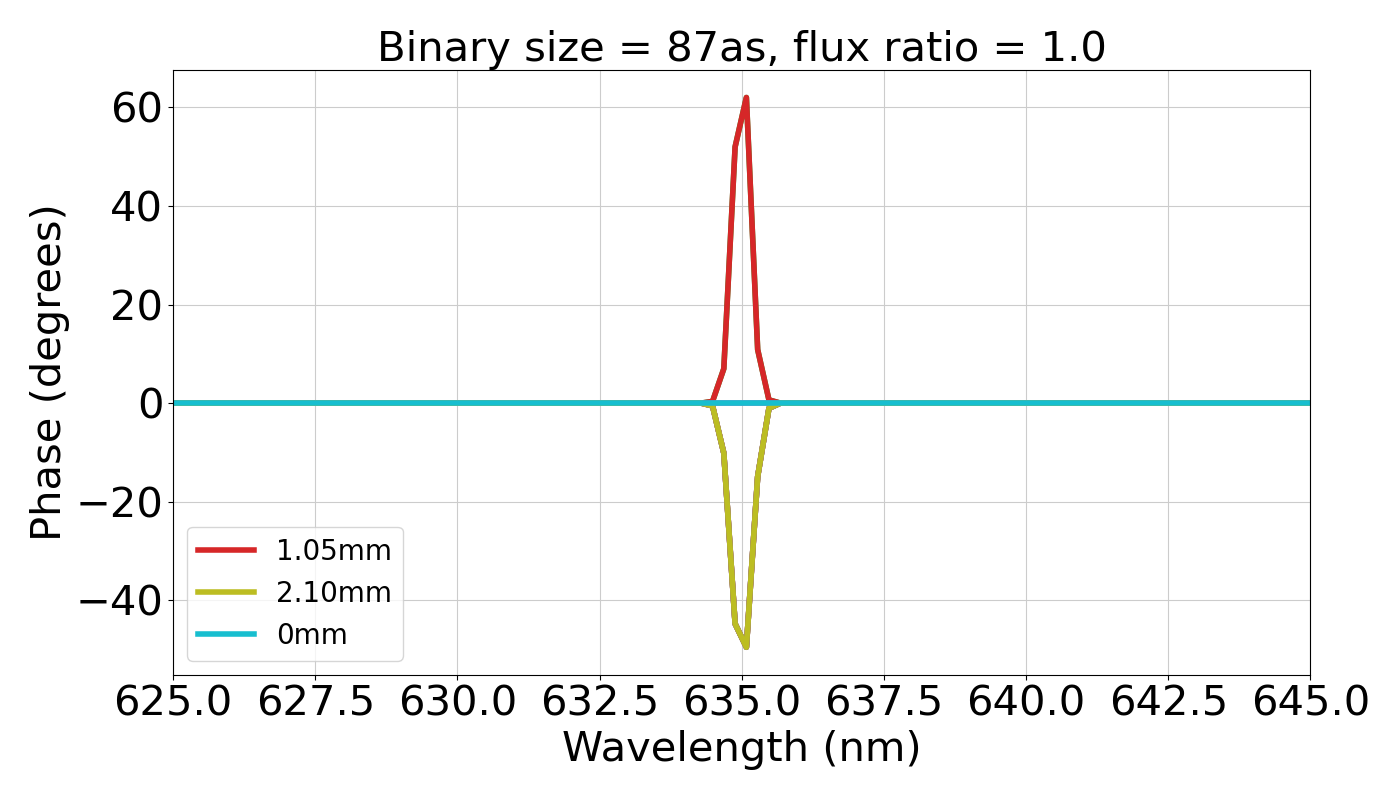}
        \caption{}
        \label{fig:PhaseDiffWaveSimuleA}
    \end{subfigure}
    \begin{subfigure}{0.5\textwidth}
        \centering
        \includegraphics[width=\textwidth]{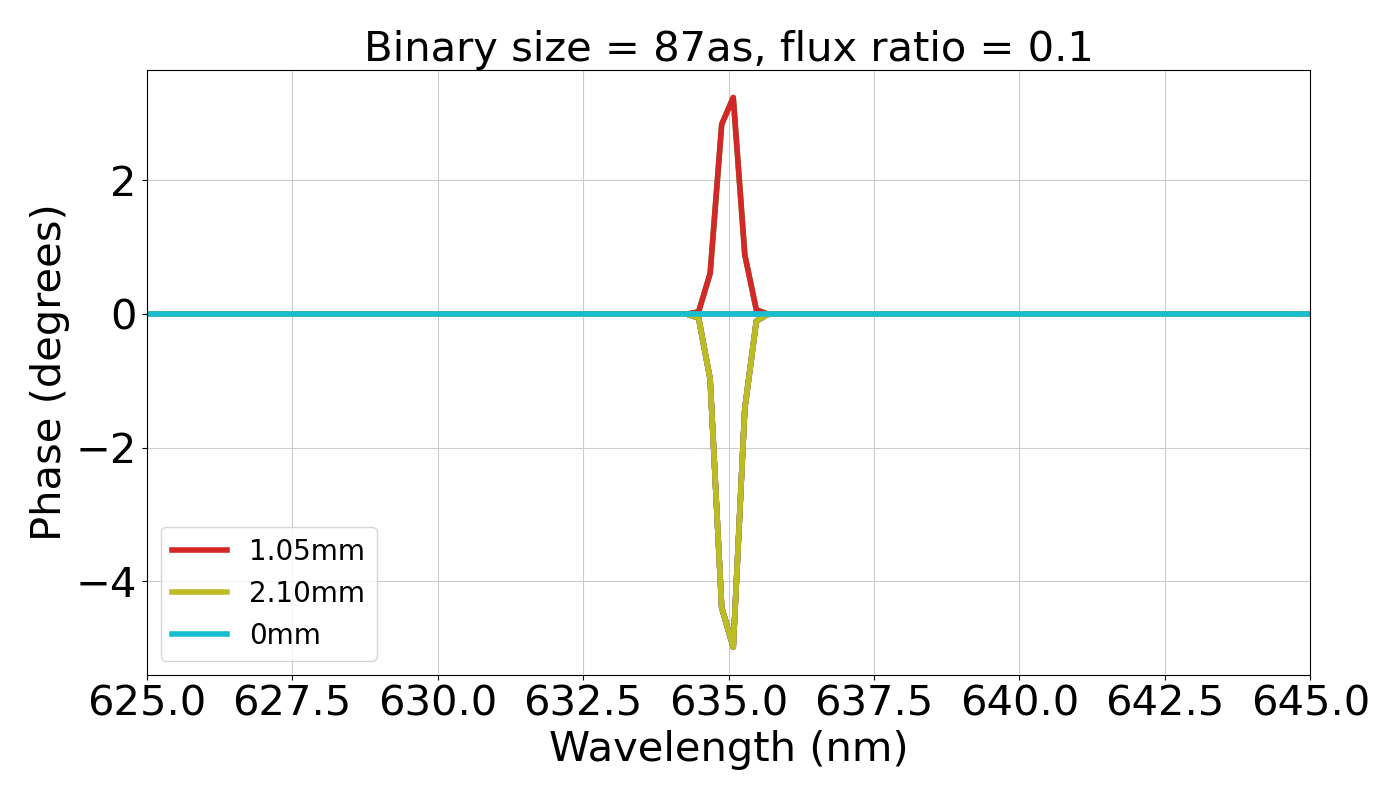}
        \caption{}
        \label{fig:PhaseDiffWaveSimuleB}
    \end{subfigure}%
    \begin{subfigure}{0.5\textwidth}
        \centering
        \includegraphics[width=\textwidth]{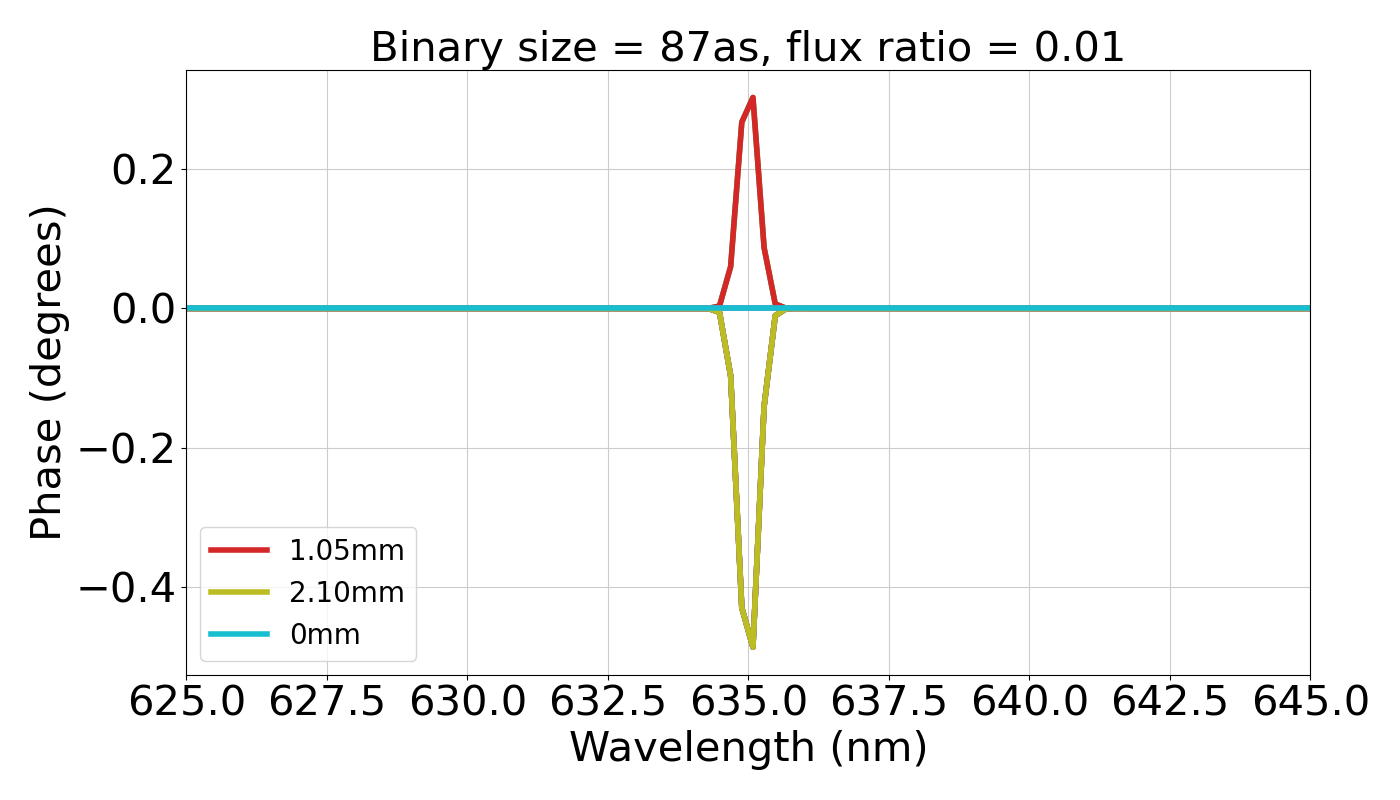}
        \caption{}
        \label{fig:PhaseDiffWaveSimuleC}
    \end{subfigure}
    \caption[Courbes de phases théoriques en fonction de la longueur d'onde pour plusieurs contrastes.]{Courbes de phases théoriques en fonction de la longueur d'onde pour les contrastes de $1$ \textit{(a)}, de $0,1$ \textit{(b)} et de $0,01$ \textit{(c)}. Le compagnon a été simulé avec une raie d'émission à $656 \,$nm, de largeur à mi-hauteur de $0,4 \,$nm et une résolution spectrale de $3\,200$.}
    \label{fig:PhaseDiffWaveSimule}
\end{figure}

Je me servirai de ce modèle pour ajuster les courbes de phases mesurées par la suite afin d'inférer le contraste du système et la séparation du compagnon. Pour cela, on souhaite ajuster aux mesures de phases différentielles, le modèle de la phase d'un système protoplanétaire en accrétion exprimé par l'équation~\ref{eq:PhaseBinaireCentree}. Pour ce faire, on souhaite minimiser la fonction $\chi^2$ définie comme suit :

\begin{equation}
    \chi^2 (\rho, \alpha, \beta) = \sum_{k}^{n_B} \frac{(\varphi_{k, mesure} - \varphi_{k, modele})^2}{\sigma^{k^2}}
    \label{eq:chi2}
\end{equation}

\noindent avec $\varphi_{k, mesure}$ l'estimation du terme de phase de la base $k$, $\varphi_{k, modele}$ le terme de phase du modèle de la base $k$, $\sigma^{k}$ l'erreur sur la mesure du terme de phase de la base $k$ et $n_B$ le nombre de bases. L'erreur $\sigma^{k}$ est estimée à partir de l'écart-type de la phase sur une centaine de canaux spectraux autour du pic d'émission du compagnon (voir ces mesures dans la section~\ref{sec:PhaseDiffMesure}). Cette minimisation est conduite sur des intervalles de contraste du système $\rho$ et des positions $(\alpha, \beta)$ qui sont les coordonnées du vecteur de séparation angulaire $\vv{\Theta}$ afin de trouver les valeurs optimales pour la longueur d'onde du pic du signal de la protoplanète.

Ensuite, à partir de la fonction $\chi^2$, on en déduit la fonction de vraisemblance suivante :

\begin{equation}
    \Like (\rho, \alpha, \beta) \propto e^{\frac{-\chi^2 (\rho, \alpha, \beta)}{2}}
\end{equation}

Chacun des trois paramètres est alors estimé par la marginalisation de la fonction de vraisemblance $\Like (\rho, \alpha, \beta)$ sur les deux autres paramètres, par exemple, selon l'expression suivante pour le paramètre $\rho$ :

\begin{equation}
    f (\rho) \propto \int_{\alpha} \int_{\beta} \Like (\rho, \alpha, \beta) d\alpha d\beta
\end{equation}

La loi de densité est obtenue après avoir normalisé numériquement cette expression et la valeur la plus probable du paramètre associé est estimée par la médiane de cette loi. Enfin, l'intervalle de confiance sur l'estimation du paramètre est pris à $68 \%$ de probabilité autour de la médiane.

%%%%%%%%%%%%%%%%
\subsection{Les phases différentielles mesurées sur une source non résolue}

Dans un premier temps, je vais présenter les mesures de phases différentielles sur une source non résolue par l'instrument. Sur le banc de test, c'est la source qui simule l'étoile centrale : la source fibrée Leukos-SM8-OEM ou super continuum (dénommé par la suite \sk), fabriquée par \textit{Leukos}\footnote{\url{https://www.leukos-laser.com/}} et qui est lumineuse sur la large bande $400 - 2\,400 \,$nm. La source est injectée dans une fibre optique PM630-HP fabriquée par \textit{Thorlabs}\footnote{\url{https://www.thorlabs.com/}} avec un \ac{MFD} égal à $4,5 \pm 0,5 \,$\um, correspondant à une taille angulaire de la source vue par le reste du banc égale à $3,1 \,$as, soit une taille angulaire de $\sim 0,05 \lambda / \text{B}$ (avec $\text{B} = 2,1 \,$mm, la longueur de la plus grande base). La figure~\ref{fig:SuperKSpectrum} montre l'intensité lumineuse normalisée de la \sk~mesurée par \ac{FIRSTv2} en fonction de la longueur d'onde, qui a une forme qui dépend de la transmission du banc. On note que le spectre présente des oscillations à hautes fréquences, qu'on appellera \wiggles~par la suite. Comme nous le verrons, ces oscillations sont problématiques pour les mesures de phases, je présenterai leur caractérisation dans la section~\ref{sec:wiggles}.

\begin{figure}[ht!]
    \centering
    \includegraphics[width=\figwidth]{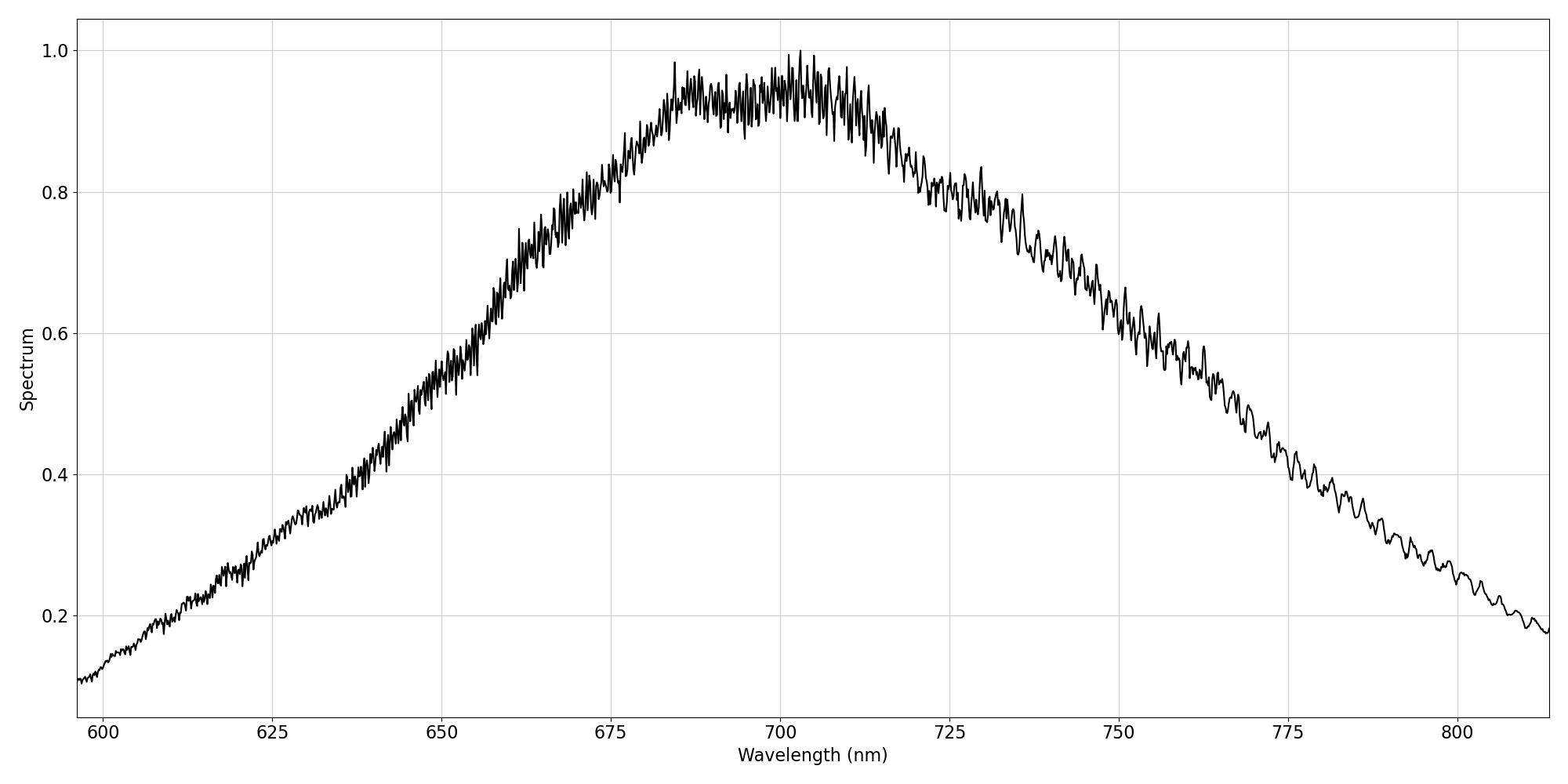}
    \caption[Intensité lumineuse en fonction de la longueur d'onde de la source de référence (\sk) mesurée par FIRSTv2.]{Intensité lumineuse en fonction de la longueur d'onde de la source \sk~de référence, simulant l'étoile centrale du système protoplanétaire, mesurée par FIRSTv2, entre $600 \,$nm et $810 \,$nm.}
    \label{fig:SuperKSpectrum}
\end{figure}

Les phases différentielles mesurées pour les dix bases sur la source \sk, selon la méthode décrite dans la section~\ref{sec:PhaseSpecDiff}, sont tracées sur le graphique de la figure~\ref{fig:PhaseDiffSuperK}. Le signal de phase différentielle sur une source unique non résolue est attendu nul. Or ici nous remarquons 1- des oscillations hautes fréquences qui font penser aux \wiggles~précédemment vus sur le spectre de la source; 2- des oscillations plus basses fréquences qui ressemblent aux spectres de transmissions relatives des puces présentés sur la figure~\ref{fig:V2PMtransmission}.

\begin{figure}[ht!]
    \centering
    \includegraphics[width=\figwidth]{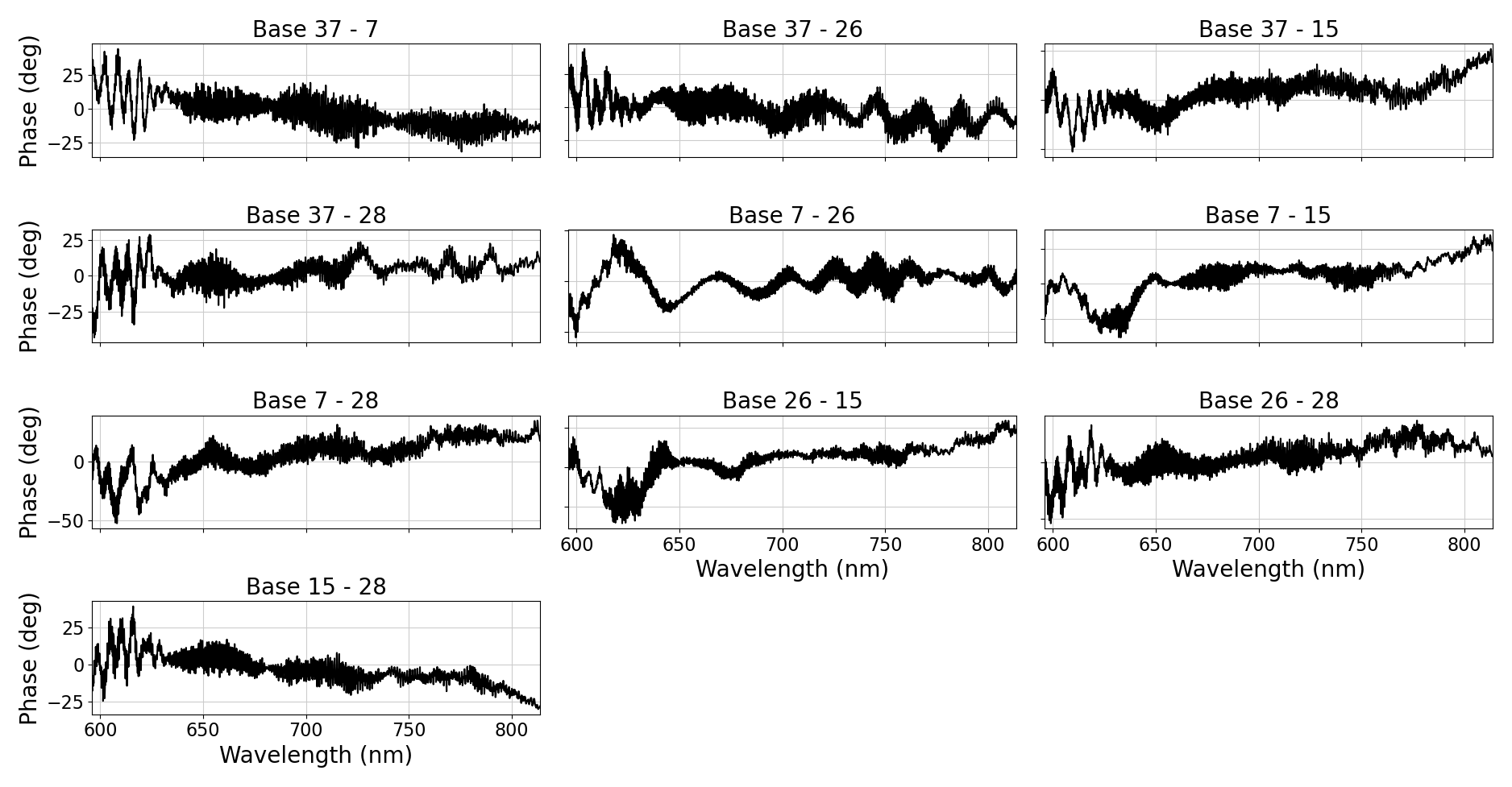}
    \caption[Phases différentielles mesurées sur la source \sk~de référence simulant l'étoile centrale.]{Phases différentielles mesurées sur la source \sk~de référence simulant l'étoile centrale, pour les dix bases de FIRSTv2. Les mesures sont présentées en fonction de la longueur d'onde entre $620 \,$nm et $660 \,$nm.}
    \label{fig:PhaseDiffSuperK}
\end{figure}

Les mesures de phases subissent donc les effets indésirables de phénomènes instrumentaux. La phase différentielle est une grandeur auto-étalonnée des termes instrumentaux qui ont une dépendance polynomiale en fonction de la longueur d'onde. Les perturbations telles que les \wiggles~ne sont donc pas étalonnées. C'est pour cette raison que je vais chercher à les corriger sur les phases différentielles de la source protoplanétaire par la soustraction des phases différentielles de la source non résolue.

%%%%%%%%%%%%%%%%
\subsection{Mesure et étalonnage des phases différentielles sur une source protoplanétaire simulée}
\label{sec:PhaseDiffMesure}

Les résultats préliminaires portant sur les mesures que je vais présenter ici ont été présentés dans la conférence \cite{barjot2022} (voir la section~\ref{sec:HypatiaProceeding}). Dans cette partie je vais exposer les mesures des phases différentielles sur la source protoplanétaire simulée sur le banc de test. Pour cela, la source simulant l'étoile centrale est la source de référence décrite dans la partie précédente et la source simulant une protoplanète en accrétion est un laser fibré, à $635 \,$nm, fabriqué par \textit{Thorlabs}. La figure~\ref{fig:BinarySpectrum} montre l'intensité lumineuse en fonction de la longueur d'onde de la source protoplanétaire simulée, mesurée par \ac{FIRSTv2}. Le graphique est une superposition de l'intensité de l'étoile centrale en trait discontinu et de l'intensité de la source du système (étoile + protoplanète) en trait plein, normalisées par l'intensité de la source centrale à $635 \,$nm, où est visible le pic du laser. La largeur spectrale (largeur à mi-hauteur) du laser est mesurée à $\sim 0,4 \,$nm. À l'aide de ces courbes, une première estimation du rapport de flux entre les deux composantes est faite à $\sim 0,39$. Je règle grossièrement ce contraste lumineux entre les deux sources en desserrant le connecteur de la fibre optique de la source laser.

\begin{figure}[ht!]
    \centering
    \includegraphics[width=\figwidth]{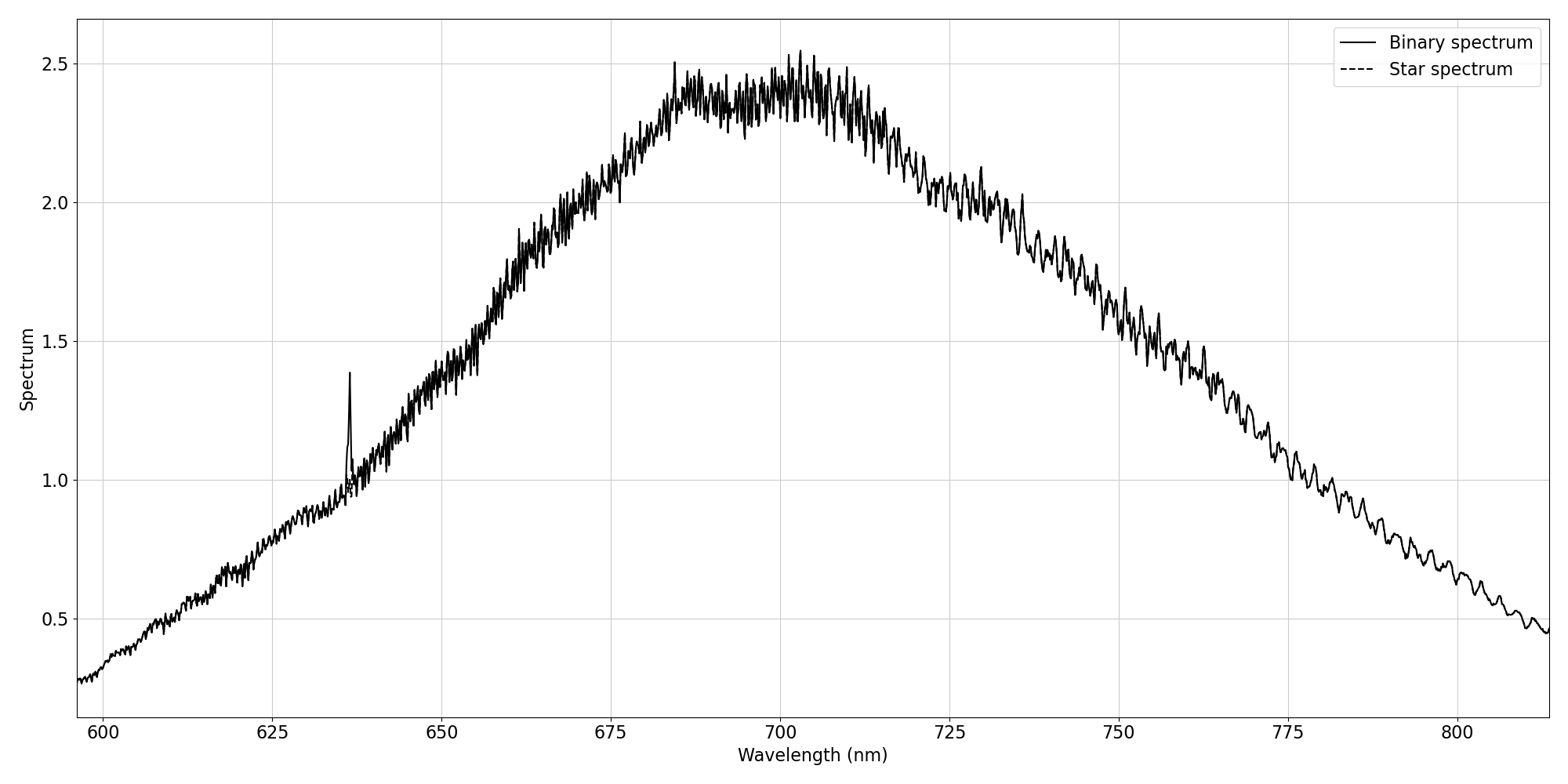}
    \caption[Intensité lumineuse en fonction de la longueur d'onde de la source protoplanétaire mesurée par FIRSTv2.]{Intensité lumineuse en fonction de la longueur d'onde de la source protoplanétaire mesurée par FIRSTv2, entre $600 \,$nm et $810 \,$nm. L'intensité de l'étoile seule est tracée en trait discontinu et l'intensité du système total (étoile + protoplanète) est tracée en trait plein. Les deux courbes sont normalisées par l'intensité de l'étoile à $635 \,$nm, où est visible le pic du laser et le rapport de flux entre les deux composantes (à cette longueur d'onde) est estimé à $\sim 0,39$.}
    \label{fig:BinarySpectrum}
\end{figure}

Les phases différentielles mesurées sur cette source, avant étalonnage par la source centrale, sont tracées sur les graphiques du haut de la figure~\ref{fig:PhaseDiffBinary}. On remarque la présence d'un pic de phase à $\sim 636 \,$nm, correspondant à l'émission du compagnon. De plus, on retrouve les mêmes perturbations instrumentales hautes fréquences qui étaient observées sur la figure~\ref{fig:PhaseDiffSuperK} des phases mesurées sur l'étoile seule. Les courbes sont la moyenne de quelques dizaines de mesures de courbe de phase et les barres d'erreur (très petites sur ces graphiques) sont estimées à partir de leur écart-type. De plus, on estime l'erreur induite par les \wiggles~sur la mesure de la phase différentielle à la longueur d'onde du pic d'émission en calculant l'écart-type $\sigma_{\lambda}$ sur une centaine de canaux spectraux de la phase autour du pic du signal de la protoplanète (le pic étant exclu), pour chaque base. Cet écart-type est indiqué dans les sous-titres au-dessus de chaque graphique et il est en moyenne égal à $8 \pm 2\degree$. De plus, on note, qualitativement, que ces perturbations sont de même amplitude, voire de plus grande amplitude pour certaines bases ($7-15$, $7-28$, $26-15$ et $26-28$), que le signal associé au compagnon, il est donc important qu'on puisse les corriger.

Avant toute chose, il faut préciser que le laser utilisé pour simuler le compagnon n'est pas stable en longueur d'onde et celle-ci varie aléatoirement sur une bande spectrale d'environ $0,6 \,$nm de largeur (ce qui correspond à environ cinq pixels de la caméra). Par conséquent, le pic du signal du compagnon se décale d'autant de pixels de manière aléatoire au fil des mesures temporelles des phases. Une première solution est d'appliquer un \textit{binning} spectral afin de sommer la valeur des pixels par paquet de $5$. J'ai préféré abandonner cette méthode car l'effet de moyenne dégrade la résolution spectrale et diminue l'amplitude du signal qu'on souhaite mesurer. Une deuxième solution, que je préfère utiliser et qui est celle montrée ici, est de sélectionner pour chaque base, un ensemble des courbes de phase pour lesquelles le pic du signal est mesuré sur un même pixel. En pratique, cela correspond à garder environ $60\%$ des données. Ensuite, ces phases différentielles sélectionnées sont moyennées avant d'être étalonnées.

\begin{figure}[ht!]
    \centering
    \begin{subfigure}{\textwidth}
        \centering
        \includegraphics[width=\textwidth]{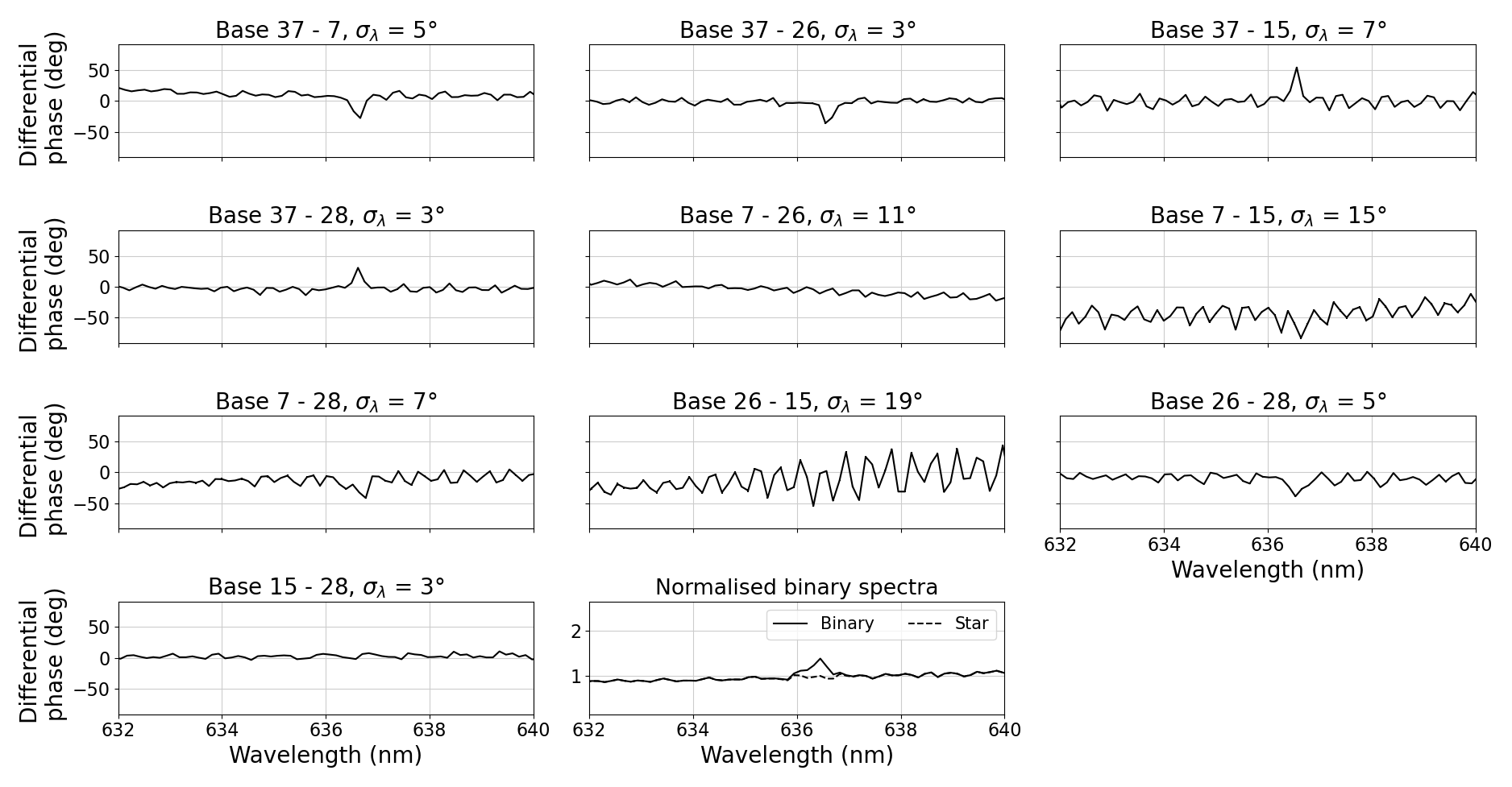}
    \end{subfigure}
    \begin{subfigure}{\textwidth}
        \centering
        \includegraphics[width=\textwidth]{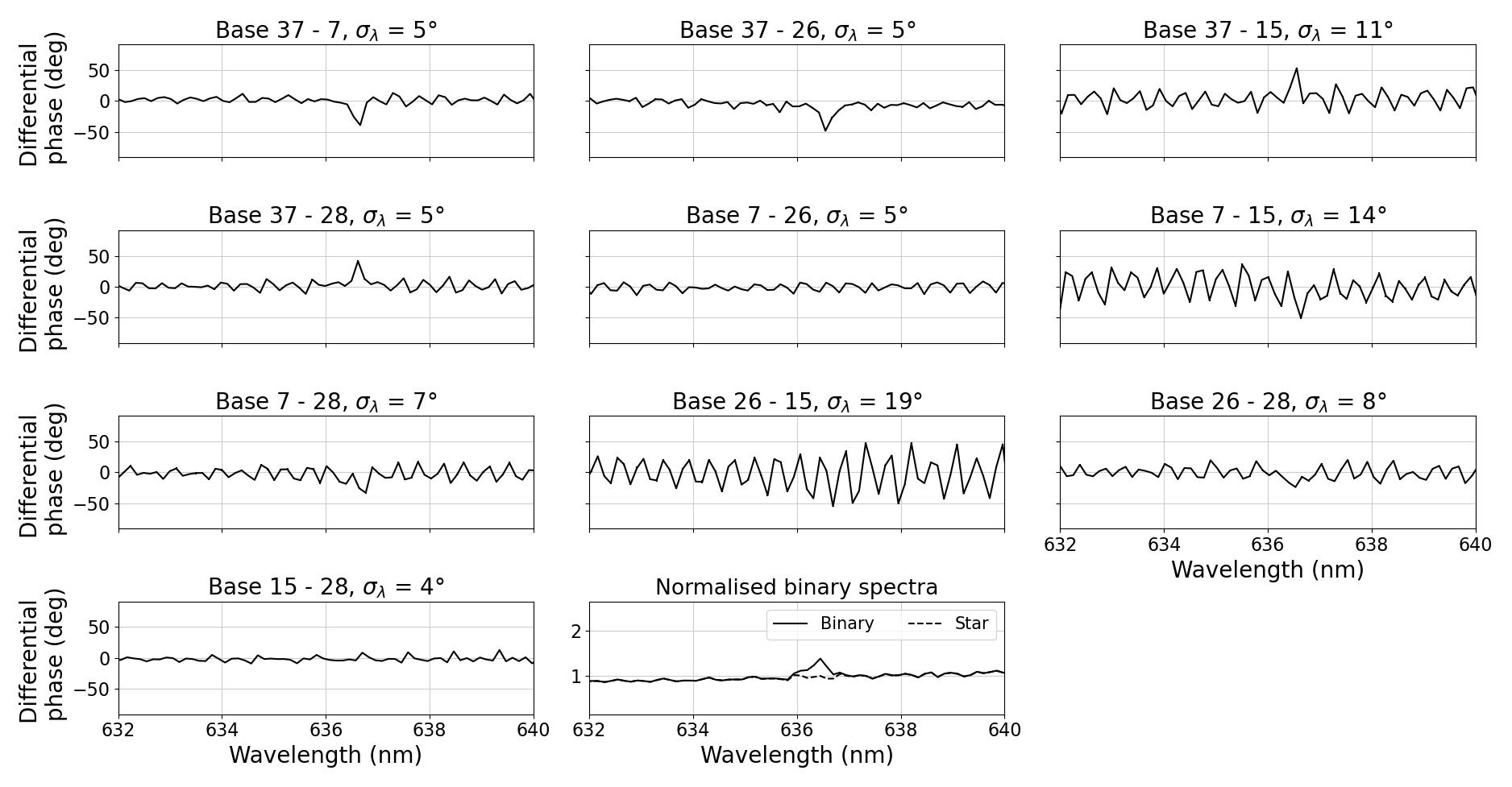}
    \end{subfigure}
    \caption[Phases différentielles de la source protoplanétaire mesurée par FIRSTv2.]{Phases différentielles de la source protoplanétaire mesurée par FIRSTv2, sur toutes les bases, en fonction de la longueur d'onde. Le graphique du haut présente les phases avant et le graphique du bas présente les phases après l'étalonnage par les phases différentielles de la source non résolue. On note que l'étalonnage n'a pas réduit l'amplitude des \wiggles~mais seulement la pente globale.}
    \label{fig:PhaseDiffBinary}
\end{figure}

Le graphique du bas de la figure~\ref{fig:PhaseDiffBinary} montre l'étalonnage par la soustraction des phases différentielles mesurées sur l'étoile seule. Les perturbations de basses fréquences (ici vues comme une pente à cause du zoom sur la bande spectrale) sont corrigées et le signal de phase est centré sur zéro. L'écart-type moyen des \wiggles~est égal à $8 \pm 1\degree$ sur toutes les bases et par comparaison entre les deux graphiques, il a parfois augmenté (pour les bases $37-26$, $7-11$, $37-28$, $26-28$ et $15-28$). Les \wiggles~n'ont donc pas été corrigés par cet étalonnage. Une telle erreur sur la phase par les \wiggles~correspond à un contraste de $0,14$, ce qui est insuffisant par rapport aux performances nécessaires pour étudier le cas scientifique des protoplanètes.

Dans le cadre d'observations sur les cibles du cas scientifique auquel cette thèse s'intéresse (e.g. la protoplanète PDS 70 b, voir la section~\ref{sec:pds70}), l'objectif est de mesurer des signaux de phase d'amplitude $\sim 1\degree$ (correspondant à un contraste de $\sim 2.10^{-2}$) ce qui est bien plus faible que les \wiggles~observés ici. Ces perturbations limitent les capacités de l'instrument et il est alors crucial d'identifier leur cause afin de réussir à les réduire. Une étude de ceux-ci est présentée plus tard dans la section~\ref{sec:wiggles}.

%%%%%%%%%%%%%%%%
\subsection{Résultats de l'ajustement du modèle de protoplanète sur les phases différentielles}

Nous continuons cependant à étalonner les phases différentielles car cela corrige les pentes globales comme nous l'avons vu dans la partie précédente. Il s'agit maintenant de les ajuster au modèle de la protoplanète. Pour ce faire, je calcule les paramètres ($\rho, \alpha, \beta$) qui minimisent la fonction de l'équation~\ref{eq:chi2}, dans le canal spectral du pic central de la raie d'émission du compagnon ($635 \,$nm). Les intervalles de ces paramètres sur lesquels le calcul de la fonction $\chi^2$ est effectué, sont définis comme suit :

\begin{itemize}
    \item le contraste entre l'étoile et le compagnon sur le canal spectral où se situe le pic d'émission, $\rho \in [0,05; 0,995]$ avec un pas de $0,005$, constituant $190$ valeurs;
    \item la coordonnée du compagnon $\alpha \in [-100; 100]\,$as avec un pas de $0,5 \,$as, constituant $400$ valeurs;
    \item la coordonnée du compagnon $\beta \in [-100; 100]\,$as avec un pas de $0,5 \,$as, constituant $400$ valeurs.
\end{itemize}

La minimisation de la fonction $\chi^2$ est donc faite dans un espace de paramètres de dimension $190 \times 400 \times 400$. De plus, j'exclus les phases différentielles des bases $7-15$ et $26-15$ au cours de cette minimisation car ce sont les deux bases pour lesquelles les \wiggles~sont très intenses. La minimisation est donc effectuée sur huit des dix bases. La figure~\ref{fig:PhaseDiffBin01LikeliMap} présente la carte des valeurs maxima de la fonction de vraisemblance $\Like$ selon la dimension du contraste, en fonction des coordonnées ($\alpha, \beta$). Le cercle noir a pour rayon la séparation théorique du système protoplanétaire ($\sim 87,3\,$as), la croix noire indique la position de la source centrale et la croix rouge indique la position trouvée après marginalisation de la fonction de vraisemblance sur les deux paramètres de positions $(\alpha; \beta)$, estimés ici à $-87\substack{+4 \\ -3} \,$as et $0\substack{+1 \\ -1} \,$as, respectivement. Le contraste de la source est estimé à $0,68\substack{+0,05 \\ -0,04}$. On remarque que deux maxima locaux principaux sont trouvés sur la carte, ce qui est cohérent avec les graphiques de la figure~\ref{fig:PhaseDiffBinSizeSimule}, sur lesquelles on peut identifier une dégénérescence du signal avec une période de $\uplambda / \text{B}_{\text{min}} = 124,7 \,$as. Ainsi, le deuxième maximum est (théoriquement) localisé à $-87,3 + 124,7 = 37,4 \,$as. Pour discriminer la séparation du compagnon sur la carte de la vraisemblance, il est nécessaire d'effectuer une mesure avec une base d'une troisième longueur, pour sonder une troisième fréquence spatiale. Étant donné que nous connaissons la séparation angulaire du système observé (indiquée par le cercle noir sur la carte), je ne considère que le minimum local qui en est le plus proche, lors de l'estimation des trois paramètres.

\begin{figure}[ht!]
    \centering
    \begin{subfigure}{0.5\textwidth}
        \centering
        \includegraphics[width=\textwidth]{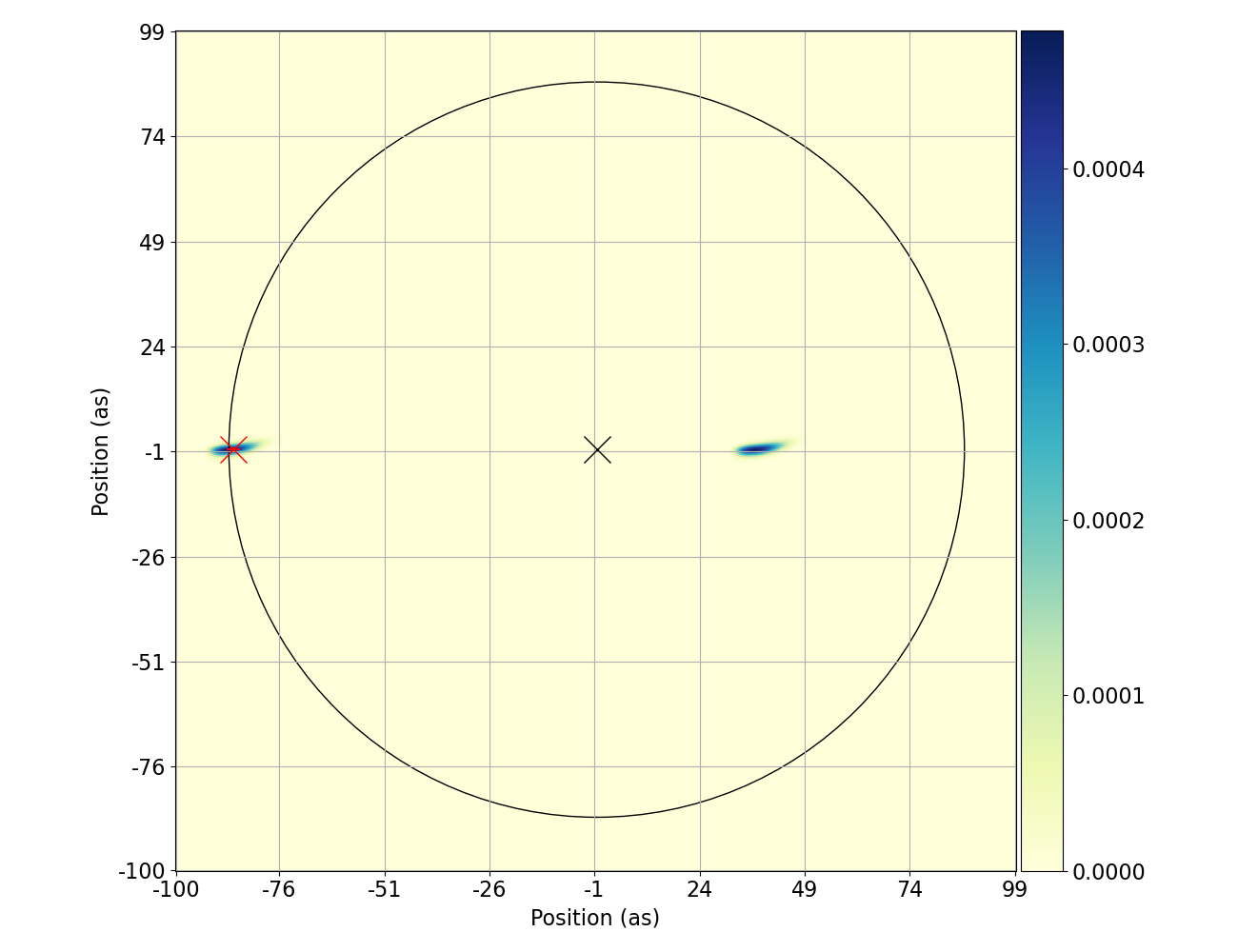}
    \end{subfigure}%
    \begin{subfigure}{0.5\textwidth}
        \centering
        \includegraphics[width=\textwidth]{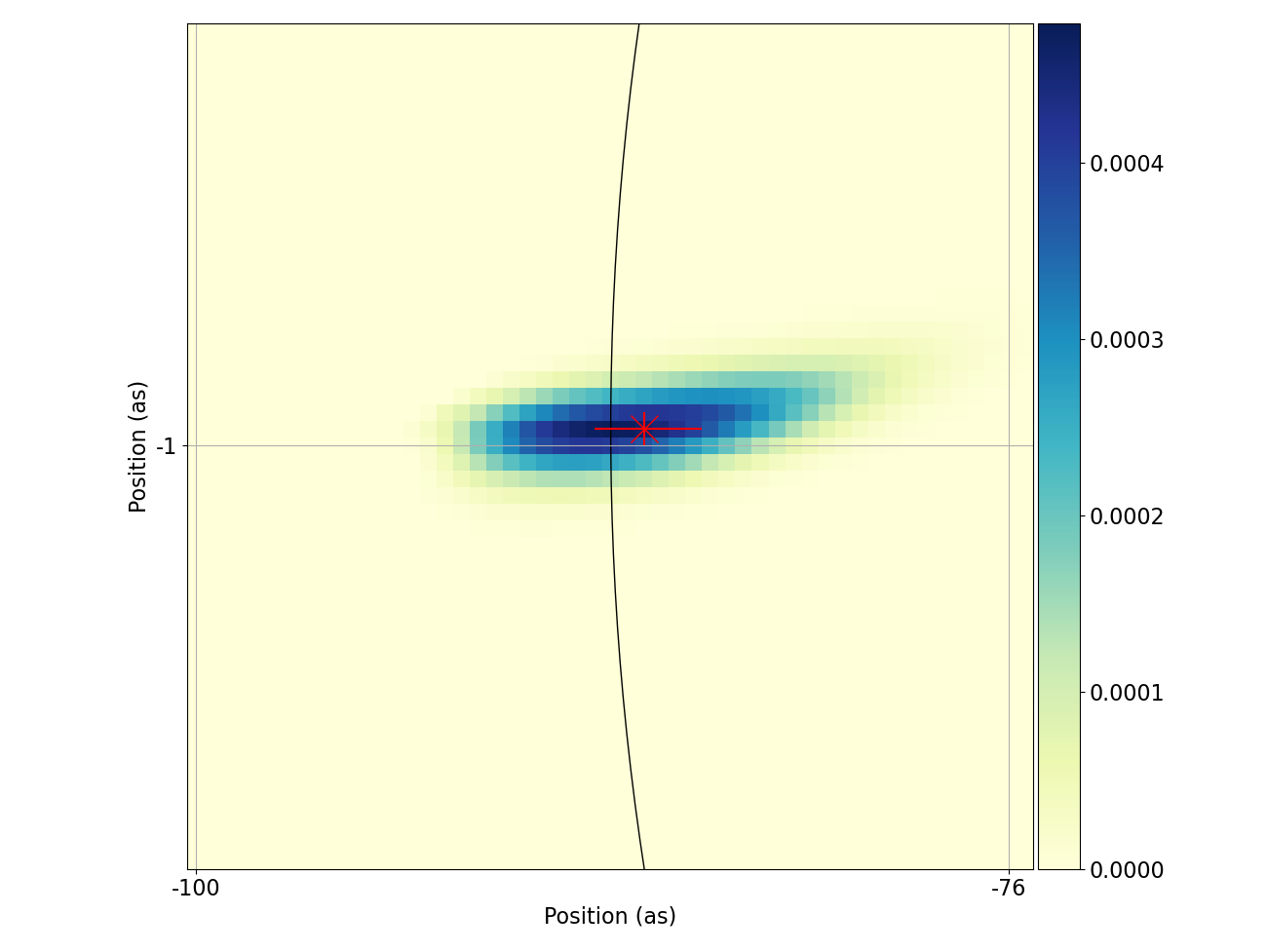}
    \end{subfigure}
    \caption[Carte de la fonction de vraisemblance calculée à partir des phases différentielles mesurées sur la puce $Y$.]{Carte de la fonction de vraisemblance (à gauche) calculée à partir des phases différentielles mesurées sur la puce $Y$, en fonction des coordonnées ($\alpha; \beta$) et zoom centré sur la croix rouge (à droite). La croix noire indique la position de la source centrale ($0; 0$) et la croix rouge indique la position la plus probable à partir de la marginalisation de la vraisemblance ($-87\substack{+4 \\ -3}; 0\substack{+1 \\ -1} \,$)as, trouvé pour un contraste de $0,68\substack{+0,05 \\ -0,04}$. Le cercle noir a pour centre la croix noire et pour rayon la séparation angulaire attendue du compagnon.}
    \label{fig:PhaseDiffBin01LikeliMap}
\end{figure}

À partir des paramètres optimaux trouvés précédemment, je trace les phases différentielles en utilisant l'équation~\ref{eq:PhaseBinaireCentree} sur la figure~\ref{fig:PhaseDiffBin01LikeliFit}, en pointillés, superposées aux phases différentielles mesurées, en trait plein. Les phases différentielles estimées des bases $7-15$ et $26-15$ exclues de la minimisation, sont laissées à zéro. La comparaison des deux montre qu'on a bien réussi à estimer les trois paramètres qui ajustent les courbes de phases mesurées.

\begin{figure}[ht!]
    \centering
    \includegraphics[width=\figwidth]{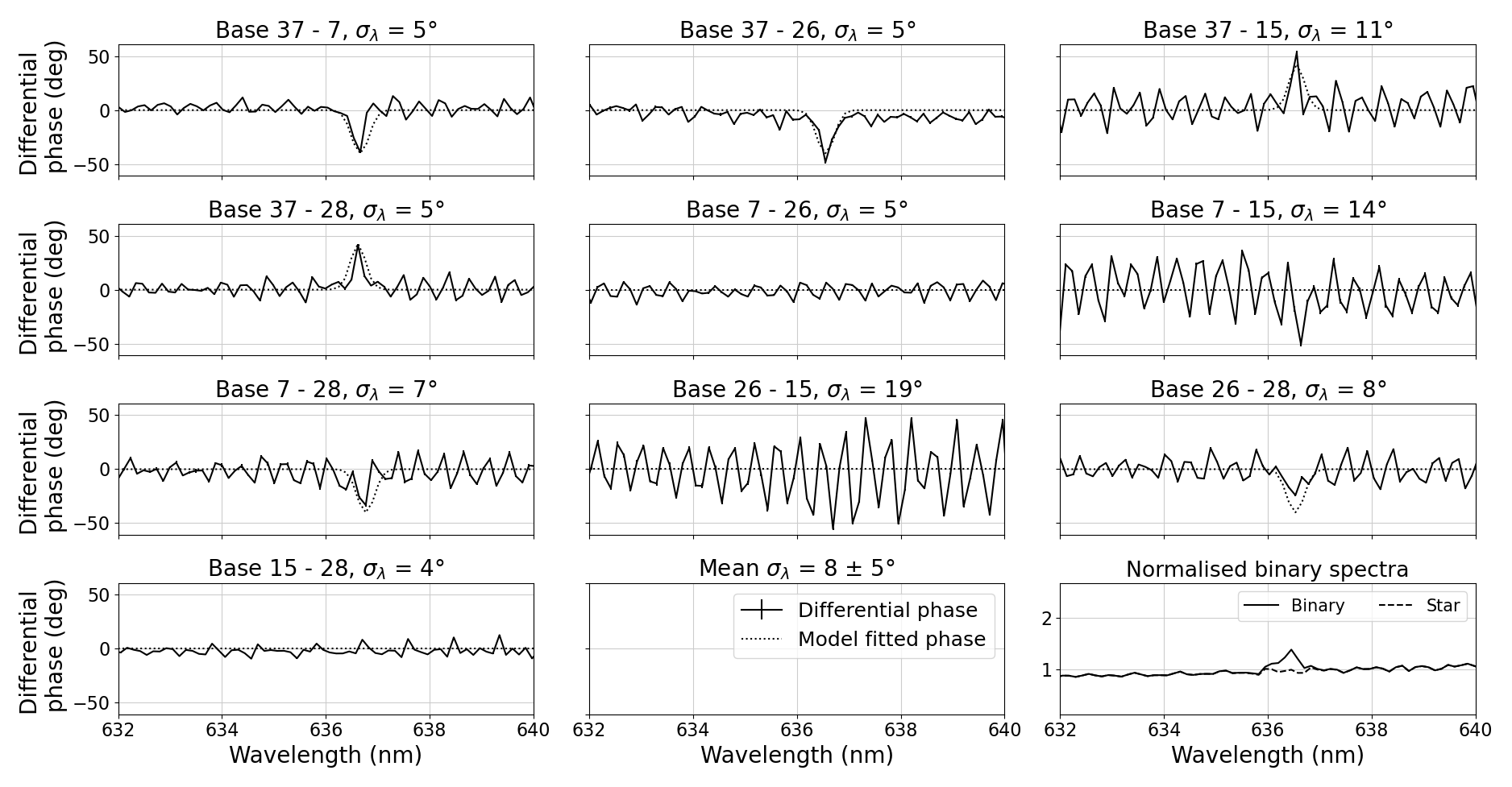}
    \caption[Graphique des phases différentielles mesurées et des phases différentielles modélisées à partir des paramètres d'ajustement du modèle de système protoplanétaire.]{Graphique des phases différentielles mesurées (en trait plein) et des phases différentielles modélisées (en pointillés) à partir des paramètres d'ajustement du modèle de système protoplanétaire. Les paramètres optimaux utilisés pour le calcul des phases à partir de ce modèle sont : ($\rho; \alpha; \beta$) = ($0,68$; $-87$; $0$).}
    \label{fig:PhaseDiffBin01LikeliFit}
\end{figure}

La même étude est faite pour des données acquises sur la source protoplanétaire avec un contraste légèrement plus élevé. La figure~\ref{fig:PhaseDiffBin02LikeliFit} présente les phases différentielles mesurées superposées au modèle calculé à partir des paramètres optimaux de la marginalisation de la fonction $\Like$. Comme précédemment, la carte résultante de la vraisemblance est montrée sur la figure~\ref{fig:PhaseDiffBin02LikeliMap}. Le contraste de cette source est estimé à $0,57 \pm 0,06$ et les coordonnées de la séparation angulaire sont estimées à ($-87\substack{+16 \\ -4}; 1\substack{+2 \\ -1} \,$)as. La carte de la vraisemblance présente deux maxima locaux en plus que sur la première carte. Cela est dû au fait que le signal du pic de phase du compagnon est plus faible, ce qui diminue le \ac{SNR} des données, entraînant une difficulté accrue de l'estimation des paramètres optimaux. Par conséquent, on peut estimer que la limite inférieure du contraste d'une source protoplanétaire détectable par le banc de test actuel de \ac{FIRSTv2} est d'environ $0,55$.

\begin{figure}[ht!]
    \centering
    \includegraphics[width=\figwidth]{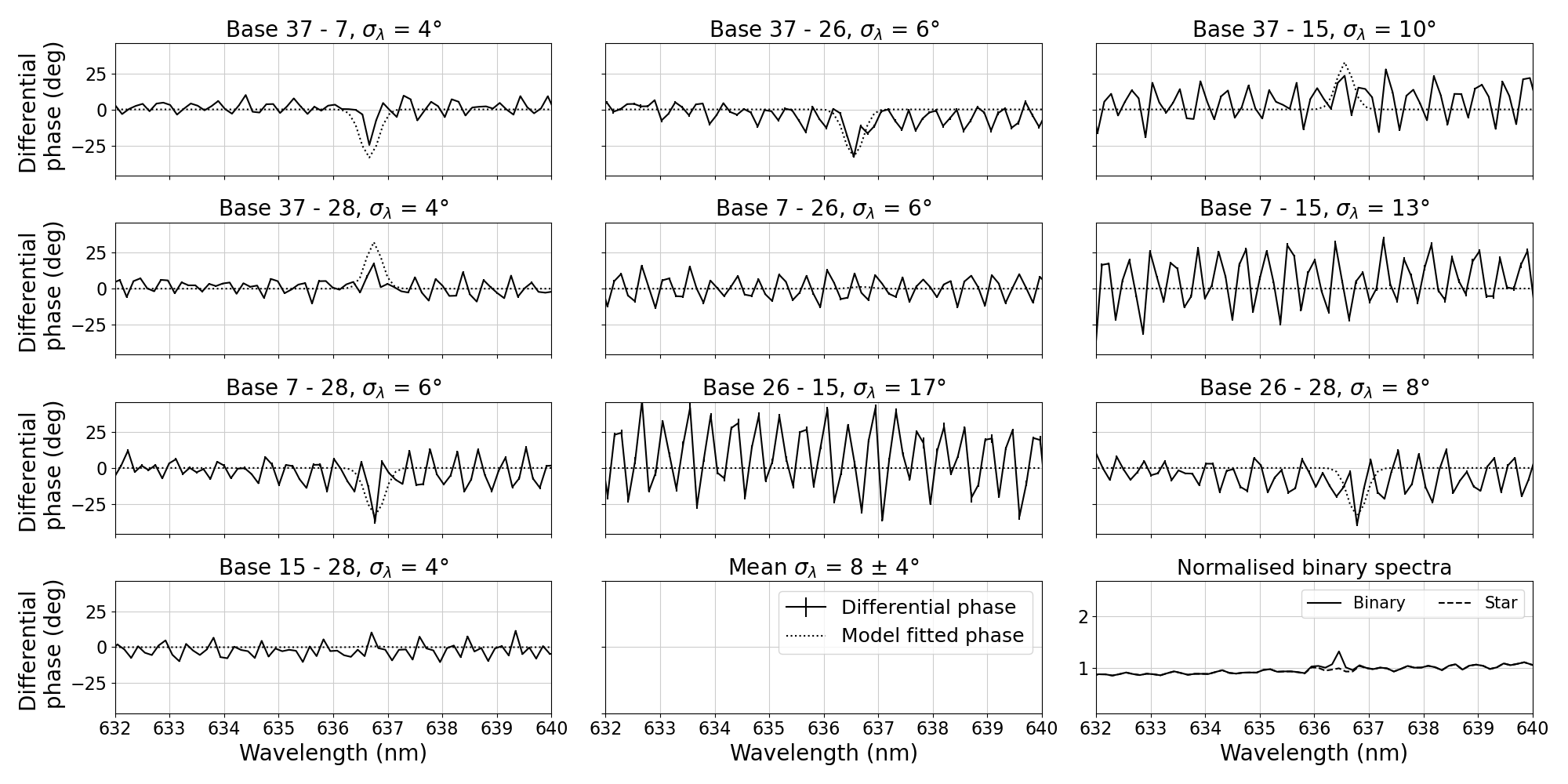}
    \caption[Graphique des phases différentielles mesurées et des phases différentielles modélisées à partir des paramètres d'ajustement du modèle de système protoplanétaire.]{Graphique des phases différentielles mesurées (en trait plein) et des phases différentielles modélisées (en pointillés) à partir des paramètres d'ajustement du modèle de protoplanète. Les paramètres optimaux utilisés pour le calcul des phases à partir de ce modèle sont : ($\rho; \alpha; \beta$) = ($0,57$; $-87$; $1$).}
    \label{fig:PhaseDiffBin02LikeliFit}
\end{figure}

\begin{figure}[ht!]
    \centering
    \begin{subfigure}{0.5\textwidth}
        \centering
        \includegraphics[width=\textwidth]{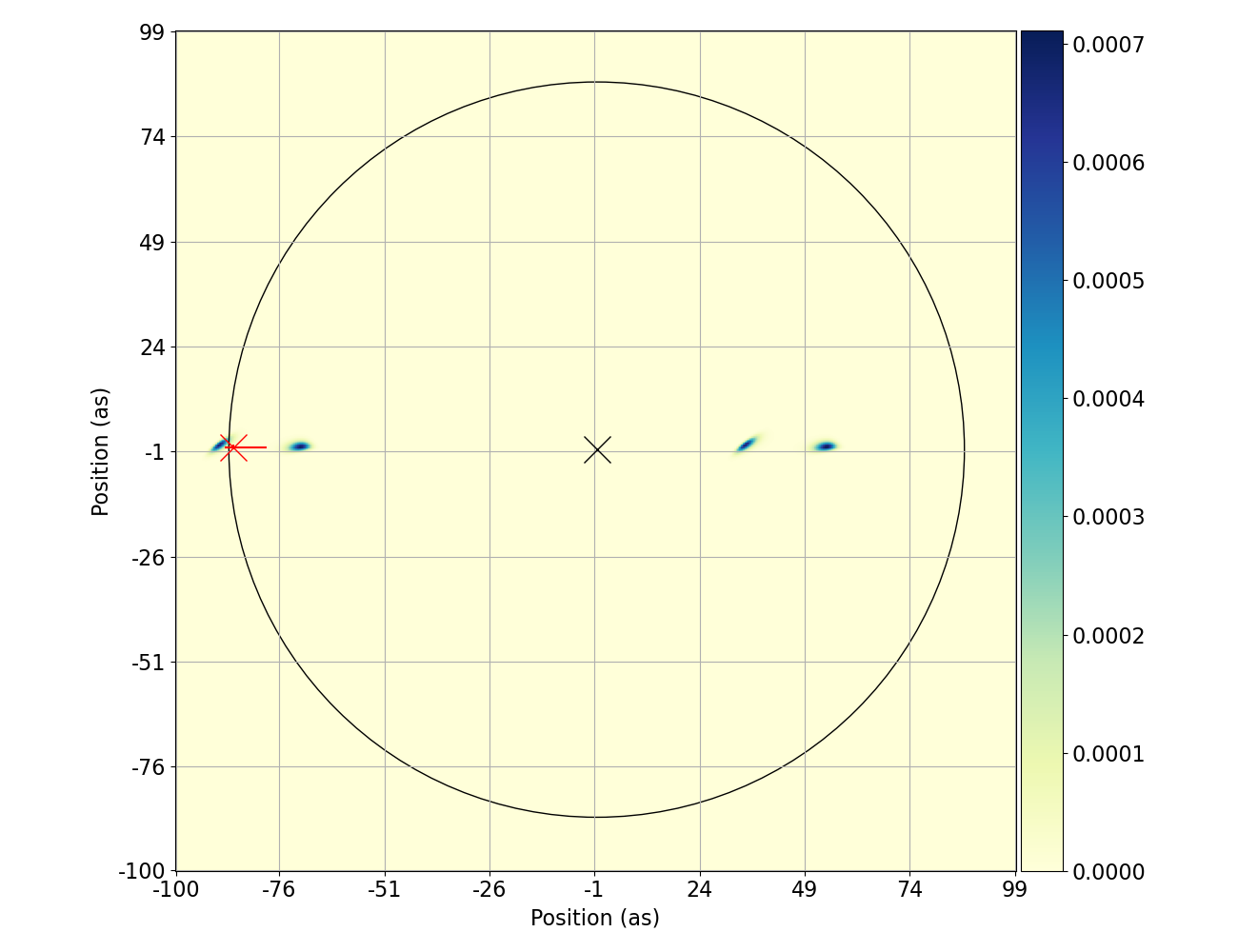}
    \end{subfigure}%
    \begin{subfigure}{0.5\textwidth}
        \centering
        \includegraphics[width=\textwidth]{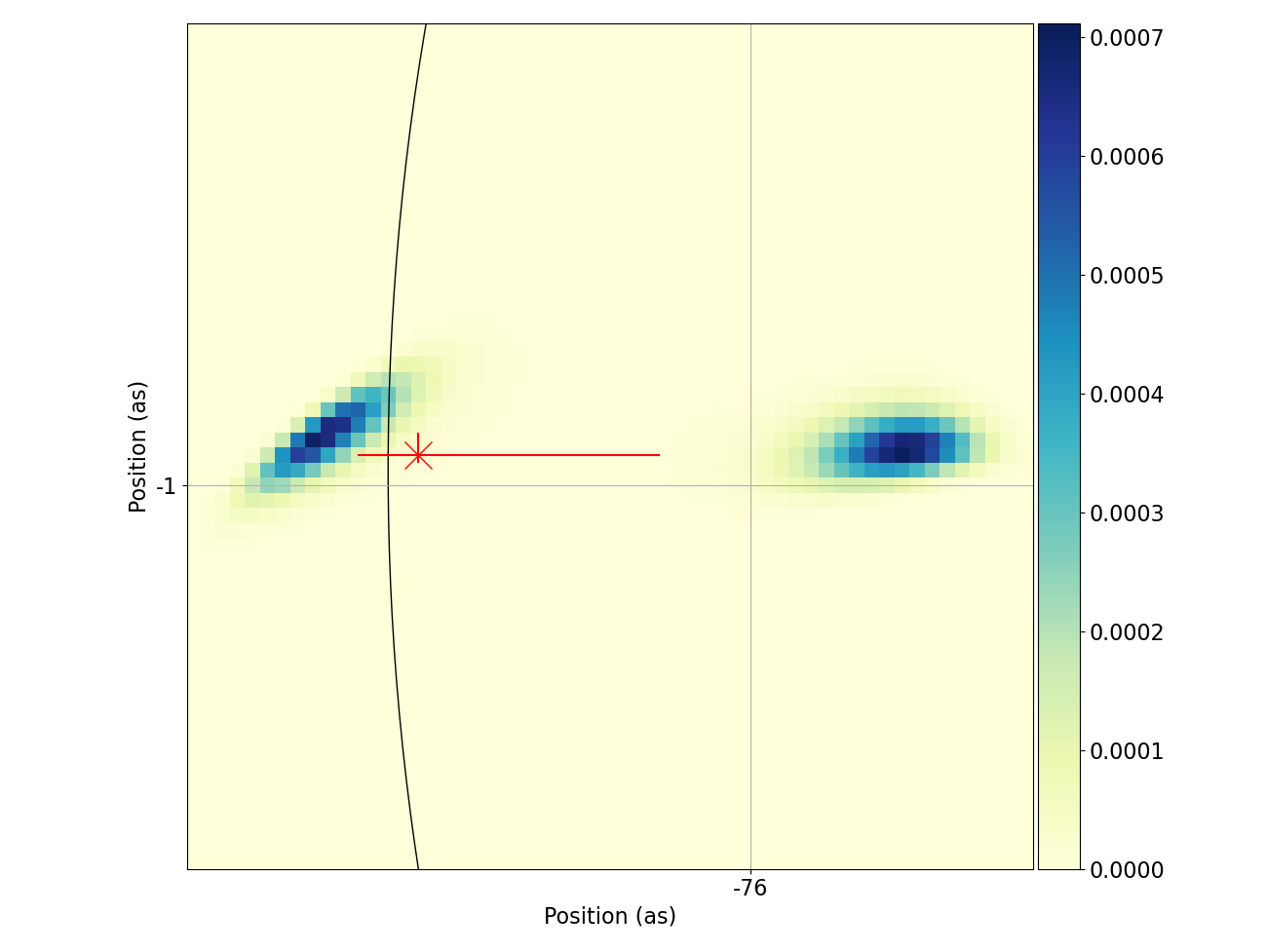}
    \end{subfigure}
    \caption[Carte de la fonction de vraisemblance calculée à partir des phases différentielles mesurées sur la puce $Y$.]{Carte de la fonction de vraisemblance (à gauche) calculée à partir des phases différentielles mesurées sur la puce $Y$, en fonction des coordonnées ($\alpha, \beta$) et zoom centré sur la croix rouge (à droite). La croix noire indique la position de la source centrale et la croix rouge indique la position la plus probable à partir de la marginalisation de la vraisemblance ($-87\substack{+16 \\ -4}; 1\substack{+2 \\ -1} \,$)as, trouvé pour un contraste de $0,57 \pm 0,06$. Le cercle noir a pour centre la croix noire et pour rayon la séparation angulaire attendue du compagnon.}
    \label{fig:PhaseDiffBin02LikeliMap}
\end{figure}

%%%%%%%%%%%%%%%%%%%%%%%%%%%%%%%%
\section{Les wiggles}
\label{sec:wiggles}

%%%%%%%%%%%%%%%%
\subsection{Les wiggles sur d'autres instruments}

Dans cette partie je présente une étude réalisée sur les \wiggles~mesurés sur les données de \ac{FIRSTv2} afin de les caractériser. Mais tout d'abord, je fais la liste de plusieurs expériences sur lesquelles ce phénomène a déjà été mesuré :

\begin{itemize}
    \item les \wiggles~sont visibles sur les phases différentielles d'\ac{AMBER} \citep{millour2008} et leur cause a été identifiée comme provenant d'une cavité Fabry-Pérot créée par la lame d'air présente dans les polariseurs \citep{malbet2008}, ce qui a permis de totalement les corriger en optant pour des polariseurs à lame mince;
    \item les \wiggles~sont aussi visibles sur les phases différentielles de \ac{GRAVITY} \citep{amorim2020} mais sont d'assez faible amplitude pour ne pas gêner les mesures. A la suite de discussions privées avec Nicolas Pourré, doctorant sur le projet \ac{GRAVITY}, il semblerait que le même phénomène soit mesuré et visible sur la norme et la phase des visibilités complexes;
    \item des franges spectrales sont aussi visibles sur les données de l'instrument \ac{MIRI} du \ac{JWST} \citep{argyriou2020} et leur cause est identifiée comme provenant d'une cavité Fabry-Pérot créée par les différentes couches électroniques de la caméra. Les \wiggles~sont corrigés au cours du traitement de données, par ajustement de la fonction de transmission Fabry-Pérot.
\end{itemize}

%%%%%%%%%%%%%%%%
\subsection{Première analyse des wiggles sur FIRSTv2}

Les \wiggles~sur le banc de test de \ac{FIRSTv2} apparaissent sur toutes les images. La figure~\ref{fig:WigglesFlat} présente le spectre mesuré sur une des sorties imagées sur la caméra, lorsque une seule entrée est illuminée. Il s'agit donc d'un spectre sans frange d'interférence, sur toute la bande spectrale transmise (en haut) et sur une bande comprise entre $660 \,$nm et $690 \,$nm (en bas). On note que les \wiggles~apparaissent sur toute la bande spectrale et leur période est estimée à $\sim 0,5 \,$nm. Une rapide analyse montre que cette période est constante en fonction du nombre d'onde $\sigma$.

\begin{figure}[ht!]
    \centering
    \includegraphics[width=\figwidth]{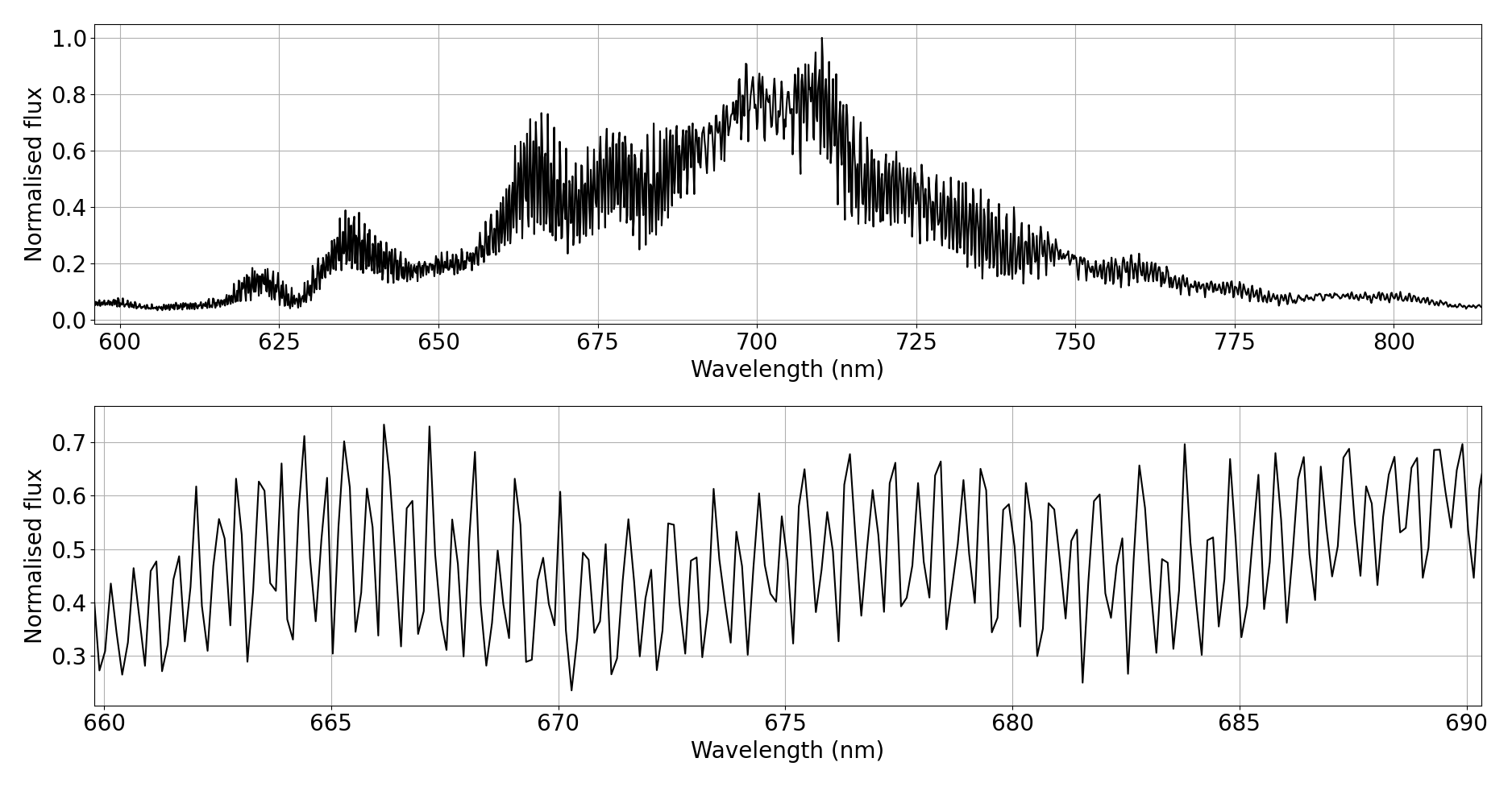}
    \caption[Spectre présentant des wiggles mesuré sur une sortie sans frange d'interférence, sur FIRSTv2.]{Spectre normalisé présentant des wiggles mesuré sur une sortie sans frange d'interférence, sur FIRSTv2. Le spectre est tracé sur toute la bande spectrale transmise par FIRSTv2 (en haut) et sur la bande comprise entre $660 \,$nm et $690 \,$nm (en bas).}
    \label{fig:WigglesFlat}
\end{figure}

Un filtrage fréquentiel pourrait être utilisé afin de corriger ces perturbations. Lors d'un traitement préliminaire dans lequel un filtrage passe-bas a été appliqué aux mesures de phases différentielles sur le système protoplanétaire simulé, il a été remarqué que le filtrage pouvait détériorer la raie du compagnon (diminuant son amplitude jusqu'à un facteur $2$). Cela parait cohérent puisque la largeur de la raie du compagnon est du même ordre de grandeur que la largeur des \wiggles. Ainsi, il a été choisi pour l'instant de ne pas appliquer de filtrage spatial sur les phases différentielles.

% Tests done at the lab to understand the cause
% Harry-Dean study:
% - tests on different polarizers with differents thickness => didn't change the wiggles frequency
% - no polar + Wolla in spectro => no wiggles
% - 2 wolla => wiggles
% - data without the chip and without polar/Wolla => no wiggles
% ==> il semble que cela vienne des fibres et du fait qu'on sélectionne la polar injectée
De récentes mesures effectuées sur le banc par Manon Lallement et Harry-Dean Kenchington Goldsmith semblent montrer que les puces ne sont pas la seule cause des \wiggles. Plusieurs tests ont été faits en injectant la source \sk~dans le montage suivant : (1) polariseur, (2) fibre optique, (3) monture pour fibre optique avec la rotation autour de l'axe $Z$ comme degré de liberté, (4) le spectro-imageur et (5) la caméra. Différents tests mènent aux observations suivantes :

\begin{itemize}
    \item en utilisant des polariseurs de différentes épaisseurs (avant ou après la fibre), la période des \wiggles~reste inchangée, ce qui montre qu'ils ne sont pas causés par une cavité Fabry-Pérot créée par le polariseur;
    \item en retirant le polariseur en entrée de fibre, les \wiggles~disparaissent;
    \item en utilisant un prisme de wollaston en entrée de fibre, les \wiggles~sont visibles;
    \item les \wiggles~disparaissent pour certaines valeurs d'angle de la fibre optique en entrée du spectro-imageur.
\end{itemize}

On peut déduire de ces observations qu'il semblerait que la sélection d'une polarisation avant l'injection des faisceaux dans les fibres soit la cause de l'apparition des \wiggles. En effet, la biréfringence des fibres optiques impose l'injection de lumière polarisée dans le sens de son axe. Autrement, la polarisation est elliptique en sortie de la fibre et interfère avec elle-même. Il semble donc nécessaire de s'assurer que l'axe de polarisation de tous les composants utilisés sur le banc de test (polariseur, fibres optiques, puce photonique, prisme de Wollaston) soient alignés. Les extrémités des fibres optiques sont montées sur les férules de sorte que leurs axes principaux de polarisation soient alignés quand on les raccorde. Cela signifie qu'il serait nécessaire de les aligner avec plus de précision à l'avenir.

De plus amples investigations sont nécessaires et sont toujours conduites afin de comprendre totalement l'origine de ces perturbations et de déterminer une solution pour contourner ce problème.

%%%%%%%%%%%%%%%%%%%%%%%%%%%%%%%%
\section{Conclusion}
% en parlant de bruit, je me demande si on ne pourrait pas te demander à quelle magnitude équivalente le signal avec lequel tu travailles correspond. Mais ce n'est pas simple comme question... Sylvestre tu as peut être une idée sur ce point ?

Dans ce chapitre j'ai présenté le simulateur de source protoplanétaire que j'ai mis en place sur le banc de test, ce qui a été primordial pour le développement de \ac{FIRSTv2} pour la caractérisation de systèmes protoplanétaires. Les mesures que j'ai présentées montrent la validité de la technique des phases différentielles mesurées sur une source de type protoplanétaire, présentant un contraste plus faible sur une bande spectrale étroite.

J'ai ainsi montré la détection d'un compagnon à une séparation de $0,68 \lambda / B$ avec un contraste égal à $\sim 0,55$, à partir du calcul et de l'analyse des phases différentielles. Le contraste atteint dans cette étude est encore assez faible comparé au contraste des protoplanètes que l'on connaît actuellement (contraste de $\sim 10^{-2} - 10^{-3}$ pour la protoplanète PDS 70 b). En effet, les mesures de phases différentielles sont limitées par les \wiggles~dont l'erreur moyenne qu'elle ajoute est estimée à $\sim 8\degree$, correspondant à un contraste de $\sim 0,14$, soit $\sim 0,56$ pour un signal de phase au moins quatre fois plus grand que les \wiggles~(\ac{SNR} $= 4$).

Une partie des efforts de l'équipe se concentre actuellement sur la compréhension de l'origine de ces \wiggles~ainsi que sur leur correction.

%%%%%%%%%%%%%%%%%%%%%%%%%%%%%%%%
\clearpage
\section*{Conférence Hypatia Colloquium}
\label{sec:HypatiaProceeding}
\phantomsection
\addcontentsline{toc}{section}{Conférence Hypatia Colloquium}

\clearpage
\includepdf[pages=-, pagecommand={}, offset=-10 -20, templatesize={0.8\textwidth}{0.8\textheight}]{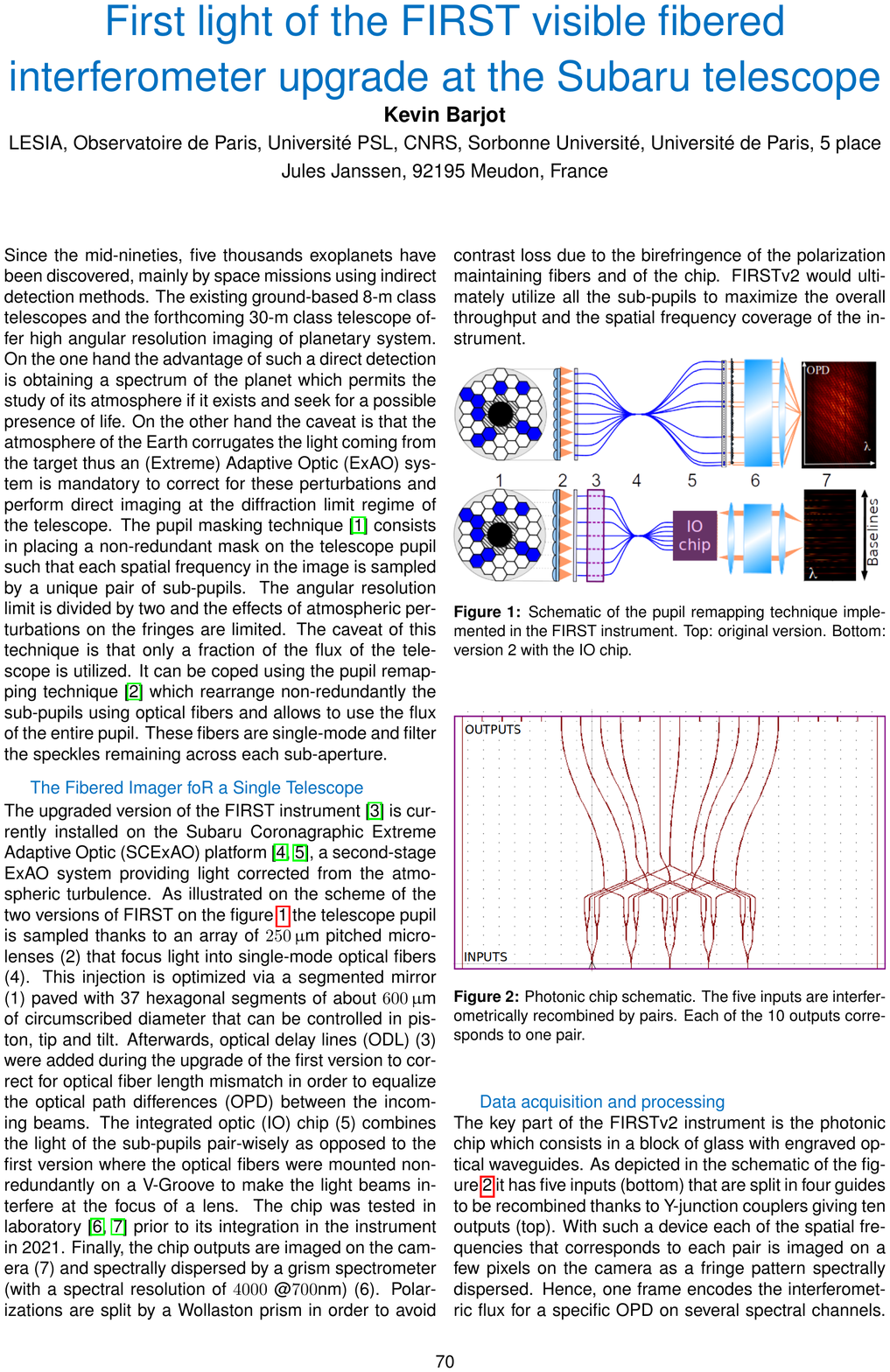}

\chapter{FIRSTv2 au télescope Subaru}
\label{sec:FIRSTv2Subaru}
\setcounter{figure}{0}
\setcounter{table}{0}
\setcounter{equation}{0}

\minitoc

\begin{figure}[ht!]
    \centering
    \includegraphics[width=0.47\textwidth]{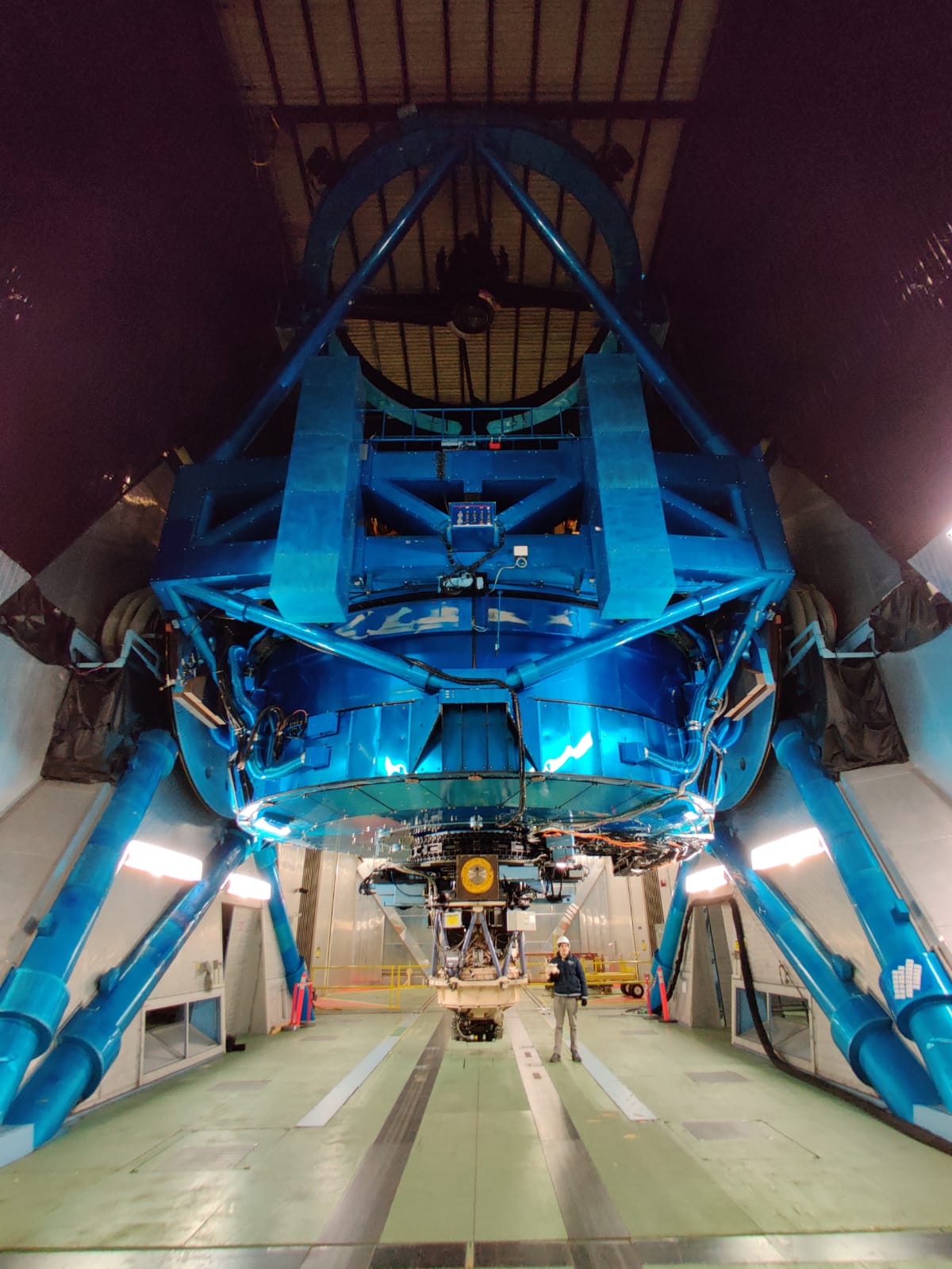}
\end{figure}

\clearpage
Durant ma thèse, j'ai participé à l'intégration de \ac{FIRSTv2} sur la plateforme \ac{SCExAO} ainsi qu'a sa première lumière le 10 septembre 2021.

Dans ce chapitre je présenterai la plateforme \ac{SCExAO} qui fournit la lumière du télescope Subaru corrigée par un système d'optique adaptative extrême (\ac{ExAO}) et permet le développement de nouveaux instruments pour l'imagerie haut contraste. à haute résolution angulaire. J'exposerai ensuite les différentes étapes d'intégration de \ac{FIRSTv2} sur la plateforme, aux côtés de \ac{FIRSTv1}. Je montre dans une troisième partie la caractérisation de \ac{FIRSTv2} sur \ac{SCExAO} et nous verrons les difficultés rencontrées. Pour finir je présenterai la première lumière et discuterai des améliorations qui ont pu être effectuées au cours des différentes nuits d'observation et tests qui ont suivi. Je conclurai enfin en présentant les développements futurs prévus sur l'instrument.

%%%%%%%%%%%%%%%%%%%%%%%%%%%%%%%%
\section{La plateforme SCExAO}

\acrfull{SCExAO} \citep{jovanovic2015} est une plateforme de R\&D pour l'imagerie \ac{HRA}, installée au foyer Nasmyth IR (voir la photographie de la figure~\ref{fig:SCExAOPhoto}) du télescope Japonais Subaru de $8,2 \,$m de diamètre. L'objectif primaire de la plateforme \ac{SCExAO} est de développer et de mettre en place un système d'optique adaptative extrême afin de fournir à plusieurs instruments un faisceau avec un haut rapport de Strehl ($80 \%$ dans la bande H) pour l'imagerie \ac{HRA}. Certains de ces instruments sont accessibles à la communauté scientifique : \acrfull{CHARIS} \citep{groff2015}, \acrfull{VAMPIRES} \citep{norris2015}, \ac{REACH} \citep{kotani2018}, \ac{MEC} \citep{walter2020} et Fast \ac{PDI} \citep{lozi2020}. D'autres sont des modules permettant le développement de nouvelles techniques pour l'imagerie \ac{HRA} : \ac{GLINT} \citep{norris2020b}, \ac{RHEA}\footnote{\url{https://www.naoj.org/Projects/SCEXAO/scexaoWEB/050devmodules.web/030rhea.web/indexm.html}} et \ac{FIRST}.

\begin{figure}[ht!]
    \centering
    \includegraphics[width=\figwidth]{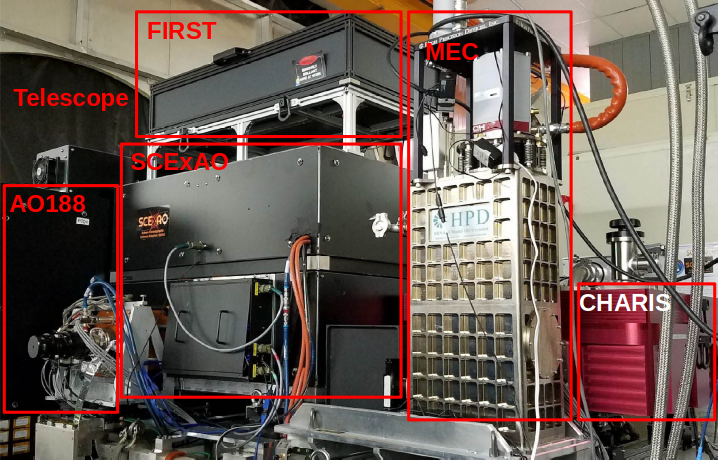}
    \caption[Photographie de la plateforme SCExAO et des modules installés sur la plateforme Nasmyth IR du télescope Subaru.]{Photographie de la plateforme SCExAO et des modules installés sur la plateforme Nasmyth IR du télescope Subaru. Crédit : équipe SCExAO.}
    \label{fig:SCExAOPhoto}
\end{figure}

La figure~\ref{fig:SCExAOScheme} présente le schéma optique global de \ac{SCExAO}. Il est divisé en quatre blocs représentant les deux bancs infrarouge et visible de la plateforme ainsi que la source de calibration et le banc de recombinaison de \ac{FIRST}. La lumière provenant du télescope subit une première correction appliquée par l'étage d'optique adaptative AO188 \citep{minowa2010} (module en amont de \ac{SCExAO}) avec un miroir déformable comportant $188$ actionneurs.

Le faisceau corrigé est ensuite injecté sur le banc \ac{IR} (\textit{IR bench}) de \ac{SCExAO} (dans le cadre du bas de la figure~\ref{fig:SCExAOScheme}) et subit un deuxième étage de correction par un miroir déformable de $2\,000$ actionneurs (optique adaptative extrême). Puis, grâce à la dichroïque il est divisé en un faisceau visible ($< 950 \,$nm) et un faisceau \ac{IR} ($< 950 \,$nm). Ce dernier est dirigé vers les instruments travaillant dans l'\ac{IR}, indiqués par les flèches oranges sur les bords du cadre. Le faisceau lumineux passe, entre autre, par des composants (\ac{PIAA} et \textit{focal plane mask}) coronographiques (voir à ce sujet la section~\ref{sec:ImagerieDirecte}).

La partie visible du faisceau est envoyée sur le banc visible (cadre du milieu nommé \textit{visible bench}) à travers un périscope. Ce faisceau est de nouveau divisé en deux sous-faisceaux. L'un d'eux ($800 - 950 \,$nm) est analysé par le senseur de front d'onde comportant une pyramide \citep{lozi2019} et permet d'appliquer la deuxième étape de correction par optique adaptative extrême à l'aide d'un miroir déformable comportant $2\,040$ actionneurs, intégré sur le banc \ac{IR}. L'autre ($< 800 \,$nm) est injecté dans les instruments travaillant dans le visible tels que \ac{FIRST}, \ac{VAMPIRES} et \ac{RHEA}.

Le faisceau est divisé entre les instruments \ac{VAMPIRES} et \ac{RHEA} d'une part et \ac{FIRST} d'autre part. Il est dirigé dans \ac{FIRST} au niveau de la partie centrale du cadre qui contient le \ac{MEMS}, les micro-lentilles et les fibres optiques monomodes qui sont connectées au banc de recombinaison. Celui-ci est représenté dans le cadre en haut à droite, nommé \textit{FIRST recombination} et il contient les deux versions de \ac{FIRST}. Il s'agit de sa configuration actuelle, qui est légèrement différente mais équivalente à celle présentée sur le schéma du bas de la figure~\ref{fig:FIRSTV1V2FinalScheme} de la section~\ref{sec:V1V2Subaru}.

Enfin, le cadre de gauche en haut de la figure~\ref{fig:SCExAOScheme} présente le banc des différentes sources internes, notamment la source super continuum ($600 - 2500 \,$nm) qui est utilisée pour simuler un point source vu par \ac{SCExAO} servant pour les tests et l'étalonnage de la \ac{P2VM}. Ces sources sont injectées via une fibre optique sur la plateforme \ac{SCExAO} par le banc \ac{IR} (cadre du bas), au niveau de l'injection du faisceau du télescope. De cette façon, la source lumineuse parcourt toute la plateforme jusqu'aux instruments.

\begin{figure}[ht!]
    \centering
    \includegraphics[width=0.89\textwidth]{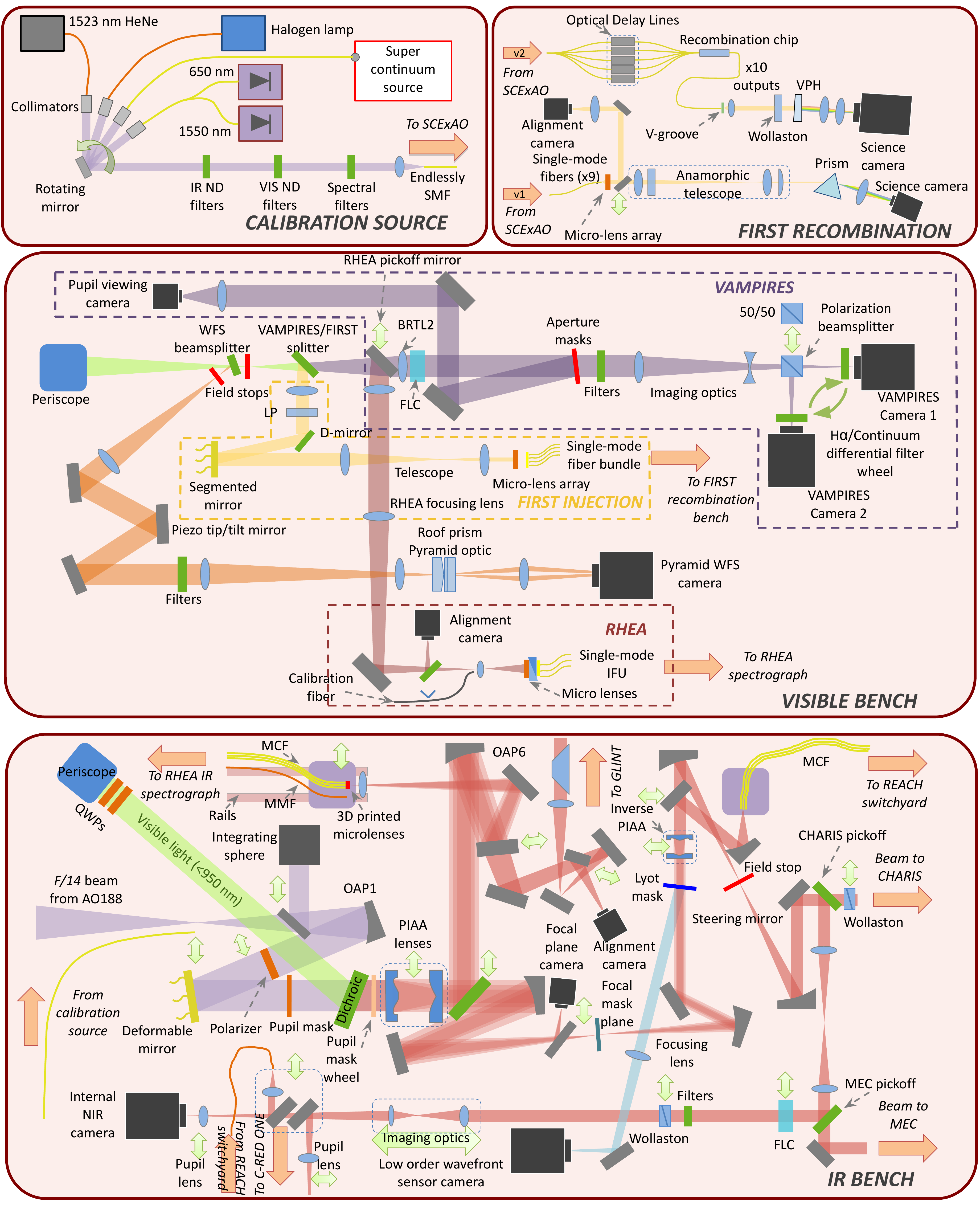}
    \caption[Schéma de principe de la plateforme SCExAO.]{Schéma de principe de la plateforme SCExAO. L'entrée se situe sur le banc infrarouge nommé \textit{IR bench} (cadre du bas). Le faisceau est divisé en une voie visible ($< 950 \,$nm) et en une voie infrarouge (IR) dirigée vers les instruments travaillant dans l'IR (GLINT, REACH, CHARIS, MEC, RHEA). La voie visible est envoyée sur le banc visible nommé \textit{visible bench} (cadre du milieu) via un périscope, est divisée et dirigée vers le senseur de front d'onde pour l'ExAO et vers les instruments travaillant dans le visible (FIRST, VAMPIRES, RHEA). Le faisceau est injecté dans FIRST (au centre) et les deux versions de recombinaison (v1 et v2) sont sur le banc nommé \textit{FIRST recombination} (cadre en haut à droite). Le banc des sources de calibration nommé \textit{calibration source} (cadre en haut à gauche) permet d'injecter différentes sources à l'entrée de SCExAO. Crédit : équipe SCExAO.}
    \label{fig:SCExAOScheme}
\end{figure}

On peut voir les différents bancs mentionnés (excepté celui de la source de calibration) sur la photo de la figure~\ref{fig:SCExAOPhoto}. En effet, ils sont montés les uns sur les autres, du banc \ac{IR} jusqu'au banc de recombinaison de \ac{FIRST}, de bas en haut.

%%%%%%%%%%%%%%%%%%%%%%%%%%%%%%%%
\section{Intégration de FIRSTv2 sur SCExAO}

Deux missions à Hawaï m'ont permis d'intégrer une partie de \ac{FIRSTv2} sur le télescope Subaru. Durant la première mission, se déroulant deux semaines en janvier 2020, j'ai déployé le nouveau logiciel de contrôle que j'avais développé en laboratoire, tout en intégrant la caméra \ac{EMCCD} Andor Ixon (voir plus loin). J'ai ainsi participé à une nuit d'observation (ma toute première) le 30 janvier 2020 permettant de tester la nouvelle installation de \ac{FIRSTv1} sur l'étoile $\upbeta$ Ori. Pendant la deuxième mission, se déroulant tout le mois de février 2022, j'ai intégré la puce $Y$ que j'ai testée au préalable sur le banc de test à Meudon. Trois nuits d'observation ont permis d'acquérir des données avec cette puce.

Le bon déroulement des intégrations et des nuits d'observation a pu être possible grâce à \textbf{l'aide et au travail précieux} de Sébastien Vievard avec qui j'ai beaucoup collaboré à distance (en partie à cause des limitations de la pandémie de COVID-19 sur les missions). Il a ainsi intégré les \acrshort{ODL}s et la puce $X$ (envoyée par colis depuis Meudon) s'assurant que la première lumière de \ac{FIRSTv2} ait bien lieu en septembre 2021.

%%%%%%%%%%%%%%%%
\subsection{Premières étapes d'installation}
\label{sec:V1V2Integration}

% Intégration de la caméra Ixion pendant ma première mission en 2020
Lors de ma première mission à Hawaï, en janvier 2020, j'ai déployé le logiciel de contrôle que j'avais commencé à développer sur le banc de test à Meudon (voir plus de détails sur le logiciel dans la section~\ref{sec:ControlSoftware}). Dans le même temps, Sébastien Vievard a intégré sur l'instrument \ac{FIRSTv1} la caméra de la gamme Andor Ixon Ultra $897$ fabriquée par \textit{Oxford Instruments}. Elle dispose d'un capteur de la technologie \acrfull{EMCCD} de $512 \,\text{px} \times 512\,\text{px}$ de taille égale à $16 \,$\um. On a fait face à des difficultés pour l'acquisition des images et, avec l'aide de Barnaby Norris, on a pu trouver une procédure qui menait à bien l'acquisition de cubes d'images. De plus, la caméra Ixon était en conflit avec la caméra photométrique du même fabricant, de la gamme Andor Luca, lorsque les deux étaient simultanément connectées à l'ordinateur. La solution a été d'alimenter les caméras par des prises dont l'alimentation est contrôlable à distance. On a fini par tester et valider la nouvelle installation de \ac{FIRSTv1} en observant l'étoile $\upbeta$ Ori lors de la nuit du 30 janvier 2020.

% Intégration des ODLs pendant l'été 2021
L'intégration de \ac{FIRSTv2} a continué par l'achat et l'intégration des lignes à retard (\acrfull{ODL}). Sébastien Vievard s'est chargé de commander auprès de \textit{Oz Optics} cinq \ac{ODL}s du même modèle que celles dont nous disposons à Meudon (voir plus de détails dans la section~\ref{sec:InstruODL}) avant de fabriquer l'interface électronique et de les intégrer sur \ac{FIRSTv1} pendant l'été 2021. La différence de sensibilité de l'instrument lors de tests sur la source interne nous a paru conséquente et il a été décidé de mesurer leur transmission, estimée à $20 - 30 \%$. Cette faible transmission est bien inférieure à celle spécifiée ($\sim 75 \%$ à $650 \,$nm) et nous avons donc décidé de conduire les mêmes mesures à Meudon (pour cela voir la section~\ref{sec:OdlThroughput}). C'est un point très préoccupant car cela va à l'encontre de notre objectif de gagner en sensibilité avec \ac{FIRSTv2}. On a ainsi commencé à envisager de redéfinir le concept de l'instrument pour se débarrasser, à terme, des \ac{ODL}s, comme j'ai pu en discuter dans la section~\ref{sec:ODLDiscussions}.

% Intégration de la puce X pendant l'été 2021
% Les nuits d'observations entre 20210910 et <20220214 sont sur Ixion, v2 seul
Dans le même temps, la puce $X$ a été envoyée à Hawaï depuis Meudon pour finir l'intégration des éléments de \ac{FIRSTv2} et la première lumière a pu avoir lieu le 10 septembre 2021. La figure~\ref{fig:FIRSTV1V2IxonPhoto} présente une photographie du banc de recombinaison de \ac{FIRST} en février 2022, après la ré-intégration en parallèle de \ac{FIRSTv1}. Dans le coin en haut à gauche se trouve la puce photonique protégée par une feuille de papier optique, sur une petite plateforme montée au dessus des \ac{ODL}s (non visibles) connectées à l'ordinateur de contrôle par les câbles blancs sur la partie haute. La puce est connectée aux fibres du V-Groove installé sur la partie centre gauche de la photographie, devant l'objectif de microscope. Le faisceau en sortie de ce dernier est tracé en bleu et traverse d'abord le prisme de Wollaston avant d'être dispersé par un prisme. Une lentille image le faisceau sur la caméra en haut à droite. Le faisceau de \ac{FIRSTv1} est tracé en rouge au niveau de l'entrée du système anamorphique et est dispersé par un autre prisme, en bas de la photographie. Le faisceau est ensuite réfléchi par deux miroirs plans et imagé par une lentille sur la même caméra que précédemment. Les images acquises contiennent alors les interférogrammes des deux versions de l'instrument (voir la figure~\ref{fig:V1V2Image}), ce qui a permis une estimation de leur différence de sensibilité, présentée plus loin dans la section~\ref{sec:V1V2Throughput}.

% Entre 20220214 et <=20220302, imagerie simultanée de v1 et v2 sur Ixion
\begin{figure}[ht!]
    \centering
    \includegraphics[width=\figwidth]{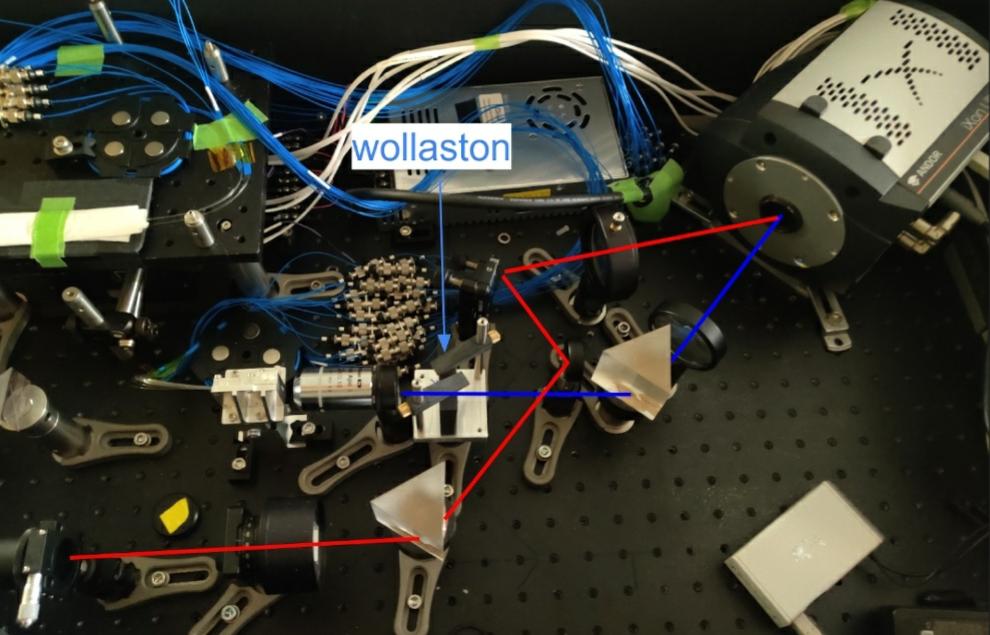}
    \caption[Photographie des deux versions de FIRST installées sur le banc de recombinaison de SCExAO, partageant la même caméra.]{Photographie des deux versions de FIRST installées sur le banc de recombinaison de SCExAO, partageant la même caméra. La puce $X$ (en haut à gauche) est enroulée dans du papier optique sur une plateforme au-dessus des ODLs et ses fibres de sortie sont connectées aux fibres du V-Groove (centre gauche). Le faisceau de FIRSTv2 est tracé en bleu en sortie de l'objectif de microscope placé devant les sorties du V-Groove, traverse le prisme de Wollaston et est dispersé par un prisme. Une lentille image ensuite le faisceau sur la caméra Andor Ixon (en haut à droite). Le faisceau de FIRSTv1 est tracé en rouge en bas à gauche, en entrée du système anamorphique et est dispersé par un prisme avant d'être réfléchi par deux miroirs plans et imagé par une lentille sur la même caméra. Crédit : Sébastien Vievard.}
    \label{fig:FIRSTV1V2IxonPhoto}
\end{figure}

\begin{figure}[ht!]
    \centering
    \includegraphics[width=\figwidth]{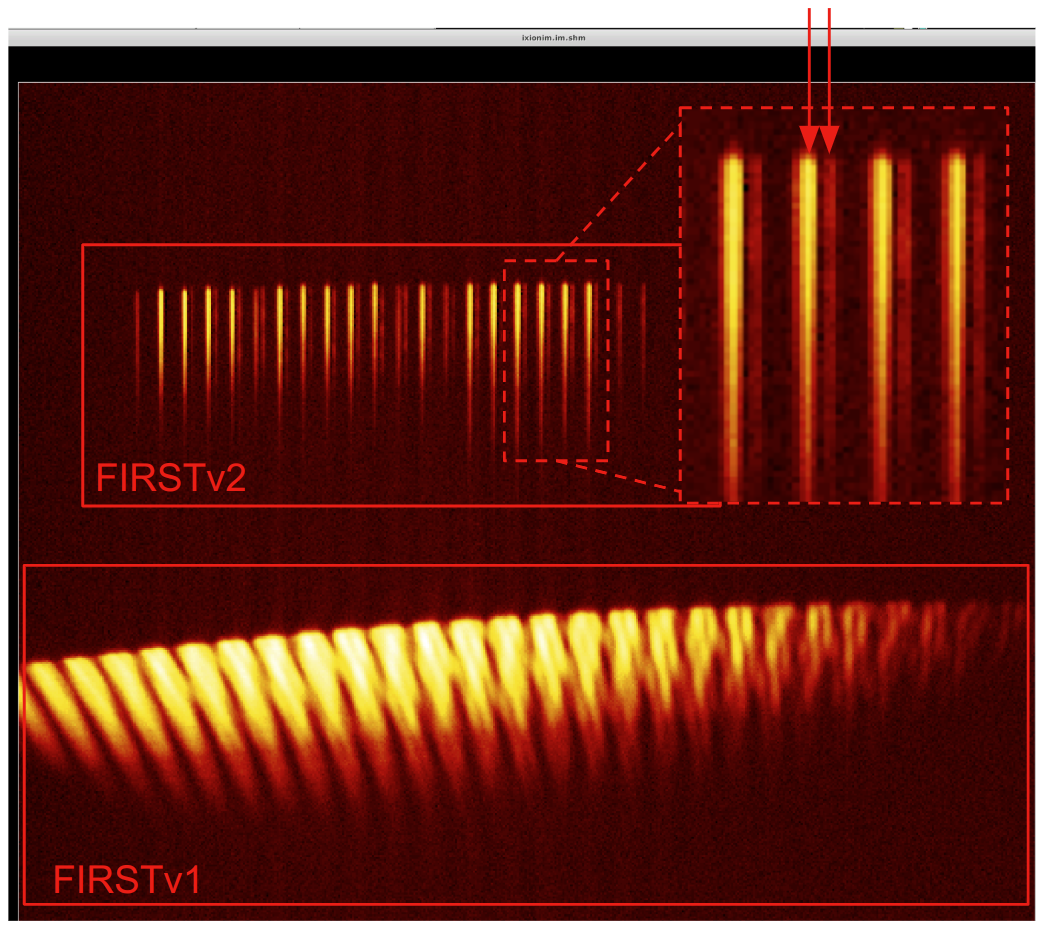}
    \caption[Images des interférogrammes de FIRSTv1 et FIRSTv2 sur la caméra Andor Ixon sur source interne.]{Images des interférogrammes de FIRSTv1 (en bas) et FIRSTv2 (en haut) sur la caméra Andor Ixon sur source interne. Les OPDs ne sont pas annulées sur les voies de FIRSTv2 et aucune frange n'apparaît sur les sorties correspondantes. Un agrandissement de quatre des sorties de V2 est montré pour mettre en évidence les sorties sélectionnées par le prisme de Wollaston dans la polarisation verticale (flèche rouge de gauche) et horizontale (flèche rouge de droite). Crédit : Sébastien Vievard.}
    \label{fig:V1V2Image}
\end{figure}

% Intégration de la puce Y pendant ma deuxième mission en février 2022
Ma deuxième mission, en février 2022, a été l'occasion d'intégrer la puce $Y$ sur le banc de recombinaison et de la tester lors de plusieurs nuits d'observations. De plus, j'ai également intégré un prisme de Wollaston ce qui nous a permis d'améliorer la qualité des données en traitant les deux polarisations séparément. Enfin, j'ai été amené à améliorer la procédure de prise de données au niveau de la synchronisation de la caméra et du \ac{MEMS}, le logiciel à Hawaï étant différent de celui à Meudon. Mais aussi, je me suis rendu compte que la procédure de prise de données n'assurait pas la connaissance de l'étape de modulation pour chaque image, ce qui est primordial pour le traitement de données. Pour ce faire, il a fallu assigner un entier à chaque étape de la séquence de modulation qui était enregistré dans un pixel d'un des coins des images. En effet, contrairement à la procédure d'acquisition à Hawaï, celle au laboratoire à Meudon est telle que la première image de chaque séquence acquise correspond nécessairement à la première étape de la séquence de modulation, ce qui dispensait de se préoccuper de ce sujet.

%%%%%%%%%%%%%%%%
\subsection{Augmentation de la résolution spectrale}
\label{sec:V1V2Subaru}

% Intégration du nouveau spectro et de la nouvelle caméra de V2 (>20220309) + ré-intégration de V1 (>=20220427)
En mars 2022, Sébastien Vievard et Manon Lallement ont procédé au réalignement des deux versions de \ac{FIRST}, à la suite de la réception d'une nouvelle caméra (présentée ci-après) et des optiques du nouveau concept de spectro-imageur, conçu par Manon Lallement (voir plus de détails dans la section~\ref{sec:InstruSpectro}). La nouvelle configuration est illustrée par le schéma de principe de la figure~\ref{fig:FIRSTV1V2FinalScheme} qui représente le banc de recombinaison de \ac{FIRST}. De manière similaire à la photographie de la figure~\ref{fig:FIRSTV1V2IxonPhoto} présentée précédemment, les fibres proviennent du banc visible par la gauche et neuf d'entre elles sont connectées sur les fibres du V-Groove de \ac{FIRSTv1} (en bas à gauche), disposé devant une matrice de micro-lentilles qui collimatent les faisceaux de sortie. Ces derniers peuvent être dirigés par un miroir plan contrôlable en position vers le système d'imagerie photométrique pour l'optimisation de l'injection du flux dans les fibres par le \ac{MEMS} (sur le banc visible), ou vers la partie recombinaison interférométrique composée du système anamorphique, de deux miroirs plans, du prisme, de la lentille d'imagerie et de la caméra de science (Andor Ixon, en haut à droite). En parallèle, cinq des fibres provenant du banc visible sont connectées aux \ac{ODL}s de \ac{FIRSTv2} (en haut à gauche) dont les fibres de sortie sont connectées à celles en entrée de la puce photonique (en haut au centre). Les fibres de sortie de cette dernière sont connectées aux fibres du V-Groove (au centre) et les faisceaux sont collimatés par l'objectif de microscope, traversent le prisme de Wollaston, sont dispersés par le nouveau spectro-imageur (composé d'un réseau holographique et de deux lentilles) et sont imagés sur la nouvelle caméra.

% Schéma final du banc de recombinaison
\begin{figure}[ht!]
    \centering
    \includegraphics[width=\figwidth]{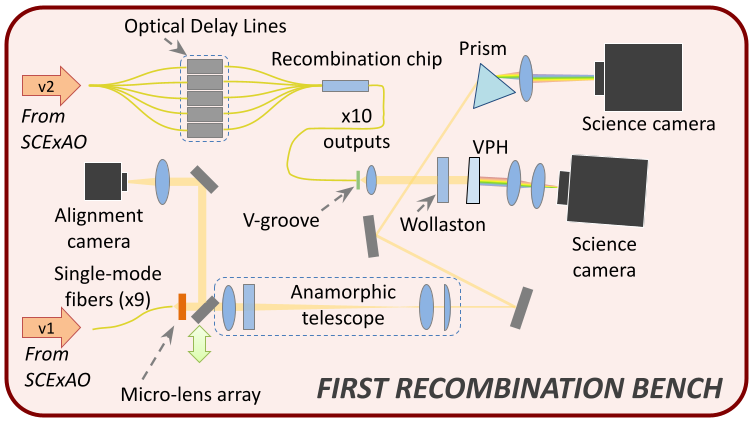}
    \caption[Schéma de principe des deux versions de FIRST installées en parallèle sur le banc de recombinaison de SCExAO.]{Schéma de principe des deux versions de FIRST installées en parallèle sur le banc de recombinaison de SCExAO. Les $9$ fibres de v1 provenant du banc visible (en bas à gauche) sont connectées aux fibres d'un V-Groove dont les faisceaux de sortie sont collimatés par une matrice de micro-lentilles, puis peuvent être orientés par un miroir plan mobile vers la voie photométrique ou vers la voie interférométrique. Celle-ci est composée (du coin en bas à gauche vers le coin en haut à droite) d'un système anamorphique, de 2 miroirs plans, d'un prisme dispersant, d'une lentille d'imagerie et de la caméra (Andor Ixon). Les 5 fibres de v2 provenant du banc visible (en haut à gauche) sont connectées aux ODLs, elles-mêmes connectées à la puce photonique, elle-même connectée au V-Groove (au centre). Puis se trouvent un objectif de microscope, un prisme de Wollaston, un réseau holographique, deux lentilles d'imagerie et la caméra (Hamamatsu). Crédit : Sébastien Vievard.}
    \label{fig:FIRSTV1V2FinalScheme}
\end{figure}

% Intégration de la nouvelle caméra Orca à partir du 9 mars 2022 (données >=20220323) en parallele de V1
% Hamamatsu Orca Quest qCMOS camera: https://www.hamamatsu.com/eu/en/product/type/C15550-20UP/index.html
La nouvelle caméra d'imagerie scientifique de \ac{FIRSTv2} est de la gamme Orca Quest qCMOS fabriquée par \textit{Hamamatsu}\footnote{\url{https://www.hamamatsu.com/}}. De même que la caméra à Meudon, elle dispose d'un capteur de la technologie \ac{CMOS}, de $4\,000 \,\text{px} \times 2\,300 \,$px de taille égale à $4,6 \,$\um. Sa bande passante est $300 - 1000 \,$nm avec une efficacité quantique de $80 - 65\%$ sur la gamme spectrale $600 - 700 \,$nm. Le bruit sur le courant d'obscurité lorsqu'elle est refroidie à l'eau (température du capteur de $-35\degree \,$C) vaut $0,006 \,\text{e}^-.px^{-1}.s^{-1}$ et le bruit de lecture est de $0,25 - 0,4 \,\text{e}^-$ suivant le mode de lecture, ce qui est mieux d'un facteur $\sim 24$ et d'un facteur $\sim 3$, respectivement, par rapport à celle de la gamme Andor Zyla à Meudon. La taille plus petite des pixels de cette caméra (d'un facteur $\sim 3,5$) permet d'obtenir la résolution spectrale estimée dans le prochain paragraphe.

% Intégration du nouveau spectro à partir du 9 mars 2022 (données >=20220323)
Dans le même temps, le spectro-imageur conçu à Meudon par Manon Lallement (pour plus de détails voir la section~\ref{sec:InstruSpectro}) est intégré à \ac{FIRSTv2} et est indiqué par la mention \acrshort{VPH} (\acrlong{VPH}) sur le schéma de la figure~\ref{fig:FIRSTV1V2FinalScheme}. Un étalonnage spectral à partir de la source d'étalonnage de l'instrument \ac{CHARIS} (dont la longueur d'onde est ajustable grâce à un filtre ajustable) permet d'estimer la résolution spectrale du spectro-imageur à $\sim 3\,000$ à $650 \,$nm. L'identification en longueur d'onde des raies dans le flux mesuré et son ajustement polynomial sont présentés sur le graphique du haut et du bas de la figure~\ref{fig:V2SpecCalSubaru}, respectivement. De même que l'étalonnage spectral effectué à Meudon et présenté dans la section~\ref{sec:EtalonnageSpectral}, la fonction polynomiale de l'ajustement de la longueur d'onde sur les pixels de la caméra est linéaire et la résolution spectrale est similaire.

\begin{figure}[ht!]
    \centering
    \begin{subfigure}{\textwidth}
        \centering
        \includegraphics[width=\textwidth]{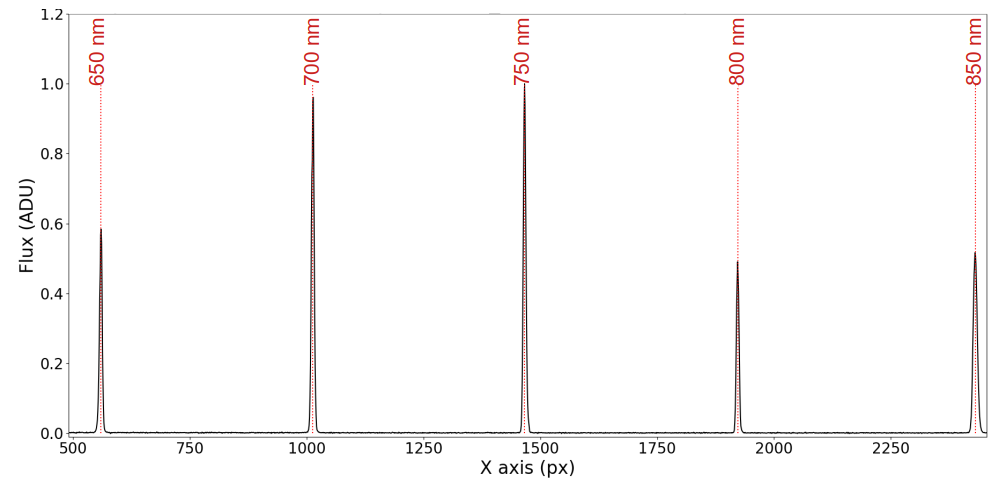}
    \end{subfigure}
    \begin{subfigure}{\textwidth}
        \centering
        \includegraphics[width=\textwidth]{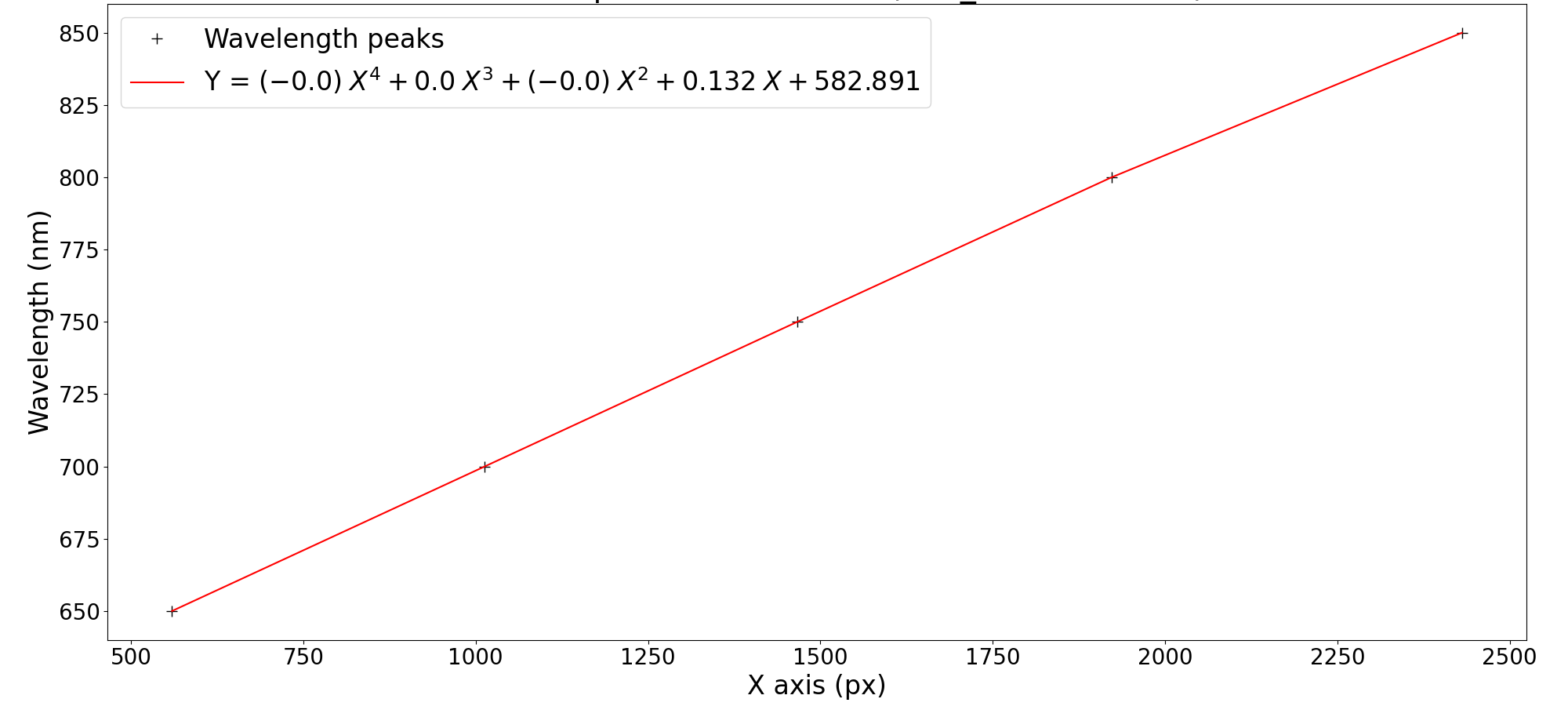}
    \end{subfigure}
    \caption[Étalonnage spectral de FIRSTv2 sur le banc de recombinaison de SCExAO.]{Étalonnage spectral de FIRSTv2 sur le banc de recombinaison de SCExAO. En haut : identification des pics de la source d'étalonnage de l'instrument CHARIS sur les pixels de la caméra pour l'une des sorties de la puce. En bas : les longueurs d’onde en fonction des positions des pics sur la caméra ainsi que l’ajustement polynomial associé.}
    \label{fig:V2SpecCalSubaru}
\end{figure}

Enfin, la figure~\ref{fig:FIRSTV1V2FinalPhoto} présente une photographie du banc de recombinaison de \ac{SCExAO} décrit précédemment par le schéma de principe de la figure~\ref{fig:FIRSTV1V2FinalScheme}. Je ne reviendrai pas sur sa description qui est similaire au premier paragraphe de cette section.

% Photo du banc de recombinaison après intégration du nouveau spectro et de la nouvelle caméra de V2
% + ré-intégration de V1 (>=20220427)
\begin{figure}[ht!]
    \centering
    \includegraphics[width=\figwidth]{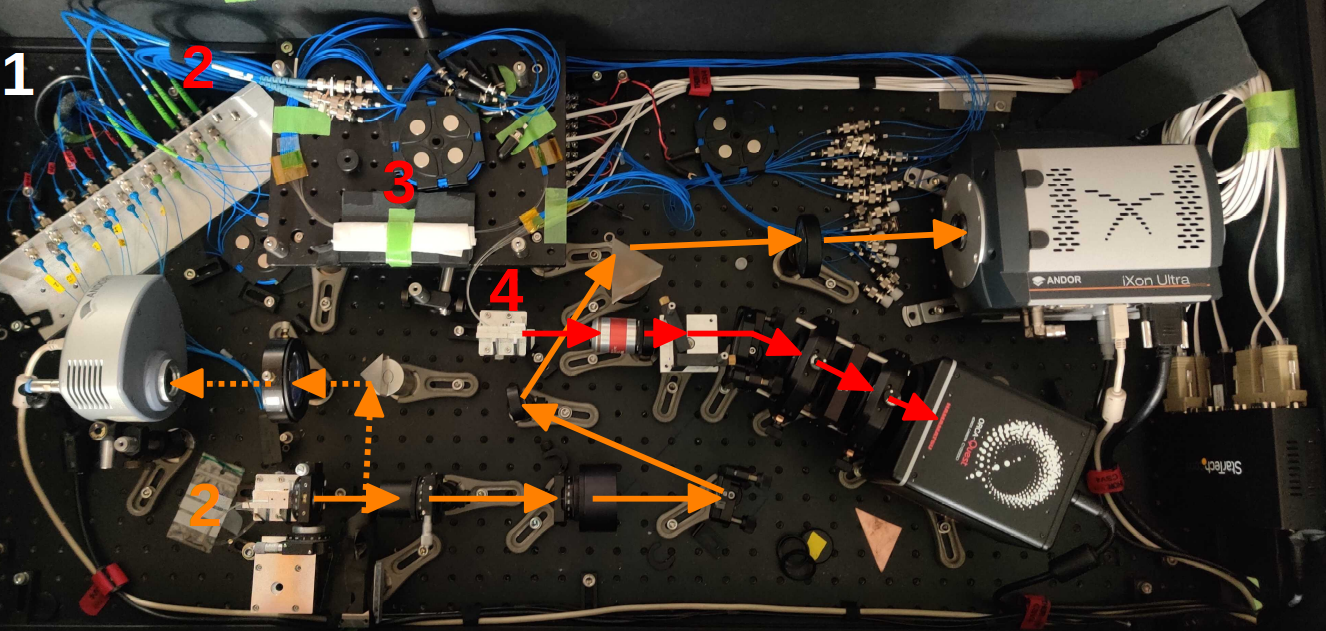}
    \caption[Photographie des deux versions de FIRST installées en parallèle sur le banc de recombinaison de SCExAO.]{Photographie des deux versions de FIRST installées en parallèle sur le banc de recombinaison de SCExAO. Les éléments et chemin optique de FIRSTv1 et FIRSTv2 sont représentés, respectivement, en orange et en rouge. Les fibres (1) proviennent du banc visible. 9 sont connectées au V-Groove de v1 (2) puis un miroir plan mobile permet de diriger les faisceaux vers la voie photométrique (flèches en pointillés) ou vers la voie interférométrique (flèches continues) composée par le système anamorphique, deux miroirs plans, un prisme dispersant, une lentille d'imagerie et une caméra (Andor Ixon). 5 des fibres d'entrée de v2 sont connectées à l'enchaînement de composants : ODLs (2), puce photonique (3) et V-Groove (4). La voie interférométrique de v2 (en rouge) est ensuite composée de l'objectif de microscope, du prisme de Wollaston, du réseau holographique, de deux lentilles d'imagerie et de la caméra (Hamamatsu). Crédit : Sébastien Vievard.}
    \label{fig:FIRSTV1V2FinalPhoto}
\end{figure}

%%%%%%%%%%%%%%%%
\subsection{La configuration des bases}

La configuration des sous-pupilles dont les faisceaux sont recombinés est présentée sur la figure~\ref{fig:SegUVSubaruA}. À la différence de celle qui est choisie sur le banc de test en laboratoire, ici l'ombre de l'obstruction centrale du télescope doit être prise en compte. Celle-ci est visible sur l'image du plan pupille en entrée de la partie de l'injection de \ac{FIRST} (du banc visible de \ac{SCExAO}) montrée en arrière plan. Ainsi, tous les segments de couleur noire ne sont jamais utilisés. Les segments de couleur beige et ceux de couleur verte sont ceux choisis pour la recombinaison de \ac{FIRSTv1} et \ac{FIRSTv2} (numérotés 22, 26, 27, 31 et 35), respectivement. Enfin, le segment rouge est défectueux et les segments bleus ne sont utilisés par aucun des instruments. Ici, la configuration optique est telle que devant chaque segment est alignée une fibre optique du toron. La figure~\ref{fig:SegUVSubaruB} présente le plan UV des fréquences spatiales échantillonnées par les sous-pupilles de \ac{FIRSTv2} précédemment choisies. Les couleurs correspondent aux longueurs d'onde sur la gamme spectrale $600 - 800 \,$nm. Enfin, seules les bases $22 - 26$ et $22 - 35$ sont redondantes, respectivement, avec les bases $35 - 31$ et $26 - 31$.

\begin{figure}[ht!]
    \centering
    \begin{subfigure}[t]{0.43\textwidth}
        \centering
        \includegraphics[width=\textwidth]{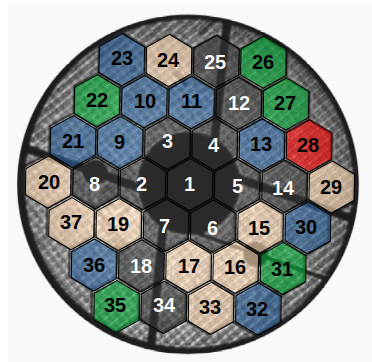}
        \caption{Configuration des sous-pupilles choisies pour FIRSTv1 (en beige) et FIRSTv2 (en vert) dans le plan pupille du MEMS de FIRST au Subaru. Les segments jamais utilisés sont en noirs (à cause de l'obstruction centrale du télescope, montrée en arrière plan), dysfonctionnel en rouge et disponibles en bleus. Crédit : Sébastien Vievard.}
        \label{fig:SegUVSubaruA}
    \end{subfigure}\hfill
    \begin{subfigure}[t]{0.55\textwidth}
        \centering
        \includegraphics[width=\textwidth]{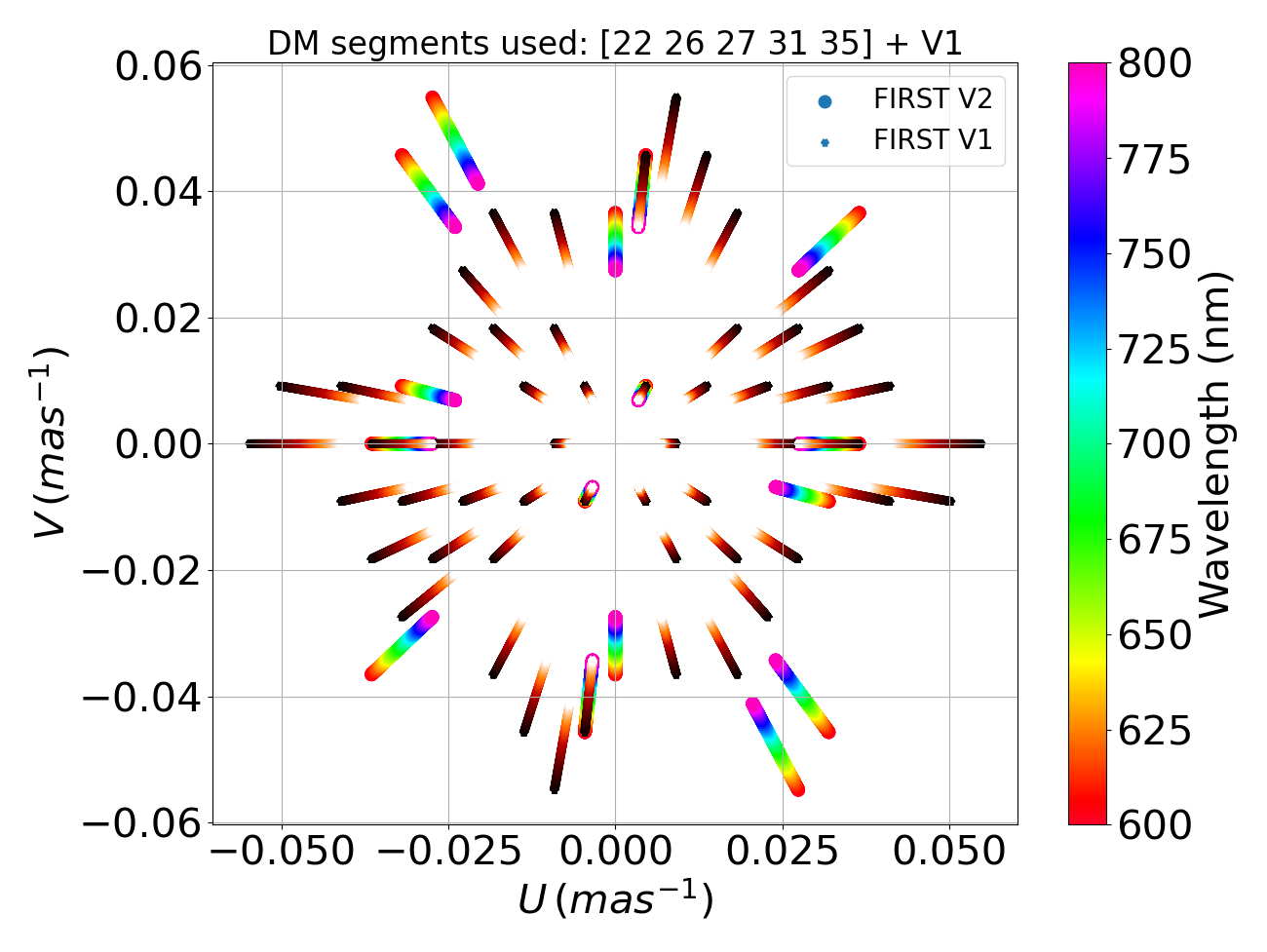}
        \caption{Les répartitions des bases choisies pour FIRSTv1 (étoile) et FIRSTv2 (point) représentées dans l'espace de Fourier, appelée aussi la couverture du plan UV des fréquences spatiales. Les couleurs représentent la longueur d'onde (les bandes spectrales sont identiques pour les deux versions).}
        \label{fig:SegUVSubaruB}
    \end{subfigure}
    \caption[Configuration des sous-pupilles et couverture du plan UV de FIRSTv1 et de FIRSTv2 sur SCExAO.]{Configuration des sous-pupilles et couverture du plan UV de FIRSTv1 et de FIRSTv2 sur SCExAO.}
    \label{fig:SegUVSubaru}
\end{figure}

%%%%%%%%%%%%%%%%%%%%%%%%%%%%%%%
\section{Caractérisation de FIRSTv2 sur SCExAO}

%%%%%%%%%%%%%%%%
\subsection{Étude de sensibilité}
\label{sec:V1V2Throughput}

% Mesures pendant la premiere integration à R = 300
On a profité du fait que les deux versions de \ac{FIRST} soient imagées simultanément sur la même caméra (Andor Ixon), comme montré sur la photographie de la figure~\ref{fig:FIRSTV1V2IxonPhoto} de la section~\ref{sec:V1V2Integration}, pour comparer leur sensibilité. La figure~\ref{fig:V1V2Vega} présente une image en échelle logarithmique acquise à un temps d'exposition de $50 \,$ms lors de la nuit d'observation du 24 février 2022 sur l'étoile Véga. La longueur d'onde est selon l'axe horizontal et l'\ac{OPD} selon l'axe vertical pour \ac{FIRSTv1}. Les interférogrammes de \ac{FIRSTv1} et de \ac{FIRSTv2} sont sur la moitié gauche et la moitié droite de l'image, respectivement. Trois sources lasers (de longueurs d'onde $785 \,$nm, $850 \,$nm et $852 \,$nm) sont injectées dans un coupleur fibré intégré juste avant le séparateur \textit{VAMPIRES/FIRST splitter} sur le banc visible (discussion dans la section~\ref{sec:V2SubaruProspectives}), dans le cadre de tests pour la mesure des perturbations de phase induites par des éléments du montage optique, tels que les fibres optiques. Les flux de celles-ci apparaissent comme des colonnes verticales sur la droite des interférogrammes de \ac{FIRSTv1} et comme des points sur la droite des interférogrammes de \ac{FIRSTv2} (indiquées à chaque fois par les flèches rouges).

\begin{figure}[ht!]
    \centering
    \includegraphics[width=0.8\textwidth]{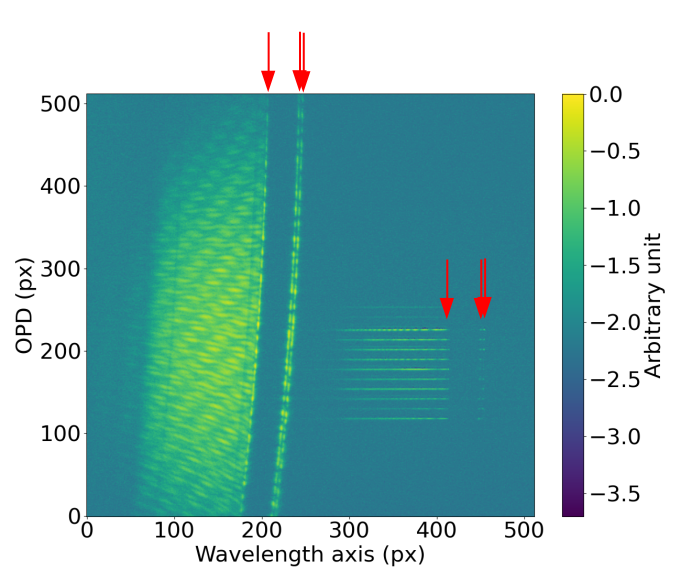}
    \caption[Image des interférogrammes des deux versions de FIRST sur la caméra Andor Ixon obtenues lors d'une observation de l'étoile Véga sur SCExAO.]{Image des interférogrammes des deux versions de FIRST sur la caméra Andor Ixon obtenues lors d'une observation de l'étoile Véga sur SCExAO. L'image est normalisée par le flux maximum et est en échelle logarithmique. Les longueurs d'onde et les OPDs (pour FIRSTv1) sont selon les axes horizontal et vertical, respectivement. Les interférogrammes de V1 et de V2 sont imagés sur la moitié gauche et sur la moitié droite de l'image, respectivement. Les flèches rouges montrent les positions des trois sources lasers injectées en entrée de FIRST.}
    \label{fig:V1V2Vega}
\end{figure}

% Estimation du rapport de transmission entre V1 et V2
À partir de cette image, le flux total des deux versions de \ac{FIRST} est calculé en sommant les pixels sur lesquels les interférogrammes sont imagés, en omettant le flux provenant des sources lasers. Ces flux sont ensuite normalisés par le nombre de faisceaux injectés : 9 pour \ac{FIRSTv1} et 5 pour \ac{FIRSTv2}. Enfin, le rapport de ces deux flux normalisés donne une estimation de la différence de transmission entre les deux versions. On estime ainsi que \ac{FIRSTv2} transmet $\sim 20$ fois moins que \ac{FIRSTv1}. De plus, dans cette configuration, les deux versions comprennent les mêmes composants exceptées les \ac{ODL}s et la puce photonique intégrées à \ac{FIRSTv2}. D'après les transmissions des \ac{ODL}s et de la puce $Y$ estimées, respectivement, à $30\%$ (section~\ref{sec:V1V2Integration}) et $15 \%$ (section~\ref{sec:IOChipThroughput}), on s'attend à une transmission de $4,5\%$ par rapport à \ac{FIRSTv1}, ce qui est cohérent avec l'estimation précédente.

Ainsi, avec le nouveau spectro-imageur intégré par la suite, de résolution spectrale égale à $\sim 3\,000$, on s'attend à ce que le flux par pixel soit divisé par 10. Cela a empêché par la suite l'acquisition de données sur ciel après l'intégration du nouveau spectro-imageur montrant bien la nécessité de se passer des \ac{ODL}s et du développement de puce d'optique intégrée plus transmissives.

%%%%%%%%%%%%%%%%
\subsection{Étude de stabilité}
\label{sec:V2StabilitySubaru}

% Perturbations visibles sur les données de fullon
La figure~\ref{fig:FullOnDataSCExAO} présente les données interférométriques acquises sur \ac{FIRSTv2}, sur la source interne de \ac{SCExAO} (la source \sk). La puce $Y$ est installée ainsi que le premier spectro-imageur de résolution spectrale égale à $300$. Chaque graphique de gauche à droite correspond à une base et les graphiques du haut et du bas représentent les interférogrammes mesurés et ajustés, respectivement, selon le processus présenté dans la section~\ref{sec:FullOnFit}. L'acquisition de ces données est effectuée avec la séquence de modulation à $20$ pas (axe horizontal sur les images de la figure~\ref{fig:FullOnDataSCExAO}), représentée par la figure~\ref{fig:ModSeq20}, avec un temps d'exposition de la caméra égal à $20 \,$ms à une fréquence de $16,45 \,$Hz. Ainsi, entre chaque point de mesure s'écoule $60 \,$ms et un interférogramme est acquis en $1,2 \,$s (correspondant à $20$ points de mesure).

\begin{figure}[ht!]
    \centering
    \includegraphics[width=\figwidth]{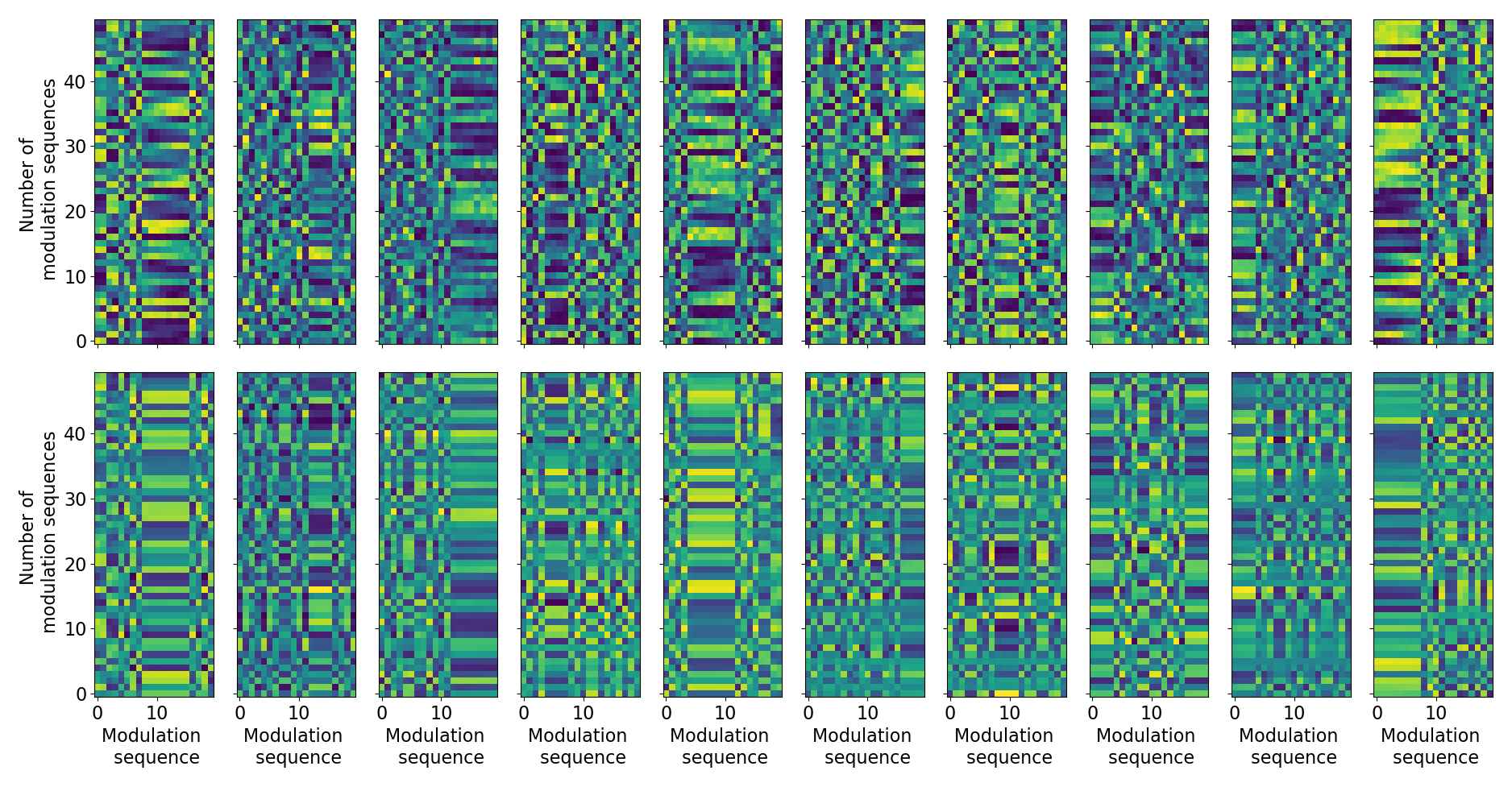}
    \caption[Images des interférogrammes mesurés et ajustés, avec la puce $Y$ sur la source interne de SCExAO.]{Images des interférogrammes mesurés (lignes du haut) et ajustés (lignes du bas) sur les $10$ bases (en colonne), avec la puce $Y$ et la source interne de SCExAO. Les axes horizontal et vertical représentent, respectivement, la séquence de modulation et le nombre de fois que la séquence de modulation a été acquise. Ces images sont pour le canal spectral $\sim 700 \,$nm.}
    \label{fig:FullOnDataSCExAO}
\end{figure}

La figure~\ref{fig:FullOnData} présente les mesures effectuées en laboratoire, équivalentes à celles montrées ici, sur la puce $Y$ (figure du haut), avec le spectro-imageur de résolution spectrale égale à $\sim 3\,200$, pour un temps d'exposition de la caméra de $100 \,$ms à une fréquence de $9,9 \,$Hz. Chaque point est donc acquis toutes les $101 \,$ms et un interférogramme toutes les $1,2 \,$s, ce qui est similaire aux données présentées sur la figure~\ref{fig:FullOnDataSCExAO}. Par comparaison visuelle des images des interférogrammes de ces deux prises de données, on remarque que ceux estimés sur \ac{FIRSTv2} intégré sur \ac{SCExAO} présentent une plus grande perturbation en fonction du temps (axe vertical). 

% Mesure de CP
On s'en rend bien compte en traçant l'\ac{OPD} estimé à partir de l'ajustement des phases des interférogrammes, en fonction du temps. En suivant la même procédure que dans la section~\ref{sec:CPStabilityMeudon}, sur les données montrées dans le paragraphe précédent, on obtient les graphiques de l'évolution de l'\ac{OPD} et des clôtures de phase en fonction du temps, présentés sur la figure~\ref{fig:OPDfitVStimeSubaru}. Cette dernière est similaire à la figure~\ref{fig:OPDfitVStime}, pour les dix bases. Les courbes d'\ac{OPD} présentent des pics qui restent présents dans les courbes de clôtures de phase. Ces pics apparaissent lorsque l'estimation de la phase échoue. En retirant ces points extrêmes on estime l'écart-type de la clôture de phase en moyenne égal à $\sim 1,3 \,$\um~rms, contre $\sim 50 \,$nm rms estimé en laboratoire.

\begin{figure}[ht!]
    \centering
    \includegraphics[width=\figwidth]{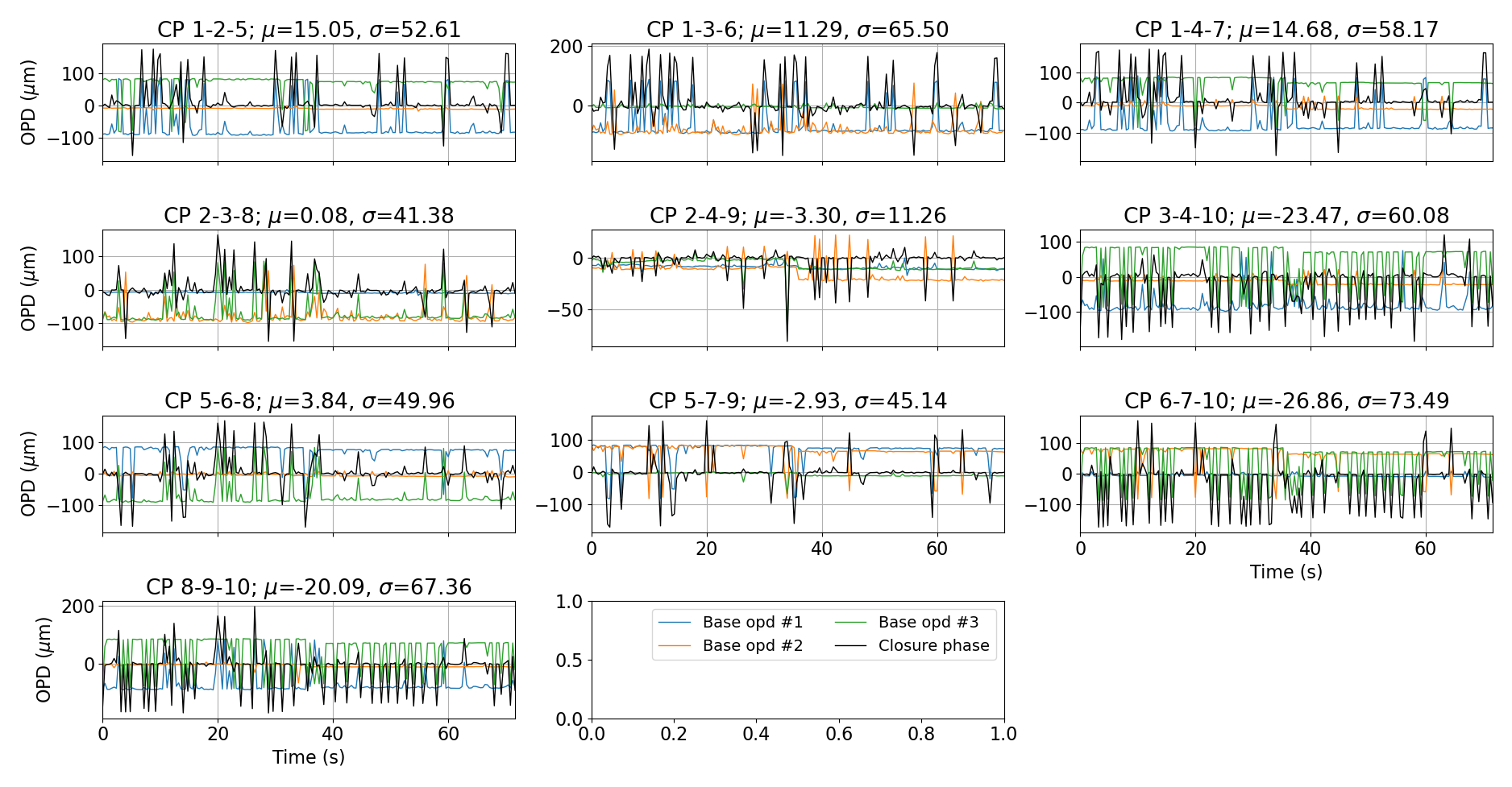}
    \caption[Graphiques de l'OPD en fonction du temps des dix triangles formés par les dix bases de FIRSTv2 sur la source interne de SCExAO avec la puce $Y$, avant la protection des fibres.]{Graphiques de l'OPD en fonction du temps (tracé en couleur) des dix triangles formés par les dix bases de FIRSTv2 sur la source interne de SCExAO avec la puce $Y$, avant la protection des fibres contre les caméras de VAMPIRES. Les clôtures de phase sont tracées en noir et leur valeur moyenne et l'écart-type sont écrits dans les sous-titres. Les courbes sont tracées pour $4\,000$ images de temps d'exposition égal à $20 \,$ms.}
    \label{fig:OPDfitVStimeSubaru}
\end{figure}

% Origines et solutions
Il est très probable que ce soit les fibres qui sont sensibles aux perturbations qui induisent ces pics observés dans les mesures d'\ac{OPD} en ajoutant un bruit trop important sur les phases estimées. Avant que \ac{FIRSTv2} soit intégré sur \ac{SCExAO}, les fibres optiques qui se trouvent dans la transition entre le banc visible et le banc de recombinaison de \ac{FIRST} (se trouvant donc à l'extérieur des coffrages) avaient déjà été protégées en les plaçant dans un tube en mousse, afin de les protéger des courants d'air se trouvant sur la plateforme Nasmyth du télescope. On pense donc que ce sont les ventilateurs des caméras de l'instrument \ac{VAMPIRES} sur le banc visible qui seraient (au moins en partie) à l'origine de ces perturbations car les fibres du toron en sont à proximité. Celles-ci ont donc aussi été protégées sur le banc visible et de nouvelles mesures, présentées sur la figure~\ref{fig:OPDfitVStimeSubaruProtect}, montrent des courbes présentant moins de pics extrêmes. Malgré cela, la stabilité n'est pas améliorée car l'écart-type moyen des clôtures de phase (après avoir retiré les pics) est estimé à $\sim 2 \,$\um~rms. De plus, l'écart-type des clôtures de phase a été estimé à $\sim 66 \,$nm sur \ac{FIRSTv1}. Ainsi, ces perturbations ont probablement lieu au niveau des \ac{ODL}s ou la méthode d'acquisition des interférogrammes par la modulation des franges en est peut-être trop sensible. D'autres tests sont ainsi nécessaires pour identifier la source de ces perturbations et améliorer la stabilité des mesures sur \ac{FIRSTv2}.

\begin{figure}[ht!]
    \centering
    \includegraphics[width=\figwidth]{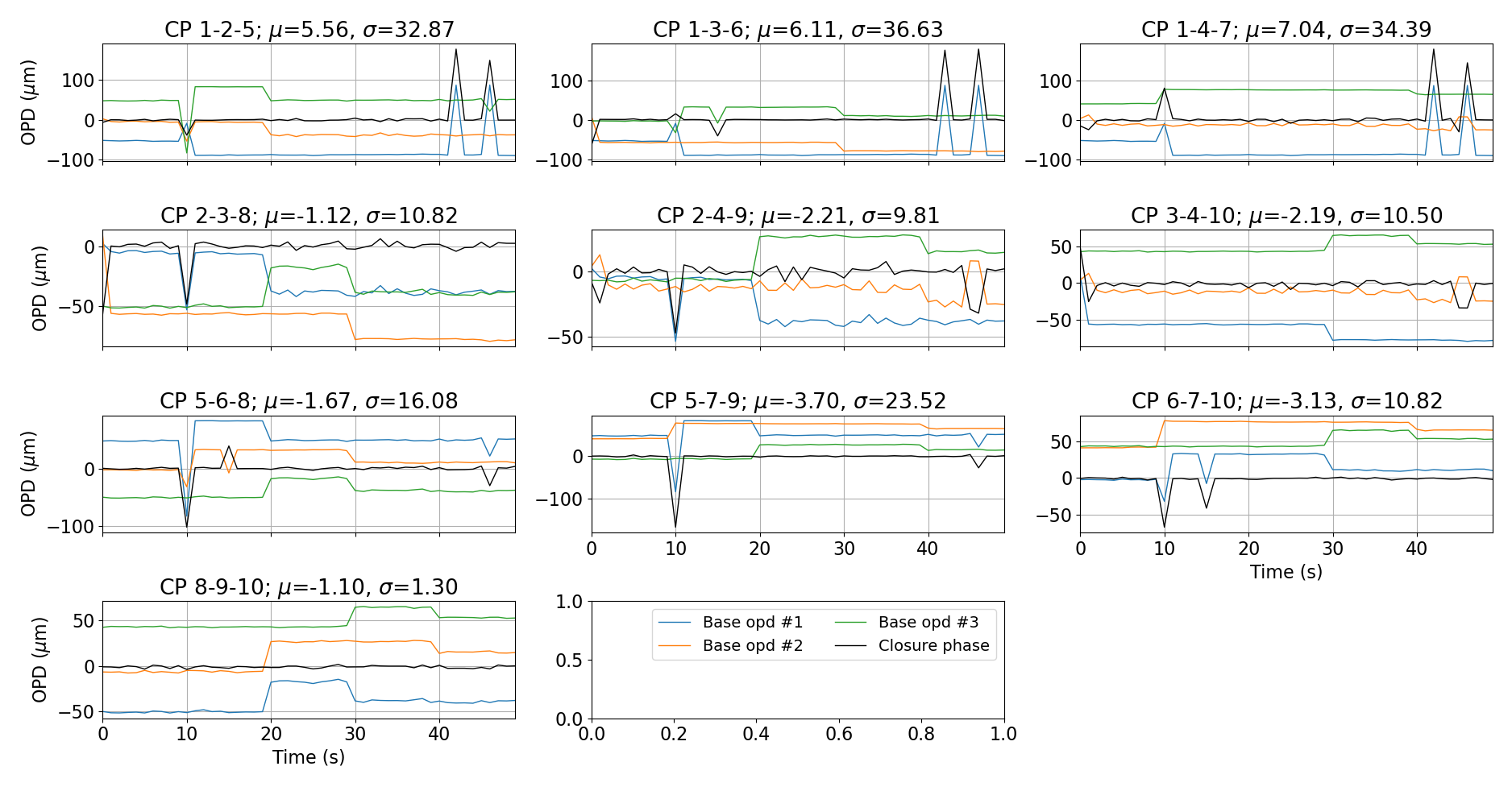}
    \caption[Graphiques de l'OPD en fonction du temps des dix triangles formés par les dix bases de FIRSTv2 sur la source interne de SCExAO avec la puce $Y$, après la protection des fibres.]{Graphiques de l'OPD en fonction du temps (tracé en couleur) des dix triangles formés par les dix bases de FIRSTv2 sur la source interne de SCExAO avec la puce $Y$, après la protection des fibres contre les caméras de VAMPIRES. Les clôtures de phase sont tracées en noir et leur valeur moyenne et l'écart-type sont écrits dans les sous-titres. Les courbes sont tracées pour $1\,000$ images de temps d'exposition égal à $50 \,$ms.}
    \label{fig:OPDfitVStimeSubaruProtect}
\end{figure}

%%%%%%%%%%%%%%%%%%%%%%%%%%%%%%%
\section{Première lumière au télescope Subaru}

% Première lumière le 10 septembre 2021 (AlphaPer)
La première lumière de \ac{FIRSTv2} a eu lieu le 10 septembre 2021, par l'observation de $\upalpha$ Per. La figure~\ref{fig:V2FirstLight} présente une image de cette première lumière, à un temps d'exposition égal à $100 \,$ms et un gain \textit{EM} de la caméra de $300$. Seuls les faisceaux des sous-pupilles de la première base (formée par les segments 26 et 27) sont injectés dans les fibres induisant la formation des franges sur la base en bas de l'image. Sur la partie gauche de l'image apparaît la lumière des trois lasers de longueurs d'onde $785 \,$nm, $850 \,$nm et $852 \,$nm, injectés sur l'instrument pour effectuer des tests sur la mesure de la perturbation de la phase sur le banc (voir plus de détails dans la section~\ref{sec:V2SubaruProspectives}).

\begin{figure}[ht!]
    \centering
    \includegraphics[width=\figwidth]{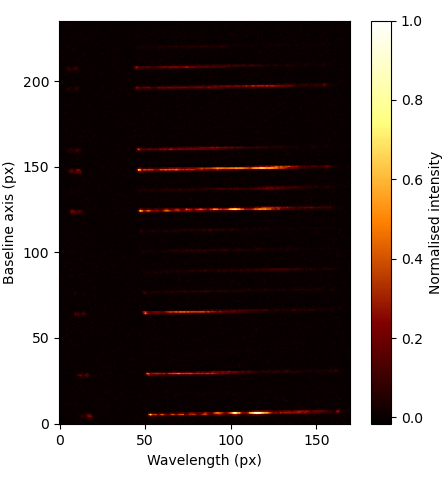}
    \caption[Image des sorties de la puce $X$ lors de la première lumière de FIRSTv2 sur SCExAO sur l'étoile Alpha Per, le 21 septembre 2021.]{Image des sorties de la puce $X$ lors de la première lumière de FIRSTv2 sur SCExAO sur l'étoile Alpha Per, le 21 septembre 2021. Seuls les faisceaux des sous-pupilles 26 et 27 formant la première base sont injectés. La base est imagée en bas de l'image et des franges d'interférence sont visibles. Le temps d'exposition est de $100 \,$ms et le gain \textit{EM} de la caméra vaut 300.}
    \label{fig:V2FirstLight}
\end{figure}

Par la suite, nous avons observé avec \ac{FIRSTv2} des étoiles brillantes telles que Véga et Capella. Les données acquises n'ont pas permis d'estimer des courbes de phase mais nous ont permis d'améliorer l'exploitation de l'instrument d'un point de vue instrumental et d'un point de vue logiciel. 

% Nécessité de protéger les fibres optiques
En effet, ce qui a été identifié comme le plus critique est sans doute l'isolation des fibres optiques des perturbations de leur environnement. Comme discuté dans la section~\ref{sec:V2StabilitySubaru}, elles sont sensibles aux mouvements d'air générés par les différentes caméras (trois sont, par exemple, installées sur le banc de recombinaison) et les vibrations probablement transmises sur le banc de recombinaison vissé sur \ac{SCExAO} ainsi qu'aux autres instruments (voir la photographie de la figure~\ref{fig:SCExAOPhoto}). Une nette amélioration a déjà été observée (figure~\ref{fig:OPDfitVStimeSubaruProtect}) après la protection des fibres de \ac{FIRST} sur le banc visible mais il reste encore à protéger les fibres présentes sur le banc de recombinaison, qui sont plus longues dans le cas de \ac{FIRSTv2}, à cause des \ac{ODL}s. Mais encore, la méthode de reconstruction des interférogrammes par modulation des franges est probablement sensible aux résidus de phase en aval du système d'optique adaptative, empêchant l'estimation des courbes de phase lors du traitement de données. D'autres solutions à ce sujet pourraient être envisagées comme l'utilisation d'une puce photonique dite \textit{ABCD} (voir la section~\ref{sec:V2SubaruProspectives}).

% Nécessité du polariseur + Wollaston (<20220214)
De plus, il est primordial de disposer d'un prisme de Wollaston en sortie du V-Groove connecté à la puce afin de séparer les deux polarisations sur la caméra et de les traiter séparément. Un tel prisme n'a pu être installé qu'en mars 2022 et toutes les données qui précèdent cette date ne sont pas exploitables.

% Nécessité de connaître le step de modulation sur chaque image acquise (<20220225)
Enfin, lors du traitement des données, l'étape de modulation des franges doit être identifiée à chaque image. La matrice des données doit contenir les franges modulées de la même façon que les franges d'étalonnage contenues dans la \ac{P2VM} lors de l'ajustement des franges décrit dans la section~\ref{sec:FullOnFit}. Durant quelques nuits d'observation, la construction du logiciel ne permettait pas de connaître les étapes de la séquence de modulation de chaque image acquises et les données ne peuvent donc pas être traitées.

%%%%%%%%%%%%%%%%%%%%%%%%%%%%%%%%
\section{Perspectives}
\label{sec:V2SubaruProspectives}

Mon projet de thèse a permis d'effectuer certains progrès sur l'implémentation de \ac{FIRSTv2} sur ciel, que j'ai exposé dans ce chapitre. J'expose ici les futurs développement prévus.

% Nessécité d'améliorer les performances en transmission (retirer les odls, utiliser de nouvelles puces transmissives)
L'aspect le plus urgent est l'augmentation de la transmission de \ac{FIRSTv2}. Comme nous l'avons vu, la configuration actuelle de l'instrument ne permet pas d'obtenir un signal avec un \ac{SNR} suffisant. Nous prévoyons ainsi de retirer les lignes à retard de l'instrument car elles sont trop peu transmissives ($\sim 30\%$). Les solutions possibles ont été discutées dans la section~\ref{sec:ODLDiscussions}. De plus, l'équipe travaille à de nouvelles puces photoniques plus transmissives comme je l'ai exposé dans la section~\ref{sec:ChipCharacDiscu}.

% Métrologie pour la mesure des perturbations de phase sur le banc afin de les compenser au traitement
Quatre sources lasers pour des mesures de métrologie sont installées afin de tirer profit des mesures d'\ac{OPD} entre les paires de sous-pupilles permises par \ac{FIRST} pour mesurer les perturbations du front d'onde incident et tenter de les corriger. De cette manière, \ac{FIRST} pourrait fournir simultanément des données scientifiques et des informations sur le front d'onde incident qui seraient utilisées par un système de correction d'optique adaptative \citep{vievard2021}. Les longueurs d'onde des quatre sources lasers sont $642 \,$nm, $785 \,$nm, $850 \,$nm et $852 \,$nm. Les deux sources émettant dans les deux dernières longueurs d'onde sont couplées dans la même fibre grâce à un coupleur $Y$ dont la sortie est combinée avec les deux autres sources grâce à un composant de multiplexage en longueur d'onde (\ac{WDM}). Les quatre sources lasers sont ainsi combinées dans une unique fibre optique qui est intégrée sur le banc visible de \ac{SCExAO} directement en amont du séparateur de faisceau entre les instruments \ac{FIRST} et \ac{VAMPIRES}, nommé \textit{VAMPIRES/FIRST splitter} sur le schéma de la figure~\ref{fig:SCExAOScheme}. Les quatre sources peuvent ainsi être imagées sur la caméra avec les franges interférences et un programme de traitement de données additionnel pourrait permettre de connaître les perturbations présentes sur le banc que subissent les mesures de phase en temps réel afin de les corriger. De premiers tests à ce sujet ont été faits durant ma thèse mais n'ont pour l'instant pas fonctionné car les courbes de phase estimées sont très repliées à cause de la difficulté d'obtenir l'\ac{OPD} nulle entre tous les faisceaux.

% Développement du nouveau logiciel de contrôle de FIRST pour l'adapter au calcul rapide synchronisé avec CACAO
Comme discuté dans le paragraphe précédent, un des projets en cours concerne l'exploitation de \ac{FIRST} comme un senseur de front d'onde. Pour cela, un ordinateur de contrôle dédié à cela a été acquis pour augmenter la cadence de calculs et la synchronisation avec les autres composants et processus de \ac{SCExAO} (e.g. avec le miroir déformable pour la correction par optique adaptative, le logiciel \ac{CACAO} \citep{guyon2020}). Dans ce but, de nouveaux développements doivent être effectués sur le logiciel de contrôle de \ac{FIRST}. L'augmentation de la puissance de calcul pourrait aussi permettre une augmentation de la fréquence de modulation des franges lors de l'acquisition des images interférométriques, à condition que l'instrument soit plus transmissif. Les interférogrammes mesurés ne seraient donc plus affectés par les résidus des perturbations atmosphériques, pour des fréquences d'acquisition supérieures à quelques centaines d'images par seconde.

% Utilisation de puce ABCD pour éviter la modulation, diminuer la quantité de données nécessaire (d'un facteur 4)
Enfin, les développements s'orientent vers la fabrication de puces photoniques dites \textit{ABCD}. Une telle puce fournit quatre sorties par base déphasées de $\uppi / 2 \,$rad et permet la mesure complète d'une frange sur une unique image. Cette technique est aussi mentionnée dans la section~\ref{sec:ChipCharacDiscu}. Elle permet ainsi de diviser par 12 la quantité de données à acquérir pour 10 bases, tout en permettant de s'affranchir des perturbations sur les mesures de phases qui ont lieu sur la version actuelle de \ac{FIRSTv2} au cours de l'application de la séquence de modulation des franges (nécessitant 12 images).

%%%%%%%%%%%%%%%%%%%%%%%%%%%%%%%%%%%%%%%%%%%%%%%%%%%%%%%%%%%%%%%%
\chapter*{Conclusion}
\addcontentsline{toc}{chapter}{Conclusion}

Mon travail de thèse s'inscrit dans la poursuite du développement de l'instrument FIRST qui implémente le concept de masquage de pupille fibré pour l'imagerie à haut contraste et à haute résolution angulaire (HRA). Il a déjà été montré que FIRST permettait la détection de compagnons stellaires à des séparations inférieures à la limite de diffraction du télescope, ce qui est un avantage considérable pour l'imagerie directe d'exoplanètes avec les télescopes de la classe des $10 \,$m de diamètre. FIRST est intégré au banc d'optique adaptative extrême SCExAO sur le télescope de $8 \,$m de diamètre Subaru. Ma thèse a permis les tests et la caractérisation de composants d'optique intégrée sur sa réplique modifiée en laboratoire (FIRSTv2) dans l'objectif d'augmenter les performances en contraste de l'instrument.

Comme je l'ai présenté dans le premier chapitre de ce manuscrit, les systèmes protoplanétaires sont des cas d'études cruciaux pour notre compréhension de la formation planétaire. Il semblerait que ces systèmes présentent une raie d'émission \ha~provenant du phénomène d'accrétion de matière, diminuant le contraste du système dans ces longueurs d'onde. Le calcul de la phase différentielle, une observable auto-étalonnée des perturbations atmosphériques et instrumentales, est particulièrement adaptée aux systèmes avec un tel spectre.

J'ai ensuite présenté l'installation de FIRSTv2 au laboratoire dans le second chapitre. Les résultats de la caractérisation des deux composants d'optique intégrée sur lesquels j'ai travaillé y sont présentés et ont fait l'objet d'un article de conférence. Les transmissions des deux composants ont été estimés à $15\%$ et $30\%$. Je présente aussi les autres composants et discute d'une possibilité de retirer les lignes à retard dans le futur car elles sont trop peu transmissives. J'expose une autre partie de ma thèse qui a consisté au développement du logiciel de contrôle du banc de test, que j'ai ensuite déployé sur ciel. Ce logiciel permet le contrôle synchronisé des différents composants clés (miroir déformable, lignes à retard et caméra) pour la mesure des interférogrammes en sortie de la puce photonique.

Un autre aspect de ma thèse a été le développement du programme de traitement et d'analyse de données. Comme je l'ai décrit dans le troisième chapitre, il s'agit d'étalonner les interférogrammes imagés sur la caméra de l'instrument par le calcul de la P2VM. Celle-ci permet d'ajuster les franges mesurées afin d'estimer la phase des visibilités complexes des bases. À partir de celle-ci est calculée la phase différentielle, qui tire profit du fait qu'on mesure la phase en fonction de la longueur d'onde. Je présente également l'étalonnage spectral nécessaire à l'analyse des données ainsi que la séquence de modulation temporelle des franges.

Dans le quatrième chapitre je présente le système simulant une source protoplanétaire que j'ai intégré sur le banc de test afin de caractériser l'instrument lors d'observations de ce type de sources. Cela m'a permis de montrer que les phases différentielles permettent de détecter un compagnon de type protoplanétaire à une séparation de $0,68 \lambda / B$, validant les composants photoniques pour l'étude de ces objets. On a pu voir que les mesures étaient limitées par une modulation de la phase, qu'on appelle \wiggles. L'amplitude de ceux-ci limite la détection de systèmes protoplanétaire à un contraste de $\sim 0,56$. Les efforts sont actuellement investis dans la recherche de la source de cette perturbation ainsi que sa diminution, car l'objectif est d'atteindre des contrastes de l'ordre de $\sim 10^{-2} - 10^{-3}$ dans le cadre d'observations de protoplanètes.

Pour finir, j'ai présenté dans le cinquième chapitre les différentes étapes de l'installation de FIRSTv2 sur la plateforme SCExAO au télescope Subaru qui ont mené à sa première lumière le 10 septembre 2021. J'ai pu intégrer l'instrument et le logiciel de contrôle développé au laboratoire lors de deux missions à Hawaï ainsi qu'à distance avec l'aide précieuse de Sébastien Vievard sur place. Nous avons dû faire face à de nouvelles perturbations sur les mesures des interférogrammes. En effet, la phase montrait des variations temporelles importantes, avant une isolation renforcée des fibres optiques. Cette expérience confirme que des fibres optiques trop longues sont à éviter et que la méthode d'acquisition des interférogrammes par modulation temporelle est limitante car trop sensibles aux perturbations présentes sur le banc (vents, vibrations, résidus en aval de l'optique adaptative).

Ainsi, ma thèse a pu montrer la capacité des phases différentielles à détecter un système de type protoplanétaire. Les prochains développements se concentreront sur de nouveaux concepts de puces photoniques (3D, ABCD, haut contraste d'indice, lanterne photonique) qui permettront d'augmenter la sensibilité de l'instrument. De nouveaux efforts sont aussi nécessaires afin de comprendre l'origine des \wiggles~et de les réduire car ils limitent la précision sur les mesures de phase à un contraste du système protoplanétaire observé de $\sim 0,56$. Enfin, un gain en stabilité sur les mesures de phase sur le banc SCExAO est encore nécessaire et deux pistes principales s'offrent à nous : la protection des fibres optiques sur le banc de recombinaison et le passage à une modulation des franges dans la puce, pour éviter la modulation temporelle.

} \endgroup

%%%%%%%%%%%%%%%%%%%%%%%%%%%%%%%%%%%%%%%%%%%%%%%%%%%%%%%%%%%%%%%%
% Print glossary
%%%%%%%%%%%%%%%%%%%%%%%%%%%%%%%%%%%%%%%%%%%%%%%%%%%%%%%%%%%%%%%%

\newpage\thispagestyle{empty}
\printglossary[type=\acronymtype, style=indexgroup, title=Acronymes, toctitle=Acronymes]

%%%%%%%%%%%%%%%%%%%%%%%%%%%%%%%%%%%%%%%%%%%%%%%%%%%%%%%%%%%%%%%%
% Print the list of figures
%%%%%%%%%%%%%%%%%%%%%%%%%%%%%%%%%%%%%%%%%%%%%%%%%%%%%%%%%%%%%%%%

\newpage\thispagestyle{empty}
\renewcommand{\listfigurename}{Liste des figures}

\phantomsection
\listoffigures

%%%%%%%%%%%%%%%%%%%%%%%%%%%%%%%%%%%%%%%%%%%%%%%%%%%%%%%%%%%%%%%%
% Print the list of tables
%%%%%%%%%%%%%%%%%%%%%%%%%%%%%%%%%%%%%%%%%%%%%%%%%%%%%%%%%%%%%%%%

\newpage\thispagestyle{empty}
\renewcommand{\listtablename}{Liste des tableaux}

\phantomsection
\listoftables

%%%%%%%%%%%%%%%%%%%%%%%%%%%%%%%%%%%%%%%%%%%%%%%%%%%%%%%%%%%%%%%%
% Print the bibliography
%%%%%%%%%%%%%%%%%%%%%%%%%%%%%%%%%%%%%%%%%%%%%%%%%%%%%%%%%%%%%%%%

\newpage\thispagestyle{empty}

\bibliographystyle{plainnat}
\bibliography{main}
% \addcontentsline{toc}{chapter}{Bibliographie}
% When compiling with LaTex on a local machine (without Overleaf), execute these commands:
% (pdf)latex main.tex
% BibTex main.aux
% (pdf)latex main.tex
% (pdf)latex main.tex

%%%%%%%%%%%%%%%%%%%%%%%%%%%%%%%%%%%%%%%%%%%%%%%%%%%%%%%%%%%%%%%
% Short abstract
%%%%%%%%%%%%%%%%%%%%%%%%%%%%%%%%%%%%%%%%%%%%%%%%%%%%%%%%%%%%%%%

%----------------------------------------------------------------------------------------
%	FRENCH
%----------------------------------------------------------------------------------------

\newpage
\thispagestyle{empty}
\chapter*{Résumé court}

\noindent\rule[2pt]{\textwidth}{0.5pt}
\begin{center}
    \large\textbf{\ttitle\\}
\end{center}
    \footnotesize J’ai travaillé pendant ma thèse dans le domaine de l’imagerie à haut contraste et à haute résolution angulaire (HRA) pour l’étude des protoplanètes. L'observation des protoplanètes est crucial pour l'étude des mécanismes de la formation planétaire. Ces jeunes planètes se caractérisent par la présence de lignes d'émission dans leur spectre. Celle qui nous intéresse est la raie Ha et le contraste du système est alors plus faible à ces longueurs d’onde et donc plus accessible. L’instrument Fibered Interferometer foR a Single Telescope (FIRST) utilise le concept de réarrangement de pupille fibré dans le visible, intégré sur le banc d’optique adaptative extrême Subaru Coronagraphic Extreme Adaptive Optics (SCExAO) du télescope Subaru. Son principe est de sous-divisé la pupille d’entrée en sous-pupilles dont les faisceaux sont injectés dans les fibres optiques monomodes. Celles-ci ont le double intérêt d’appliquer un filtrage spatial du front d’onde supprimant ainsi les aberrations optiques à l’échelle des sous-pupilles, en même temps que le réarrangement des sous-pupilles. Les interférogrammes résultant de l’interférence des faisceaux de ces dernières sont ensuite mesurés. FIRST a démontré qu’il permettait l’imagerie de binaires stellaires à une résolution en-deçà de la limite de diffraction du télescope avec ce concept. La réplique d’une nouvelle version de cet instrument (FIRSTv2) a été construite en laboratoire afin de permettre les développements de la technologie d’optique intégrée. Mon travail de thèse est d’évaluer ses performances et sa faisabilité pour l’imagerie HRA, dans le but d’améliorer les performances en contraste de l’instrument. La recombinaison de chaque paire de sous-pupilles est ainsi codée sur une sortie de la puce photonique avant d’être dispersée et imagée sur quelques pixels de la caméra. L’échantillonnage des franges se fait par modulation temporelle des chemins optiques des faisceaux par les segments du miroir déformable. De plus, j’ai continué le développement du banc de test, aussi bien au niveau du montage optique que du logiciel de contrôle, de mettre au point une procédure spécifique d’acquisition des données et de développer un programme de traitement et d’analyse de données. Cette analyse des données s’effectue via le calcul de la phase différentielle spectrale, qui est une observable auto-étalonnée des perturbations atmosphériques et instrumentales. Pour cela, les mesures de phase des visibilités complexes de chaque base sont étalonnées par le signal du continuum mettant en évidence un possible signal dans la raie d’intérêt. Ainsi, à partir d’un simulateur de source protoplanétaire que j’ai intégré sur le banc, j’ai pu démontrer que FIRSTv2 pouvait détecter un compagnon de type protoplanètaire à une séparation équivalente à 0.7 l / B de la source centrale, avec un contraste d’environ 0.5. Enfin, j’ai participé à l’intégration et à la première lumière de FIRSTv2 sur le banc SCExAO pour laquelle le logiciel développé en laboratoire a été déployé. Les tests et les données acquises lors de quelques nuits d’observations montrent qu’une isolation des fibres optiques sur le banc est nécessaire et que la méthode d’acquisition des interférogrammes par modulation temporelle doit probablement être changée pour une modulation dans la puce en changeant la technologie de recombinaison pour une ABCD.

\vspace{0.2cm}
\noindent{\normalsize\textbf{Mots clés :}} \keywordnamesfr

\noindent\rule[2pt]{\textwidth}{0.5pt}

%----------------------------------------------------------------------------------------
%	ENGLISH
%----------------------------------------------------------------------------------------

\chapter*{Short abstract}

\noindent\rule[2pt]{\textwidth}{0.5pt}
\begin{center}
    \large\textbf{Characterization and deployement of Photonic Integrated Circuits for the FIRST fibered interferometer at the Subaru Telescope in the context of accreting protoplanets studies\\}
\end{center}
    \footnotesize During my thesis, I worked in the field of high contrast and high angular resolution (HRA) imaging for the study of protoplanets. The observation of protoplanets is key to constrain the mechanisms of planetary formation. These young planets are characterized by the presence of emission lines in their spectrum. The one we are interested in is the Ha line and the contrast of the system is then weaker at these wavelengths and thus more accessible. The Fibered Interferometer foR a Single Telescope (FIRST) instrument uses the concept of fibered pupil remapping in the visible, integrated on the Subaru Coronagraphic Extreme Adaptive Optics (SCExAO) telescope. Its principle is to sub-divide the entrance pupil into sub-pupils whose beams are injected into single mode optical fibers. The latest have the double advantage of applying a spatial filtering on the wavefront thus suppressing the optical aberrations at the scale of the sub-pupils, at the same time as the remapping of the sub-pupils. The interferograms resulting from the interference of the sub-pupil beams are then measured. FIRST has demonstrated that it can detect stellar binaries at a resolution below the diffraction limit of the telescope with this concept. The replica of a new version of this instrument (FIRSTv2) has been built in laboratory to allow the development of the integrated optics technology. My thesis work is to evaluate its performance and its feasibility for HRA imaging, in order to improve the contrast performance of the instrument. The recombination of each pair of sub-pupils is encoded on an output of the photonic chip then dispersed and imaged on few pixels of the camera. The fringe sampling is performed by temporal modulation of the sub-pupil optical paths by the segments of the deformable mirror. In addition, I continued the development of the testbed, both in terms of the optical assembly and the control software, to develop a specific procedure for data acquisition and to develop a program for data reduction and analysis. The goal of this data analysis  is to calculate the spectral differential phase, which is a self-calibrated observable of the atmospheric and instrumental perturbations. In that purpose, the phase measurements of the complex visibilities of each baseline are calibrated by the continuum signal highlighting a possible signal in the line of interest. Thus, from a protoplanetary source simulator that I integrated on the testbed, I was able to demonstrate that FIRSTv2 could detect a protoplanetary companion at a separation equivalent to 0.7 l / B from the central source, with a contrast of about 0.5. Finally, I participated in the integration and the first light of FIRSTv2 on the SCExAO bench for which the software developed in laboratory was deployed. The tests and data acquired during several observation nights show that an isolation of the optical fibers on the bench is necessary and that the method of acquisition of the interferograms by temporal modulation may need to be changed for an on-chip modulation by changing the recombination technology using an ABCD chip.

\vspace{0.2cm}
\noindent{\normalsize\textbf{Keywords\string:}} \keywordnamesen

\noindent\rule[2pt]{\textwidth}{0.5pt}

\end{document}